\documentclass[final]{aa}

\usepackage{graphicx}
\usepackage{txfonts}
\usepackage{natbib}
\usepackage{soul}
\usepackage{longtable}
\bibpunct{(}{)}{;}{a}{}{,} 
\usepackage{lscape}
\usepackage{colortbl}
\usepackage[normalem]{ulem}

\newcommand{\Nstars}{102~}     
     
\newcommand{\NstarsFour}{98~}     
\newcommand{\Vmax}{14~}     
\newcommand{\vsinimax}{6.5~}   
\newcommand{\Nspectra}{1965~}   
\newcommand{\Nhours}{460~} 
\newcommand{\dmax}{11~}
\newcommand{\NplanetsAnnouncedByUs}{9~}
\newcommand{\NstarsPerioVarNoPl}{11~}

\begin{document}
	\title{The HARPS search for southern extra-solar planets\thanks{Based on observations made with the HARPS instrument on the ESO 3.6-m telescope at La Silla Observatory under programme ID 072.C-0488(E)}} 
	\subtitle{XXXI. The M-dwarf sample}

	\author{X. Bonfils \inst{1,2}
	         \and X. Delfosse \inst{1}
	         \and S. Udry \inst{2}
	         \and T. Forveille \inst{1}
	         \and M. Mayor \inst{2}
	         \and C. Perrier \inst{1}
	         \and F. Bouchy \inst{3,4}
	         \and M. Gillon \inst{5,2}
	         \and C. Lovis \inst{2}
	         \and F. Pepe \inst{2}
	         \and D. Queloz \inst{2}
	         \and N. C. Santos \inst{6}
	         \and D. S\'egransan \inst{2}
	         \and J.-L. Bertaux \inst{7}
		}

	\offprints{X. Bonfils\\ \email{Xavier.Bonfils@obs.ujf-grenoble.fr}}
	
	\institute{ UJF-Grenoble 1 / CNRS-INSU, Institut de Plan\'etologie et d'Astrophysique de Grenoble (IPAG) UMR 5274, Grenoble, F-38041, France
	          \and  Observatoire de Gen\`eve, 51 ch. des Maillettes, 1290 Sauverny, Switzerland
        \and
        Institut d'Astrophysique de Paris, CNRS, 
	  Universit\'e Pierre et Marie Curie,
	  98bis Bd Arago, 75014 Paris, France
	\and Observatoire de Haute-Provence, CNRS/OAMP, 04870 Saint-Michel-lÕObservatoire, France
		\and Institut dÕAstrophysique et de G\'eophysique, Universit\'e de Li\`ege, All\'ee du 6 Ao\^{u}t 17, Bat. B5C, 4000 Li\`ege, Belgium
        \and
        Centro de Astrof{\'\i}sica, Universidade do Porto, Rua das Estrelas,
	  P4150-762 Porto, Portugal
		        \and
        Service d'A\'eronomie du CNRS, BP 3, 91371
	  Verri\`eres-le-Buisson, France
		}

	\date{Received/Accepted}
	
	\abstract
	{Searching for planets around stars with different masses probes the outcome of planetary formation for different initial conditions. The low-mass M dwarfs are also the most frequent stars in our Galaxy and potentially therefore, the most frequent planet hosts.}
	{This drives observations of a sample of \Nstars southern nearby M dwarfs, using a fraction of our guaranteed time on the ESO/HARPS spectrograph. We observed  \Nhours hours and gathered \Nspectra precise ($\sim1-3$ m/s) radial velocities, spanning the period from Feb. 11th, 2003 to Apr. 1st 2009.}
	{This paper makes available the sample's time series, presents their precision and variability. We apply systematic searches for long-term trends, periodic signals and Keplerian orbits (from 1 to 4 planets). We analyze the subset of stars with detected signals and apply several diagnostics to discriminate whether the observed Doppler shifts are caused by stellar surface inhomogeneities or by the radial pull of orbiting planets. To prepare for the statistical view of our survey we also compute the limits on possible unseen signals, and derive a first estimate of the frequency of planets orbiting M dwarfs.} 
	{We recover the planetary signals corresponding to \NplanetsAnnouncedByUs planets already announced by our group (Gl\,176 b, Gl\,581 b, c, d~\&~e, Gl\,674 b, Gl\,433 b, Gl\,667C b and Gl\,667C c). We present radial velocities that confirm GJ 849 hosts a Jupiter-mass planet, plus a long-term radial-velocity variation. We also present RVs that precise the planetary mass and period of Gl\,832b. We detect long-term RV changes for Gl\,367, Gl\,680 and Gl\,880 betraying yet unknown long-period companions. We identify candidate signals in the radial-velocity time series of \NstarsPerioVarNoPl other M dwarfs. Spectral diagnostics and/or photometric observations demonstrate however that they are most probably caused by stellar surface inhomogeneities. Finally, we find our survey sensitive to few Earth-mass planets for periods up to several hundred days. We derive a first estimate of the occurrence of M-dwarf planets as a function of their minimum mass and orbital period. In particular, we find that giant planets ($m \sin i = 100-1,000~\mathrm{M_\oplus}$) have a low frequency (e.g. $f\lesssim1\%$ for $P=1-10$ d and $f=0.02^{+0.03}_{-0.01}$ for  $P=10-100$ d), whereas super-Earths ($m \sin i = 1-10~\mathrm{M_\oplus}$) are likely very abundant ($f=0.36^{+0.25}_{-0.10}$ for $P=1-10$ d and $f=0.35^{+0.45}_{-0.11}$ for $P=10-100$ d). We also obtained $\eta_\oplus=0.41^{+0.54}_{-0.13}$, the frequency of habitable planets orbiting M dwarfs ($1\le m\sin i \le 10~\mathrm{M_\oplus}$). For the first time, $\eta_\oplus$ is a direct measure and not a number extrapolated from the statistic of more massive and/or shorter-period planets. }
	{}


\titlerunning{The HARPS M-dwarf sample}
\authorrunning{X. Bonfils et al.}

\maketitle

\section{Introduction}
M dwarfs are the dominant stellar population of our Galaxy \citep[e.g.][]{Chabrier:2000}. Compared to our Sun, they are cooler, smaller and lower-mass stars. These characteristics ease the detection of planets for many techniques and M dwarfs have therefore been included early in planet-search samples. While the first claimed detections \citep[e.g. around Barnard's star -- ][]{Kamp:1963} were later found incorrect \citep{Gatewood:1973, Gatewood:1995}, targeting M dwarfs has proven more successful since.

At the forefront of planet discoveries, the radial-velocity (RV) technique was first to unveil a candidate giant planet orbiting an M dwarf. Three years after the discovery of \object{51Peg\,b} \citep{Mayor:1995}, the detection of a giant planet orbiting the M dwarf \object{GJ 876} \citep{Delfosse:1998b, Marcy:1998} proved that M dwarfs could form planets too. \object{GJ 876} was actually one of a few tens of M stars monitored by radial-velocity surveys, and its detection made the early impression that giant planets could be common around late-type stars. Today, only 6 M dwarfs are known to host a planet with a minimum mass $> 0.5~\mathrm{M_{Jup}}$ (see Table~\ref{tab:planets}) and the view has progressively shifted toward a low rate of occurrence for giant planets, compared to sun-like stars \citep{Bonfils:2006, Butler:2004, Endl:2006, Johnson:2007, Cumming:2008}. 

Improvements of the RV technique has led to the discovery of lower-mass planets down to $m \sin i \simeq 1.9~\mathrm{M_\oplus}$ \citep{Mayor:2009}. Below 25~$\mathrm{M_\oplus}$, there are 8 known M-dwarf hosts and they altogether host 12 such low-mass exoplanets. Hence, despite an adverse detection bias, planets of low-mass already appear more frequent than giant planets \citep{Bonfils:2007b}. Among them, GJ\,581d and GJ\,667Cc are noticeably interesting because they have $m \sin i < 10~\mathrm{M_\oplus}$ and receive closely the amount of light received by Earth in our Solar System \citep[][Delfosse et al., in prep.]{Udry:2007a, Mayor:2009}. Depending on their atmosphere (thickness, albedo and chemistry) liquid water may flow on their surface -- the standard criterium to define a {\it habitable planet} \citep{Kasting:1993, Selsis:2007}.

The transit technique has also been successful in detecting two planets transiting an M dwarf. One is \object{GJ 436\,b}, a Neptune-mass planet initially detected with Doppler measurements \citep{Butler:2006, Maness:2007} and subsequently seen in transit \citep{Gillon:2007a}. Finding that \object{GJ 436\,b} undergoes transits has enabled a wealth of detailed studies such as the determinations of the planet's true mass and radius and measurements of its effective temperature and  orbital eccentricity \citep{Gillon:2007b, Demory:2007, Deming:2007}. Most recently, the {\it Mearth} project, a search for transiting planets dedicated to late M dwarfs \citep{Nutzman:2008}, has unveiled a $\sim6~\mathrm{M_\oplus}$ planet transiting the nearby M4.5 dwarf GJ\,1214 \citep{Charbonneau:2009}. Like GJ\,436b, it has a favorable planetary to stellar radius ratio and is well suited to in-depth characterizations with current observatories. Both planets are considered {\it Rosetta stones} to the physics of low-mass planets. 

\begin{table*}
\caption{\label{tab:planets}Known exoplanets orbiting M dwarfs and their basic parameters}
\begin{tabular}{llccccccccc}
\multicolumn{11}{c}{Radial-Velocity detections}\\
\hline
Star	& \multicolumn{1}{c}{Pl.}	&	\multicolumn{2}{c}{$m_p \sin i^\dagger$}			&	i			&	P	&	a	&	e	& ref.                                      & ref.			& in\\
	&					&	[$\mathrm{M_\oplus}$] & [$\mathrm{M_{Jup}}$]	&[$^{\mathrm{o}}$]	&	[d]	&    [AU]   	&		& discovery$^{\ddagger}$ & param.$^{\ddagger}$ &sample\\
\hline
GJ\,176	& b	& 8.4		&       0.026        & -  		& 8.7(8)		&	0.07		&	0. (fixed)		& 25 & 25 & y\\
GJ\,317    	& b  	& 380 	&      1.2 		& -		& 69(2)		&	0.95		&	0.19 (006)		& 17 & 17 &n\\
GJ\,433	& b	& 6.41	&       0.0202	& -		& 7.36(5)		&	0.06		&	0. (fixed)		& 37 & 37 & y\\
GJ\,581    	& b	& 15.7 	&	0.0492	& $>$ 40.	& 5.368(7)	&	0.04		&	0.      (fixed)	& 8	& 23	&y\\
		& c	& 5.4 	& 	0.017	& $>$ 40.	& 12.9(3)		&	0.07		&	0.17	(0.07)	& 13	& 23	&y\\
		& d	& 7.1 	&	0.022	& $>$ 40.	& 66.(8) 		&	0.22		&	0.38 (0.09)	& 13,23& 23	& y\\	
		& e	& 1.9 	&	0.0060	& $>$ 40.	& 3.1(5)		&      0.03		&	0.	(fixed)	& 23	& 23	& y\\
GJ\,649	& b	& 106	&	0.333	& -		& 59(8)		&      1.14		&	0.30 (0.08)	& 28& 28&n\\
GJ\,667C	& b	& 6.0		&       0.019        & -		& 7.20(3)		&	0.05		&	0. (fixed)		& 37	& 37 &y\\
           	& c	& 3.9		&	0.012	& -		& 28.1(5)		&       0.28		&	0. (fixed)		& 37	& 37 & y\\ 
GJ\,674	& b	& 11		&      0.034	& -		& 4.6(9)		&	0.04		&	0.20 (0.02)	& 12	& 12	&y\\
GJ\,676A	& b	& 1300	&       4.0       	& -  		& 98(9)		&       1.61		&	0.29 (0.01)	& 36 & 36 &n\\
GJ\,832	& b	& 200   	&       0.64     	& -    		& 3(416)		&	3.4		&	0.12 (0.11)	& 20 & 39 &y\\
GJ\,849 	& b	& 310	&	0.99          	& -		& 18(52)		&      2.35		&	0.04 (0.02)	& 11 & 39 &y\\
GJ\,876	& b	& 839	&	2.64     	& 48.(9)	& 61.0(7)		&	0.211	&	0.029 (0.001)	& 1,2	&29 & y\\
 		& c	& 180    	&      0.83       	& 4(8)  	& 30.2(6)		&	0.132	& 	0.266 (0.003)	& 3	&29 & y\\
		& d	& 6.3     	&      0.020	& 50 (fix.)	& 1.9378(5) 	&	0.021	&	0.139 (0.032)	& 6	&29 & y\\
GJ\,3634  & b   & 6.6          &      0.021         & -              & 2.645(6)         &      0.028        &      0.08 (0.09)         & 36 & 36 & n\\
HIP\,12961& b	& 110	&	0.35		& -		& 57.4(3)		&	0.13		&       0.16 (0.03)	& 35 & 35 & n\\
HIP\,57050& b  & 40		&       0.3		& -		& 41.(4)		&	0.16		&	0.3 (0.1)		& 34 & 34 & n\\
HIP\,79431& b	& 350	&	1.1		& -		& 111.(7)		&	0.36		&	0.29 (0.02)	& 30& 30&n\\
\hline
\\
\end{tabular}
\begin{tabular}{llcccccccccc}
\multicolumn{12}{c}{Transit detections}\\
\hline
Star	& \multicolumn{1}{c}{Pl.}	&	\multicolumn{2}{c}{$m_p^\dagger$}			&	i			&	P	&	a	&	e	&	$R_p$	 	& ref.                                      & ref.			& in\\
	&					&	[$\mathrm{M_\oplus}$] & [$\mathrm{M_{Jup}}$]	&[$^{\mathrm{o}}$]	&	[d]	&    [AU]   	&		&	[$R_\oplus$]	& discovery$^{\ddagger}$ & param.$^{\ddagger}$ &sample\\
\hline

GJ\,436    					& b    &      22.6  &      0.0711     & 85.(9)		         & 2.643(9) 	&	0.029	&	0.14 (0.01)	 &4.(2) 	&5,14 & 15, 16&n\\
GJ\,1214					& b	&      6.5	&       0.020     & 88.(6)			& 1.5803(9)	& 	0.014(3)		&	$<0.27$		& 2.(7)		& 26& 26 &n\\
\hline
\\
\end{tabular}
\begin{tabular}{llccccccc}
\multicolumn{8}{c}{Microlensing detections}\\
\hline
Star	& \multicolumn{1}{c}{Pl.}	&	\multicolumn{2}{c}{$m_p$}			&		&	a	& $M_\star$     &ref.                                      & ref.		\\
	&					&	[$\mathrm{M_\oplus}$] & [$\mathrm{M_{Jup}}$]	&		&    [AU]   	& [M$_\odot$] & discovery$^{\ddagger}$ & param.$^{\ddagger}$ \\
\hline
OGLE235-MOA53			& b	&      830$^{+250}_{-190}$ & 2.61$^{+0.79}_{-0.60}$       & & 4.3$^{+2.5}_{-0.8}$ & 0.67$\pm$0.14 & 4 & 22\\
MOA-2007-BLG-192-L		& b	&	3.8$^{+5.2}_{-1.8}$    & 0.012$^{+0.016}_{-0.057}$& & 0.66$^{+0.11}_{-0.06}$ & 0.084$^{+0.015}_{-0.012}$ & 19 & 32\\
MOA-2007-BLG-400-L		& b	&	260$^{+160}_{-99}$   &	0.83$^{+0.49}_{-0.31}$   & & 0.72$^{+0.38}_{-0.16}$ $|$ 6.5$^{+3.2}_{-1.2}$ $^{\dagger\dagger}$&0.30$^{+0.19}_{-0.12}$ & 24&24\\
OGLE-2007-BLG-368-L		& b	&	20$^{+7}_{-8}$	    &	0.06$^{+0.02}_{-0.03}$            & & 3.3$^{+1.4}_{-0.8}$ & 0.64$^{1.4}_{-0.8}$ &31&31\\
MOA-2008-BLG-310-L		& b	&	74$\pm$17	    &	0.23$^\pm0.05$		        & & 1.25$\pm$0.10& 0.67$\pm$0.14 &27&27\\
OGLE-06-109L				& b	&	226$\pm$25         &	0.711$\pm$0.079	                 & & 2.3$\pm$0.2 &0.50$\pm$0.05 & 18& 22\\
						& c	&	86$\pm$10	    &	0.27$\pm$0.03	                          & & 4.6$\pm$0.5 &                            & 18& 22\\
OGLE-05-169L				& b	&	13$^{+4}_{-5}$	    &	0.041$^{+0.013}_{-0.016}$     & & 3.2$^{+1.5}_{-1.0}$ & 0.49$^{+0.14}_{-0.18}$ &10&22\\
OGLE-05-390L				& b 	&	5.5$^{+5.5}_{-2.7}$ & 0.017$^{+0.017}_{-0.008}$   & & 2.6$^{+1.5}_{-0.6}$ & 0.22$^{+0.21}_{-0.11}$ &9&124\\
OGLE-05-071L				& b	&	1200$\pm$100    & 3.8$\pm$0.4 		                 & & 2.1$\pm0.1$ $|$ 3.6$\pm$0.2 $^{\dagger\dagger}$&0.46$\pm$0.04&7&21\\
MOA-2009-BLG-319-L               & b   &       50$^{+44}_{-24}$     & 0.2$\pm$0.1 & & 2.4$^{+1.2}_{-1.6}$ & 0.38$^{+0.34}_{-0.18}$ & 33 & 33 \\
MOA-2009-BLG-387-L               & b   &       830$^{+1300}_{-510}$     & 2.6$^{+4.1}_{-1.6}$ & & 1.8$^{+0.9}_{-0.7}$ & 0.19$^{+0.30}_{-0.12}$ & 38 & 38$^{\dagger\dagger\dagger}$ \\
\hline
\end{tabular}
\\
$^\dagger$ The true mass ($m_p$) is reported for GJ\,876\,$b$, $c$, for the transiting planets GJ\,436b and GJ\,1214b and for all microlensing detections. The masses given for $GJ\,876d$ assumes a 50$^\mathrm{o}$ orbital inclination. We give minimum masses GJ581b, c, d and e, and dynamical consideration restrict coplanar systems to $i>40^{\mathrm{o}}$. Usually, uncertainties in planetary masses do not include the stellar mass uncertainty.\\
$^{\dagger\dagger}$ degenerated solution\\
$^{\dagger\dagger\dagger}$ instead of $1\sigma$ uncertainties, we quote 90\% confidence intervals from \citet{Batista:2011}\\
$^\ddagger$ (1)  \citet{Delfosse:1998b};  (2) \citet{Marcy:1998}; (3) \citet{Marcy:2001}; (4) \citet{Bond:2004}; (5) \citet{Butler:2004};  (6) \citet{Rivera:2005}; (7) \citet{Udalski:2005}; (8) \citet{Bonfils:2005b}; (9) \citet{Beaulieu:2006}; (10) \citet{Gould:2006}; (11) \citet{Butler:2006}; (12) \citet{Bonfils:2007b}; (13) \citet{Udry:2007a}; (14) \citet{Gillon:2007a}; (15) \citet{Gillon:2007b}; (16) \citet{Demory:2007}; (17) \citet{Johnson:2007}; (18) \citet{Gaudi:2008}; (19) \citet{Bennett:2008}; (20) \citet{Bailey:2009}; (21) \citet{Dong:2009b};  (22) \citet{Bennett:2009}; (23) \citet{Mayor:2009}; (24) \citet{Dong:2009a}; (25) \citet{Forveille:2009}; (26) \citet{Charbonneau:2009}; (27) \citet{Janczak:2010}; (28) \citet{Johnson:2010b}; (29) \citet{Correia:2010b};  (30) \citet{Apps:2010}; (31) \citet{Sumi:2010}; (32) \citet{Kubas:2010};  (33) \citet{Miyake:2010}; (34) \citet{Haghighipour:2010} (35) \citet{Forveille:2011}; (36) \citet{Bonfils:2011a}; (37) Delfosse (2011, in prep.); (38) \citet{Batista:2011}; (39) This paper.\\
\end{table*}

Anomalies in gravitational microlensing light curves can reveal planetary systems kiloparsecs away from our Sun. Most frequently, the lenses are low-mass stars of masses $\lesssim~0.6~\mathrm{M_\odot}$ and of spectral types M and K. Up to now, the technique has found 12 planets in 11 planetary systems. Among those, 7 are giant planets and 5 fall in the domain of Neptunes and Super-Earths (Tab.~\ref{tab:planets}). The technique is mostly sensitive to planets at a few AUs from their host which, for M dwarfs, is far beyond the stellar habitable zone. The microlensing technique probes a mass-separation domain complementary to the RV and transit techniques and has shown evidences that, at large  separations, low-mass planets outnumber giant planets \citep{Gould:2006}. 

Ground-based astrometry applied to planet searches has been cursed by false positives, of which Van de Kamp's attempts around BarnardÕs star are probably the most famous examples \citep{Kamp:1963, Gatewood:1973}. Fifty years ago, Van de Kamp first claimed that a $1.6~\mathrm{M_{Jup}}$ planet orbits Barnard's star every 24 years. Over the following decades, he continued to argue for a planetary system around the star \citep{Kamp:1982}, despite growing evidence of systematics in the data \citep[e.g. ][]{Gatewood:1973, Hershey:1973}. Radial-velocity and astrometric data have now completely excluded the van de Kamp planets \citep{Gatewood:1995, Kurster:2003, Benedict:2002}, but Barnard's star has been far from the only target with false astrometric detections. 

Nevertheless, astrometry has proven useful to confirmed the planetary nature of a few radial-velocity detections, and to remove from the planet sample the low-mass stars seen with an unfavorable inclination \citep[e.g.][]{Halbwachs:2000}. Moreover, thanks to HST/FGS astrometric observations,  GJ\,876b has been the second exoplanet with a true mass determination \citep{Benedict:2002}, soon after the detection of the transiting planet \object{HD 209458\,b} \citep{Charbonneau:2000, Mazeh:2000, Henry:2000}.

To complete the view of planetary-mass objects formed at the lower end of the main sequence, let us mention the $\sim5~\mathrm{M_{Jup}}$ companion detected with RV measurements around the young brown dwarf  \object{Cha H$\alpha$ 8} \citep{Joergens:2007} and the $\sim5~\mathrm{M_{Jup}}$ companion imaged around another young brown dwarf \object{2M1207} \citep{Chauvin:2004}. The protoplanetary disks of both brown dwarfs were most likely not massive enough to form such massive objects, since observations show that protoplanetary disk masses scale at most linearly with the mass of the star.
Both 2M1207b and Cha H$\alpha$ 8b therefore probably formed like stars rather than in protoplanetary disks.

Table~\ref{tab:planets} lists the known M-dwarf hosts and their procession of planets. For each planet, it gives the basic characteristics and a reference to the discovery papers. In total, there are 35 planets in 28 planetary systems. 

Planets orbiting M dwarfs formed in a different environment than those around solar-type stars, and therefore reflect a different outcome of the planetary formation mechanism. The mass of the proto-planetary disk, its temperature and density profiles, gravity, the gas-dissipation timescale, etc... all change with stellar mass \citep[e.g.~][]{Ida:2005}. For the construction of the \textsc{Harps} spectrograph for ESO, our consortium has been granted 500 observing nights with the instrument spread over 6 years. We chose to dedicate 10\% of that guaranteed time to characterize the planetary population for stars with masses $< 0.6~\mathrm{M_\odot}$. 

This paper reports on our 6-year RV search for planets around M dwarfs and the outline is as follow. We first describe our sample (Sect.~\ref{sect:sample}) and present the RV dataset collected (Sect.~\ref{sect:observations}). We next perform a systematic analysis for variability, long-term trend and periodic signals (Sect.~\ref{sect:signal}). We close that section by an automated Keplerian multi-planet search. For all signals detected with enough confidence, we apply a suite of diagnostics to disentangle Doppler shifts caused by bona fide planets from Doppler shifts caused by stellar surface inhomogeneities (Sect.~\ref{sect:activity}). Sect. 6 presents the detection limits for individual stars of our sample and Sect. 7 pools them together to estimate both the survey sensitivity and the frequency of planets orbiting M dwarfs. Sect. 8 summarizes our results and presents our conclusions. 
\section{\label{sect:sample}Sample}

Our search for planets orbiting M dwarfs originates from RV observations started in 1995 with the 1.93m/\textsc{Elodie} spectrograph (CNRS/OHP, France). This former program aimed at determining the stellar multiplicity  for very-low-mass stars \citep{Delfosse:1999} as well as the detection of their most massive planets \citep{Delfosse:1998b}. In 1998, we started to run a similar program in the southern hemisphere, with the 1.52m/\textsc{Feros} spectrograph (La Silla, Chile). With \textsc{Feros}'s early settings, we were unsuccessful to improve on its nominal precision of $\sim$ 50 m/s. Nevertheless, we benefitted from these observations to start with a better sample on \textsc{Harps}, cleaned from spectroscopic binaries and fast rotators for which precision radial-velocity is more difficult.

Our \textsc{Harps} sample corresponds to a volume limited of M dwarfs closer than \dmax  pc, with a declination $\delta < + 20^\mathrm{o}$, brighter than V = \Vmax ~mag and with a projected rotational velocity $v \sin i \lesssim \vsinimax$~km/s. We removed known spectroscopic binaries and visual pairs with separation $<5''$ (to avoid light contamination from the unwanted component). We have however encountered a few spectroscopic binaries and fast rotators which were not known before our observations. We list them in Table~\ref{tab:sb} and discard them from the sample presented here. Note that we also dismiss GJ\,1001, Gl\,452.1 and LHS\,3836. The first two stars were initially counted in our volume limited sample \citep[with $\pi=103$ and $96~\mathrm{mas}$, respectively --][]{Gliese:1991} and had their parallax revised since, now placing them beyond 11~pc \citep[with $\pi=76.86\pm3.97$ and $88.3\pm3.7~\mathrm{mas}$ --][]{Henry:2006, Smart:2010}. LHS\,3836 was initially included based on its V magnitude in \citet{Gliese:1991}'s catalog but our first measurements were indicative of a much lower brightness.

Table~\ref{tab:sample} lists the \Nstars  stars selected for the sample. Their coordinates, proper motions and parallaxes are primarily retrieved from the revised Hipparcos catalog \citep{Leeuwen:2007}. A fraction of the parallaxes,  unavailable in the Hipparcos database, were obtained from \citet{Altena:1995}, \citet{Henry:2006}, \citet{Reid:1995b} and the 4th Catalog of Nearby Stars (CNS4 -- Jahreiss, priv. comm.). V-band magnitudes are taken from Simbad and infrared J- and K-band magnitudes from 2MASS \citep{Cutri:2003}. We used the empirical \citet{Delfosse:2000}'s mass-luminosity relationship together with parallaxes and K-band photometry to compute the mass of each star. Infrared K-band photometry and (J$-$K) colors are converted to luminosities with \citet{Leggett:2001}'s bolometric correction. 

\begin{table*}
\caption{\label{tab:sb} Spectroscopic binaries and fast rotators discarded {\it a posteriori} from the sample}
\begin{tabular}{lr@{:}c@{:}lr@{:}c@{:}lcccccl}
\hline\hline
Name & \multicolumn{3}{c}{$\alpha$ (2000)}&\multicolumn{3}{c}{$\delta$ (2000)}& V [mag]&Comment\\
\hline
L\,225-57	&02	&34	&21	&$-53$	&05	&35	&7.3	& SB2\\  
CD-\,44-836B	&02	&45	&14	&$-43$	&44	&06	&12.7& Fast Rotator / SB2	\\ 
LHS\,1610	&03	&52	&42	&$+17$	&01	&06	&13.7&SB2\\
LHS\,6167	&09	&15	&36	&$-10$	&35	&47	&14.7&SB2 $^\dagger$	\\
G\,161-71	&09	&44	&55	&$-12$	&20	&53	&13.8&Fast Rotator\\ 
GJ\,1154A	&12	&14	&17	&$+00$	&37	&25	&13.7	& Unresolved SB2$^\ddagger$ \\
LHS\,3056	&15	&19	&12	&$-12$	&45	&06	&12.8& SB\\
L\,43-72	&18	&11	&15	&$-78$	&59	&17&12.5	& Unresolved SB2$^\ddagger$ \\
LTT\,7434	&18	&45	&57	&$-28$	&55	&53	&12.6	&Unresolved SB2$^\ddagger$\\ 
Gl\,867B	&22	&38	&45	&$-20$	&36	&47	&11.4	&Fast Rotator or unresolved SB2\\\hline
\end{tabular}\\
$^\dagger$ previously detected by \citet{Montagnier:2006}\\
$\ddagger$ SB with variable spectral-line width

\end{table*}

We also indicate in Table~\ref{tab:sample} the inner and outer limits for the distance of the Habitable Zone using the recent-Venus and early-Mars criterions, respectively, and Eq. (2) and (3) from \citet{Selsis:2007}. The boundaries of the Habitable Zone are uncertain and depend on the planet's atmospheric composition. Extra-solar planets found close to these edges have therefore to meet more stringent conditions to be inhabitable. For more detailed considerations, we refer the reader to more comprehensive models \citep[e.g.][]{Selsis:2007}.

Our sample is composed of the closest neighbors to the Sun. Nearby stars tend to have large proper motions and the projection of their velocity vector may change over time, up to few m/s/yr \citep{Schlesinger:1917, Kurster:2003}. We therefore report the value of their secular acceleration in Table~\ref{tab:sample}.

\begin{figure}[t]
\centering
\includegraphics[width=1.1\linewidth]{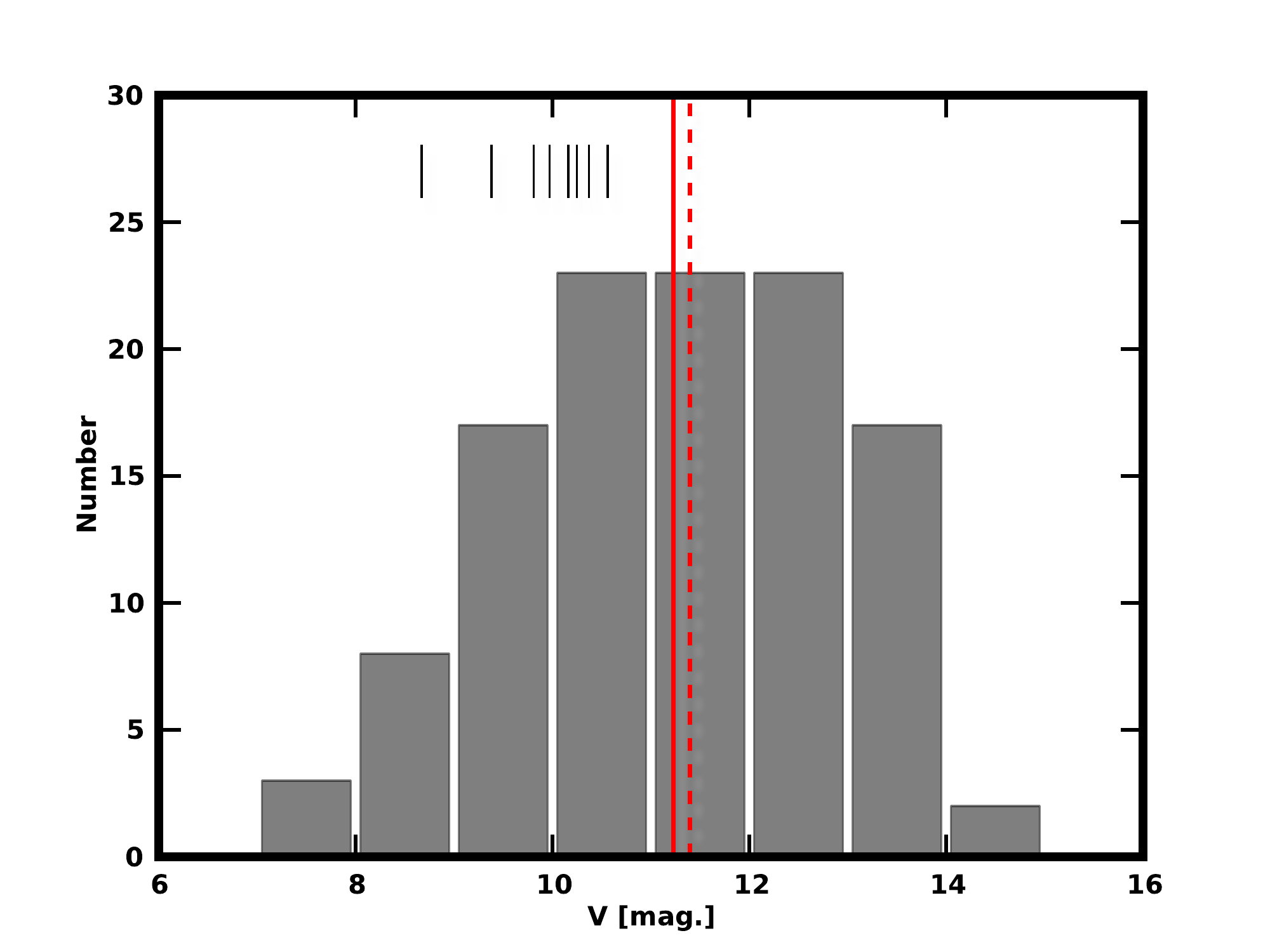}\\
\includegraphics[width=1.1\linewidth]{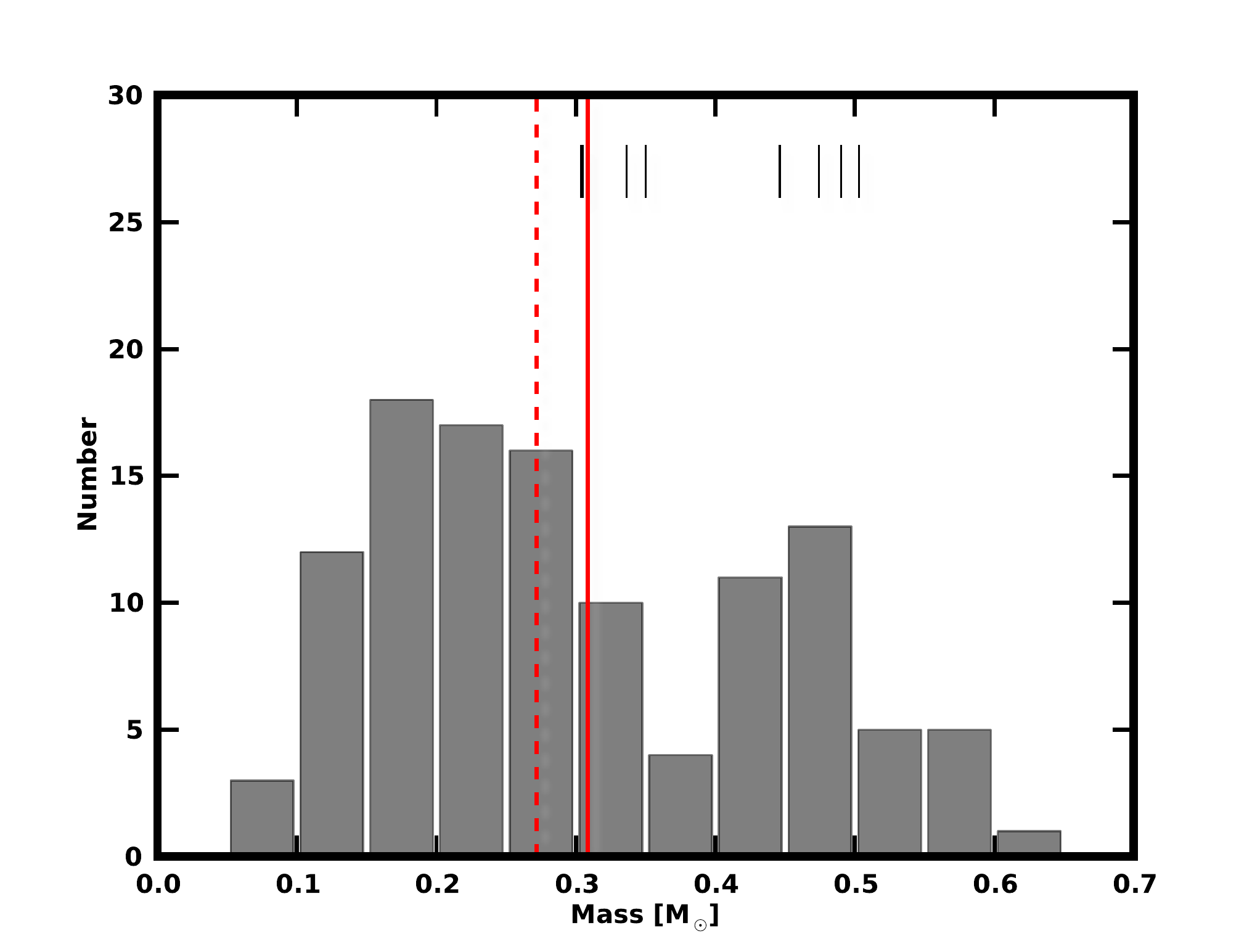}
\caption{\label{fig:hist} Sample distributions for V magnitudes and stellar masses. The vertical dashed and plain lines locate the median and averaged values, respectively. The small ticks are explained in Sect.~\ref{sect:discussion}.}
\end{figure}

To portray our sample, we show its V-mag. and mass distributions in Fig.~\ref{fig:hist}. For both distributions, the average (resp. median) value is plotted with a vertical straight (resp. dashed) line. The magnitudes and masses of planet hosts are also marked with vertical ticks on top of the histograms. The target brightness spans V$=$7.3 to \Vmax mag. with a mean (resp. median) value of 11.25 mag. (resp. 11.43 mag.). The stellar mass ranges from 0.09 to 0.60~$\mathrm{M_\odot}$ with an average (resp. median) value of 0.30 M$_\odot$ (resp. 0.27~M$_\odot$). The smaller count seen in the 0.35-0.40 $\mathrm{M_\odot}$ bin is unexplained but from statistical fluctuations. Interestingly, one can note that our sample covers a factor of $\sim$6 in stellar mass, while the mass step between our typical M dwarf ($\sim 0.27~\mathrm{M_\odot}$) and the typical Sun-like star ($\sim 1~\mathrm{M_\odot}$) corresponds to a factor of less than 4. This means that planetary formation processes depending on stellar mass could lead to larger observable differences across our sample than between our M-dwarf sample and Sun-like stars.

There are overlaps between our sample and others that similarly targets M dwarfs to search for planets. Among them, we found published RV time-series for \object{Gl\,1}, \object{Gl\,176}, \object{Gl\,229}, \object{Gl\,357}, \object{Gl\,551}, \object{Gl\,682}, \object{Gl\,699}, \object{Gl\,846} and \object{Gl\,849} in \citet[][hereafter E06]{Endl:2006}, for \object{Gl\,1}, \object{Gl\,229}, \object{Gl357}, \object{Gl\,433}, \object{Gl\,551}, \object{Gl\,682}, \object{Gl\,699}, \object{Gl\,846} and \object{Gl\,849} in \citet[][hereafter Z09]{Zechmeister:2009b}, and few others in detection papers, as for \object{Gl\,176} \citep{Endl:2008a}, \object{Gl\,832} \citep{Bailey:2009} and \object{Gl\,849} \citep{Butler:2006}. When possible, we compare our results to these time series and, for completeness, additional comparison is given in Appendix~\ref{app:comparison}.

\section{\label{sect:observations}Observations} 
To gather RV observations for the sample described above we used the \textsc{Harps} instrument \citep{Mayor:2003, Pepe:2004}, a spectrograph fiber fed with  the ESO/3.6-m telescope (La Silla, Chile). It covers the 3800$-$6800~$\AA$ wavelength domain, has a resolution $R\sim115\,000$ and an overall throughput (instrument$+$telescope) greater than $6\%$. It is enclosed in a vacuum vessel, is pressure controlled to $\pm 0.01$~mbar and thermally controlled to $\pm 0.01$~K. That ensures a minimum instrumental shift on the position of the spectrum image on the CCD or, practically, a RV drift $\lesssim0.5$ m/s/night.  To reference the origin of the instrumental RV drift, the fiber can be illuminated with a Thorium-Argon (ThAr) lamp at any time. \textsc{Harps} also offers a second fiber that can be illuminated with ThAr light {\it simultaneously} while the scientific fiber receives star light. This mode avoids the need to record frequent calibrations between scientific exposures and can correct the small instrumental drift that occurs while the stellar spectrum is recorded. Since the instrumental drift during the night is small, this mode is only used when a sub-m/s precision is required. Our observational strategy for M dwarfs aims to achieve a precision of $\sim 1$ m/s per exposure for the brightest targets. We chose therefore not to use the second fiber and relied on a single calibration done before the beginning of the night. As the science and calibration spectral orders are interlaced, avoiding the second fiber eludes light cross-contamination between science and calibration spectra. This can be a source of noise for the blue-most spectral orders, where we only reach low signal-to-noise for M dwarfs. In particular, we were interested in clean \ion{Ca}{ii} H\&K lines because they are a useful diagnostic of stellar activity (see Sect.~\ref{sect:activity}).

From the first light on \textsc{Harps} on February, 11th 2003 to the end of our guaranteed time program on April, 1st 2009, we have recorded \Nspectra  spectra for the M-dwarf sample, for a total of \Nhours  h of integration time. 

We computed RV by cross-correlating the spectra with a numerical weighted mask following \citet{Baranne:1996} and \citet{Pepe:2002}. Our numerical mask was generated from the addition of several exposures taken on \object{Gl 877}, a bright M2 star of our sample. Co-addition of spectra requires knowing their relative Doppler shifts. We computed RVs for Gl\,877 spectra with a first iteration of the template and re-built the template more precisely. We obtained convergence just with a few iterations. The numerical mask counts almost 10,000 lines and most of the Doppler information in the spectrum. No binning is done.

The RV uncertainties were evaluated from the Doppler content of individual spectra, using the linear approximation of a Doppler shift \citep[Eq. 12 -- ][]{Bouchy:2001}. This formula gives more weight to spectral elements with higher derivative because they are more sensitive to phase shifts and contribute more to the total Doppler content. It is important to note that we do not sum the Doppler content of individual spectral elements over the whole spectrum. The derivative of the spectrum has a higher variability against noise than the spectrum itself and, for low signal-to-noise ratio, doing so would over-estimate the RV precision. Instead, to mitigate the effect and compute more realistic uncertainties, we applied the formula directly to the cross-correlation profile, which has $\sim30$ times higher signal-to-noise than the individual spectral lines \citep[see appendix A in][]{Boisse:2010}. To account for the imperfect guiding ($\sim$30 cm/s) and wavelength calibration ($\sim$50 cm/s) we quadratically added 60 cm/s to the Doppler uncertainty.

\begin{figure}[t]
\centering
\includegraphics[width=1.1\linewidth]{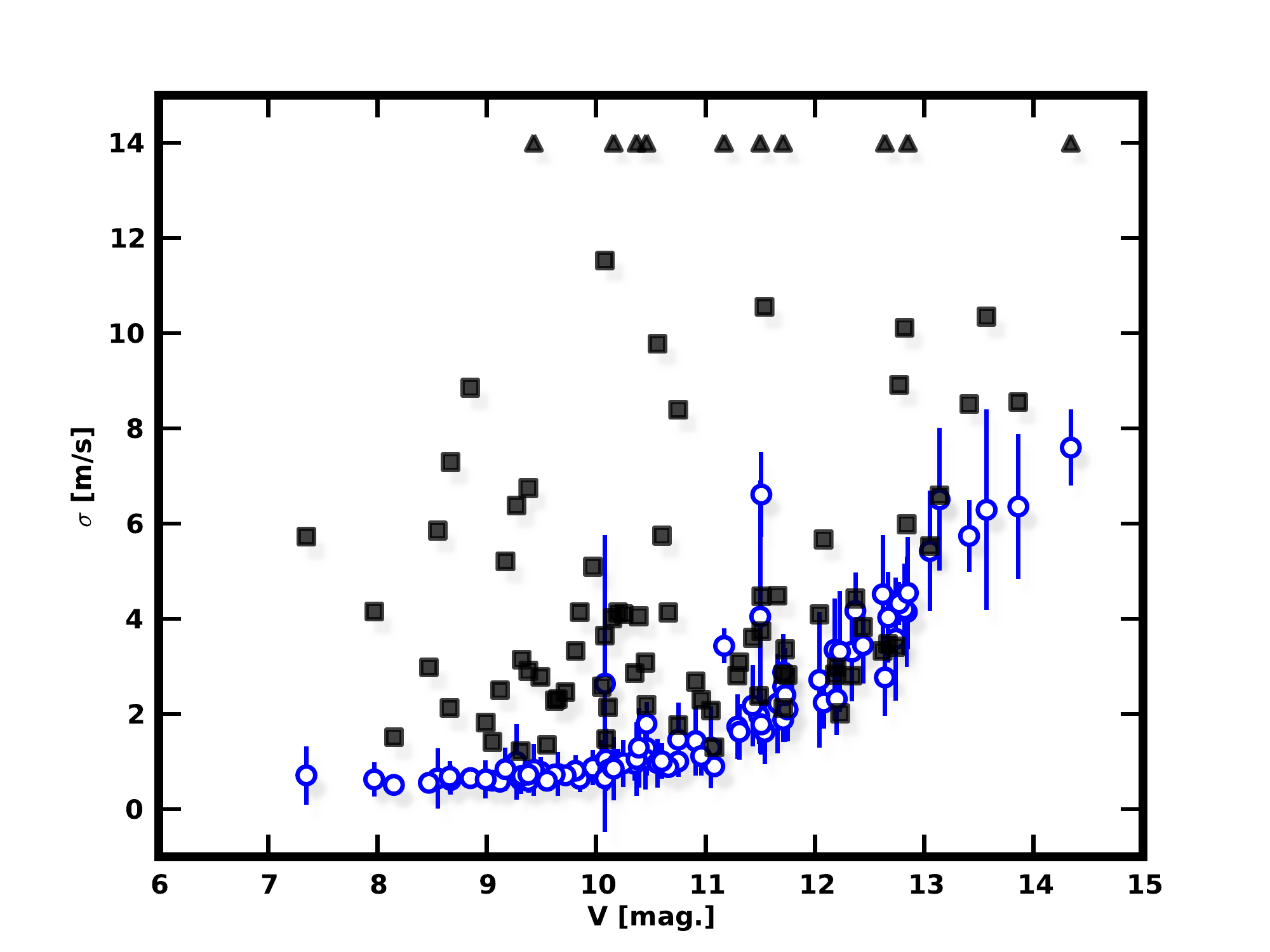}
\caption{\label{fig:SigV} Internal ($\sigma_i$) and external ($\sigma_e$) errors as a function of V-band magnitudes, for stars with 6 or more measurements.}
\end{figure}

As a trade-off between exposure time and precision we chose to fix the integration time to 900~s for all observations. We obtained a precision\footnote{As opposed to our precision, our measurement accuracy is poor. Absolute radial velocities given in this paper may not be accurate to $\pm$ 1 km/s.}
 $\sigma_i\sim 80$ cm/s from V$^{mag}=7-10$ stars and $\sigma_i\sim2.5^{(10-V)/2}$ m/s for V$^{mag}=10-14$. Our internal errors $\sigma_i$ (composed of photon noise $+$ instrumental errors) are shown in Fig.~\ref{fig:SigV} where we report, for all stars with more than 6 measurements, the mean $\sigma_i$ (blue filled circle) as a function of the star's magnitude, with error bars corresponding to $\sigma_i$'s dispersion. For comparison, Fig.~\ref{fig:SigV} shows observed dispersions $\sigma_e$ for all stars with more than 6 measurements (black squares, changed to triangles for clipped values). The $\sigma_e$ are the observed weighted r.m.s. and are related to the $\chi^2$ value :
\begin{equation}
\sigma_e^2 = \frac{ \sum{(RV-<RV>)^2/\sigma_i^2}} {  \sum{1/\sigma_i^2} }= \frac{\chi_{constant}^2}{\sum{1/\sigma_i^2}}.
\end{equation}
We discuss the difference between internal and external errors in Section~\ref{sect:signal}. 

\begin{figure*}[hhh]
\hspace{-1.5cm}
\includegraphics[width=1.2\linewidth]{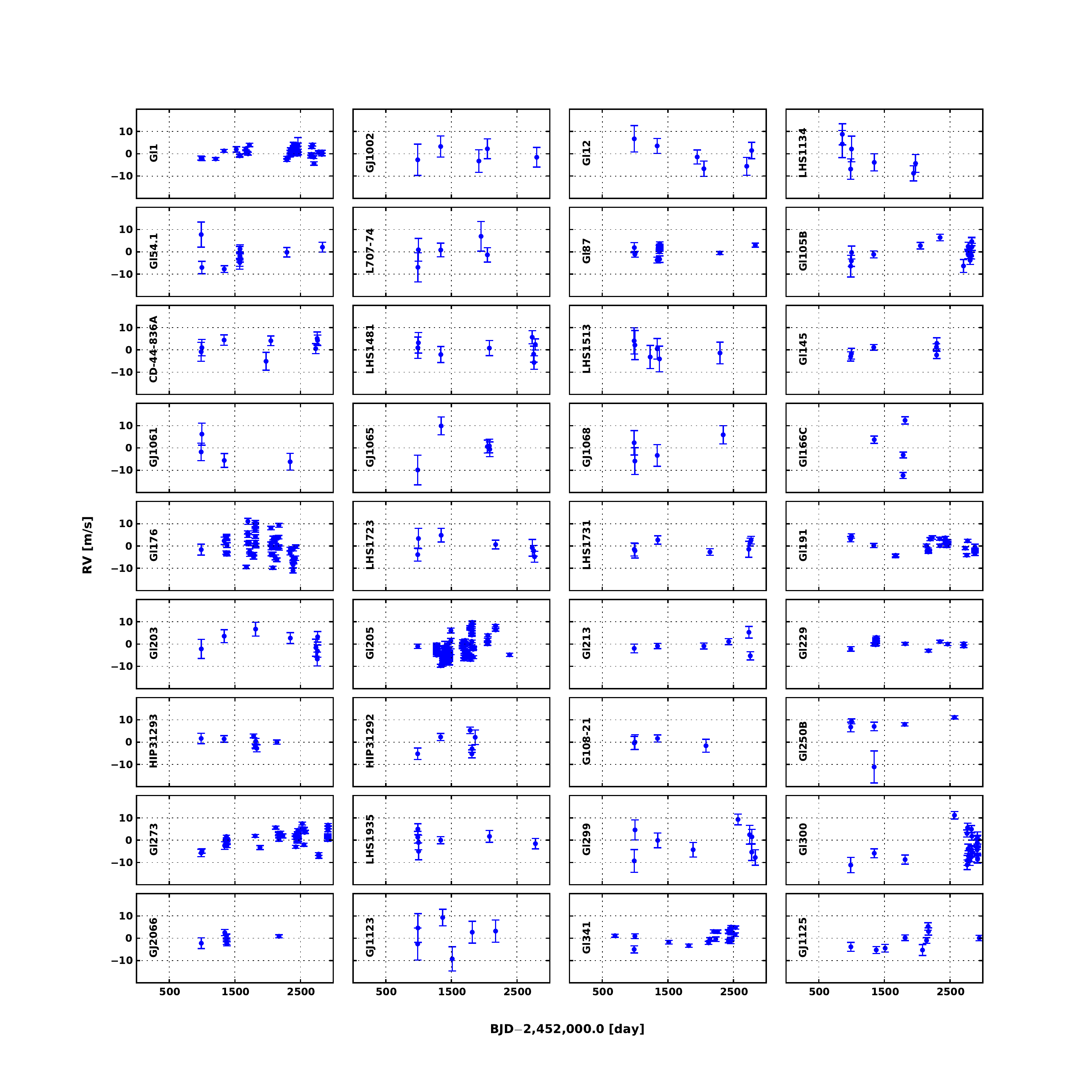}
\caption{\label{fig:series} Radial-velocity time series.}
\end{figure*}

Our RV time series are given in the Solar System barycentric reference frame and can be retrieved online from CDS. Prior to their analysis, the time series are corrected from the secular RV changes reported in Tab.~\ref{tab:sample}. We also show the time series in Fig.~\ref{fig:series} (only available on-line), after subtraction of half the min$+$max value for a better readability.

\addtocounter{figure}{-1}
\begin{figure*}
\hspace{-1.5cm}
\includegraphics[width=1.2\linewidth]{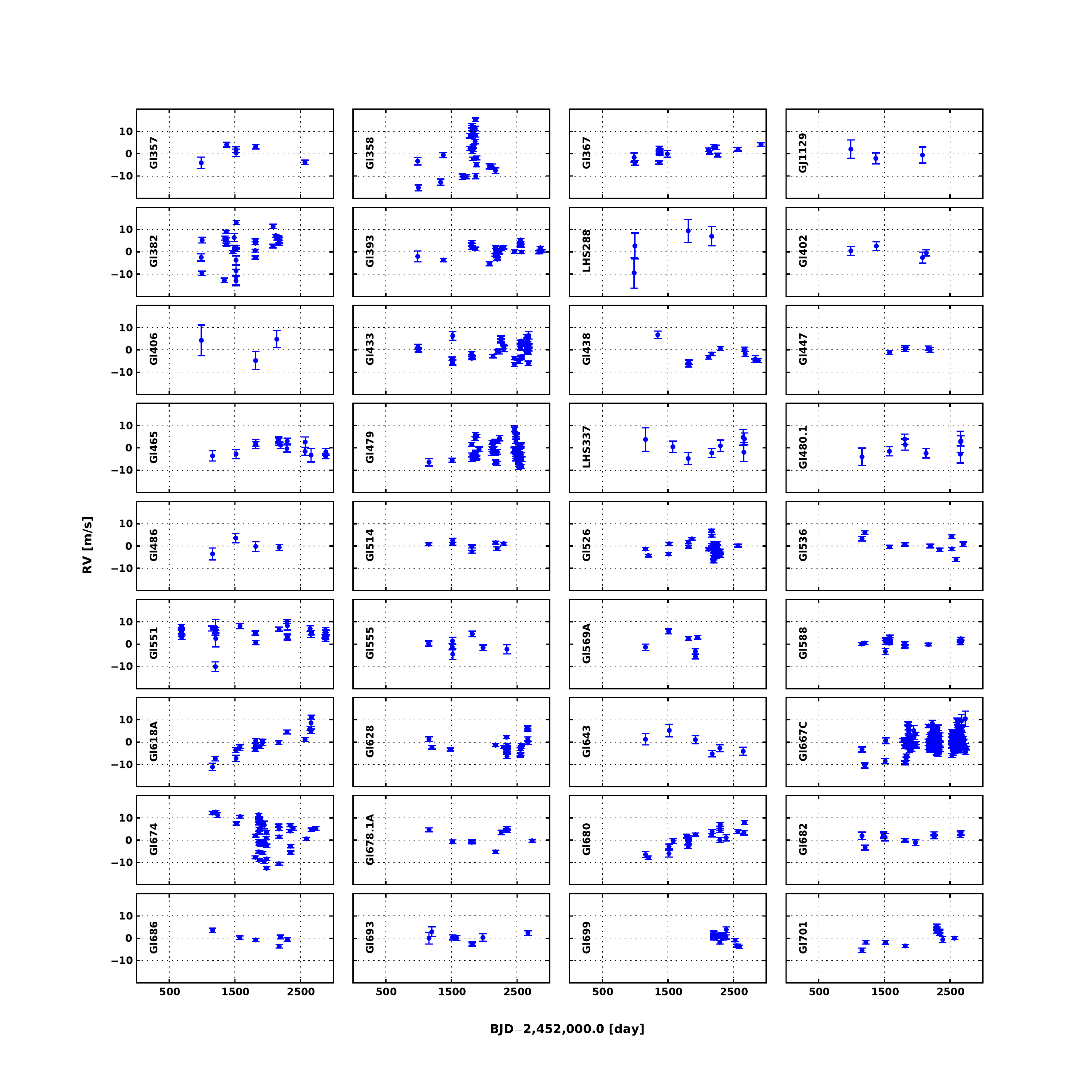}
\caption{Radial-velocity time series (continued).}
\end{figure*}

\addtocounter{figure}{-1}
\begin{figure*}
\caption{Radial-velocity time series (continued).}
\hspace{-1.5cm}
\includegraphics[width=1.2\linewidth]{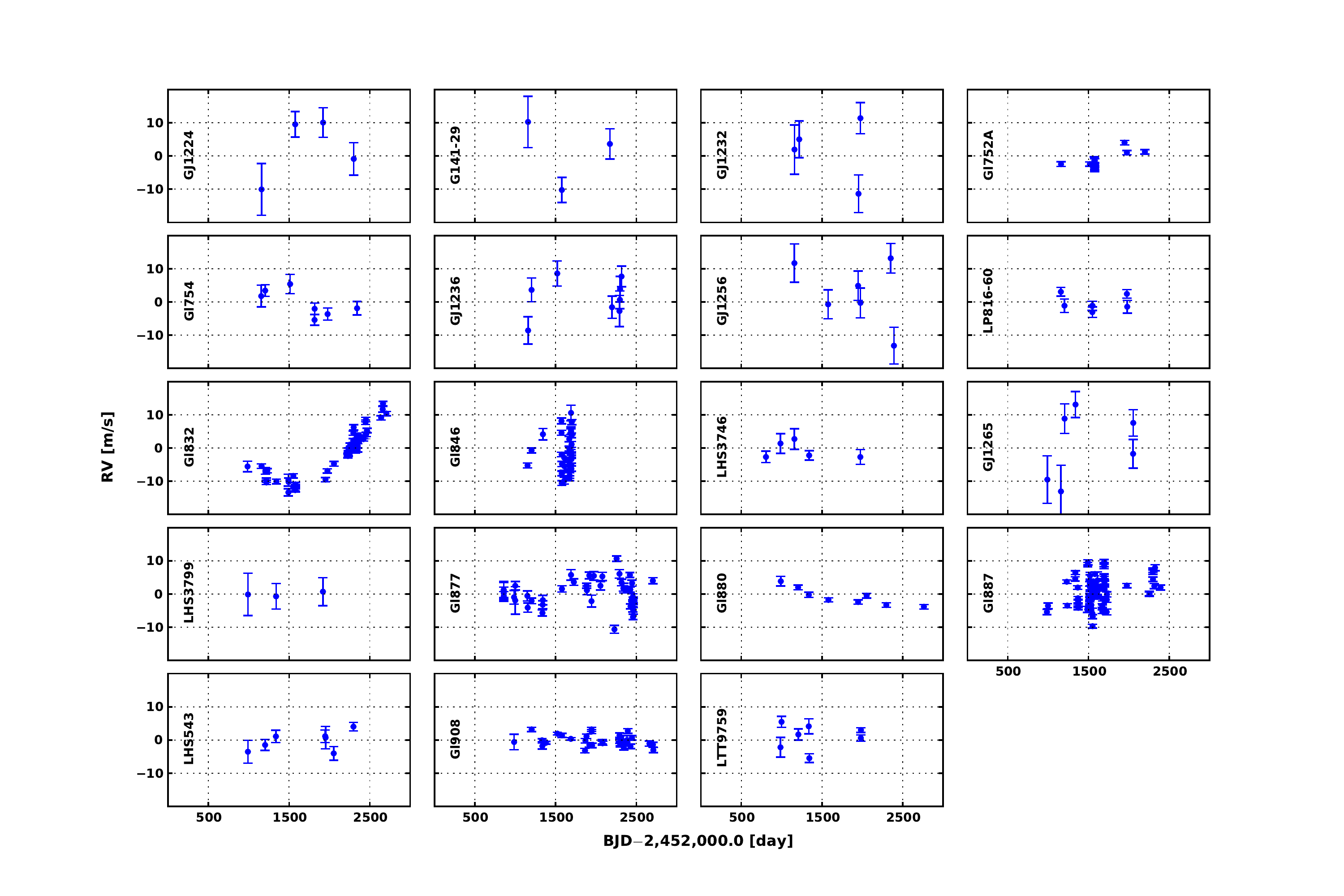}
\vspace{-6.5cm}
\hspace{-1.5cm}
\includegraphics[width=1.2\linewidth]{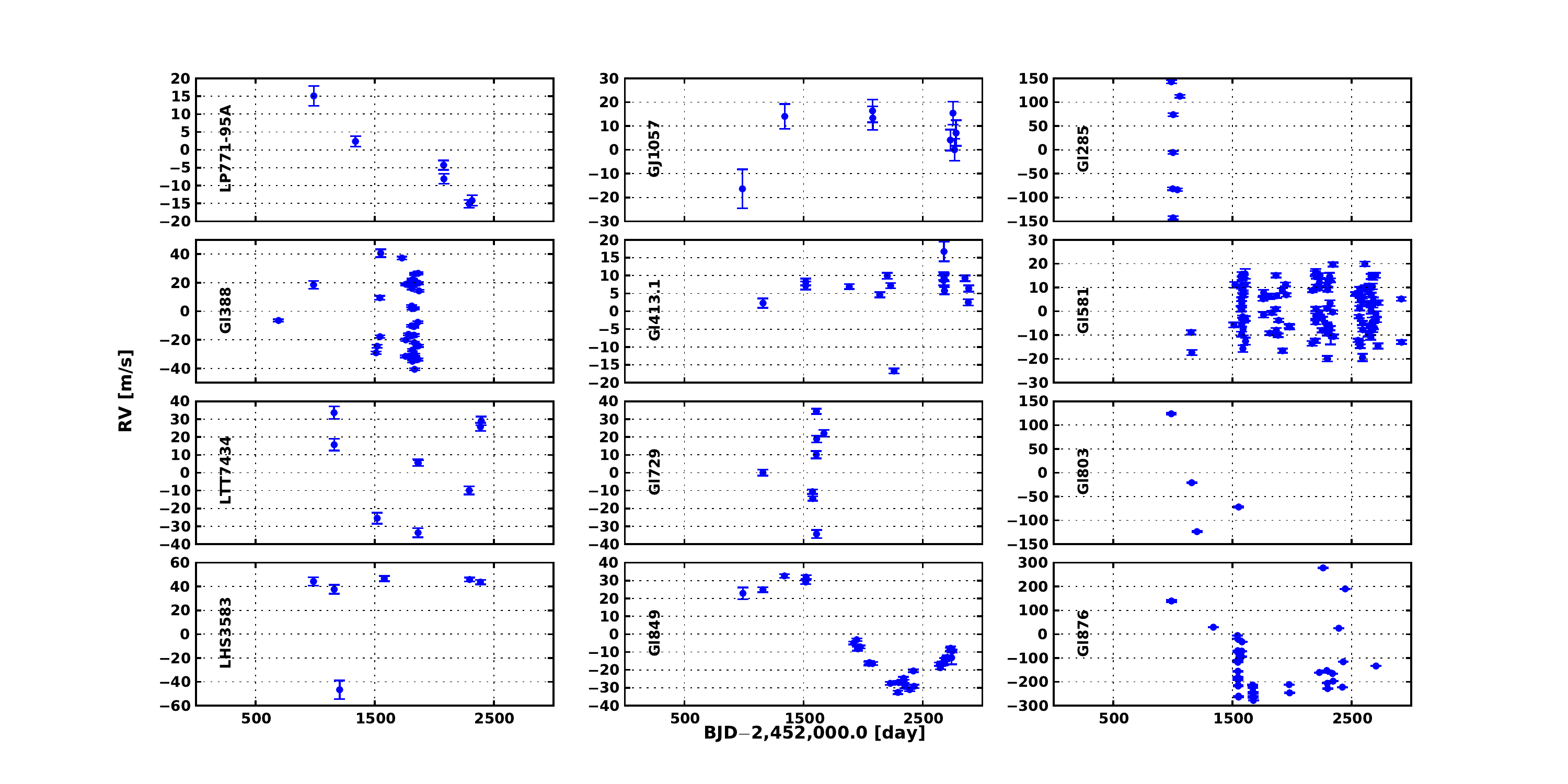}
\end{figure*}

\section{\label{sect:signal}Data analysis}
Several planet-search programs have presented a statistical analysis for their survey \citep[e.g.][]{Murdoch:1993, Walker:1995, Cumming:1999, Cumming:2008, Endl:2002, Zechmeister:2009b}. Often, statistical tests are applied to the time series in order to appraise the significance of trends or variability. Then, the time series are searched for periodicities and, if a significant periodicity is found, the corresponding period is used as a starting point for a Keplerian fit. Again, statistical tests are applied to decide whether a sinusoidal or a Keplerian model is a good description for the time series. In this section, we follow the same strategy and add a heuristic method based on genetic algorithms to the systematic search for Keplerian signals.

	\subsection{Excess variability}
Once we have computed the RVs, the Doppler uncertainties ($\sigma_i$) and the RV dispersion ($\sigma_e$) for each star, the first step is to test our time series for variability. In Fig.~\ref{fig:SigV}, we have reported both $\sigma_i$ and $\sigma_e$ as a function of stellar magnitudes for all stars with more than 6 measurements. Apart from one star (\object{Gl 447}, which has 6 RVs only), we have $\sigma_e \gtrsim \sigma_i$. The $\sigma_e$ have a lower envelope that matches the $\sigma_i$'s envelope for V$^{mag}=10-14$ and that is slightly above ($\sim$ 2 m/s) in the brighter range V$^{mag}=7-10$. 

To test whether the observed RVs vary in excess of our internal errors, we first compare the expected variance $<\sigma_i^2>$ (the mean internal error) to the measured variance $\sigma_e^2$ by applying a F-test, which gives a probability $P(F)$ to the F-value $F=\frac{\sigma_e^2}{<\sigma_i>^2}$ \citep[e.g.][]{Zechmeister:2009b}. As another test for variability, we also compute the $\chi_{constant}^2$ for the {\it constant} model and $P(\chi_{constant}^2)$, the probability of having $\chi_{constant}^2$ given the $\sigma_i$. For both the F-test and the $\chi^2$-test, a low probability means that photon noise, calibration uncertainty and guiding errors are not sufficient to explain the observed variability. In such a case, one has to invoke an additional source for variability, from unaccounted noise to planetary companions.

We report $\sigma_i$, $\sigma_e$, $P(F)$, $\chi^2$ and $P(\chi^2)$ in Table~\ref{tab:var} and change the $P(F)$ and the $P(\chi^2)$ values to boldface when smaller than 1\%, i.e. when they indicate a confidence level for variability higher than 99\%.  Using this criterion, the F probabilities (resp. the $\chi^2$ probabilities) indicate that $45\%$ (resp. 63\%) of our sample displays an excess variability. When focusing on stars with V$^{mag}=7-10$,  all stars but two are found more variable than expected according to $P(F)$, and all stars according to $P(\chi^2)$. The reason is that our external error never reaches the $\sim$70 cm/s threshold estimated for the brighter range of our sample (but rather 1.5$-$2~m/s), and is dominated by photon noise for a third of the sample in the V$^{mag}=10-14$ range.

\subsection{Trends}

Next, we examine the time series for possible trends. They may correspond to incomplete orbits and betray the presence of long-period companions. For each star, we adjust a slope $\alpha$ ($RV=\alpha t + \beta$) to the RV data and evaluate the $\chi_{slope}^2$ value of that model. 

To know whether the {\it slope} is a better description to the data than the {\it no slope} model, we use two statistical tests. First, we use the F-test to gauge whether a lower $\chi^2_{slope}$ compared to $\chi^2_{constant}$ is a sufficient improvement to justify an additional free parameter (2 for the slope model against 1 for the constant model). 

Also, because the F-test statistics is ill-behaved for non normally distributed uncertainties, we use a less-sensitive test based on bootstrap randomization.
That time, we generate virtual RV time series by shuffling the original data, i.e., we keep the same observing dates and attribute to each date a measurement randomly chosen among the other dates, without repeating twice a given measurement. On each virtual time series, we adjust a slope and compute its $\chi_{virtual}^2$ value. Then, the fraction of virtual data sets with $\chi^2_{virtual}<\chi_{slope}^2$ gives us the false-alarm probability ($FAP$) for the slope model. For all stars, but the ones with 6 measurements or less, we generate 1,000 virtual time series. And because that method probes FAPs greater than $O(1/N!)$ only,  we limit the number of trials to $N!$ for stars with fewer measurements.

Table~\ref{tab:var} gives the slope coefficient $\alpha$ as well as $P(F)$ and $FAP$ values for all time series. The $P(F)$ and the $FAP$ values are reported in boldface when below a threshold of 1\%, to point high confidence level in the slope model.   

Among our sample and according to the $FAP$ values, we find that 15 stars have time series better described by a slope than a simple constant. They are Gl\,1, LP\,771-95A, Gl\,205, Gl\,341, Gl\,382, Gl413.1, Gl\,618A, Gl\,667C, Gl\,680, Gl\,699, Gl\,701, Gl\,752A, Gl\,832, Gl\,849 and Gl\,880. We also see that, while LP\,771-95A, Gl\,367, Gl618A, Gl\,680 and Gl\,880 displays smooth RV drifts, the other 10 stars seems to obey a more complex variability. According to $P(F)$ values, we find that the same 15 stars plus 8 more have a significant chi squared improvements when we fit a slope. They are Gl54.1, Gl\,250B, Gl\,273, Gl\,367, Gl\,433, Gl\,551, Gl\,674 and Gl\,887.

\subsection{\label{subsec:periodo}Periodicity}

In our search for planets, variability selects the stars to focus on and trends reveal yet uncompleted orbits. Our next step in the search for planetary candidates is to look for periodic signals. The classical diagnostic for periodic coherent signals in unevenly spaced time series is the Lomb-Scargle periodogram \citep{Lomb:1976, Scargle:1982} or, to account for the unknown system's mean velocity, its generalized version, the floating-mean periodogram \citep{Cumming:1999}. 

We therefore compute generalized Lomb-Scargle periodograms for all our time series with at least 6 measurements. We follow the formalism developed in \citet{Zechmeister:2009a}, and choose a power normalization where 1.0 means that a sinusoidal fit is perfect ($\chi^2=0$) and where 0.0 means that a sinusoidal fit does not improve $\chi^2$ over a constant model. We calculate false-alarm probabilities very similarly to our trend analysis, with a bootstrap randomization (preferred to a F-test). We create 1,000 virtual time series by shuffling the original data set. For each individual data set we compute a periodogram and locate the highest power. Considering all periodograms from all virtual data sets, we compute the distribution of the power maxima. The power value that is above 99\% of all trial powers is then considered as the 1\% $FAP$. More generally, we attribute a $FAP$ to the maximum power value found in the original data set by counting the fraction of the simulated power maxima that have a larger value. 

We show all periodograms in Fig.~\ref{fig:periodos} (only available in electronic format). The periods corresponding to the highest power, the corresponding signal semi-amplitude, the $\chi^2$, as well as the associated $FAPs$ are reported in Table~\ref{tab:periodos}. As previously, we boldface the significant signal detections, i.e. the periods with a power excess that has a $FAP<1\%$.

In ideal cases, long-term variability only affects the long-period range of the periodogram. However, the sparse sampling of our time series can cause power excess from a long-period signal to leak to shorter periods. Removing the long-term significant trends cleans the periodogram and may reveal periodic signal at shorter periods. Thus, we also compute the periodogram for the time series {\it after} removing the slope adjusted in the previous section, as well as corresponding false-alarm probabilities ($FAP_2$ in Table~\ref{tab:periodos}). We record a noticeable change for Gl\,618A and Gl\,680 only. No power excess remains in their periodograms after subtraction of the slope, meaning their RV variation are mostly linear drifts. 

Our periodicity analysis identifies 14 stars with power excess with $FAP_2<1\%$ are : Gl\,176, G\,205, Gl\,273, Gl\,358, Gl\,388, Gl\,433, Gl\,479, Gl\,581, Gl\,667C, Gl\,674, Gl\,832, Gl\,846, Gl\,849 and Gl\,876. The star Gl\,887 additionally has $FAP$ approaching our 1\% threshold.

After identifying possible periodicities, we use the candidate periods as starting values for Keplerian fits. 
We search the residuals for periodic signals again, computing periodograms (see Fig.~\ref{fig:periodos2} in the online material) and locating power excesses with $FAP<1\%$. We find a probable second periodic signal for Gl\,176, Gl\,205, Gl\,581, Gl\,674 and Gl\,876 (Table \ref{tab:per2}). We notice that Gl\,667C, Gl\,832 and Gl\,846 have FAP approaching our 1\% threshold. Although not shown here, a third iteration only reveals coherent signals for Gl\,581 and, possibly, Gl667C. A fourth iteration found signal for Gl\,581 only.

\subsection{\label{subsec:kepler}Keplerian analysis}

Even in its generalized form, the Lomb-Scargle periodogram is optimal for sinusoidal signals only. Eccentric orbits spread the spectral power on harmonics, decreasing the power measured at the fundamental period, and fitting a Keplerian function is an obvious improvement. Compared to a periodogram search, it can detect planets with high eccentricity against a higher noise. It is often not used because it is non-linear in some of the orbital parameters. Traditional non-linear minimizations can only converge to a solution close to a starting guess, possibly far from the global optimum. Applying non-linear minimization from many starting guesses becomes quickly impracticable when the number of planets increases. Keplerian functions are indeed transcendent and evaluating the radial velocity at a given time therefore need an integration which is computationally expensive. Finally, the sequential approach outlined above requires a higher signal-to-noise for systems with several planets on commensurable orbits (i.e. with RV pulls of similar amplitudes).

To work around these shortcomings, we make use of a hybrid method based on both on a fast non-linear algorithm (Levenberg-Marquardt) and genetic operators (breeding, mutations, cross-over). The algorithm has been integrated by one of us (D.S.) in an orbit analysis software named {\it Yorbit}. We give a brief overview of this software here, but defer its detail description to a future paper (S\'egransan et al. in prep.). While a population of typically 1,\,000 solutions evolves, the top layer of $Yorbit$ evaluates the performances of the different minimization methods and tunes their time allocation in real time. Genetic algorithms efficiently explore the parameter space on large scales. They thus score well and are given more CPU time during the early phase of the minimization. Once promising solutions are found, non-linear methods converge much more efficiently toward the local optima. Hence, when new solutions arise outside of local minima, the non-linear methods are given more CPU time. This heuristic approach has proved very efficient and the solution to multi-planet systems is found in few minutes only, on common desktop computers.

We search for planets using {\it Yorbit} on all stars with more than 12 measurements, neglecting planet-planet interactions at this point. We scaled the complexity of the tested models with the number of measurements. Although in principle 5N+1 RVs are enough to obtain a Keplerian fit to an N-planet system, we wish to minimize the number of spurious solutions and arbitrary require at least 12 RVs per planet in the model. Hence we use a 1-planet model for 12 RVs, both a 1-planet and a 2-planet model for stars with more than 24 RVs, and 4 different models (1$-$4 planets) for stars with more than 48 RVs. To complement those models we also use the same suite of models with the addition of a linear drift. We allow {\it Yorbit} to run for 150 seconds per planet in the model.  

Just as for the periodicity analysis, evaluating the credence of the model is essential. The $\chi^2$ of solutions necessarily improves for more complex models as the number of degree of freedom increases, and we want to evaluate whether this improvement is statistically significant or occurs by chance. As previously, we generate virtual datasets by shuffling the original data and retaining the dates. We create 1,000 virtual datasets and refit all tested models 1,000 times with {\it Yorbit}. For each star and model, we obtain 1,000 $\chi^2$ values and count how many are below the $\chi^2$ measured on the original data. This gives the FAP for that model, compared to the no-planet hypothesis. A model is considered to improve $\chi^2$ significantly when less than 1\% of the virtual trials give as low a $\chi^2$. 

Once we find a significant model, more complex models are then evaluated against that model, and not against simpler models anymore. We consider signals in that model are {\it detected} (i.e., for instance, we assume the system is composed of 2 planets if that model is a 2-planet model). To generate the virtual datasets we use the residuals around the best solution for the new reference model (i.e., in our example, the residuals around the 2-planet model). Shuffling the residuals (and retaining the dates), we again create 1,000 virtual datasets and fit the more complex models using {\it Yorbit}. How many $\chi^2$ from these virtual trials are lower than the best $\chi^2$ obtained on the actual RVs then gives the FAP for the complex models, compared to the simpler model.

We report the parameters of the best solutions for each star and model in Table~\ref{tab:paramKep}. In Tables~\ref{tab:fapKepd0} and~\ref{tab:fapKepd1}, we report the FAP of the models, compared to selected simpler models. In both tables, we boldface the models with FAP$<$1\%. Of the 43 stars with more than 12 measurements, 19 have RVs well modeled by 1 or more planets. Among them, we recover all stars with probable RV periodicity except Gl\,680 that has less than 12 measurements and was not tested. In the following section, we discuss the interpretation of these candidate signals.

\section{\label{sect:activity}Interpretation}
The above analysis reveal Keplerian signals in our time series, but Doppler shifts do not always correspond to planets. To vet a RV variability against stellar activity, we make use of several diagnostics. The shape of the cross-correlation function (CCF) in particular is often informative. While its barycenter measures the radial-velocity, its bisector inverse slope \citep[BIS --][]{Queloz:2001} and its full width half maximum (FWHM) can diagnose inhomogeneities of the stellar surface. Alternatively, spectral indices built on \ion{Ca}{ii} H\&K or H$_\alpha$ lines can also diagnose inhomogeneities in the stellar chromosphere and photosphere, respectively \citep{Bonfils:2007b}. Finally, we obtained photometric observations of a few stars to check whether plages or spots could produce the observed Doppler changes.

\subsection{\label{subsec:plsys}Planetary systems}
Among the stars with clear periodic/Keplerian signals, we of course recover several planets that were previously known. In total, 14 planets are known to orbit 8 M dwarfs of our sample. Nine were found by this program (Gl\,176\,b, Gl\,433\,b, Gl\,581\,b, c, d \&e, Gl\,667C\,b and Gl\,674\,b), one more was found in 1998 by our former program on ELODIE (Gl\,876\,b), 2 were detected by concurrent programs and already confirmed by this program (Gl\,876\,c \& d) and 2 were detected by concurrent programs and confirmed in this paper (Gl\,832\,b and Gl\,849\,b).

\paragraph{$\bullet$ Gl 176 :} From HET radial-velocity data, it was proposed that it hosts a $m \sin i= 24~\mathrm{M_\oplus}$ planet in 10.2-d orbit \citep{Endl:2008a}. However, we found our HARPS data incompatible with that planet  \citep{Forveille:2009}. Instead, we have shown evidence for a lower-mass planet with a shorter period ($m\sin i=8~\mathrm{M_\oplus}$; $P=8$~d). Much like Gl\,674 (see below), Gl\,176 is a moderately active M dwarf. We also observe a second periodic signal ($P\sim40$ d), that has marginally higher power than the 8-d signal in our periodogram. Thanks to photometric observations and to spectroscopic indices measured on the same spectra, we identified the 40-d signal as due to a spot rather than a second planet \citep[see]{Forveille:2009}. We note that our systematic Keplerian search for planets converges to solutions with different periods and very high eccentricities. This is mostly because the signal associated with the 40-d period might be poorly described by a Keplerian function with a large eccentricity. The method converges to the same periods as the periodogram analysis when we restrict the range of eccentricities to $<0.6$.

\paragraph{$\bullet$ Gl 433 :} This nearby M2V dwarf is rather massive ($M_\star=0.48~\mathrm{M_\odot}$) for our sample. The periodogram of our HARPS RVs shows a clear power excess at a 7.2-d period. We failed to find a counter part to that signal in our activity indicators. Also, based on the intensity rather than the variability of H$\alpha$ and \ion{Ca}{ii} lines, the star seems to have a weak magnetic activity, and most probably a low rotational velocity. It is therefore likely that a planet revolves around Gl\,433 every 7.2 days. A $\chi^2$ minimization of Harps RVs lead to a minimum mass of $m \sin i = 6~\mathrm{M_\oplus}$ for that planet. We note that \citet{Zechmeister:2009b} have reported UVES radial velocities for Gl\,433  and found no significant periodicity. The semi-amplitude of our solution 3.5$\pm$0.4 m/s translates to an r.m.s. of 5.0$\pm$0.6 m/s that is nonetheless compatible with the 4.4 m/s r.m.s. reported in Z09. Also, we found Z09's RVs compatible with our data, provided that we use a model composed of a Keplerian plus a low order polynomial to fit the merged data sets. We refer the reader to Delfosse et al. (in prep.) for a detailed description.

\paragraph{$\bullet$ Gl 581 :} That system counts at least four planets, which we reported in three papers : \citet{Bonfils:2005b, Udry:2007a, Mayor:2009}. We also performed a stability analysis for the system in \citet{Beust:2008}, updated in \citet{Mayor:2009}. Composed of one Neptune-mass planet and three super-Earths, the system is remarkable because it includes both the first possibly habitable exoplanets \citep[Gl\,581c \&d --][]{Selsis:2007, Bloh:2007} and the lowest-mass planet known to date (Gl\,581e - $m \sin i = 1.9~\mathrm{M_\oplus}$ ). Besides, we found that stability constrains its configurations. In the coplanar cases, inclinations lower than $\sim40^\mathrm{o}$ induce too strong interactions between ``b'' and ``e'',  and ``e'' is ejected in a few hundred thousands years. A lower bound on the inclinations translates to upper bounds for planetary masses. Gl\,581e, for instance, would not be more massive than $\sim3~\mathrm{M_\oplus}$ if the system is coplanar. In 2010, \citeauthor{Vogt:2010} have proposed that 2 more planets orbit Gl\,581, one having $m \sin i = 3.1~\mathrm{M_\oplus}$ and being in the middle of the habitable zone, between Gl581 c and d. However, we most recently demonstrate those planets do not exist (Forveille et al., submitted).

\paragraph{$\bullet$ Gl 667C :} This is a M2V dwarf we have intensively observed. We find several coherent signals in RV data and possibly identify the rotation period on FWHM measurements of the cross-correlation function. Fitting a 1-planet model plus a $\sim 1.8~\mathrm{m\,s^{-1}\,yr^{-1}}$ linear drift to account for the A$+$B stellar-binary companion to Gl\,667C, converges toward a minimum mass of a super-Earth ($m \sin i = 5.9 \mathrm{M_\oplus}$) on a short-period orbit (7.2 day). Adding one more planet to the model makes the fit converge toward a $\sim180$ day period and a very eccentric solution, while the power excess identified in the periodogram of the first model residuals is located around 90 day. Finer analysis actually interprets that second signal as a possible harmonic of a (half-)yearly systematic affecting few data points (Delfosse et al., in prep.). Filtering the signals of 2 planets plus a linear drift reveals another candidate planet ($P_3=28$ d; $m \sin i = 3.4~\mathrm{M_\oplus}$). This candidate receives about 90\% the amount of light received by Earth in our Solar System and we speculate the planet is a habitable canidate (see Delfosse et al., in prep., for a detailed description).

\paragraph{$\bullet$ Gl 674 :} Only 4.5 pc away from our Sun, this M2.5 dwarfs hosts at least one low-mass planet \citep[$m \sin i = 11~\mathrm{M_\oplus}$; $P=4.7$~d --][]{Bonfils:2007b}. Although a second periodic signal exists for Gl\,674 ($P_2\sim35$ d), analysis of spectroscopic indices and photometric observations shows that this additional signal originated from stellar surface inhomogeneities. Today and with additional measurements, subtracting the 4.7-d periodic signal and computing a periodogram of the residuals shows power excess at a period of $\sim25$ d instead of 35. If due to a planet, it would correspond to a super-Earth in Gl\,674's habitable zone. However, the semi-amplitude $K$ of the Keplerian orbit of that second planet is $\sim3.8$~m/s, significantly above the residuals around the 2007 combined fit (r.m.s.$\sim$80 cm/s). There are two different interpretations for this apparent inconsistency :  either the 2007's solution excludes the present solution and today's 25-d periodicity is spurious or, the 2007's fit absorbed both the 35- and the 25-d signals, simultaneously. 

\begin{figure}[t]
\centering
\resizebox{\hsize}{!}{\includegraphics{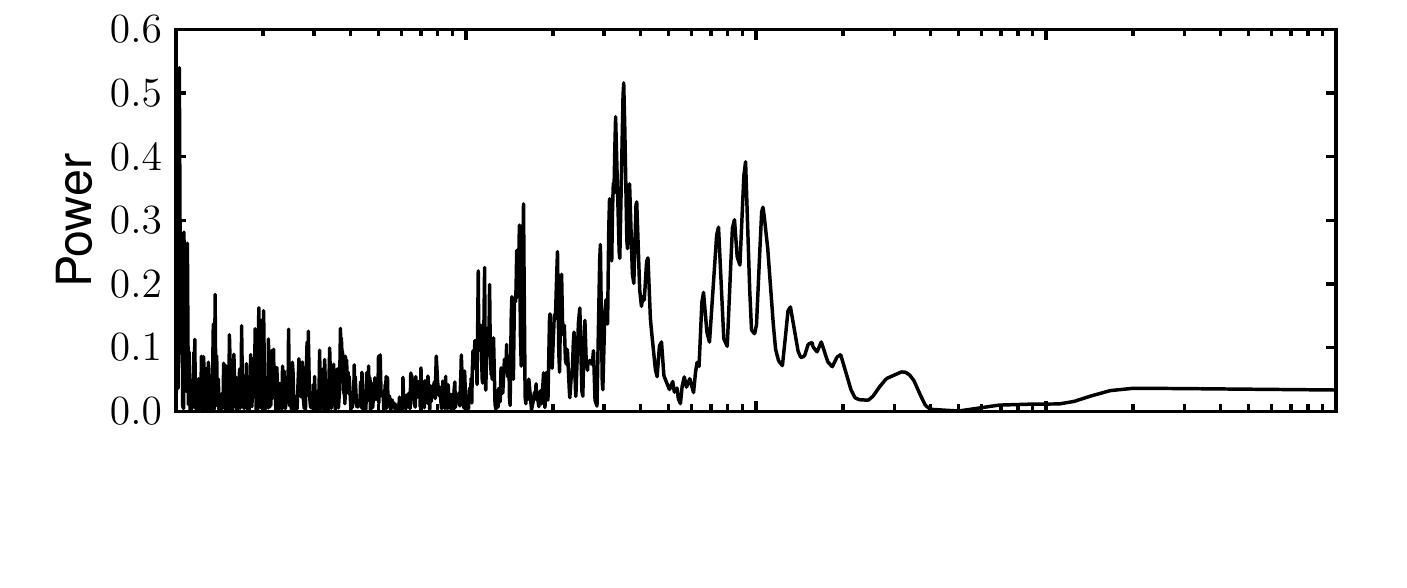}}\\\vspace{-1.cm}
\resizebox{\hsize}{!}{\includegraphics{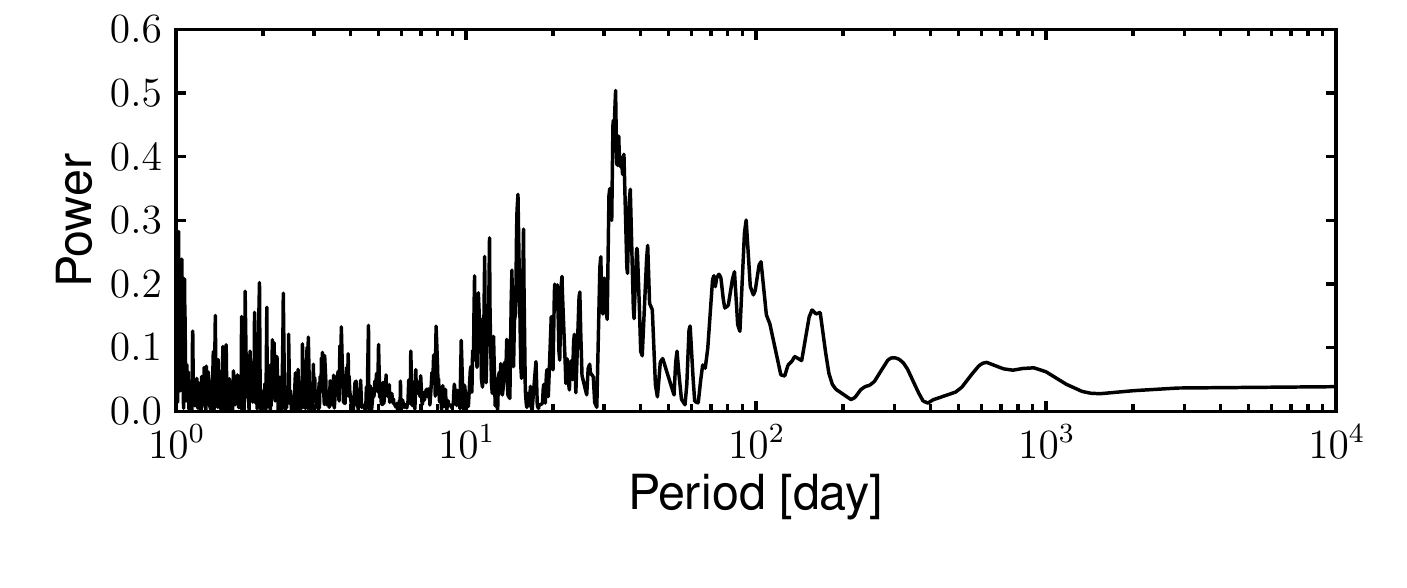}}\\
\caption{\label{fig:Gl674_caii} Periodograms of Ca II H$+$K ({\it Top}) and H$\alpha$ ({\it Bottom}) indexes for Gl 674.
   }
\end{figure}

The Keplerian analysis presented in Sect.~\ref{subsec:kepler} find a lower significance of only $\sim94.6\%$ for the second signal. We nevertheless apply further diagnostics. Restricting the data set to the 2007' RVs, we try a {\it 1 planet$+$sine wave} model instead of 2 Keplerians and found almost equally low residuals (r.m.s $\sim1.1$ m/s). That strongly opposes the presence of an additional planet with a 3.8 m/s semi-amplitude with short or moderate orbital period. On the other hand, periodograms of the H$\alpha$ and \ion{Ca}{ii} H$+$K indexes continue to peak at a $\sim35$-d period (Fig.~\ref{fig:Gl674_caii}), indicating that the signal remains coherent for these indicators. A decorrelation between spectral indexes and RVs is then hard to explain. We conclude that the case for an additional planet is not strong enough with the present data set, and that data gathered after Apr. 1st 2009, will be necessary to conclude.
	
\begin{figure}[t]
\centering
\includegraphics[width=1.\linewidth]{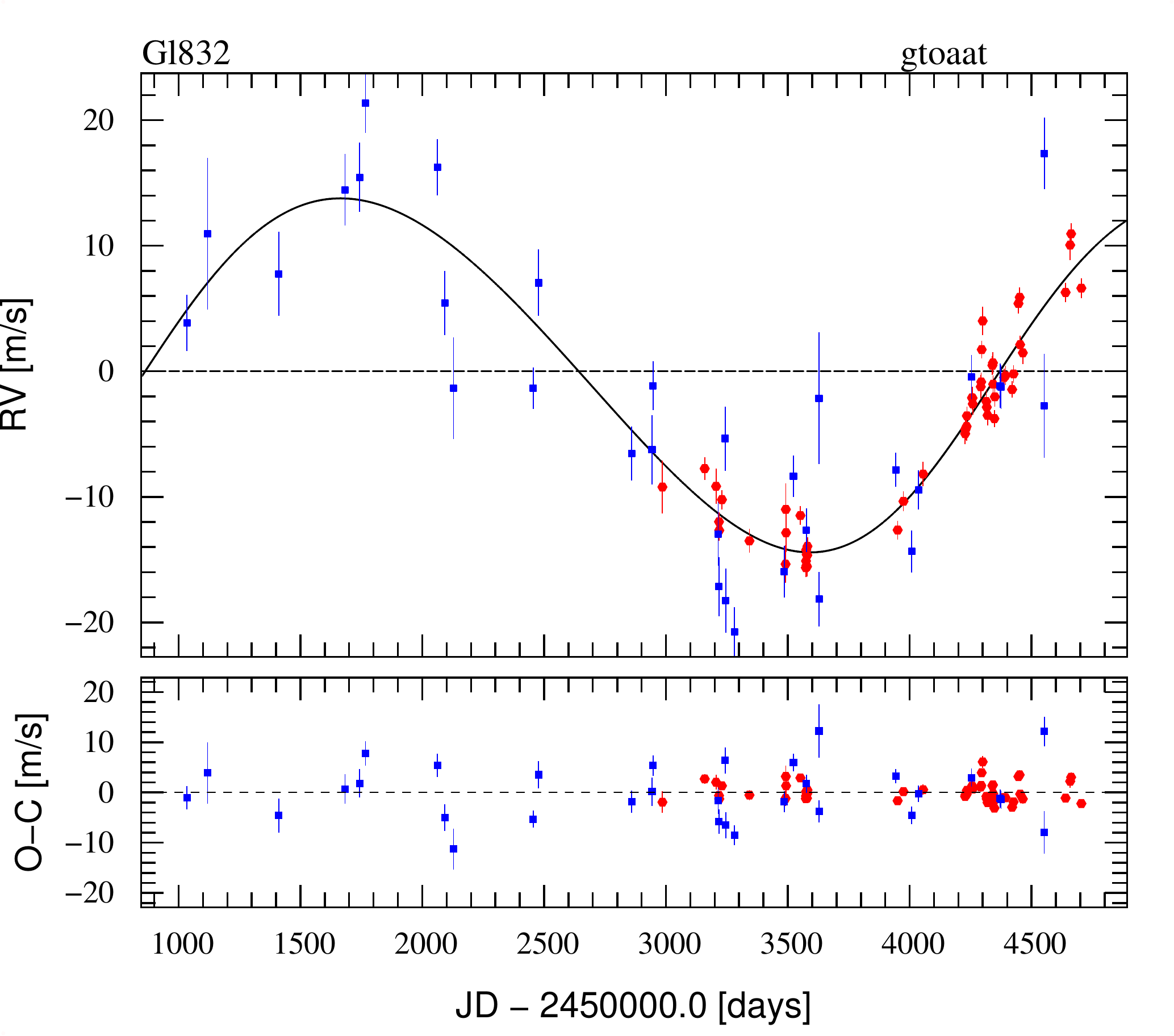}
\caption{\label{fig:Gl832rv} Best solution for the model {\it 1 planet}, with AAT (blue) and HARPS (red) RV for Gl\,832.
   }
\end{figure}

\paragraph{$\bullet$ Gl\,832 :} A decade long RV campaign with the Anglo Australian Telescope (AAT) has revealed that Gl\,832 hosts a long-period companion with a period almost as long as that survey \citep[$\sim9.5$ yr --][]{Bailey:2009}. The best fit to AAT data lead to a minimum mass $m \sin i = 0.64~\mathrm{M_{Jup}}$. Our HARPS data do not span as much time and, while they do confirm with a high confidence level the long-period RV variation, they can not confirm the planetary nature of Gl\,832b by themselves. Together with the AAT data\footnote{One RV point (with Julian Date $=$ 2\,453\,243.0503) has different values in Table 1. and Fig. 2 of \citet{Bailey:2009}. Bailey and collaborators kindly informed us of the correct value ($-2.1\pm2.5$ m/s)}, our HARPS RVs refine the orbit of Gl\,832b. A Keplerian fit using $Yorbit$ converges to $m\sin i = 0.62\pm 0.05~\mathrm{M_{Jup}}$ and $P=3507\pm 181$ d (Fig.~\ref{fig:Gl832rv}). Our Keplerian analysis (\S~\ref{subsec:kepler}) finds a possible second signal with a 35-d period. That second signal however, does not reach a 99\% significance level in either our periodicity nor our Keplerian analysis. Some power excess is seen around $\sim$40 d in the BIS periodogram, uncomfortably close to the possible 35-d periodicity. We keep measuring Gl\,832 to clarify whether a second periodic signal is present and assess its true nature.

\paragraph{$\bullet$ Gl\,849 :} This M3V dwarf is known since 2006 to host a Jupiter-mass companion \citep{Butler:2006}. The RV variation is clearly seen in our HARPS observations, and has no counter part in our activity indicators (based on the shape of the cross-correlation function or H$\alpha$ and \ion{Ca}{ii} spectral indices). Our observations confirm that Gl\,849 hosts a Jupiter-mass companion. 

\begin{figure}[t]
\centering
\includegraphics[width=1.\linewidth]{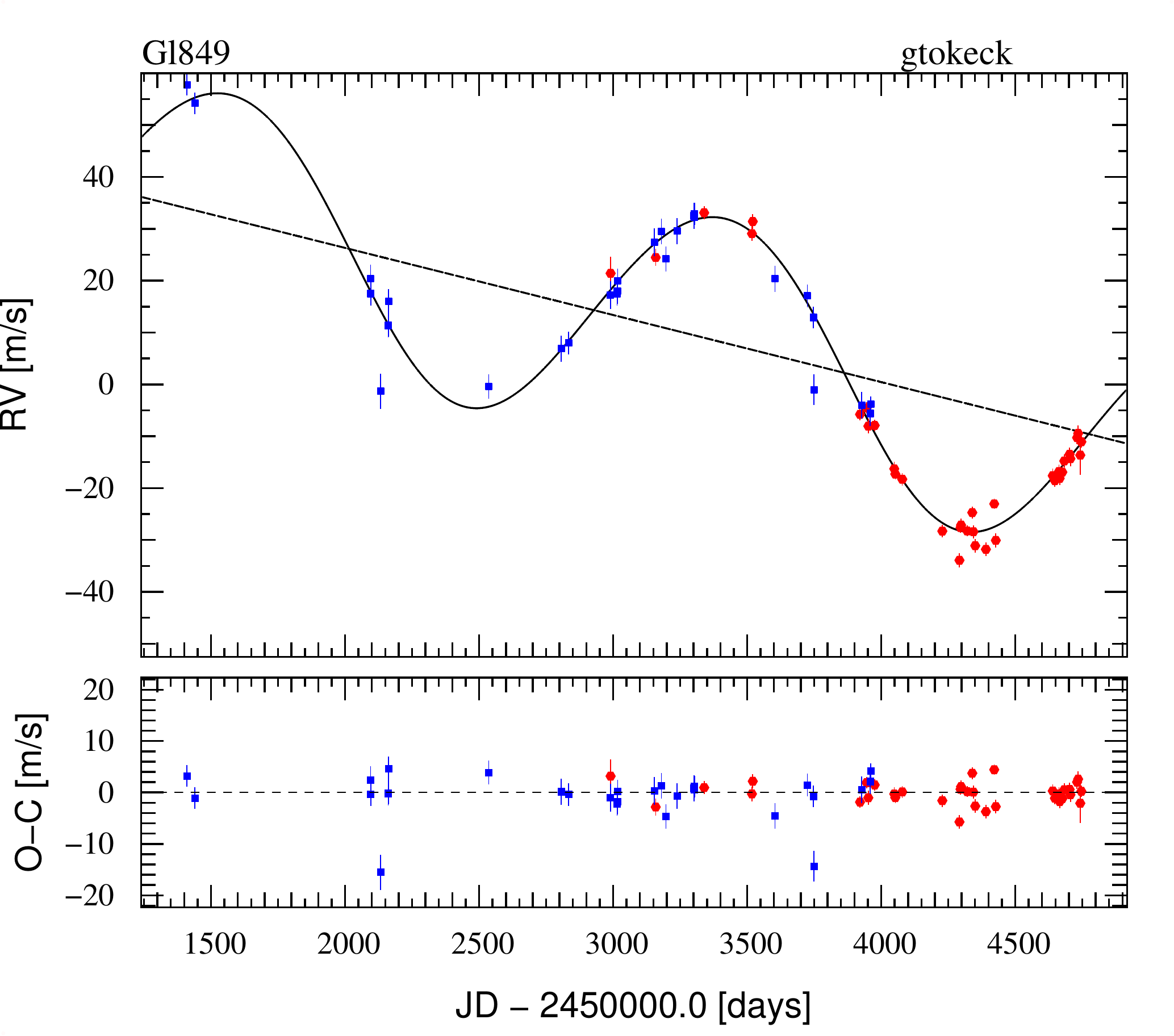}
\caption{\label{fig:Gl849rv} Best solution for the model {\it 1 planet$+$drift}, with Keck (blue) and HARPS (red) RV for Gl\,849.
   }
\end{figure}

Fitting a Keplerian orbit to HARPS observations alone converges toward a minimum mass of $m \sin i = 1.17\pm0.06~\mathrm{M_{Jup}}$ and a period $P=2165\pm132$ d. Together with the Keck RVs however, one planet is not sufficient to explain all the RV motion. As already suspected from Keck data, a long-term change is superimposed on the first periodic signal. We therefore fit the merged data set with a {\it 1 planet$+$drift}, a {\it 2 planets} and a {\it 2 planets$+$drift} model and calculate their respective FAPs. We find that a model more complex than {\it 1 planet$+$a drift} is not justified. For that model, our best solution ($\sqrt{\chi_r^2}=1.96$) corresponds to a Jupiter-mass planet ($m \sin i = 0.99\pm 0.02~\mathrm{M_{Jup}}$; $P=1852\pm 19$~d;  $e=0.04\pm0.02$) plus, a RV drift with a slope of $-3.84$ m/s/yr (Fig.~\ref{fig:Gl849rv}; Tab.~\ref{TabOrb}). Because Gl\,849 is a nearby star (d$=8.77\pm0.16$ pc), the long-period massive companion makes it an excellent target for astrometric observations and direct imaging with high angular resolution. 

\addtocounter{table}{+7}
\begin{table*}
\caption{\label{TabOrb}Fitted orbital solutions for the Gl\,849 (1planet $+$ a linear drift) and Gl\,832 (1 planet). 
  }

\begin{tabular}{l l l c c}
  \hline\hline
  \multicolumn{2}{l}{\bf Parameter} &\hspace*{2mm} 
  & \bf GJ\,849\,b  & \bf GJ\,832\,b  \\
  \hline
  $P$ & [days] & & 1845 $\pm$ 15 & 3507 $\pm$ 181 \\
  $T$ & [JD-2400000] & & 54000 $\pm$ 150 &  54224 $\pm$ 302  \\
  $e$ & & & 0.05 $\pm$ 0.03 &0.08 $\pm$   0.05\\
  $\omega$ & [deg] & & 298 $\pm$ 29 & 254 $\pm$  35 \\
  $K$ & [m s$^{-1}$] & & 24.4 $\pm$ 0.7 & 14.1 $\pm$     1.1  \\
  $V(Keck/AAT)$ & [km s$^{-1}$] & &  $-$0.0192 $\pm$ 0.0012 &0.0033 $\pm$ 0.0013 \\
 $V(Harps)$ & [km s$^{-1}$] & &  $-$15.0896 $\pm$ 0.0005 &13.3471 $\pm$ 0.0013 \\
 Slope           & [m s$^{-1}$ yr$^{-1}$] &  &$-4.76\pm$0.33&-\\
  $f(m)$ & [10$^{-9} M_{\odot}$] & & 2.77 &1.00\\
  $m_2 \sin{i}$ & [$M_{\oplus}$] & & 0.91 & 0.62 \\
  $a$ & [AU] & & 2.32 & 3.46\\
  \hline
  $\sigma$ (O-C) (AAT/Keck) &[ms$^{-1}$] && 3.64& 4.70\\
$\sigma$ (O-C) (Harps) &[ms$^{-1}$] && 2.08&1.77\\
  $\sqrt{\chi^2_{\rm red}}$ &                  & &  1.83 &2.36\\
  \hline
\end{tabular}

\end{table*}

\paragraph{$\bullet$ Gl\,876 :} That system was known to harbor planets before our observations started and at that time, was even the only planetary system centered on an M dwarf. The first giant planet found to orbit GJ\,876 was detected simultaneously by members of our team using the ELODIE and CORALIE spectrographs \citep{Delfosse:1998b} and by the HIRES/Keck search for exoplanets  \citep{Marcy:1998}. The system was later found to host a second giant planet in a 2:1 resonance with GJ\,876b \citep{Marcy:2001}. The third planet detected around GJ\,876 was the first known super-Earth, \object{Gl\,876d} \citep{Rivera:2005}. Because the 2:1 configuration of the two giant planets leads to strong interactions, the orbits differ significantly from Keplerian motions. To model the radial velocities, one has to integrate the planet movement with a N-body code. This lifts the $\sin i$ degeneracy and measures the masses of the giant planets. A full N-body analysis was first performed for GJ\,876 by \citet{Laughlin:2001} and \cite{Rivera:2001}. From an updated set of Keck RVs, \citet{Rivera:2005} found the third planet only because planet-planet interactions were properly accounted in the fitting procedure. Those authors still had to assume coplanar orbits to assign a mass to each planet. \citet{Bean:2009b} then combined the Keck RVs with HST astrometry to both measure the masses in the coplanar case and to measure the relative inclination between planets ``b'' and ``c''. Most recently, we used our HARPS data and the published Keck measurements to model the system and measure the relative inclination of both giant planets ($<1^\mathrm{o}$), relying on RVs only. The paper \citep{Correia:2010b} also analyzed the dynamical stability and show that the librations amplitude are smaller than $2^\mathrm{o}$ thanks to a damping process acting during the planet formation.

\subsection{\label{subsec:act}Activity dominated variations}
We group in this section the ``active'' cases. They are stars tested positively for periodicity and/or Keplerian signal, and their measurement variability correlates with an activity indicator. We do not show all diagnostics for each star, but rather pick the most illustrative. A statistical discussion of all activity indicators will be presented in a separate paper (Bonfils et al. 2010, in prep.). 

\begin{figure}[t]
\centering
\resizebox{\hsize}{!}{\includegraphics{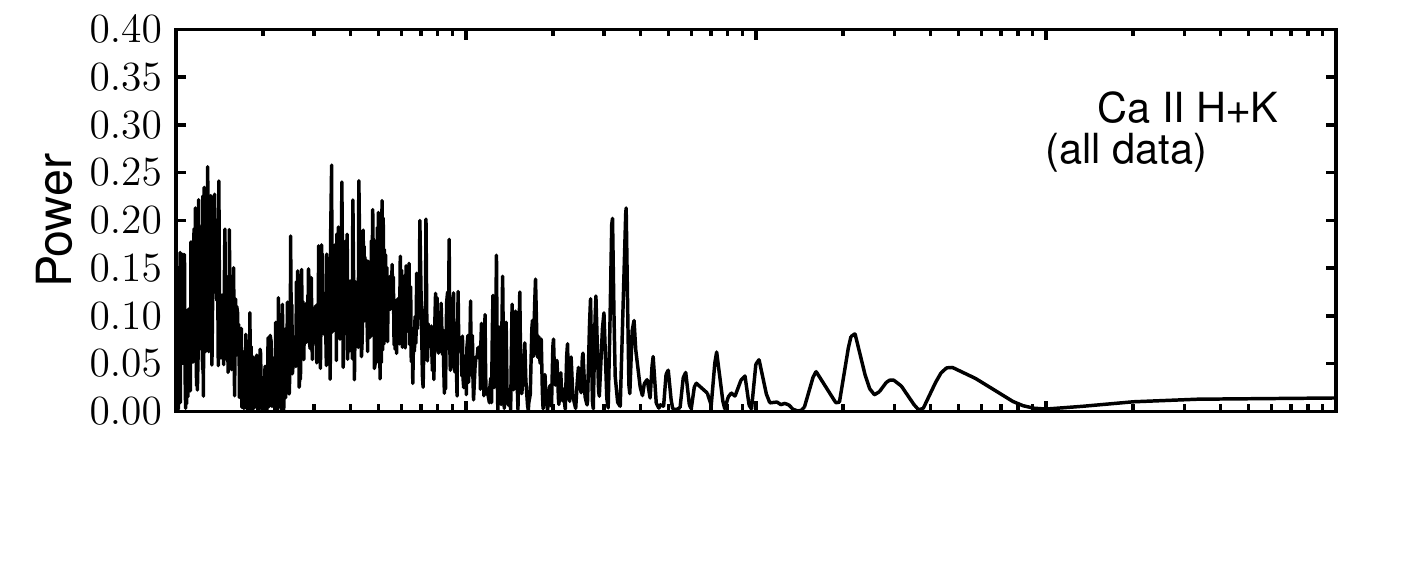}}\vspace{-1.cm}
\resizebox{\hsize}{!}{\includegraphics{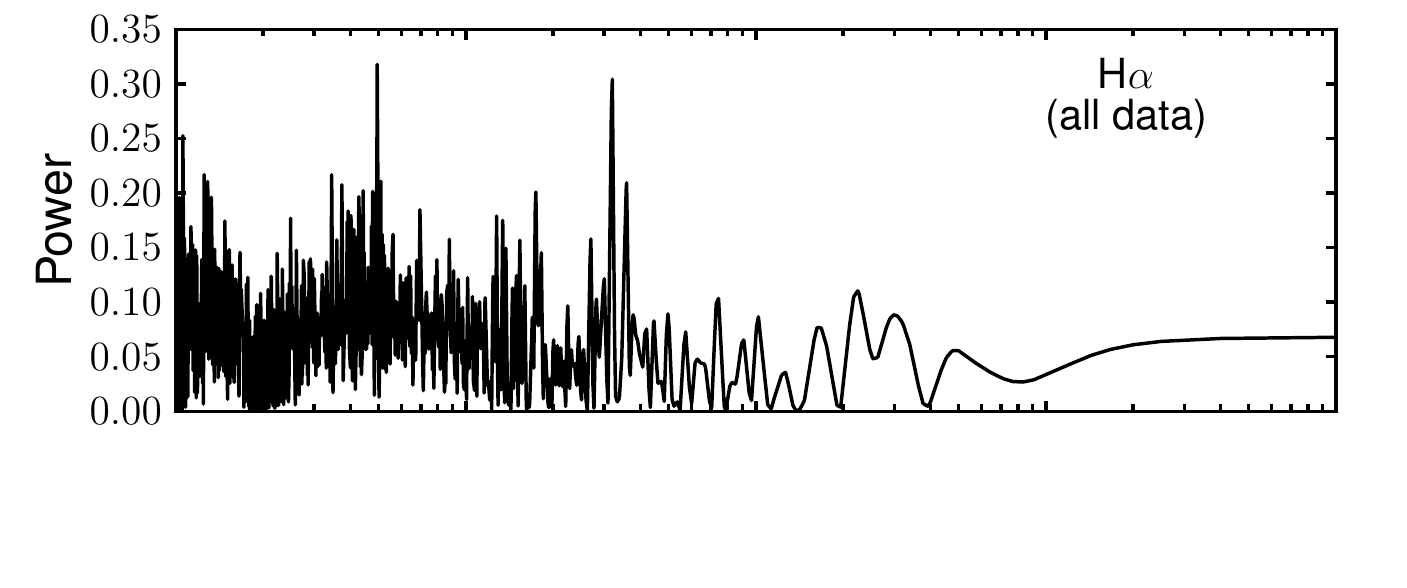}}\vspace{-1.cm}
\resizebox{\hsize}{!}{\includegraphics{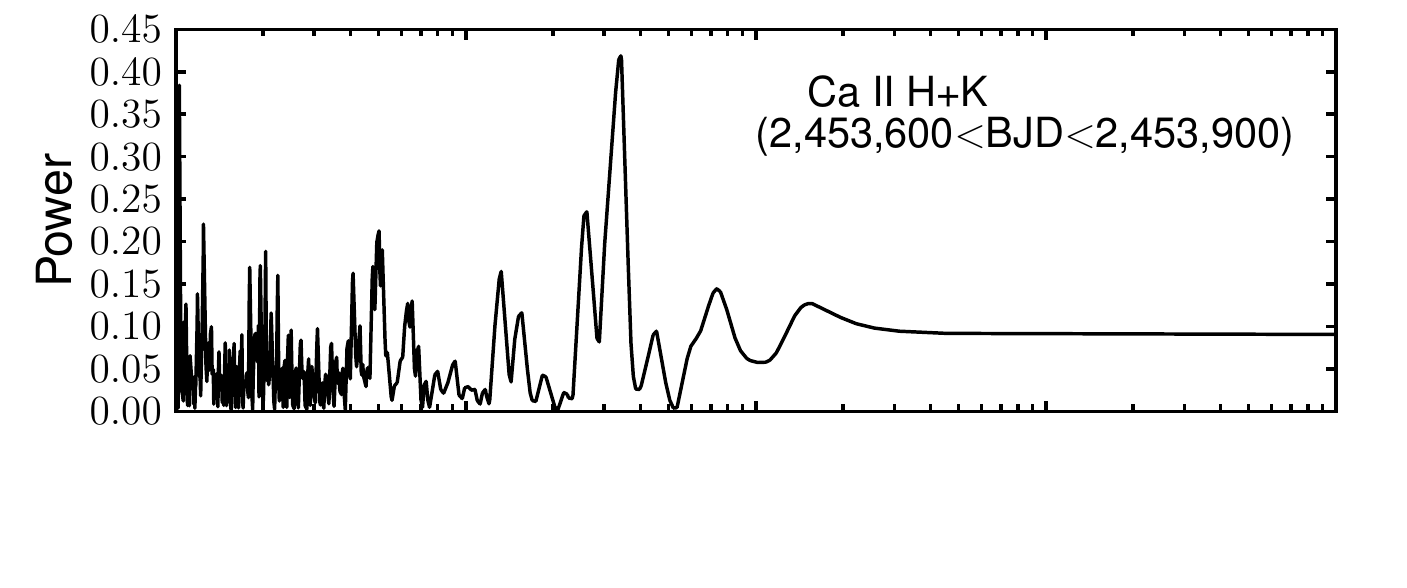}}\vspace{-1.cm}
\resizebox{\hsize}{!}{\includegraphics{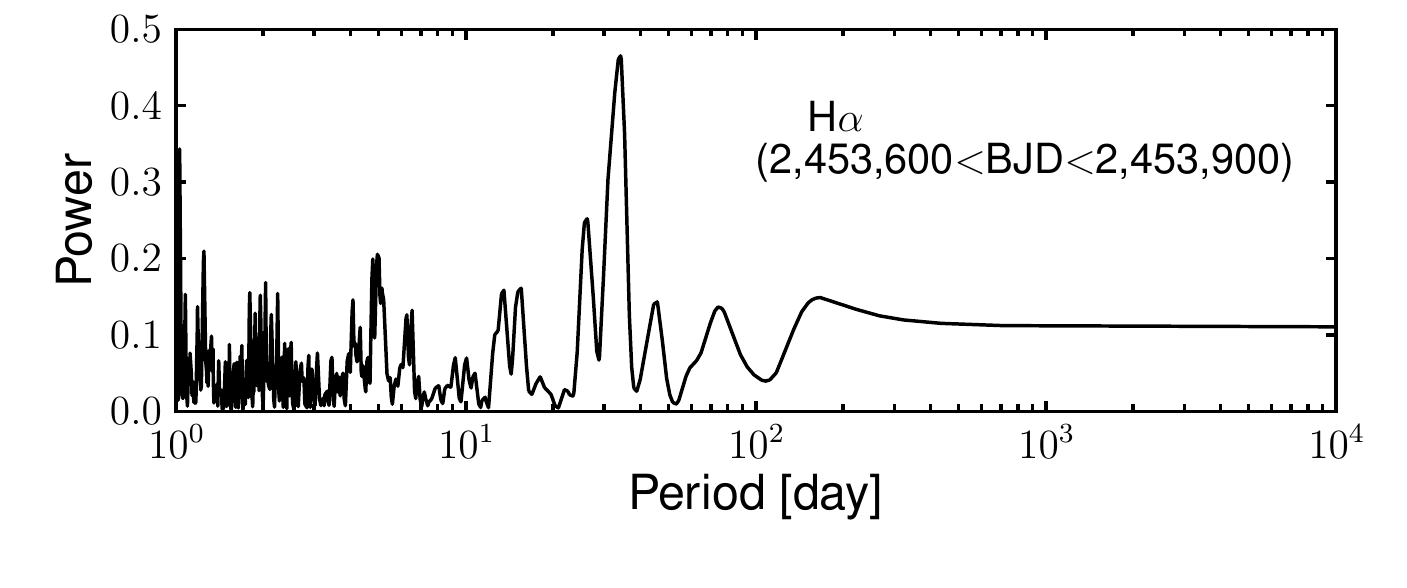}}\vspace{-0.cm}
\caption{\label{fig:IndexGl205}For Gl\,205, the top 2 panels show the periodograms for Ca II H$+$K index and H$\alpha$ index, including all data.  The bottom 2 panels show the periodograms  for Ca II H$+$K index and H$\alpha$ index, restricting the dataset to one observational season ($2,453,600<BJD<2,543,900$).} 
\end{figure}

\paragraph{$\bullet$ Gl\,205 :} A periodogram of the velocities identifies excess power around 32.8 d. Our Keplerian search with {\it Yorbit}, for a single planet or for a 1 planet$+$drift model, converges toward either a similar period or 0.970 d, an alias with the 1-d sampling. We find indications that the variation is intrinsic to the star from both spectral indices and photometric observations. Considering the whole dataset, we identify excess power around 33 d in periodograms of H$\alpha$ and \ion{Ca}{ii} H$+$K indices, though, those peaks are not the highest. Restricting instead the dataset to one observational season, a strong power excess around the period 33 d dominates the periodogram (Fig.~\ref{fig:IndexGl205}). With additionally note a high correlation between H$\alpha$ and \ion{Ca}{ii} H$+$K indices (their Pearson correlation coefficient is 0.97), suggesting both variations originate from the same surface homogeneity. Also, photometric monitoring of Gl\,205 reports a similar period for the stellar rotation \citep[33.61 d --][]{Kiraga:2007}. The observed RV modulation is most probably due to surface inhomogeneities, which remains stable over one season but not over several. 

\begin{figure}[t]
\centering
\includegraphics[width=1.\linewidth]{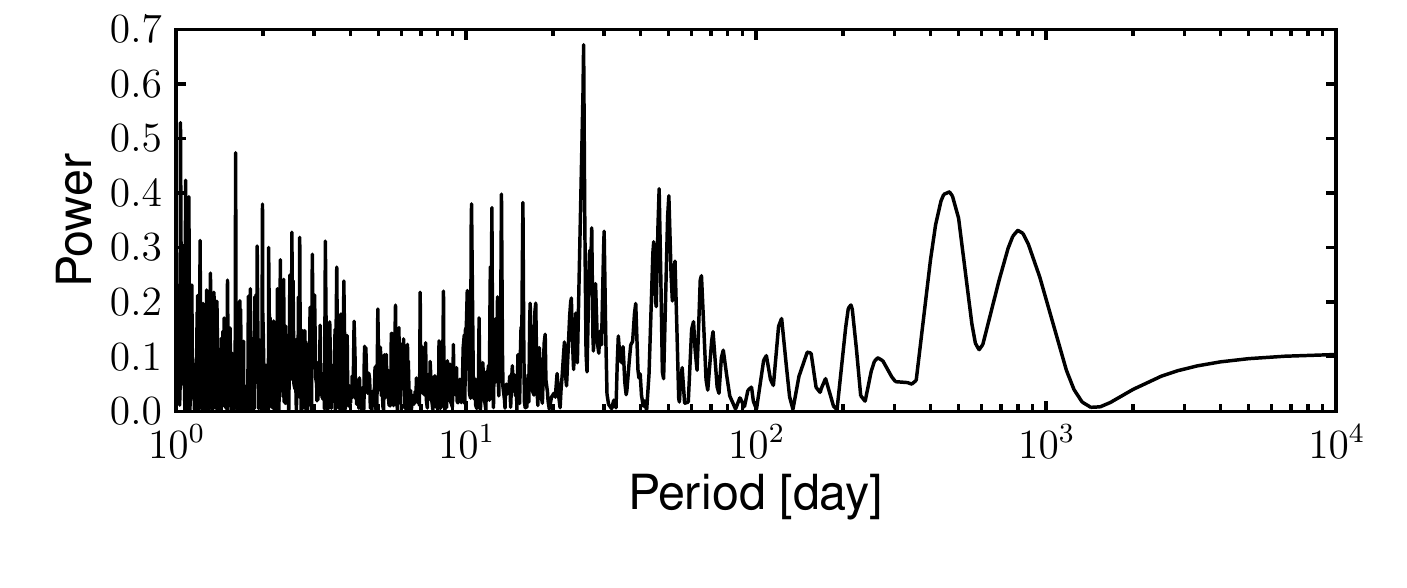}\\
\includegraphics[width=1.\linewidth]{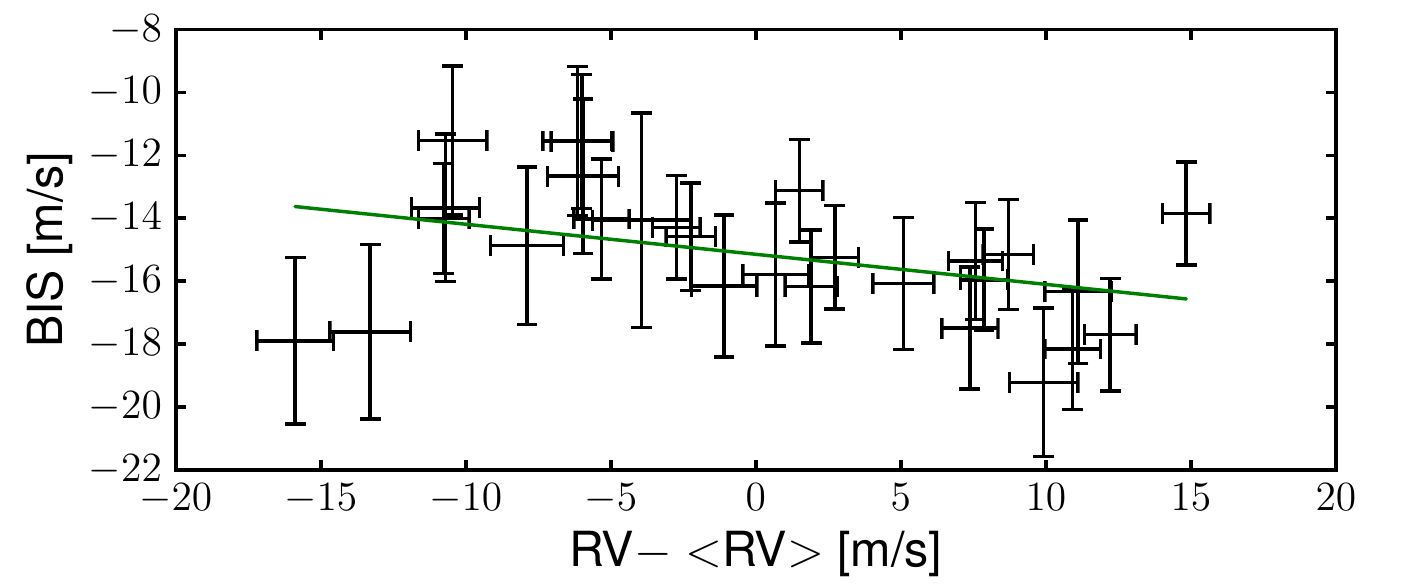}\\
\caption{\label{fig:Gl358_bis_rv} {\it Top :} Periodogram for the FWHM of CCFs for Gl\,358. {\it Bottom~:} Possible correlation between bisector spans and RV data for Gl\,358.
   }
\end{figure}

\paragraph{$\bullet$ Gl\,358 :} We have gathered 28 measurements for Gl\,358. They show significant variability with a periodicity of $\sim26$ d. In the RV time series periodogram, we also identify power excess at the first harmonic of that period ($\sim13$ d). The RV modulation is well described by a Keplerian orbit. However, we also observe similar variability in the FWHM of the CCF as well as a possible anti-correlation between RV and spectral lines asymmetry (see Fig~\ref{fig:Gl358_bis_rv}), as measured by the CCF bisector span \citep{Queloz:2001}. The Pearson's correlation for BIS and RV is $-0.40$, and rises to $-0.67$ for the 2007 measurement subset. Photometric monitoring by \citet{Kiraga:2007} found a rotational period of 25.26 d for Gl\,358. All this favors a scenario where a stellar surface inhomogeneity such as a spot or a plage produces the RV change, rather than a planet.

\begin{figure}[t]
\centering
\resizebox{\hsize}{!}{\includegraphics{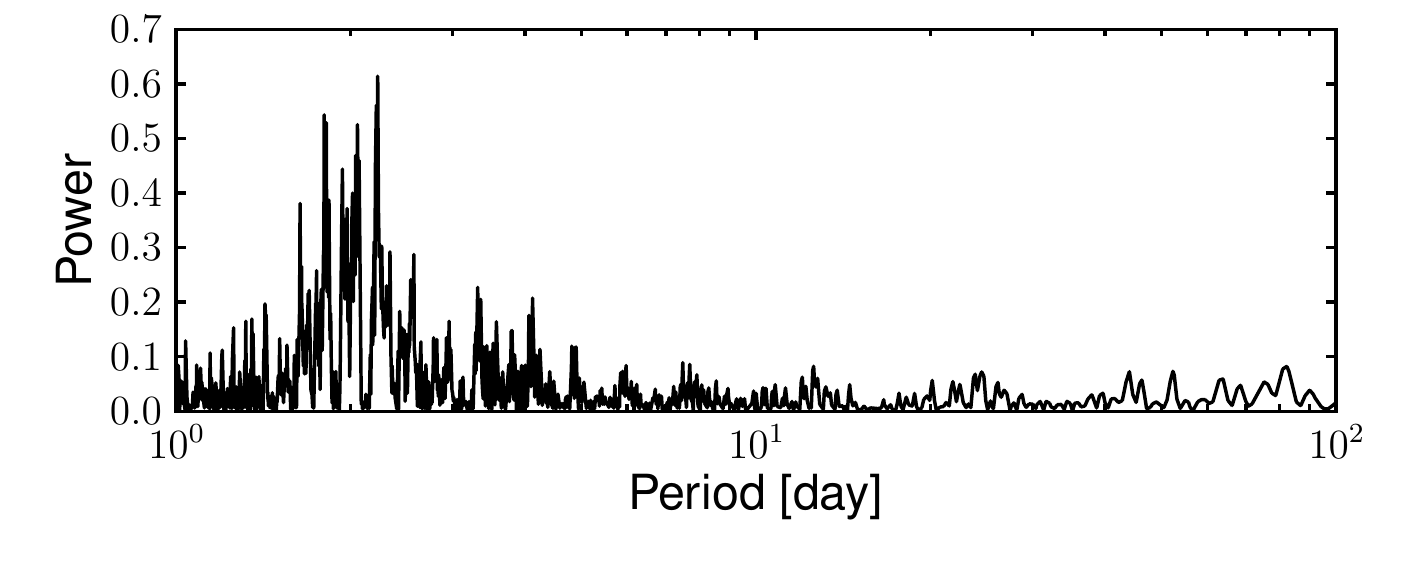}}\\\vspace{-0.cm}
\resizebox{\hsize}{!}{\includegraphics{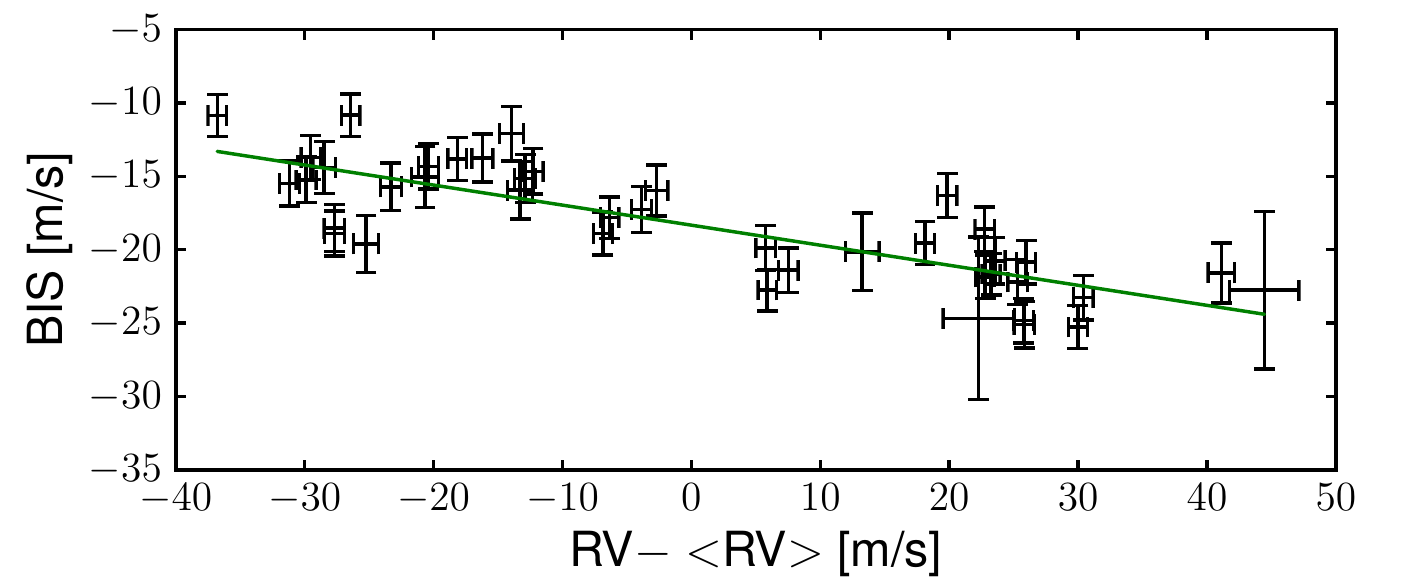}}\\
\caption{\label{fig:Gl388_bis_rv} {\it Top :} Periodogram for the CCFs bisector span for Gl\,388. {\it Bottom~:} Strong correlation between bisector spans and RV data for Gl\,388.
   }
\end{figure}

\paragraph{$\bullet$ Gl\,388 (AD Leo) :} The periodogram of its RV time series shows an important power excess at short-period, with two prominent peaks at 1.8 and 2.22 d, consistent with 2.24-d rotational period reported by \citet{Morin:2008}. These are 1-d aliases of each other, and the later is slightly stronger. Here, the bisector span demonstrates that stellar activity is responsible for the variation. Its periodogram shows a broad power excess at short period, and it is strongly anti-correlated to RV (with a Person's correlation coefficient of $-0.81$ -- see Fig.~\ref{fig:Gl388_bis_rv}). Correcting from the BIS-RV correlation by subtracting a linear fit does decrease the r.m.s from 24 to 14 m/s, but leaves some power excess around $\sim 2$ d.

\paragraph{$\bullet$ Gl\,479} We observe significant power excesses in the RV time series at two periods, $\sim$11.3 and $23-24$ d, with the shorter period roughly half the longer one. The RVs vary with an amplitude of $\sim$27 m/s and an r.m.s. of 4.13 m/s. Modeling that RV variability with Keplerians converges toward 2 planets with very similar periods (23.23 and 23.40 d), which would clearly be an unstable system. Gl\,479 shares its M3 spectral type with Gl\,674 and Gl581, which we use as benchmarks for their moderate and weak magnetic activity, respectively. From a spectral index built on the \ion{Ca}{ii} H\&K lines, we find that Gl\,479 has a magnetic activity intermediate to Gl\,674 and Gl\,581. Neither the bisector nor the spectral indices show any significant periodicity or correlation with RVs. However, we have complemented our diagnostic for stellar activity with a photometric campaign with Euler Telescope (La Silla). The periodogram of the photometry shows a maximum power excess for the period $23.75$ d, similar to the RV periodicity. The photometry phase folded to the 23.75-d period varies with a peak-to-peak amplitude of 5\% and complex patterns. We cannot ascertain the rotational period of Gl\,479, but a 5\% variability can explain the observed RV variations down to very slow rotation. We cannot correlate the photometry with RV variation because they were not taken at the same time but, phasing both to a 23 d period, we found a 0.24-phase shift consistent with a spot. The observed RV variability is therefore probably due to Gl\,479 magnetic activity rather than to planets. 
\begin{figure}[t]
\centering
\includegraphics[width=1.\linewidth]{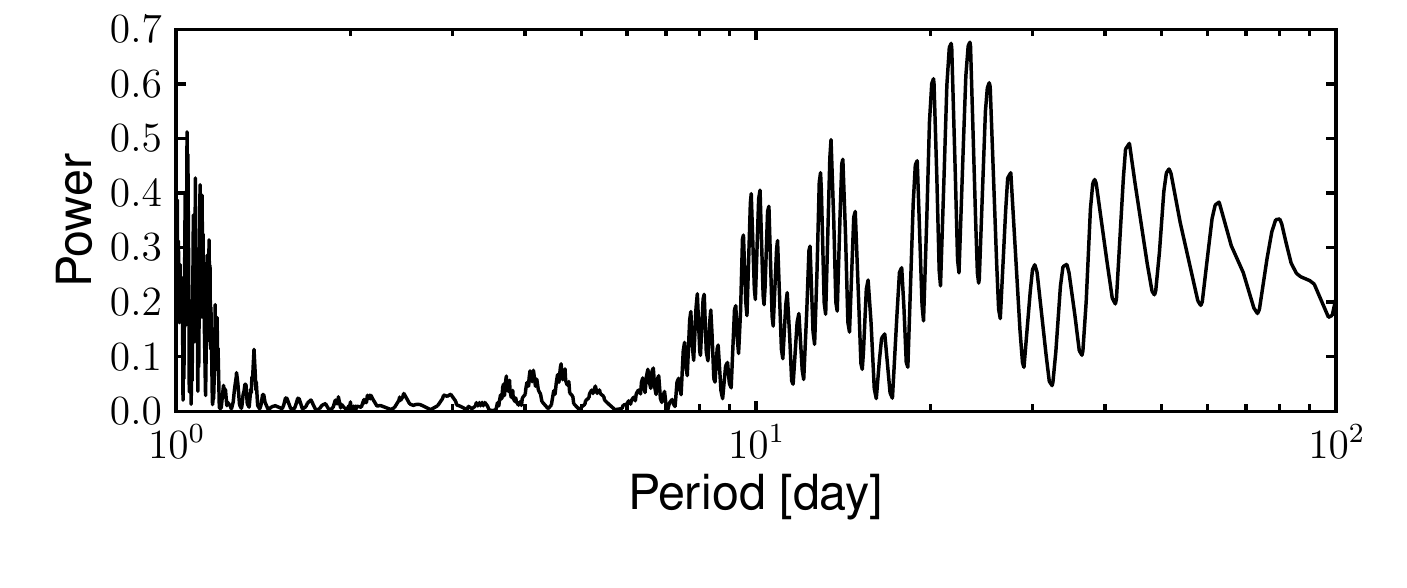}\\
\includegraphics[width=1.\linewidth]{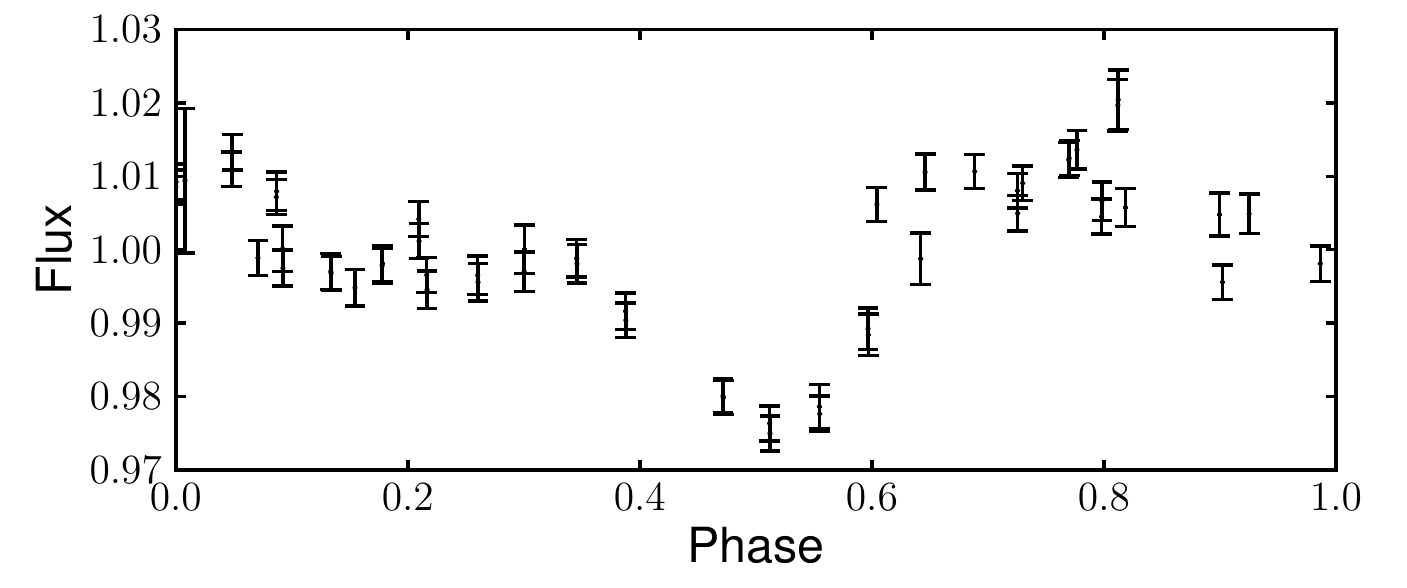}
\caption{\label{fig:Gl479_phot} {\it Top :} Periodogram of Gl\,479 photometry. {\it Bottom :} Phase-folded to the 23.75-d period.
   }
\end{figure}

\begin{figure}[t]
\centering
\includegraphics[width=1.\linewidth]{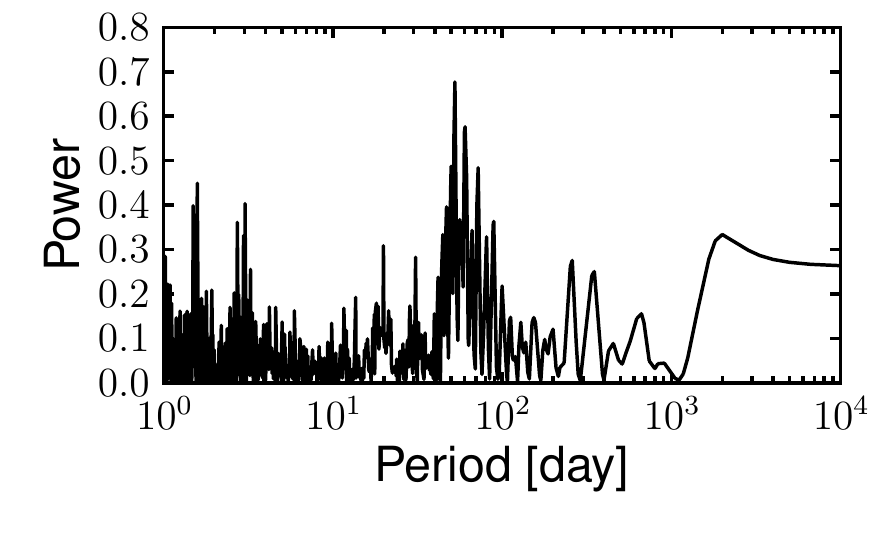}\\\vspace{-0.cm}
\includegraphics[width=1.\linewidth]{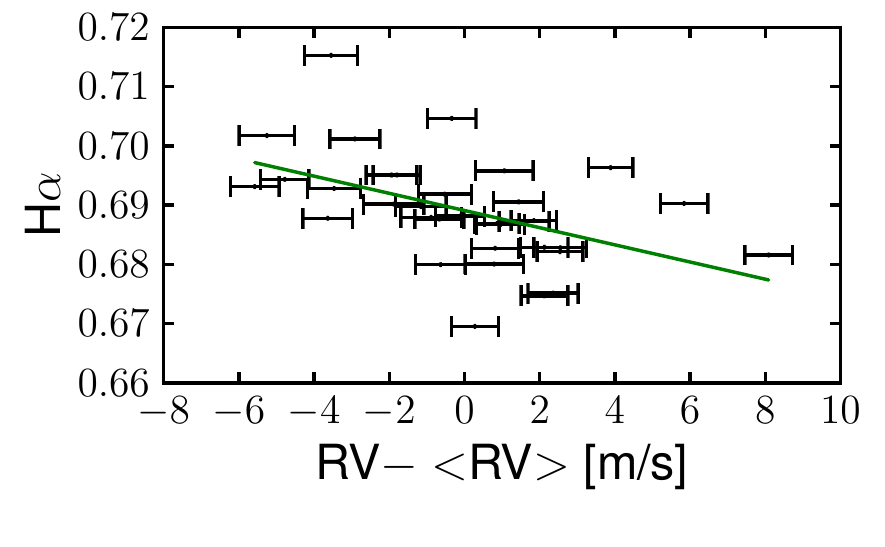}\\
\caption{\label{fig:Gl526} {\it Top :} Periodogram for the Ca II H$+$K index for Gl\,526. {\it Bottom~:} Correlation between H$\alpha$ index and RV data for Gl\,526.
   }
\end{figure}

\paragraph{$\bullet$ Gl\,526} We observe RV periodicity with power excess close to 50 d and a FAP approaching 1\%. As expected for a moderate or long period, we do find corresponding changes neither in BIS nor FWHM. As for Gl\,674, spectral indices or photometry are then more informative. For Gl\,526, we do find that RV is weakly correlated to H$\alpha$ and that Ca II H$+$K index varies with a clear 50-d period (Fig.~\ref{fig:Gl526}). Because the observed period is similar for the calcium index and the RV shift, we interpret that RV changes as due to magnetic activity rather than a planet. 

\begin{figure}[t]
\centering
\includegraphics[width=1.\linewidth]{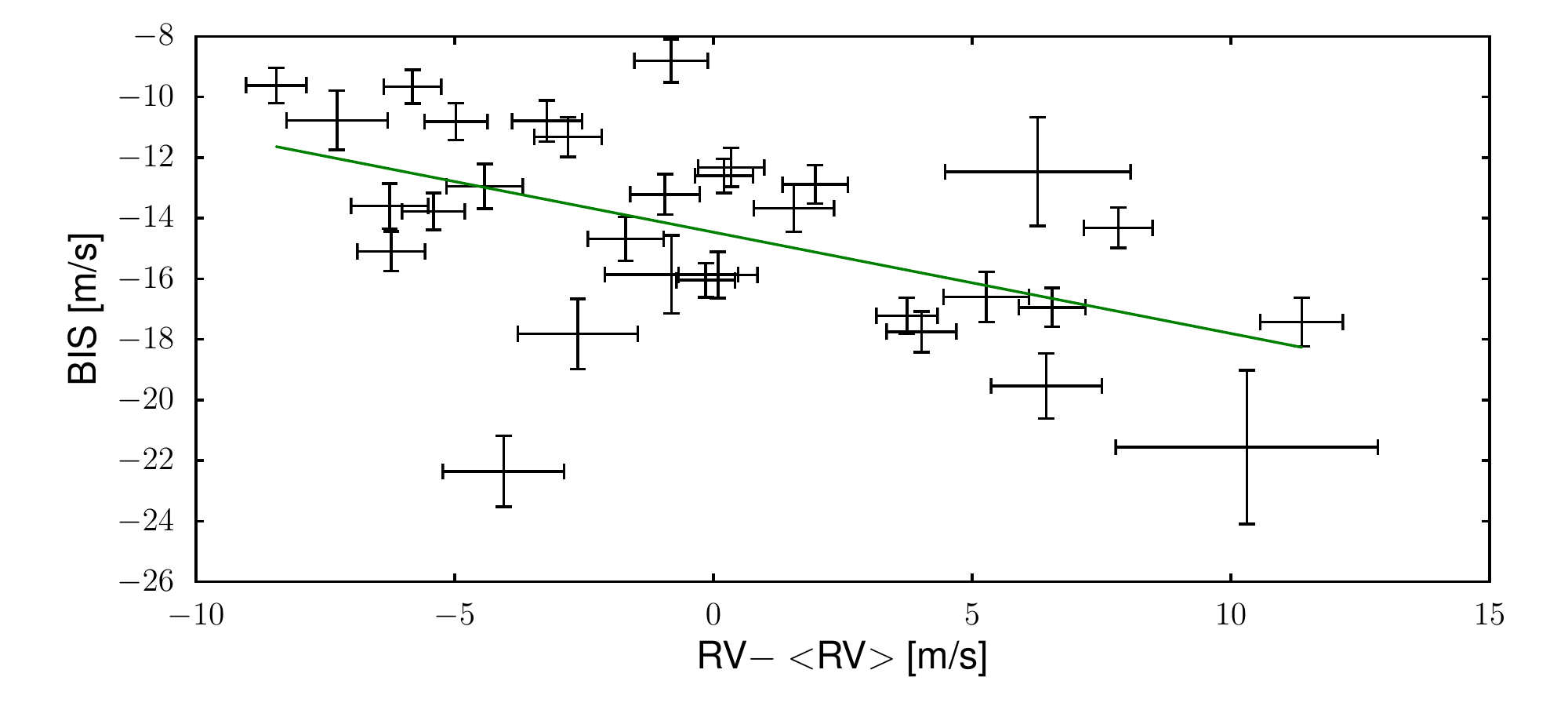}\\
\caption{\label{fig:Gl846_bis_rv} Correlation between bisector inverse slope and RV data for Gl\,846.
   }
\end{figure}

\paragraph{$\bullet$ Gl\,846 :} We observe RV variability with significant power excesses in the periodogram at several periods (7.4, 7.9 and 10.6 d), plus their aliases with the 1-d sampling, near 1-d. The bisector inverse slope is well correlated with the RV (Fig.~\ref{fig:Gl846_bis_rv}). Like Gl\,358 and Gl\,388, Gl\,846 is also a clear case of stellar intrinsic variability rather than planetary-companion Doppler shift.

\subsection{Unclear cases}
\paragraph{$\bullet$ Gl\,273} This M3.5 dwarf shows RV variability and tests positively for a RV drift according to the $\chi^2$-probability test (and approaches FAP=1\% for the permutation test). The RV periodogram indicates significant power excess at a period of $\sim434$~d, and $Yorbit$ find a good solution for a planet with $P\sim440$ d, with or without a supplementary drift. However, good solutions are found with uncomfortably high eccentricity and, most importantly, a poor phase coverage. Among activity indicators, only the H$\alpha$ index has power excess for long periods but with a different period ($\sim500-600$ d). Once a linear drift is subtracted, 2 RV points stand out (with BJD=2,454,775 and BJD=2,454,779). They have a value $8-10$ m/s lower than the residual mean. If we fit a drift again on the original data, considering all but those 2 points, the 440-d power excess disappear from the periodogram of the residuals. This suggest that, if these 2 particular points were outliers, the 440-d period could be an alias between a long-term RV drift and the 1-yr sampling. However, a direct inspection of Gl\,273 spectra, cross-correlation functions, and spectral indexes give no reason to exclude those values. We conclude that a firm conclusion on Gl\,273 would be premature and will obtain more data.

\paragraph{$\bullet$ Gl\,887} Formally significant models are found with 1 and 3 planets but all have converged to solutions with very high or unrealistic eccentricities. Most probably, the RV variability of Gl\,887 does not match a Keplerian motion, and our automatic search got confused. Also, we do not find significant periodicity or correlation with RVs in our diagnostics.

\section{\label{sect:limits}Detection limits}
While the previous sections focuses on signal detection, the present one aims at giving upper limits on the signals we were not able to detect. For individual stars, the upper limit translates into which planet, as a function of its mass and period, can be ruled out given our observations. For the sample, all upper limits taken together convert into a survey efficiency and can be used to measure statistical properties (Sect.~\ref{sect:freq}).

To derive a period-mass limit above which we can rule out the presence of a planet, given our observations, we start with the periodogram analysis presented in \S\ref{subsec:periodo}. We make the hypothesis the time series consists of noise only. Like for the FAP calculations, we evaluate the noise in the periodogram by generating virtual data sets. The virtual data are created by shuffling the time series while retaining the observing dates (i.e. by bootstrap randomization). And for each trial we compute a periodogram again. This time however, we do not look for the most powerful peak. We rather keep all periodograms and build a distribution of powers, at each period. For a given period then, the power distribution tells us the range of power values compatible with no planet, i.e. compatible with our hypothesis that the time series consists of noise only. In the same manner as for the FAP computation, we can also evaluate the probability that a given power value occurred by chance just by counting the fraction of the power distribution with lower values.

Once we know the power distribution compatible with no planet, we inject trial planetary orbits into the data. In this paper, we restrict our analysis to circular orbits and therefore fix the eccentricity and argument of periastron to zero. We nevertheless note that eccentricities as high as 0.5 do not affect much the upper limit on planet detection \citep{Endl:2002, Cumming:2009}. We thus add sine waves, choosing a period $P$, a semi-amplitude $K$ and a phase $T$. We explore all periods computed in the periodograms, from 1.5 to 10,000 day, with a linear sampling in frequency of step 1/20,000 day$^{-1}$,  and for 12 equi-spaced phases. At each trial period, we start with the semi-amplitude of the best sine fit to the original time series $K_{obs}$, and compute the new periodogram power $p_{sim}$ for that period. We next increase the semi-amplitude $K$ until $p_{sim}$ reaches a value with a probability as low as or lower than 1\%, if the data were noise only. On one hand, we impose that power threshold for {\it all} our 12 trial phases and obtain a {\it conservative} detection limit. On the other hand, we {\it averaged} that power threshold over our 12 trial phases and obtain a {\it statistical} (or {\it phase-averaged}) detection limit. Eventually, for both detection limits, we convert the $K$ semi-amplitude to planetary mass\footnote{$m \sin i  = K M_\star^{2/3} (P/2\pi G)^{1/3}$}, and orbital period to orbital semi-major axis\footnote{$a  = (P/2\pi)^{2/3} (G M_\star)^{1/3}$}, using the stellar mass listed in Table~\ref{tab:sample}.

The method described above has been applied before us by \citet{Cumming:1999, Cumming:2008} and \citet{Zechmeister:2009b}. We note a small difference between Cumming et al's and Zechmeister et al's approaches. On one hand, Cumming and collaborators add the trial sine wave to normally distributed noise and choose as variance the r.m.s around the best sine wave to the observed data. On the other hand, Zechmeister and collaborators choose not to make trial versions for the noise and considered the observed data as the noise itself (to which they add the trial orbit). We choose Zechmeister et al.'s approach because we believe it is more conservative when, in some cases, the noise departs from a normal distribution.

\begin{figure*}[thh]
\centering
\includegraphics[width=.9\linewidth]{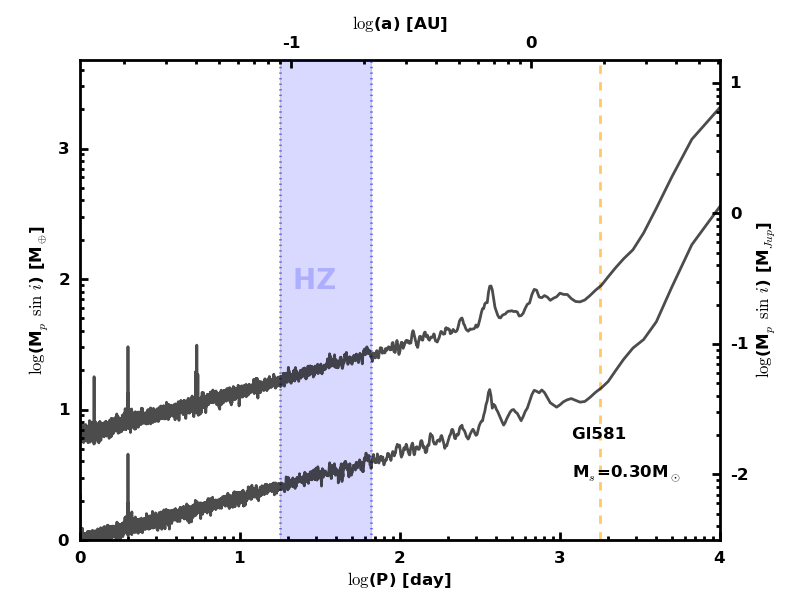}
\caption{\label{fig:limGl581} Conservative detection limit applied to Gl\,581. Planets with minimum mass above the limit are excluded with a 99\% confidence level for {\it all} 12 trial phases. The upper curve shows the limit before any planetary signal is remove to the RV time series. The sharp decrease in detection sensitivity around the period 5.3 day is caused by the RV signal of Gl\,581b. The lower curve shows the limit after the best 4-planet Keplerian fit has been subtracted. The sharp decrease in sensitivity around the period 2 day is due to sampling. Venus and Mars criterion delineate the habitable zone, shown in blue. The vertical yellow dashed line marks the duration of the survey.
   }
\end{figure*}

\begin{figure*}[thh]
\centering
\includegraphics[width=.9\linewidth]{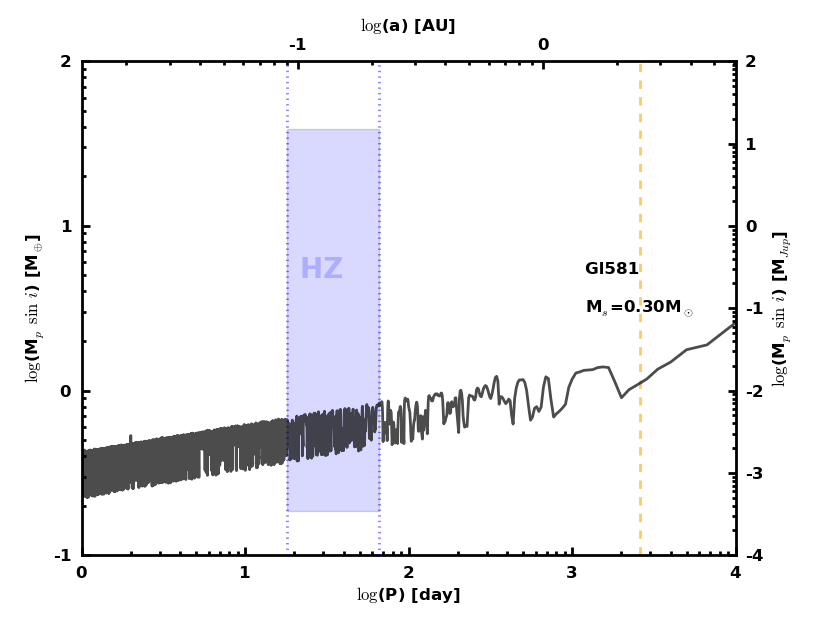}
\caption{\label{fig:limGl581b} Phase-averaged detection limit applied to Gl\,581. Planets with minimum mass above the limit are excluded with a 99\% confidence level for {\it half} our 12 trial phases. The upper curve shows the limit before any planetary signal is remove to the RV time series. The sharp decrease in detection sensitivity around the period 5.3 day is caused by the RV signal of Gl\,581b. The lower curve shows the limit after the best 4-planet Keplerian fit has been subtracted. The sharp decrease in sensitivity around the period 2 day is due to sampling. Venus and Mars criterion delineate the habitable zone, shown in blue. The vertical yellow dashed line marks the duration of the survey.
   }
\end{figure*}

We report both conservative and phase-averaged upper limits for the \NstarsFour time series with more than 4 measurements (an example is shown in Fig.~\ref{fig:limGl581} and \ref{fig:limGl581b} for Gl\,581 and we group the figures of all stars in Fig.~\ref{fig:limites} and ~\ref{fig:limitesb}, which is only available online). When a periodic variation has been attributed to stellar magnetic activity, we know the main variability is not due to a planet. We therefore apply a first order correction to the RV data by subtracting the best sine fit. We choose a simple sine wave rather than a more complex function (like a Keplerian) because the fundamental Fourier term is the least informative choice, and therefore the most conservative. When instead, we have identified the RV variability is due to one or more planets, we are interested in the upper-limit imposed by the residuals around the solution. We therefore subtract the best Keplerian solution to the time series before computing the upper limit. For the multi-planetary systems Gl\,581, Gl\,667C and Gl\,876, we compute the periodogram once with raw time series and once with the residuals around the full orbital solution (with all detected planets). Because the giant planets in Gl\,876 system undergo strong mutual interactions, we use a N-body integration to compute the residuals \citep{Correia:2010b}. For Gl\,674 and Gl\,176 (that show both planet- and activity-induced variation), we use a 1 Keplerian$+$sine model to fit the RVs. At last, when we observed no variability, variability without periodicity, or periodic variability without a well identified cause (planet {\it vs.} magnetic activity), we use the raw time series to compute periodograms and upper limits.

\begin{figure*}[thh]
\centering
\includegraphics[width=1.\linewidth]{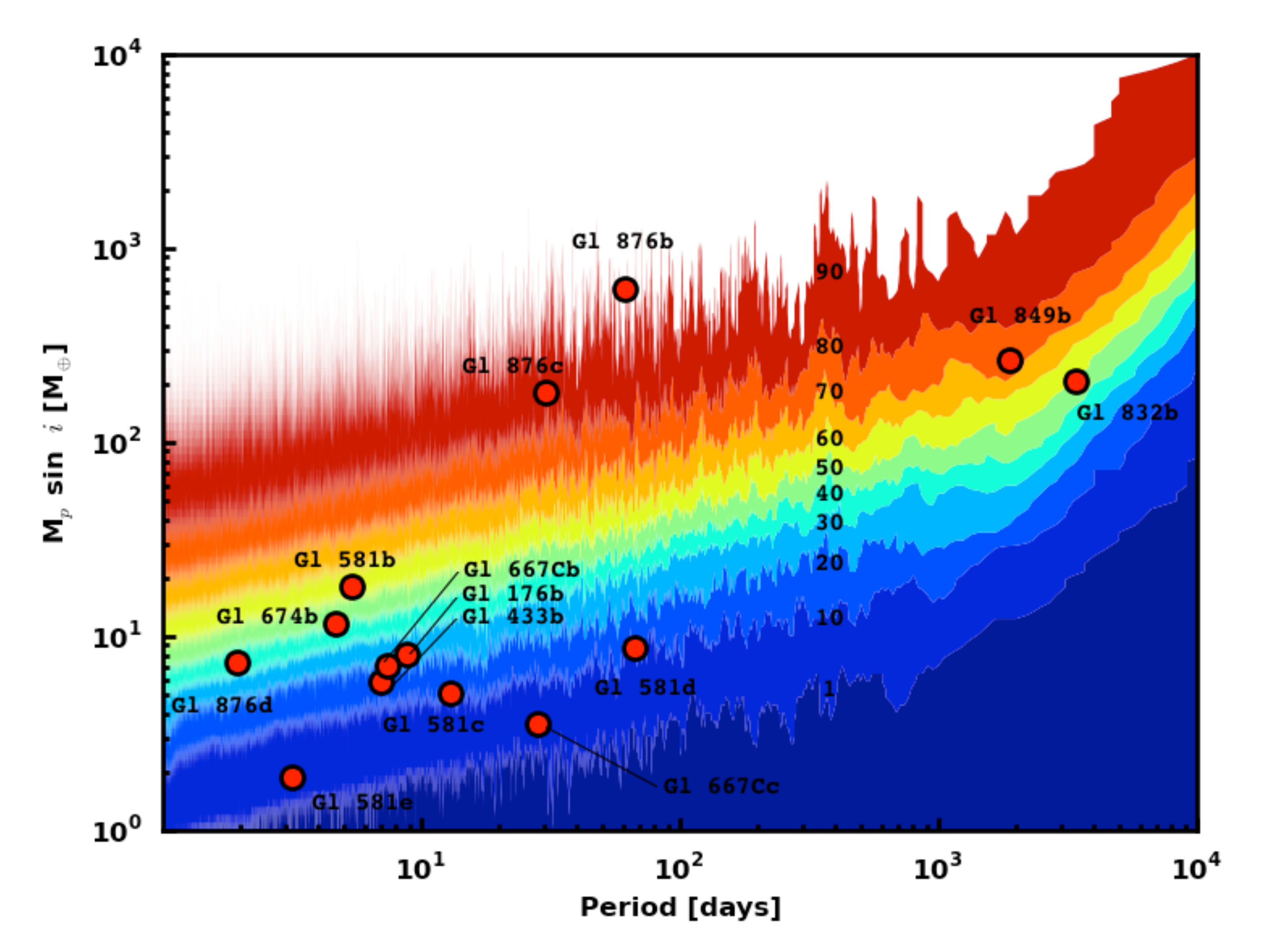}
\caption{\label{fig:all_lim2} Survey sensitivity derived from the combined phase-averaged detection limits on individual stars. Iso-contours are shown for 1, 10, 20, 30, 40, 50, 60, 70, 80 and 90 stars. Planet detected or confirmed by our survey are reported by red circles and labeled by their names.
   }
\end{figure*}

\section{\label{sect:freq}Planet occurrence}

All together, the phase-averaged detection limits calculated for individual stars give the survey efficiency, which is eventually used to correct for the detection incompleteness and derive precise occurrence of planets. Although the statistical analysis of our survey is the purpose of a companion paper (Bonfils et al., in prep.) we give here a first aper\c{c}u.

In an $m \sin i$-period diagram, we pool together the phase-averaged detection limits computed in Sect.~\ref{sect:limits} and, for each period, count the number of ruled out planets more massive than a given mass (with our 99\% criterion). This synthesis is shown in Fig.~\ref{fig:all_lim2} together with iso-contours for 1, 10, 20, 30, 40, 50, 60, 70, 80, and 90 stars. We also overlay the planet detections including all planets of multi-planet systems. Such a diagram is especially useful to compare the planet occurrence for different domains of masses and periods. For instance, for periods $P<100$ day, our sample counts 1 host star (Gl\,876) with two giant planets ($m\sin i = 0.5-10~\mathrm{M_{Jup}}$) but 7 super-Earths ($m \sin i = 1-10~\mathrm{M_\oplus}$). Whereas, our survey is sensitive to Gl\,876b-like planets for 92 stars, short-period super Earths could be detected for only $\sim5-20$ stars. It is therefore obvious that super-Earths are much more frequent than Gl\,876b-like giants.

For more precise estimates we delineate regions in the mass-period diagram and approximate the frequency of planets by the ratio $f=N_d/N_{\star,eff}$, where $N_d$ is the number of planets detected in that region and where $N_{\star,eff}$ is the number of stars whose detection limits confidently exclude planets with similar mass and period. We evaluate $N_{\star,eff}$ with Monte-Carlo sampling : we draw random mass and period within the region delineated (assuming a log-uniform probability for both quantities), use the $m \sin i$-period diagram of Fig.~\ref{fig:all_lim2} to give a local estimate of $N_{\star,eff}$ and, 
with many trials, compute an averaged $N_{\star,eff}$ value. Table.~\ref{tab:occurrence} reports the number of detections, averaged values for $N_{\star,eff}$ and corresponding occurrence of planets for different regions chosen in Fig.~\ref{fig:all_lim2}. 

Numbers of Table.~\ref{tab:occurrence} do confirm that planets are increasingly abundant toward lower-mass and longer-period planets.

At last, we estimate $\eta_\oplus$, the frequency of habitable planets orbiting M dwarfs.  For each star, we use the locations for both the inner ($a_{HZ,in}$) and outer ($a_{HZ,out}$) edges of the habitable zone computed in Sect.~\ref{sect:sample} and we consider that habitable planets have $m\sin i$ between 1 and 10 M$_\oplus$. To evaluate the sensitivity of our survey to habitable planets, we compute a new $N_{\star,eff}$ for the habitable zone with a Monte-Carlo approach again. We draw random masses between 1-10 M$_\oplus$ and random semi-major axis between $a_{HZ,in}$ and $a_{HZ,out}$, choosing a log-uniform probability for both mass and semi-major axis. We screened the detection limits computed previously and increment $N_{\star,eff}$ when the trial fall above that threshold. After normalizing  $N_{\star,eff}$ by the number of trials we found $N_{\star,eff}=4.84$. As among our detections two planets (Gl\,581d and Gl\,667Cc) falls in the habitable zone, we have $N_d=2$ and therefore $\eta_\oplus=0.41^{+0.54}_{-0.13}$.

Alternatively, we measure that 11 (resp. 3) stars of our sample have time-series precise enough to detect planet with same mass and period as Gl\,581d (resp. Gl667Cc), which lead to a very similar estimate of $\eta_\oplus$ ($\sim42\%$).
 
\begin{table*}
\caption{\label{tab:occurrence} Occurrence of planets around M dwarfs for various regions of the $m \sin i$-period diagram.}
\begin{tabular}{l|cccc}
\hline \hline
                  & \multicolumn{4}{c}{Period} \\
\multicolumn{1}{c|}{$m\sin i$}  &  \multicolumn{4}{c}{[day]} \\
\multicolumn{1}{c|}{[M$_\oplus$]} &  1$-$10  & $10-10^2$  & $10^2-10^3$  & $10^3-10^4$ \\
        \hline
$10^3-10^4$& $N_d = 0$& $N_d = 0$& $N_d = 0$& $N_d = 0$\\
$  -  $& $N_{eff} = 96.83$& $N_{eff} = 95.83$& $N_{eff} = 94.29$& $N_{eff} = 87.99$\\
$  -  $& $f<0.01 (1 \sigma)$& $f<0.01 (1 \sigma)$& $f<0.01 (1 \sigma)$& $f<0.01 (1 \sigma)$\\
$10^2-10^3$& $N_d = 0$& $N_d = 2$& $N_d = 0$& $N_d = 2$\\
$  -  $& $N_{eff} = 92.18$& $N_{eff} = 88.54$& $N_{eff} = 81.51$& $N_{eff} = 53.77$\\
$  -  $& $f<0.01 (1 \sigma)$& $f=0.02^{+0.03}_{-0.01}$& $f<0.01 (1 \sigma)$& $f=0.04^{+0.05}_{-0.01}$\\
$10-10^2$& $N_d = 2$& $N_d = 0$& $N_d = 0$& $N_d = 0$\\
$  -  $& $N_{eff} = 69.76$& $N_{eff} = 51.96$& $N_{eff} = 30.75$& $N_{eff} = 9.47$\\
$  -  $& $f=0.03^{+0.04}_{-0.01}$& $f<0.02 (1 \sigma)$& $f<0.04 (1 \sigma)$& $f<0.12 (1 \sigma)$\\
$1-10$& $N_d = 5$& $N_d = 3$& $N_d = 0$& $N_d = 0$\\
$  -  $& $N_{eff} = 13.97$& $N_{eff} = 5.79$& $N_{eff} = 1.53$& $N_{eff} = 0.03$\\
$  -  $& $f=0.36^{+0.24}_{-0.10}$& $f=0.52^{+0.50}_{-0.16}$& -& -\\
\hline\hline
\end{tabular}
\end{table*}

\section{\label{sect:discussion}Conclusion}
We have reported on HARPS guaranteed-time observations for a volume-limited sample of nearby M dwarfs. The paper develops on a systematic analysis of the time series for \Nstars M dwarfs. It analyzes their variability, look for possible trends, and search them for periodicities and Keplerian signals.

We find significant periodic signals for 14 stars and linear trends for 15. We recover the signal for 14 known planets in 8 systems. In particular, we confirm the detection of 2 giant planets, Gl\,849\,b and Gl\,832\,b, and we confirm that an additional long-period companion is probably orbiting Gl\,849. We analyze the RV periodicity against stellar diagnostics for 8 other stars, and find evidences the observed RV variation originate from stellar surface inhomogeneities for all but one (Gl\,273). We find periodic RV variation in Gl\,273 time series without counter part in activity indicators, though the phase coverage is too poor for a robust detection.

Our search for planets with HARPS has detected 9 planets in that sample alone, and a total of 11 planets when counting 2 M dwarfs from another complementary sample. Our detections includes the lowest-mass planet known so far and the first prototype of habitable super-Earths. They are the fruit of slightly less than 500 h of observing time on a 3.6-m telescope, now-days considered as a modest size telescope.

Beyond individual detections, we have also reported a first statistical analysis of our survey. We derived the occurrence of M-dwarf planets for different regions of the  $m\sin i$-period diagram. In particular, we find that giant planets ($m \sin i = 100-1,000 ~\mathrm{M_\oplus}$) have a low frequency (e.g. $f<1$\% for P = $1-10$~d and $f = 0.02^{+0.03}_{-0.01}$ for P = $10-100$ d), whereas super-Earths ($m\sin i = 1-10~\mathrm{M_\oplus}$) are likely very abundant ( $f = 0.36^{+0.25}_{-0.10}$ for P = $1-10$ d and $f = 0.35^{+0.47}_{-0.11}$ for P = $10-100$ d). We also obtained $\eta_\oplus= 0.41^{+0.54}_{-0.13}$, the frequency of habitable planets orbiting M dwarfs. Considering M dwarfs dominate the stellar count in the Galaxy, that estimate weighs considerably in the measure of the frequency of habitable planets in our Galaxy. Also, for the first time, $\eta_\oplus$ is a direct measure and not a number extrapolated from the statistic of more massive planets.

Of course, much refinements are possible. For instance, back to Fig.~\ref{fig:hist} we indicate with vertical ticks (above the histograms) the V and mass values for the known planet-host stars included in our sample. It is striking that all planet-host stars are found in the brightest and more massive halves of the two distributions. This is reminiscent of what we observe between solar type stars and M dwarfs. Solar-like stars have many more detected planets, but observational advantages and disadvantages compared to M dwarfs are difficult to weight. On the one hand solar-type stars are brighter and have a lower jitter level \citep[e.g.][]{Hartman:2011}, but on the other hand they are more massive and the reflex motion induced by a given planet is weaker. To know whether we face an observational bias or a true stellar mass dependance in the formation of planets, we need to evaluate the detection efficiency for all mass ranges, which is the purpose of a forthcoming companion paper (Bonfils et al., in prep.). 

\begin{acknowledgements}
We express our gratitude to Martin Kuerster who serve as referee to our paper. His comments were most useful and significantly improved the manuscript.
\end{acknowledgements}

\bibliographystyle{aa}
\bibliography{mybib}

\longtabL{3}{
\begin{landscape}
\begin{longtable}{lr@{:}c@{:}lr@{:}c@{:}lccr@{$\pm$}lclllllllll}

\caption{\label{tab:sample} Sample}\\
\hline\hline 
Name & \multicolumn{3}{c}{$\alpha$ (2000)} & \multicolumn{3}{c}{$\delta$ (2000)} & $\mu_\alpha$ & $\mu_\delta$ & \multicolumn{2}{c}{$\pi$}  & $\pi_{\mathrm{src}}^a$ & Sp. Typ. &  V         & J		& K        & M$_\star$     & L$_\star$      &  $a_{HZ,in}$ & $a_{HZ,out}$ & d$v$/d$t$\\ 
            & \multicolumn{3}{c}{}                              & \multicolumn{3}{c}{}                             &  \multicolumn{2}{c}{[arcsec]}    & \multicolumn{2}{c}{[mas]} &              &                 &  [mag] & [mag]	& [mag] & [M$_\odot$] & [L$_\odot$] &					&	& [m/s/yr]\\ \hline \endfirsthead
\caption{continued.}\\ \hline\hline
Name & \multicolumn{3}{c}{$\alpha$ (2000)} & \multicolumn{3}{c}{$\delta$ (2000)} & $\mu_\alpha$ & $\mu_\delta$ & \multicolumn{2}{c}{$\pi$}   & $\pi_{\mathrm{src}}^a$ & Sp. Typ. &  V         & J		& K        & M$_\star$     & L$_\star$   &  $a_{HZ,in}$ & $a_{HZ,out}$ & d$v$/d$t$\\
            & \multicolumn{3}{c}{}                              & \multicolumn{3}{c}{}                             &  \multicolumn{2}{c}{[arcsec]}    & \multicolumn{2}{c}{[mas]}  &              &                 &  [mag] & [mag]	& [mag] & [M$_\odot$] & [L$_\odot$] &					&	& [m/s/yr]\\ \hline \endhead \hline \endfoot
Gl\,1	&00	&05	&25	&$-37$	&21	&23	&$-5.635$	&$-2.334$	& 230.4	&0.9	&H	&M3   V	&8.6	&5.3	&4.5	&0.39	&0.021		&0.11	&0.30	&3.71\\
GJ\,1002	&00	&06	&44	&$-07$	&32	&23	&$+0.815$	&$-1.874$	& 213.0	&3.6	&H	&M5.5 V	&13.8	&8.3	&7.4	&0.11	&0.001	&0.03	&0.08	&0.45\\
Gl\,12	&00	&15	&49	&$+13$	&33	&17	&$-0.620$	&$+0.314$	&  88.8	&3.5	&H	&M3	&12.6	&8.6	&7.8	&0.22	&0.007		&0.06	&0.17	&0.12\\
LHS\,1134	&00	&43	&26	&$-41$	&17	&36	&$+0.501$	&$-0.553$	& 101.0	&16.0	&R	&M3	&13.1	&8.6	&7.7	&0.20	&0.005		&0.06	&0.15	&0.13\\
Gl\,54.1	&01	&12	&31	&$-17$	&00	&00	&$-1.210$	&$+0.646$	& 271.0	&8.4	&H	&M4.5 V	&12.0	&7.3	&6.4	&0.13	&0.003		&0.04	&0.10	&0.16\\
L\,707-74	&01	&23	&18	&$-12$	&56	&23	&$+0.013$	&$+0.332$	&  97.8	&13.5	&Y	&m	&13.0	&9.2	&8.3	&0.15	&0.003		&0.04	&0.12	&0.03\\
Gl\,87	&02	&12	&21	&$+03$	&34	&30	&$+1.760$	&$-1.854$	&  96.0	&1.7	&H	&M1.5	&10.1	&6.8	&6.1	&0.45	&0.032		&0.14	&0.37	&1.56\\
Gl\,105B	&02	&36	&16	&$+06$	&52	&12	&$-1.806$	&$+1.442$	& 139.3	&0.5	&H	&M3.5 V	&11.7	&7.3	&6.6	&0.25	&0.010		&0.07	&0.20	&0.88\\
CD-\,44-836A	&02	&45	&11	&$-43$	&44	&30	&$-0.071$	&$-0.403$	& 113.9	&38.7	&C	&M4	&12.3	&8.1	&7.3	&0.22	&0.006		&0.06	&0.16	&0.03\\
LHS\,1481	&02	&58	&10	&$-12$	&53	&06	&$-0.319$	&$+0.566$	&  95.5	&10.9	&H	&M2.5	&12.7	&9.0	&8.2	&0.17	&0.005		&0.05	&0.14	&0.10\\
LP\,771-95A	&03	&01	&51	&$-16$	&35	&36	&$+0.367$	&$-0.277$	& 146.4	&2.9	&H06	&M3	&11.5	&7.3	&6.5	&0.24	&0.009		&0.07	&0.19	&0.03\\
LHS\,1513	&03	&11	&36	&$-38$	&47	&17	&$-0.839$	&$-0.201$	& 130.0	&20.0	&R	&M3.5	&11.5	&9.8	&9.0	&0.09	&0.001		&0.03	&0.07	&0.13\\
GJ\,1057	&03	&13	&23	&$+04$	&46	&30	&$-1.705$	&$+0.098$	& 117.1	&3.5	&H	&M5	&13.9	&8.8	&7.8	&0.16	&0.003		&0.04	&0.12	&0.57\\
Gl\,145	&03	&32	&56	&$-44$	&42	&06	&$+0.312$	&$+0.131$	&  93.1	&1.9	&H	&M2.5	&11.5	&7.7	&6.9	&0.32	&0.014		&0.09	&0.24	&0.03\\
GJ\,1061	&03	&36	&00	&$-44$	&30	&48	&$-0.730$	&$-0.407$	& 271.9	&1.3	&H	&M5.5 V	&13.1	&7.5	&6.6	&0.12	&0.002		&0.03	&0.09	&0.06\\
GJ\,1065	&03	&50	&44	&$-06$	&05	&42	&$+0.416$	&$-1.368$	& 105.4	&3.2	&H	&M4   V	&12.8	&8.6	&7.8	&0.19	&0.005		&0.05	&0.15	&0.45\\
GJ\,1068	&04	&10	&28	&$-53$	&36	&06	&$+0.851$	&$-2.444$	& 143.4	&1.9	&H	&M4.5	&13.6	&8.7	&7.9	&0.13	&0.002		&0.04	&0.10	&1.07\\
Gl\,166C	&04	&15	&22	&$-07$	&39	&23	&$+2.239$	&$-3.421$	& 200.6	&0.2	&H	&M4.5 V	&11.2	&6.7	&6.0	&0.23	&0.008		&0.07	&0.18	&1.91\\
Gl\,176	&04	&42	&56	&$+18$	&57	&29	&$-0.660$	&$-1.114$	& 106.2	&2.5	&H	&M2.5	&10.0	&6.5	&5.6	&0.50	&0.033		&0.14	&0.38	&0.36\\
LHS\,1723	&05	&01	&57	&$-06$	&56	&47	&$+0.542$	&$-0.550$	& 187.9	&1.3	&H	&M3.5 V	&12.2	&7.6	&6.7	&0.17	&0.004	&0.05	&0.12	&0.07\\
LHS\,1731	&05	&03	&20	&$-17$	&22	&23	&$+0.226$	&$-0.448$	& 108.6	&2.7	&H	&M3.0 V	&11.7	&7.8	&6.9	&0.27	&0.009	&0.07	&0.20	&0.05\\
Gl\,191	&05	&11	&40	&$-45$	&01	&06	&$-6.497$	&$-5.727$	& 255.3	&0.9	&H	&M1 p V	&8.8	&5.8	&5.0	&0.27	&0.011		&0.08	&0.22	&6.75\\
Gl\,203	&05	&28	&00	&$+09$	&38	&36	&$+0.194$	&$-0.760$	& 113.5	&5.0	&H	&M3.5 V	&12.4	&8.3	&7.5	&0.19	&0.006		&0.06	&0.16	&0.12\\
Gl\,205	&05	&31	&27	&$-03$	&40	&42	&$-0.762$	&$-2.093$	& 176.8	&1.2	&H	&M1.5 V	&8.0	&5.0	&4.0	&0.60	&0.043	&0.16	&0.43	&0.65\\
Gl\,213	&05	&42	&09	&$+12$	&29	&23	&$-1.997$	&$-1.572$	& 171.6	&4.0	&H	&M4   V	&11.5	&7.1	&6.4	&0.22	&0.008		&0.07	&0.18	&0.87\\
Gl\,229	&06	&10	&34	&$-21$	&51	&53	&$+0.137$	&$-0.714$	& 173.8	&1.0	&H	&M1   V	&8.2	&5.1	&4.2	&0.58	&0.041		&0.15	&0.42	&0.07\\
HIP\,31293	&06	&33	&43	&$-75$	&37	&47	&$+0.290$	&$+0.279$	& 110.9	&2.2	&H	&M3 V	&10.5	&6.7	&5.9	&0.43	&0.024		&0.12	&0.32	&0.03\\
HIP\,31292	&06	&33	&47	&$-75$	&37	&30	&$+0.309$	&$+0.256$	& 114.5	&3.2	&H	&M3/4 V	&11.4	&7.4	&6.6	&0.31	&0.012		&0.08	&0.23	&0.03\\
G\,108-21	&06	&42	&11	&$+03$	&34	&53	&$-0.040$	&$-0.270$	& 103.1	&8.5	&H	&M3.5	&12.1	&8.2	&7.3	&0.23	&0.008		&0.07	&0.18	&0.02\\
Gl\,250B	&06	&52	&18	&$-05$	&11	&24	&$+0.544$	&$-0.003$	& 114.8	&0.4	&H	&M2.5 V	&10.1	&6.6	&5.7	&0.45	&0.026		&0.12	&0.33	&0.06\\
Gl\,273	&07	&27	&24	&$+05$	&13	&30	&$-0.572$	&$-3.694$	& 263.0	&1.4	&H	&M3.5 V	&9.8	&5.7	&4.9	&0.29	&0.011		&0.08	&0.22	&1.22\\
LHS\,1935	&07	&38	&41	&$-21$	&13	&30	&$-0.455$	&$-0.474$	&  94.3	&3.3	&H	&M3	&11.7	&7.8	&7.1	&0.29	&0.013		&0.09	&0.23	&0.11\\
Gl\,285	&07	&44	&40	&$+03$	&33	&06	&$+0.345$	&$-0.451$	& 167.9	&2.3	&H	&M4   V	&11.2	&6.6	&5.7	&0.31	&0.012		&0.08	&0.22	&0.04\\
Gl\,299	&08	&11	&57	&$+08$	&46	&23	&$-1.145$	&$-5.082$	& 146.3	&3.1	&H	&M4   V	&12.8	&8.4	&7.7	&0.14	&0.003		&0.04	&0.12	&4.26\\
Gl\,300	&08	&12	&41	&$-21$	&33	&12	&$-0.031$	&$-0.704$	& 125.8	&1.0	&H	&M3.5 V	&12.1	&7.6	&6.7	&0.26	&0.008		&0.07	&0.19	&0.09\\
GJ\,2066	&08	&16	&08	&$+01$	&18	&11	&$+0.375$	&$+0.060$	& 109.6	&1.5	&H	&M2	&10.1	&6.6	&5.8	&0.46	&0.027		&0.13	&0.34	&0.03\\
GJ\,1123	&09	&17	&05	&$-77$	&49	&17	&$-0.655$	&$-0.806$	& 110.9	&2.0	&H	&M4.5 V	&13.1	&8.3	&7.4	&0.21	&0.005		&0.06	&0.15	&0.22\\
Gl\,341	&09	&21	&38	&$-60$	&16	&53	&$+0.839$	&$+0.181$	&  95.6	&0.9	&H	&M0 V	&9.5	&6.4	&5.6	&0.55	&0.042		&0.16	&0.42	&0.18\\
GJ\,1125	&09	&30	&44	&$+00$	&19	&18	&$+0.570$	&$-0.552$	& 103.5	&3.9	&H	&M3.0 V	&11.7	&7.7	&6.9	&0.29	&0.012		&0.08	&0.22	&0.14\\
Gl\,357	&09	&36	&02	&$-21$	&39	&42	&$-0.137$	&$-0.990$	& 110.8	&1.9	&H	&M3   V	&10.9	&7.3	&6.5	&0.33	&0.014		&0.09	&0.24	&0.21\\
Gl\,358	&09	&39	&47	&$-41$	&04	&00	&$+0.526$	&$+0.357$	& 105.6	&1.6	&H	&M3.0 V	&10.8	&6.9	&6.1	&0.42	&0.023		&0.11	&0.31	&0.09\\
Gl\,367	&09	&44	&30	&$-45$	&46	&36	&$+0.466$	&$-0.586$	& 101.3	&3.2	&H	&M1	&10.1	&6.6	&5.8	&0.49	&0.031		&0.14	&0.37	&0.13\\
GJ\,1129	&09	&44	&48	&$-18$	&12	&48	&$+1.604$	&$-0.163$	&  90.9	&3.8	&H	&M3.5 V	&12.5	&8.1	&7.3	&0.28	&0.010		&0.08	&0.21	&0.66\\
Gl\,382	&10	&12	&17	&$-03$	&44	&47	&$+0.153$	&$-0.243$	& 127.1	&1.9	&H	&M2   V	&9.3	&5.9	&5.0	&0.54	&0.039		&0.15	&0.41	&0.01\\
Gl\,388	&10	&19	&36	&$+19$	&52	&12	&$-0.501$	&$-0.043$	& 204.6	&2.8	&H	&M4.5	&9.4	&5.4	&4.6	&0.42	&0.023		&0.12	&0.31	&0.03\\
Gl\,393	&10	&28	&55	&$+00$	&50	&23	&$+0.603$	&$-0.731$	& 141.5	&2.2	&H	&M2   V	&9.7	&6.2	&5.3	&0.44	&0.024		&0.12	&0.32	&0.15\\
LHS\,288	&10	&44	&32	&$-61$	&11	&35	&$+0.334$	&$+1.627$	& 209.7	&2.7	&H	&M5.5	&13.9	&8.5	&7.7	&0.10	&0.001		&0.03	&0.08	&0.30\\
Gl\,402	&10	&50	&52	&$+06$	&48	&30	&$+0.853$	&$-0.818$	& 147.9	&3.5	&H	&M4   V	&11.7	&7.3	&6.4	&0.26	&0.007		&0.07	&0.18	&0.22\\
Gl\,406	&10	&56	&29	&$+07$	&00	&54	&$+3.829$	&$-2.711$	& 419.1	&2.1	&H	&M6   V	&13.4	&7.1	&6.1	&0.10	&0.001		&0.03	&0.07	&1.21\\
Gl\,413.1	&11	&09	&31	&$-24$	&36	&00	&$+0.797$	&$-0.447$	&  93.0	&1.7	&H	&M2	&10.4	&6.9	&6.1	&0.46	&0.028		&0.13	&0.35	&0.21\\
Gl\,433	&11	&35	&27	&$-32$	&32	&23	&$+0.070$	&$-0.853$	& 112.6	&1.4	&H	&M2.0 V	&9.8	&6.5	&5.6	&0.47	&0.030		&0.13	&0.36	&0.15\\
Gl\,438	&11	&43	&20	&$-51$	&50	&23	&$-0.674$	&$-0.532$	& 119.0	&10.2	&R	&M0	&10.4	&7.1	&6.3	&0.33	&0.015		&0.09	&0.25	&0.14\\
Gl\,447	&11	&47	&44	&$+00$	&48	&16	&$-0.605$	&$-1.219$	& 299.6	&2.2	&H	&M4	&11.1	&6.5	&5.7	&0.17	&0.004		&0.05	&0.13	&0.14\\
Gl\,465	&12	&24	&53	&$-18$	&14	&30	&$-1.096$	&$-2.307$	& 113.0	&2.5	&H	&M3   V	&11.3	&7.7	&7.0	&0.26	&0.010		&0.08	&0.20	&1.33\\
Gl\,479	&12	&37	&53	&$-52$	&00	&06	&$+1.034$	&$+0.031$	& 103.2	&2.3	&H	&M3   V	&10.7	&6.9	&6.0	&0.43	&0.025		&0.12	&0.33	&0.24\\
LHS\,337	&12	&38	&50	&$-38$	&22	&53	&$+0.667$	&$-1.326$	& 156.8	&2.0	&H	&M4.5 V	&12.7	&8.2	&7.4	&0.15	&0.003		&0.04	&0.12	&0.32\\
Gl\,480.1	&12	&40	&46	&$-43$	&34	&00	&$+0.782$	&$+0.694$	& 128.5	&3.9	&H	&M3.0 V	&12.2	&8.2	&7.4	&0.18	&0.005		&0.05	&0.14	&0.20\\
Gl\,486	&12	&47	&57	&$+09$	&45	&12	&$+1.008$	&$-0.461$	& 119.5	&2.7	&H	&M3.5	&11.4	&7.2	&6.4	&0.32	&0.014		&0.09	&0.24	&0.24\\
Gl\,514	&13	&30	&00	&$+10$	&22	&36	&$-1.128$	&$-1.074$	& 130.6	&1.1	&H	&M1   V	&9.1	&5.9	&5.0	&0.53	&0.037		&0.15	&0.40	&0.43\\
Gl\,526	&13	&45	&44	&$+14$	&53	&30	&$-1.778$	&$-1.456$	& 185.5	&1.1	&H	&M1.5 V	&8.5	&5.2	&4.4	&0.50	&0.039		&0.15	&0.40	&0.65\\
Gl\,536	&14	&01	&03	&$-02$	&39	&18	&$+0.824$	&$-0.599$	&  98.3	&1.6	&H	&M1	&9.7	&6.5	&5.7	&0.52	&0.038		&0.15	&0.40	&0.24\\
Gl\,551	&14	&29	&43	&$-62$	&40	&47	&$+3.777$	&$+0.768$	& 771.6	&2.6	&H	&M5.5	&11.1	&5.4	&4.4	&0.12	&0.002		&0.03	&0.08	&0.44\\
Gl\,555	&14	&34	&17	&$-12$	&31	&06	&$+0.358$	&$+0.595$	& 165.0	&3.3	&H	&M3.5 V	&11.3	&6.8	&5.9	&0.28	&0.009		&0.07	&0.20	&0.07\\
Gl\,569A	&14	&54	&29	&$+16$	&06	&04	&$-0.276$	&$-0.122$	& 101.9	&1.7	&H	&M2.5	&10.2	&6.6	&5.8	&0.49	&0.031		&0.13	&0.36	&0.02\\
Gl\,581	&15	&19	&26	&$-07$	&43	&17	&$+1.225$	&$-0.099$	& 160.9	&2.6	&H	&M2.5 V	&10.6	&6.7	&5.8	&0.30	&0.011		&0.08	&0.22	&0.22\\
Gl\,588	&15	&32	&13	&$-41$	&16	&36	&$+1.178$	&$-1.028$	& 168.7	&1.3	&H	&M2.5 V	&9.3	&5.6	&4.8	&0.47	&0.027		&0.13	&0.34	&0.33\\
Gl\,618A	&16	&20	&04	&$-37$	&31	&41	&$+0.728$	&$+0.992$	& 119.8	&2.5	&H	&M3   V	&10.6	&6.8	&6.0	&0.39	&0.020		&0.11	&0.29	&0.29\\
Gl\,628	&16	&30	&18	&$-12$	&39	&47	&$+0.093$	&$-1.185$	& 233.0	&1.6	&H	&M3   V	&10.1	&6.0	&5.1	&0.30	&0.011	&0.08	&0.22	&0.14\\
Gl\,643	&16	&55	&25	&$-08$	&19	&23	&$+0.814$	&$-0.895$	& 148.9	&4.0	&H	&M3.5 V	&11.8	&7.6	&6.7	&0.21	&0.006	&0.06	&0.16	&0.23\\
Gl\,667C	&17	&18	&58	&$-34$	&59	&42	&$-1.159$	&$-0.114$	& 146.3	&9.0	&H	&M2   V	&10.2	&6.8	&6.0	&0.30	&0.013	&0.09	&0.23	&0.21\\
Gl\,674	&17	&28	&40	&$-46$	&53	&42	&$-0.574$	&$-0.880$	& 220.2	&1.4	&H	&M3   V	&9.4	&5.7	&4.9	&0.35	&0.015	&0.09	&0.26	&0.12\\
Gl\,678.1A	&17	&30	&22	&$+05$	&32	&53	&$-0.028$	&$-0.248$	& 100.2	&1.1	&H	&M1 V	&9.3	&6.2	&5.4	&0.57	&0.048	&0.17	&0.45	&0.01\\
Gl\,680	&17	&35	&13	&$-48$	&40	&53	&$-0.085$	&$+0.456$	& 102.8	&2.8	&H	&M1.5	&10.2	&6.7	&5.8	&0.47	&0.030	&0.13	&0.36	&0.05\\
Gl\,682	&17	&37	&03	&$-44$	&19	&11	&$+0.710$	&$-0.938$	& 196.9	&2.1	&H	&M4.5 V	&11.0	&6.5	&5.6	&0.27	&0.008	&0.07	&0.19	&0.16\\
Gl\,686	&17	&37	&53	&$+18$	&35	&30	&$-0.927$	&$+0.983$	& 123.0	&1.6	&H	&M1	&9.6	&6.4	&5.6	&0.45	&0.029	&0.13	&0.35	&0.34\\
Gl\,693	&17	&46	&35	&$-57$	&19	&11	&$+1.119$	&$-1.353$	& 171.5	&2.3	&H	&M3.5 V	&10.8	&6.9	&6.0	&0.26	&0.009	&0.07	&0.20	&0.41\\
Gl\,699	&17	&57	&49	&$+04$	&41	&36	&$+0.795$	&$+10.338$	& 549.0	&1.6	&H	&M4   V	&9.6	&5.2	&4.5	&0.16	&0.004	&0.05	&0.14	&4.50\\
Gl\,701	&18	&05	&07	&$-03$	&01	&53	&$-0.570$	&$-0.333$	& 128.9	&1.4	&H	&M0   V	&9.4	&6.2	&5.3	&0.48	&0.030	&0.13	&0.36	&0.08\\
GJ\,1224	&18	&07	&33	&$-15$	&57	&47	&$+0.562$	&$-0.352$	& 132.6	&3.7	&H	&M4.5 V	&13.6	&8.6	&7.8	&0.14	&0.003	&0.04	&0.11	&0.08\\
G\,141-29	&18	&42	&44	&$+13$	&54	&17	&$+0.043$	&$+0.309$	&  93.3	&11.5	&H	&M4	&12.8	&8.4	&7.6	&0.23	&0.008	&0.07	&0.18	&0.02\\
Gl\,729	&18	&49	&49	&$-23$	&50	&12	&$-0.638$	&$-0.192$	& 336.7	&2.0	&H	&M3.5 V	&10.5	&6.2	&5.4	&0.17	&0.004	&0.05	&0.13	&0.03\\
GJ\,1232	&19	&09	&51	&$+17$	&40	&07	&$-0.637$	&$-0.425$	&  93.6	&2.8	&H	&M4.5	&13.6	&8.8	&7.9	&0.20	&0.005	&0.05	&0.14	&0.14\\
Gl\,752A	&19	&16	&55	&$+05$	&10	&05	&$+0.579$	&$-1.331$	& 170.4	&1.0	&H	&M3   V	&9.1	&5.6	&4.7	&0.48	&0.028	&0.13	&0.35	&0.28\\
Gl\,754	&19	&20	&48	&$-45$	&33	&30	&$-0.668$	&$-2.869$	& 169.2	&1.6	&H	&M4.5	&12.2	&7.7	&6.8	&0.18	&0.005		&0.05	&0.14	&1.18\\
GJ\,1236	&19	&22	&03	&$+07$	&02	&36	&$+0.736$	&$-0.393$	&  92.9	&2.5	&H	&M3	&12.4	&8.5	&7.7	&0.22	&0.007		&0.06	&0.17	&0.17\\
GJ\,1256	&20	&40	&34	&$+15$	&29	&57	&$-1.324$	&$+0.660$	& 102.0	&2.2	&H	&M4.5	&13.4	&8.6	&7.7	&0.19	&0.005	&0.05	&0.14	&0.49\\
Gl\,803$^\dagger$	&20	&45	&10	&$-31$	&20	&30	&$-0.280$	&$-0.360$	& 100.9	&1.1	&H	&M0 V e	&8.8	&5.4	&4.5	&0.75	&0.091	&0.23	&0.62	&0.05\\
LHS\,3583	&20	&46	&37	&$-81$	&43	&12	&$-0.540$	&$-0.544$	&  77.1	&21.2	&C	&M2.5	&11.5	&7.7	&6.8	&0.40	&0.020	&0.11	&0.29	&0.17\\
LP\,816-60	&20	&52	&33	&$-16$	&58	&30	&$+0.307$	&$+0.031$	& 175.0	&3.4	&H	&m	&11.4	&7.1	&6.2	&0.23	&0.007		&0.06	&0.17	&0.01\\
Gl\,832	&21	&33	&34	&$-49$	&00	&36	&$+0.046$	&$-0.818$	& 201.9	&1.0	&H	&M1 V	&8.7	&5.3	&4.5	&0.45	&0.026		&0.12	&0.33	&0.08\\
Gl\,846	&22	&02	&10	&$+01$	&24	&00	&$+0.455$	&$-0.280$	&  97.6	&1.5	&H	&M0.5 V	&9.2	&6.2	&5.3	&0.60	&0.050		&0.17	&0.46	&0.07\\
LHS\,3746	&22	&02	&29	&$-37$	&04	&54	&$-0.807$	&$-0.227$	& 134.3	&1.3	&H	&M3.5	&11.8	&7.6	&6.7	&0.24	&0.007		&0.06	&0.18	&0.12\\
Gl\,849	&22	&09	&40	&$-04$	&38	&30	&$-1.135$	&$-0.020$	& 109.9	&2.1	&H	&M3   V	&10.4	&6.5	&5.6	&0.49	&0.028	&0.13	&0.35	&0.27\\
GJ\,1265	&22	&13	&42	&$-17$	&41	&12	&$-0.826$	&$-0.331$	&  96.0	&3.9	&H	&M4.5	&13.6	&9.0	&8.1	&0.17	&0.004	&0.05	&0.13	&0.19\\
LHS\,3799	&22	&23	&07	&$-17$	&36	&23	&$-0.290$	&$-0.726$	& 134.4	&4.9	&H	&M4.5 V	&13.3	&8.2	&7.3	&0.18	&0.004	&0.05	&0.13	&0.10\\
Gl\,876	&22	&53	&17	&$-14$	&15	&48	&$-0.961$	&$-0.675$	& 213.3	&2.1	&H	&M3.5 V	&10.2	&5.9	&5.0	&0.34	&0.013	&0.09	&0.24	&0.15\\
Gl\,877	&22	&55	&46	&$-75$	&27	&36	&$+1.028$	&$-1.061$	& 116.1	&1.2	&H	&M2.5	&10.4	&6.6	&5.8	&0.43	&0.025	&0.12	&0.33	&0.43\\
Gl\,880	&22	&56	&35	&$+16$	&33	&12	&$-1.033$	&$-0.283$	& 146.1	&1.0	&H	&M1.5V	&8.7	&5.4	&4.5	&0.58	&0.050	&0.17	&0.46	&0.18\\
Gl\,887	&23	&05	&52	&$-35$	&51	&12	&$-6.767$	&$+1.328$	& 303.9	&0.9	&H	&M2   V	&7.3	&4.3	&3.5	&0.47	&0.028	&0.13	&0.35	&3.60\\
LHS\,543	&23	&21	&37	&$+17$	&17	&25	&$-0.537$	&$-1.385$	&  91.0	&2.9	&H	&M4	&11.7	&7.4	&6.5	&0.40	&0.019	&0.10	&0.29	&0.56\\
Gl\,908	&23	&49	&13	&$+02$	&24	&06	&$-0.996$	&$-0.968$	& 167.3	&1.2	&H	&M1   V	&9.0	&5.8	&5.0	&0.42	&0.026	&0.12	&0.33	&0.26\\
LTT\,9759	&23	&53	&50	&$-75$	&37	&53	&$-0.243$	&$-0.378$	& 100.1	&1.1	&H	&M...	&10.0	&6.5	&5.5	&0.54	&0.037	&0.15	&0.40	&0.05\\
\end{longtable}
$^a$ (H) revised Hipparcos catalog \citep{Leeuwen:2007}; (R) \citep{Reid:1995b} ; (H06): \citet{Henry:2006};(Y) \citet{Altena:1995}; (C) CNS4 catalog (Jahreiss, private comm.)\\
$^\dagger$ Gl\,803 (AU Mic) is a young ($\sim20$ Myr) M dwarfs with a circumstellar disk \citep{Kalas:2004}. The calibration used to determine its mass may not be adapted for this age.
\end{landscape}
}

\longtab{4}{\begin{longtable}{lcccccccccc}
\caption{\label{tab:var} Test for variability}\\
\hline\hline
Name    & N     & $\sigma_i$    & $\sigma_e$    & $P(F)$                    & $\chi_{constant}^2$      & $P(\chi_{constant}^2)$   & Slope       & $\chi^2_{slope}$      & $P(F_{slope})$      & FAP   \\
        &       & [m/s]          & [m/s]          &                           &                           &                           & [m/s/yr]    &                       &                     &       \\ \hline \endfirsthead
\caption{continued.}\\
\hline\hline
Name    & N     & $\sigma_i$    & $\sigma_e$    & $P(F)$                    & $\chi_{constant}^2$      & $P(\chi_{constant}^2)$   & Slope       & $\chi^2_{slope}$      & $P(F_{slope})$      & FAP   \\
        &       & [m/s]          & [m/s]          &                           &                           &                           & [m/s/yr]    &                       &                    &       \\ \hline \endhead \hline \endfoot
Gl1	& 45	&  0.6	&  2.0	& {\boldmath $<10^{-9}$}	& 517	& {\boldmath $<10^{-9}$}	& 0.332	& 487	& {\boldmath $0.001$}	& {\boldmath$0.006$}\\
GJ1002	& 5	&  6.0	&  2.0	& $0.993$	&  0.6	& $0.959$	& -0.375	&  0.6	& $0.929$	& $0.530$\\
Gl12	& 6	&  4.5	&  3.4	& $0.851$	&  3.7	& $0.595$	& -0.862	&  3.1	& $0.741$	& $0.244$\\
LHS1134	& 7	&  5.4	&  5.6	& $0.622$	&  9.0	& $0.176$	& -2.506	&  5.4	& $0.152$	& $0.219$\\
Gl54.1	& 12	&  2.7	&  4.1	& $0.155$	& 37.5	& {\boldmath $9.5~10^{-5}$}	& 1.745	& 22.7	& {\boldmath $0.004$}	& $0.048$\\
L707-74	& 5	&  5.9	&  5.7	& $0.702$	&  3.7	& $0.444$	& 0.074	&  2.9	& $0.755$	& $0.185$\\
Gl87	& 15	&  0.9	&  1.5	& $0.078$	& 50.5	& {\boldmath $5.1~10^{-6}$}	& -0.039	& 49.3	& $0.990$	& $0.348$\\
Gl105B	& 17	&  2.2	&  3.6	& $0.051$	& 60.7	& {\boldmath $3.9~10^{-7}$}	& 0.418	& 56.8	& $0.566$	& $0.301$\\
CD-44-836A	& 8	&  3.3	&  2.8	& $0.794$	&  5.4	& $0.614$	& 0.111	&  4.8	& $0.767$	& $0.201$\\
LHS1481	& 8	&  4.0	&  3.5	& $0.773$	&  8.0	& $0.333$	& -0.087	&  7.9	& $1.000$	& $0.850$\\
LP771-95A	& 6	&  1.6	& 10.6	& {\boldmath $9.4~10^{-4}$}	& 246	& {\boldmath $<10^{-9}$}	& -7.343	& 24.0	& {\boldmath $0.001$}	& {\boldmath$0.005$}\\
LHS1513	& 6	&  6.6	&  3.8	& $0.949$	&  2.1	& $0.831$	& -1.382	&  1.7	& $0.655$	& $0.525$\\
GJ1057	& 8	&  6.4	&  8.1	& $0.398$	&  8.9	& $0.257$	& 0.282	&  8.4	& $0.949$	& $0.152$\\
Gl145	& 6	&  2.0	&  2.4	& $0.512$	&  7.7	& $0.173$	& 0.244	&  7.3	& $0.977$	& $0.185$\\
GJ1061	& 4	&  4.5	&  4.2	& $0.749$	&  3.6	& $0.308$	& -1.827	&  1.8	& $0.473$	& $0.084$\\
GJ1065	& 5	&  4.6	&  6.3	& $0.448$	&  7.6	& $0.106$	& -2.482	&  6.0	& $0.747$	& $0.130$\\
GJ1068	& 4	&  6.0	&  3.9	& $0.904$	&  1.7	& $0.645$	& 1.309	&  1.0	& $0.657$	& $0.474$\\
Gl166C	& 4	&  1.5	&  9.0	& $0.017$	& 146	& {\boldmath $<10^{-9}$}	& -5.703	& 127	& $0.944$	& $0.514$\\
Gl176	& 57	&  0.9	&  5.2	& {\boldmath $<10^{-9}$}	& 2438	& {\boldmath $<10^{-9}$}	& -1.023	& 2365	& $0.033$	& $0.383$\\
LHS1723	& 7	&  3.3	&  2.9	& $0.780$	&  6.6	& $0.359$	& -0.650	&  5.3	& $0.575$	& $0.410$\\
LHS1731	& 7	&  2.6	&  2.8	& $0.577$	& 17.5	& {\boldmath $0.008$}	& 0.383	& 16.7	& $0.985$	& $0.964$\\
Gl191	& 30	&  0.7	&  2.4	& {\boldmath $<10^{-9}$}	& 442	& {\boldmath $<10^{-9}$}	& -0.113	& 435	& $0.986$	& $0.158$\\
Gl203	& 8	&  3.4	&  3.9	& $0.518$	& 11.0	& $0.140$	& -1.009	&  8.7	& $0.422$	& $0.187$\\
Gl205	& 103	&  0.6	&  3.9	& {\boldmath $<10^{-9}$}	& 6224	& {\boldmath $<10^{-9}$}	& 3.186	& 4371	& {\boldmath $<10^{-9}$}	& {\boldmath$0.002$}\\
Gl213	& 6	&  1.8	&  4.0	& $0.099$	& 22.5	& {\boldmath $4.3~10^{-4}$}	& 0.238	& 20.9	& $0.957$	& $0.114$\\
Gl229	& 15	&  0.5	&  1.5	& {\boldmath $2.2~10^{-4}$}	& 157	& {\boldmath $<10^{-9}$}	& -0.257	& 148	& $0.784$	& $0.617$\\
HIP31293	& 8	&  1.3	&  2.2	& $0.153$	& 38.1	& {\boldmath $2.9~10^{-6}$}	& -0.656	& 37.0	& $0.993$	& $0.687$\\
HIP31292	& 6	&  2.2	&  3.6	& $0.251$	& 24.6	& {\boldmath $1.7~10^{-4}$}	& 0.090	& 23.9	& $0.996$	& $0.774$\\
G108-21	& 4	&  2.9	&  1.4	& $0.961$	&  1.4	& $0.696$	& -0.958	&  0.6	& $0.408$	& $0.299$\\
Gl250B	& 6	&  2.6	& 11.5	& {\boldmath $0.006$}	& 204	& {\boldmath $<10^{-9}$}	& 0.664	& 18.4	& {\boldmath $0.001$}	& $0.045$\\
Gl273	& 49	&  0.6	&  3.0	& {\boldmath $<10^{-9}$}	& 1221	& {\boldmath $<10^{-9}$}	& 0.628	& 1108	& {\boldmath $2.3~10^{-7}$}	& $0.040$\\
LHS1935	& 7	&  2.9	&  2.1	& $0.873$	&  3.2	& $0.778$	& -0.273	&  2.5	& $0.503$	& $0.123$\\
Gl285	& 7	&  3.4	& 102	& {\boldmath $3.2~10^{-8}$}	& 5855	& {\boldmath $<10^{-9}$}	& 59.513	& 5844	& $1.000$	& $0.577$\\
Gl299	& 9	&  4.2	&  4.4	& $0.580$	& 16.8	& $0.032$	& 0.447	& 15.3	& $0.815$	& $0.854$\\
Gl300	& 24	&  2.2	&  5.6	& {\boldmath $3.8~10^{-5}$}	& 201	& {\boldmath $<10^{-9}$}	& 1.167	& 187	& $0.194$	& $0.787$\\
GJ2066	& 8	&  1.0	&  1.5	& $0.296$	& 13.6	& $0.059$	& 0.024	& 13.5	& $1.000$	& $0.084$\\
GJ1123	& 6	&  6.5	&  6.5	& $0.664$	&  7.9	& $0.165$	& 0.169	&  7.1	& $0.927$	& $0.669$\\
Gl341	& 23	&  0.8	&  2.6	& {\boldmath $4.0~10^{-7}$}	& 273	& {\boldmath $<10^{-9}$}	& 0.934	& 208	& {\boldmath $4.5~10^{-5}$}	& {\boldmath$0.001$}\\
GJ1125	& 8	&  1.9	& 144	& {\boldmath $<10^{-9}$}	& $4.8~10^{+4}$	& {\boldmath $<10^{-9}$}	& 29.684	& $4.5~10^{+4}$	& $0.922$	& $0.287$\\
Gl357	& 6	&  1.4	&  2.7	& $0.162$	& 38.3	& {\boldmath $3.3~10^{-7}$}	& -1.682	& 15.2	& $0.067$	& $0.081$\\
Gl358	& 28	&  1.0	&  8.4	& {\boldmath $<10^{-9}$}	& 2130	& {\boldmath $<10^{-9}$}	& 3.344	& 1944	& $0.016$	& $0.041$\\
Gl367	& 19	&  0.8	&  2.0	& {\boldmath $3.6~10^{-4}$}	& 139	& {\boldmath $<10^{-9}$}	& 0.741	& 97.8	& {\boldmath $1.1~10^{-4}$}	& $0.048$\\
GJ1129	& 3	&  3.8	&  0.4	& $0.999$	&  0.0	& $0.978$	& 0.246	&  0.0	& $0.755$	& $0.329$\\
Gl382	& 33	&  1.0	&  6.4	& {\boldmath $<10^{-9}$}	& 1581	& {\boldmath $<10^{-9}$}	& 1.037	& 1259	& {\boldmath $7.0~10^{-8}$}	& {\boldmath$<~10^{-3}$}\\
Gl388	& 41	&  0.8	& 23.7	& {\boldmath $<10^{-9}$}	& $4.2~10^{+4}$	& {\boldmath $<10^{-9}$}	& 2.616	& $4.2~10^{+4}$	& $1.000$	& $0.068$\\
Gl393	& 29	&  0.7	&  2.3	& {\boldmath $7.2~10^{-8}$}	& 347	& {\boldmath $<10^{-9}$}	& 0.371	& 332	& $0.406$	& $0.188$\\
LHS288	& 4	&  6.5	&  7.3	& $0.638$	&  4.4	& $0.218$	& 3.392	&  1.6	& $0.308$	& $0.176$\\
Gl402	& 4	&  2.0	&  1.0	& $0.956$	&  0.9	& $0.813$	& -0.073	&  0.9	& $0.998$	& $0.535$\\
Gl406	& 3	&  5.7	&  5.7	& $0.745$	&  3.9	& $0.141$	& -0.054	&  3.7	& $0.995$	& $0.663$\\
Gl413.1	& 17	&  1.1	&  3.0	& {\boldmath $1.4~10^{-4}$}	& 93.1	& {\boldmath $<10^{-9}$}	& 0.206	& 73.9	& {\boldmath $0.008$}	& {\boldmath$0.001$}\\
Gl433	& 50	&  0.8	&  3.3	& {\boldmath $<10^{-9}$}	& 985	& {\boldmath $<10^{-9}$}	& 0.885	& 897	& {\boldmath $2.4~10^{-7}$}	& $0.055$\\
Gl438	& 12	&  1.0	&  3.0	& {\boldmath $6.8~10^{-4}$}	& 73.2	& {\boldmath $<10^{-9}$}	& -0.656	& 58.7	& $0.118$	& $0.039$\\
Gl447	& 6	&  0.9	&  1.3	& $0.370$	& 12.2	& $0.032$	& 0.889	& 10.1	& $0.754$	& $0.415$\\
Gl465	& 15	&  1.7	&  2.2	& $0.269$	& 32.4	& {\boldmath $0.004$}	& -0.811	& 26.7	& $0.055$	& $0.153$\\
Gl479	& 58	&  0.9	&  4.1	& {\boldmath $<10^{-9}$}	& 1272	& {\boldmath $<10^{-9}$}	& -0.240	& 1269	& $1.000$	& $0.137$\\
LHS337	& 8	&  3.6	&  3.3	& $0.721$	&  9.1	& $0.243$	& 1.159	&  7.0	& $0.327$	& $0.188$\\
Gl480.1	& 8	&  3.3	&  1.9	& $0.960$	&  3.6	& $0.822$	& 0.422	&  3.2	& $0.716$	& $0.286$\\
Gl486	& 4	&  2.2	&  2.7	& $0.580$	&  4.9	& $0.180$	& 0.740	&  3.7	& $0.824$	& $0.229$\\
Gl514	& 8	&  0.6	&  1.5	& $0.025$	& 50.6	& {\boldmath $1.1~10^{-8}$}	& -0.383	& 46.6	& $0.888$	& $0.368$\\
Gl526	& 29	&  0.6	&  2.9	& {\boldmath $<10^{-9}$}	& 887	& {\boldmath $<10^{-9}$}	& 0.104	& 881	& $1.000$	& $0.893$\\
Gl536	& 12	&  0.7	&  2.7	& {\boldmath $1.4~10^{-4}$}	& 165	& {\boldmath $<10^{-9}$}	& -1.049	& 117	& $0.025$	& $0.054$\\
Gl551	& 32	&  1.3	&  2.1	& {\boldmath $0.007$}	& 136	& {\boldmath $<10^{-9}$}	& -0.234	& 125	& {\boldmath $0.006$}	& $0.352$\\
Gl555	& 7	&  1.6	&  3.1	& $0.127$	& 30.5	& {\boldmath $3.2~10^{-5}$}	& -0.105	& 27.1	& $0.838$	& $0.313$\\
Gl569A	& 6	&  1.0	&  4.1	& {\boldmath $0.007$}	& 121	& {\boldmath $<10^{-9}$}	& -1.704	& 109	& $0.910$	& $0.408$\\
Gl581	& 121	&  1.0	&  9.8	& {\boldmath $<10^{-9}$}	& $1.5~10^{+4}$	& {\boldmath $<10^{-9}$}	& -0.440	& $1.5~10^{+4}$	& $1.000$	& $0.021$\\
Gl588	& 21	&  0.6	&  1.1	& {\boldmath $0.009$}	& 63.1	& {\boldmath $2.3~10^{-6}$}	& 0.093	& 61.9	& $0.992$	& $0.041$\\
Gl618A	& 19	&  1.0	&  5.4	& {\boldmath $2.6~10^{-9}$}	& 543	& {\boldmath $<10^{-9}$}	& 3.681	& 78.1	& {\boldmath $<10^{-9}$}	& {\boldmath$<~10^{-3}$}\\
Gl628	& 23	&  0.6	&  3.6	& {\boldmath $<10^{-9}$}	& 646	& {\boldmath $<10^{-9}$}	& 0.616	& 618	& $0.622$	& $0.068$\\
Gl643	& 6	&  2.1	&  3.0	& $0.347$	& 15.8	& {\boldmath $0.007$}	& -1.753	&  6.7	& $0.082$	& $0.042$\\
Gl667C	& 143	&  1.0	&  4.0	& {\boldmath $<10^{-9}$}	& 2984	& {\boldmath $<10^{-9}$}	& 0.916	& 2801	& {\boldmath $<10^{-9}$}	& {\boldmath$0.002$}\\
Gl674	& 44	&  0.6	&  6.8	& {\boldmath $<10^{-9}$}	& 6588	& {\boldmath $<10^{-9}$}	& -2.166	& 6124	& {\boldmath $2.0~10^{-4}$}	& $0.142$\\
Gl678.1A	& 11	&  0.7	&  3.1	& {\boldmath $5.5~10^{-5}$}	& 237	& {\boldmath $<10^{-9}$}	& -0.216	& 229	& $0.978$	& $0.597$\\
Gl680	& 22	&  0.9	&  4.0	& {\boldmath $2.2~10^{-9}$}	& 507	& {\boldmath $<10^{-9}$}	& 3.203	& 116	& {\boldmath $<10^{-9}$}	& {\boldmath$<~10^{-3}$}\\
Gl682	& 12	&  1.1	&  2.2	& $0.027$	& 52.1	& {\boldmath $2.6~10^{-7}$}	& 0.685	& 42.5	& $0.149$	& $0.081$\\
Gl686	& 6	&  0.7	&  2.6	& $0.016$	& 64.6	& {\boldmath $<10^{-9}$}	& -1.693	& 28.8	& $0.097$	& $0.043$\\
Gl693	& 8	&  1.5	&  1.6	& $0.579$	& 26.8	& {\boldmath $3.7~10^{-4}$}	& 0.710	& 20.5	& $0.329$	& $0.638$\\
Gl699	& 22	&  0.6	&  1.5	& {\boldmath $7.7~10^{-5}$}	& 124	& {\boldmath $<10^{-9}$}	& -3.043	& 58.5	& {\boldmath $1.1~10^{-9}$}	& {\boldmath$<~10^{-3}$}\\
Gl701	& 12	&  0.7	&  2.8	& {\boldmath $9.3~10^{-5}$}	& 187	& {\boldmath $<10^{-9}$}	& 1.692	& 76.1	& {\boldmath $1.1~10^{-4}$}	& {\boldmath$0.001$}\\
GJ1224	& 4	&  6.3	&  8.0	& $0.553$	&  6.8	& $0.079$	& -0.880	&  5.2	& $0.838$	& $0.283$\\
G141-29	& 3	&  6.4	&  5.9	& $0.785$	&  3.4	& $0.181$	& 0.927	&  2.5	& $0.914$	& $0.330$\\
Gl729	& 8	&  1.8	& 20.9	& {\boldmath $1.4~10^{-6}$}	& 1105	& {\boldmath $<10^{-9}$}	& 5.628	& 1087	& $0.999$	& $0.473$\\
GJ1232	& 4	&  7.0	&  6.3	& $0.773$	&  4.2	& $0.241$	& 1.267	&  3.9	& $0.978$	& $0.832$\\
Gl752A	& 13	&  0.6	&  2.4	& {\boldmath $2.5~10^{-5}$}	& 246	& {\boldmath $<10^{-9}$}	& 2.646	& 113	& {\boldmath $9.1~10^{-5}$}	& {\boldmath$0.008$}\\
Gl754	& 7	&  2.3	&  3.6	& $0.242$	& 21.3	& {\boldmath $0.002$}	& -2.550	&  9.9	& $0.048$	& $0.074$\\
GJ1236	& 8	&  4.2	&  4.3	& $0.606$	&  8.9	& $0.257$	& 1.240	&  7.5	& $0.592$	& $0.302$\\
GJ1256	& 6	&  5.7	&  8.7	& $0.315$	& 12.2	& $0.032$	& -0.883	& 12.0	& $0.999$	& $0.377$\\
Gl803	& 4	&  1.3	& 89.1	& {\boldmath $1.3~10^{-5}$}	& $1.6~10^{+4}$	& {\boldmath $<10^{-9}$}	& -41.444	& $1.2~10^{+4}$	& $0.829$	& $0.188$\\
LHS3583	& 6	&  4.0	& 63.9	& {\boldmath $1.2~10^{-5}$}	& 1005	& {\boldmath $<10^{-9}$}	& 2.901	& 287	& $0.025$	& $0.034$\\
LP816-60	& 7	&  1.6	&  1.9	& $0.486$	& 13.0	& $0.042$	& 0.330	& 12.6	& $0.991$	& $0.913$\\
Gl832	& 54	&  0.6	&  7.2	& {\boldmath $<10^{-9}$}	& 9240	& {\boldmath $<10^{-9}$}	& 5.198	& 2092	& {\boldmath $<10^{-9}$}	& {\boldmath$<~10^{-3}$}\\
Gl846	& 31	&  0.8	&  5.2	& {\boldmath $<10^{-9}$}	& 1424	& {\boldmath $<10^{-9}$}	& 1.109	& 1383	& $0.735$	& $0.098$\\
LHS3746	& 5	&  2.5	&  2.3	& $0.733$	&  5.6	& $0.233$	& -0.380	&  3.5	& $0.470$	& $0.294$\\
Gl849	& 35	&  1.1	& 18.2	& {\boldmath $<10^{-9}$}	& $1.1~10^{+4}$	& {\boldmath $<10^{-9}$}	& -9.616	& 6602	& {\boldmath $<10^{-9}$}	& {\boldmath$<~10^{-3}$}\\
GJ1265	& 6	&  6.3	& 10.3	& $0.257$	& 19.0	& {\boldmath $0.002$}	& -0.482	& 15.2	& $0.667$	& $0.275$\\
LHS3799	& 3	&  5.7	&  2.9	& $0.942$	&  0.7	& $0.717$	& 1.878	&  0.2	& $0.538$	& $0.175$\\
Gl876	& 52	&  0.9	& 120	& {\boldmath $<10^{-9}$}	& $1.1~10^{+6}$	& {\boldmath $<10^{-9}$}	& 14.304	& $1.1~10^{+6}$	& $0.058$	& $0.018$\\
Gl877	& 43	&  1.3	&  4.0	& {\boldmath $<10^{-9}$}	& 882	& {\boldmath $<10^{-9}$}	& 0.020	& 881	& $1.000$	& $0.885$\\
Gl880	& 8	&  0.7	&  2.4	& {\boldmath $0.004$}	& 107	& {\boldmath $<10^{-9}$}	& -1.270	& 27.9	& {\boldmath $0.001$}	& {\boldmath$0.004$}\\
Gl887	& 75	&  0.7	&  4.3	& {\boldmath $<10^{-9}$}	& 4422	& {\boldmath $<10^{-9}$}	& 1.489	& 4061	& {\boldmath $<10^{-9}$}	& $0.057$\\
LHS543	& 7	&  2.4	&  2.9	& $0.466$	& 11.3	& $0.079$	& 1.105	&  7.3	& $0.206$	& $0.094$\\
Gl908	& 33	&  0.6	&  1.8	& {\boldmath $2.9~10^{-8}$}	& 385	& {\boldmath $<10^{-9}$}	& -0.352	& 366	& $0.125$	& $0.444$\\
LTT9759	& 7	&  1.7	&  4.4	& $0.033$	& 56.0	& {\boldmath $<10^{-9}$}	& 0.447	& 53.3	& $0.978$	& $0.326$\\
\end{longtable}}

\begin{figure*}[t]
\hspace{-1.5cm}
\includegraphics[width=1.2\linewidth]{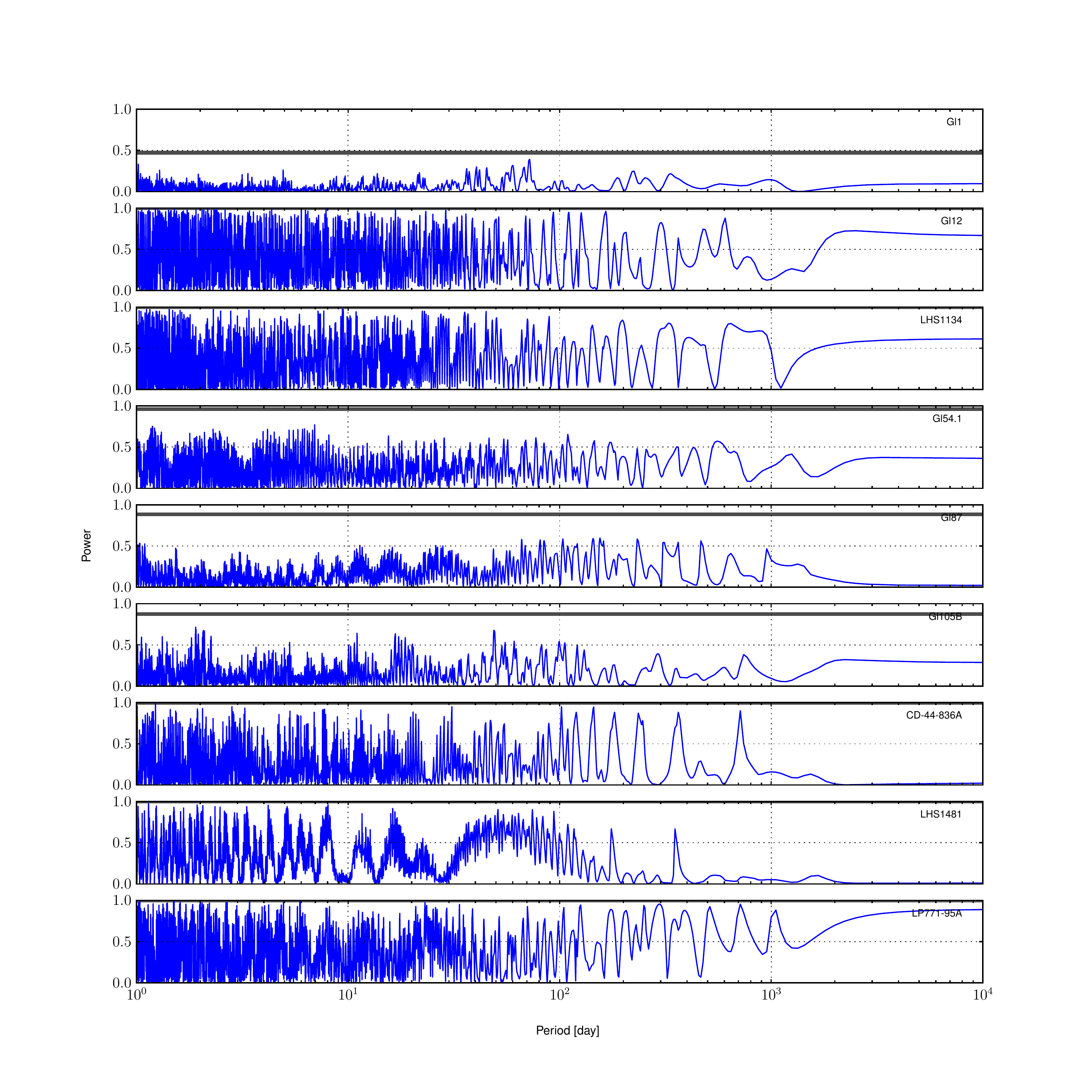}
\caption{\label{fig:periodos} Periodograms for RV time series with more than 6 measurements.}
\end{figure*}
\addtocounter{figure}{-1}
\begin{figure*}[t]
\hspace{-1.5cm}
\includegraphics[width=1.2\linewidth]{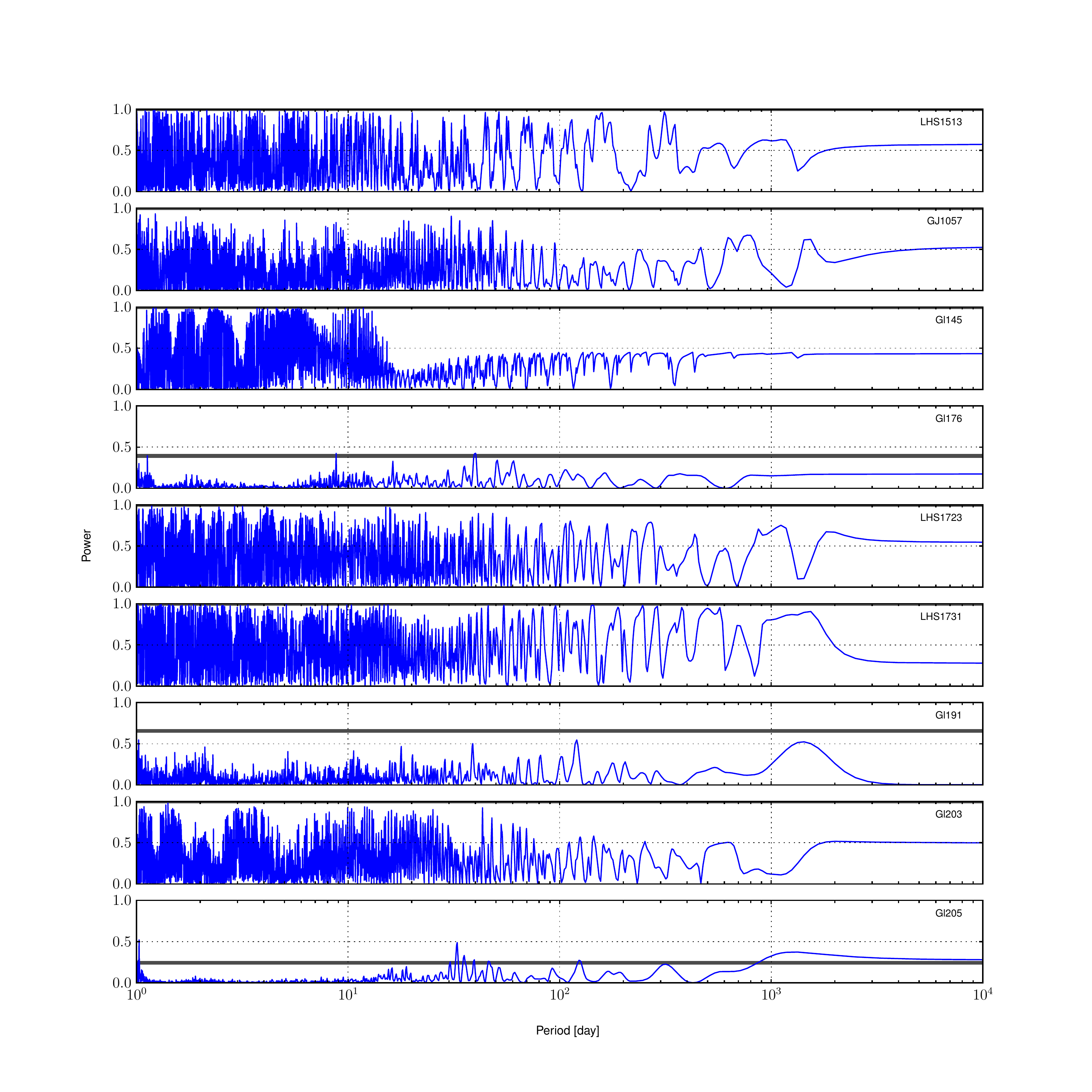}
\caption{Periodograms for RV time series (continued).}
\end{figure*}
\addtocounter{figure}{-1}
\begin{figure*}[t]
\hspace{-1.5cm}
\includegraphics[width=1.2\linewidth]{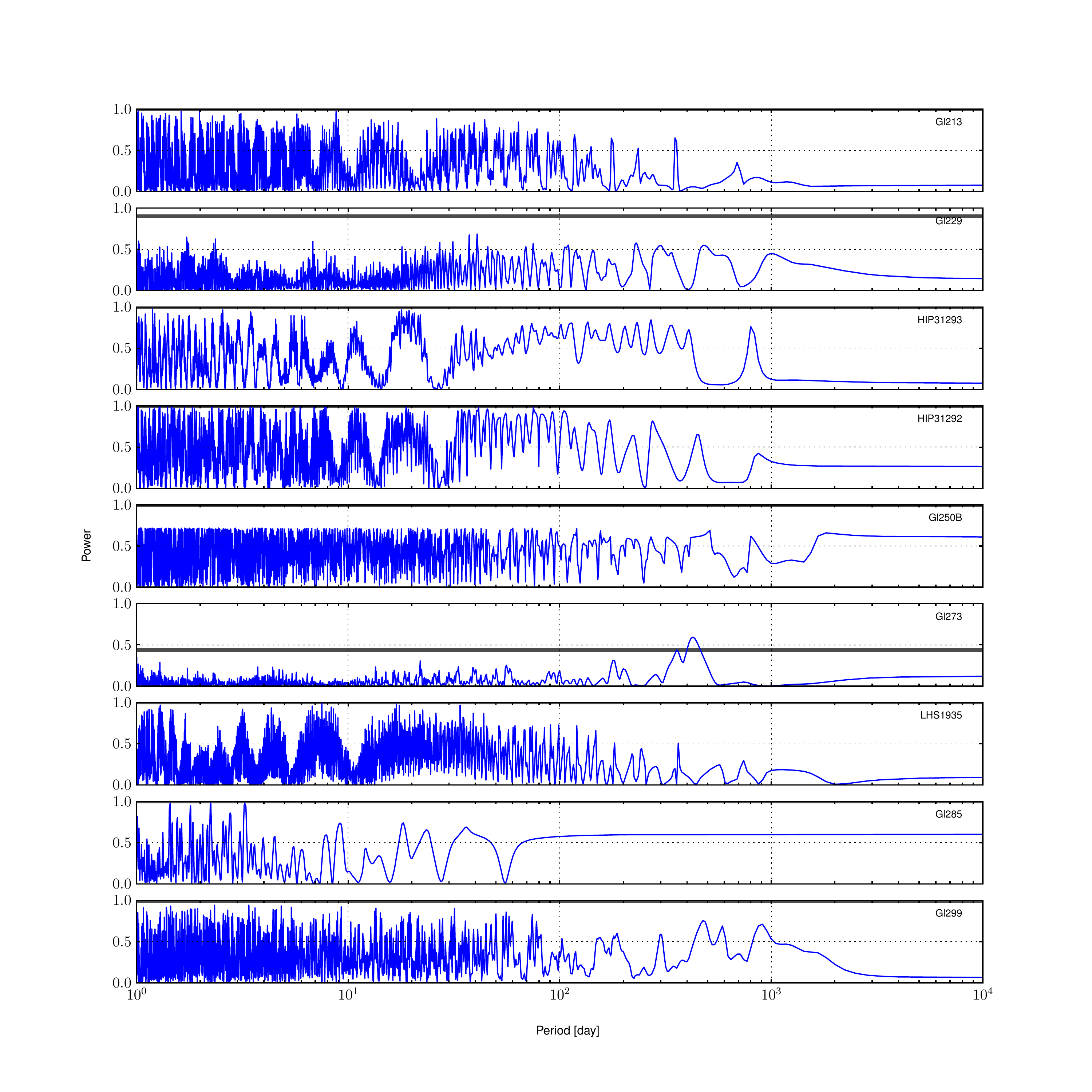}
\caption{Periodograms for RV time series (continued).}
\end{figure*}
\addtocounter{figure}{-1}
\begin{figure*}[t]
\hspace{-1.5cm}
\includegraphics[width=1.2\linewidth]{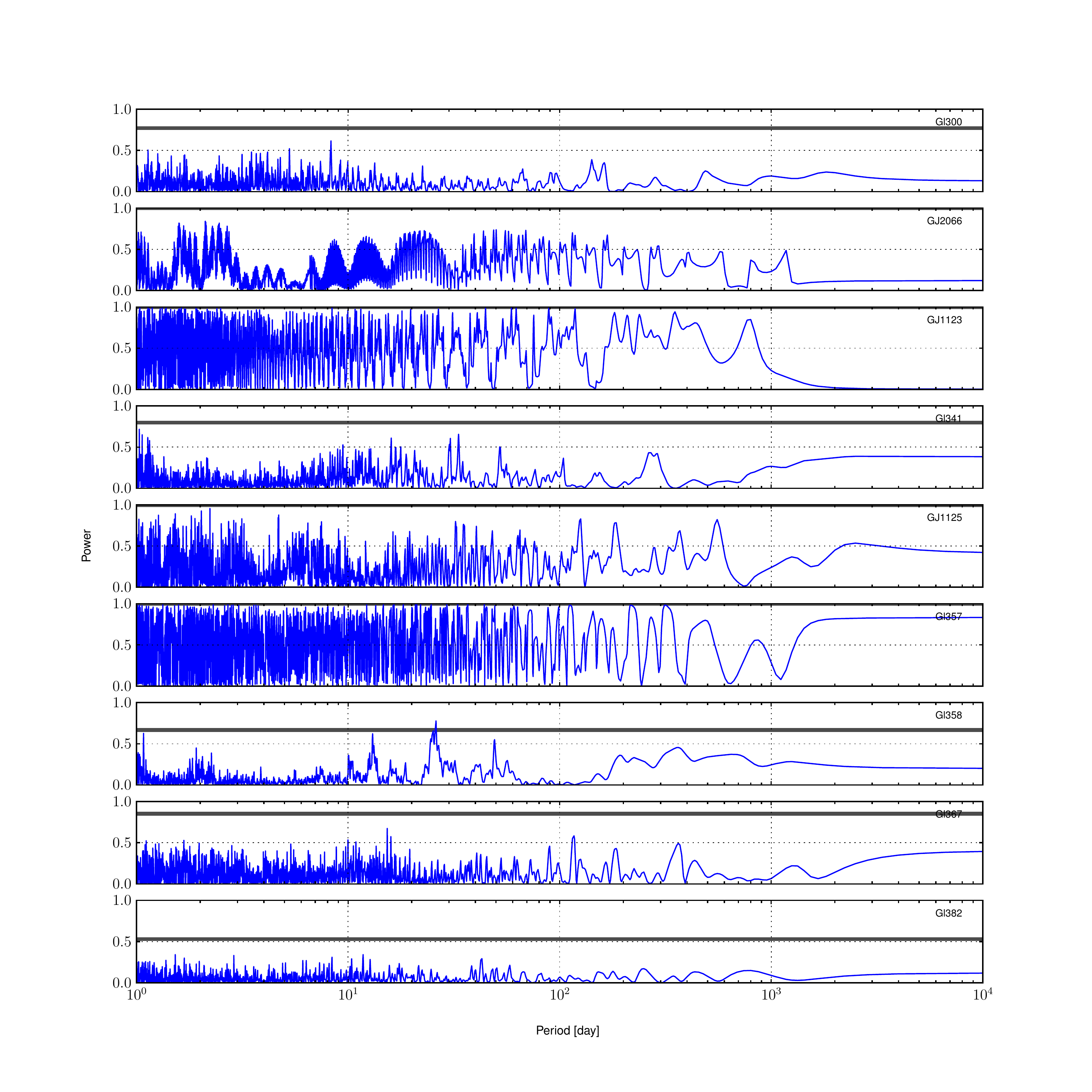}
\caption{Periodograms for RV time series (continued).}
\end{figure*}\addtocounter{figure}{-1}
\begin{figure*}[t]
\hspace{-1.5cm}
\includegraphics[width=1.2\linewidth]{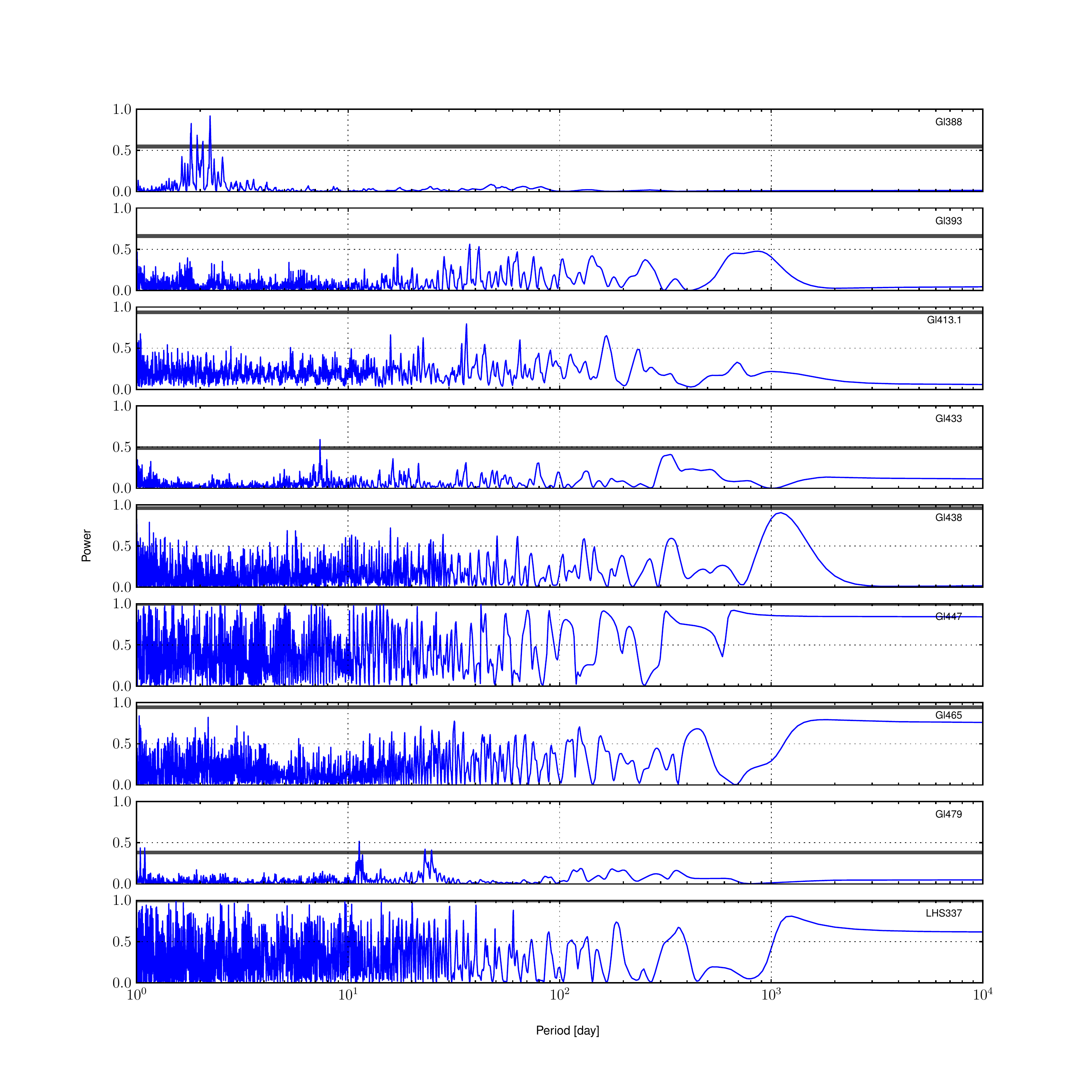}
\caption{Periodograms for RV time series (continued).}
\end{figure*}
\addtocounter{figure}{-1}
\begin{figure*}[t]
\hspace{-1.5cm}
\includegraphics[width=1.2\linewidth]{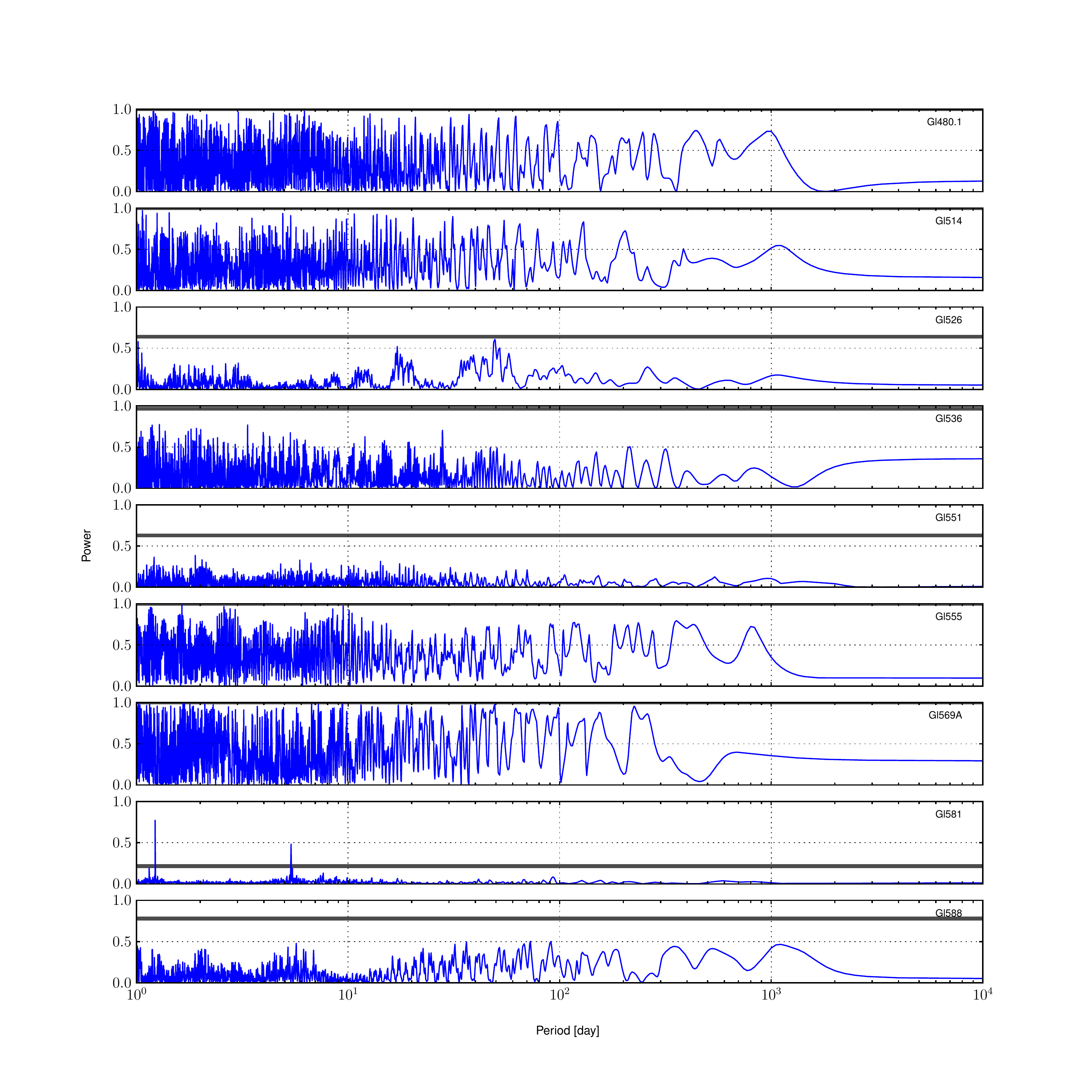}
\caption{Periodograms for RV time series (continued).}
\end{figure*}\addtocounter{figure}{-1}
\begin{figure*}[t]
\hspace{-1.5cm}
\includegraphics[width=1.2\linewidth]{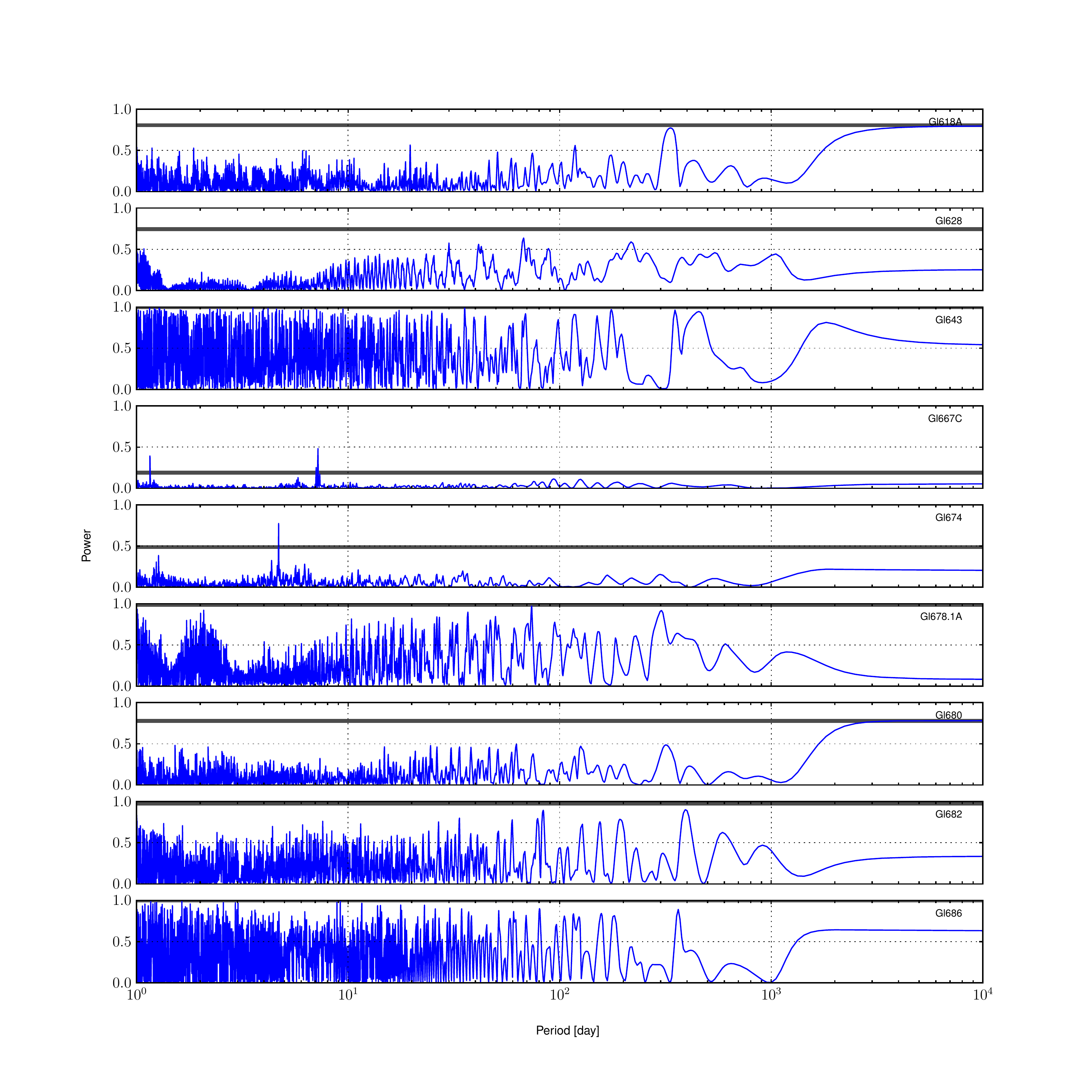}
\caption{Periodograms for RV time series (continued).}
\end{figure*}
\addtocounter{figure}{-1}
\begin{figure*}[t]
\hspace{-1.5cm}
\includegraphics[width=1.2\linewidth]{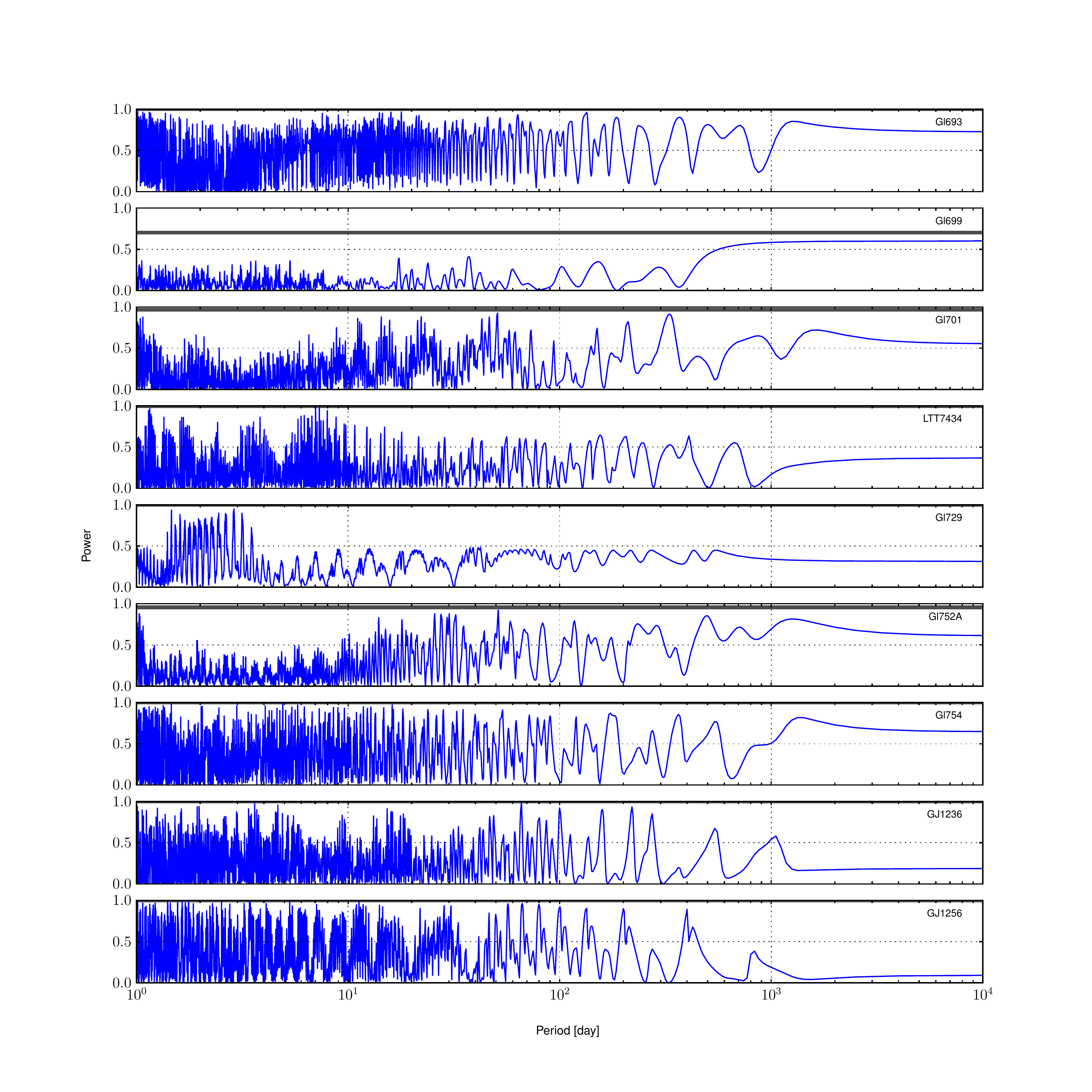}
\caption{Periodograms for RV time series (continued).}
\end{figure*}\addtocounter{figure}{-1}
\begin{figure*}[t]
\hspace{-1.5cm}
\includegraphics[width=1.2\linewidth]{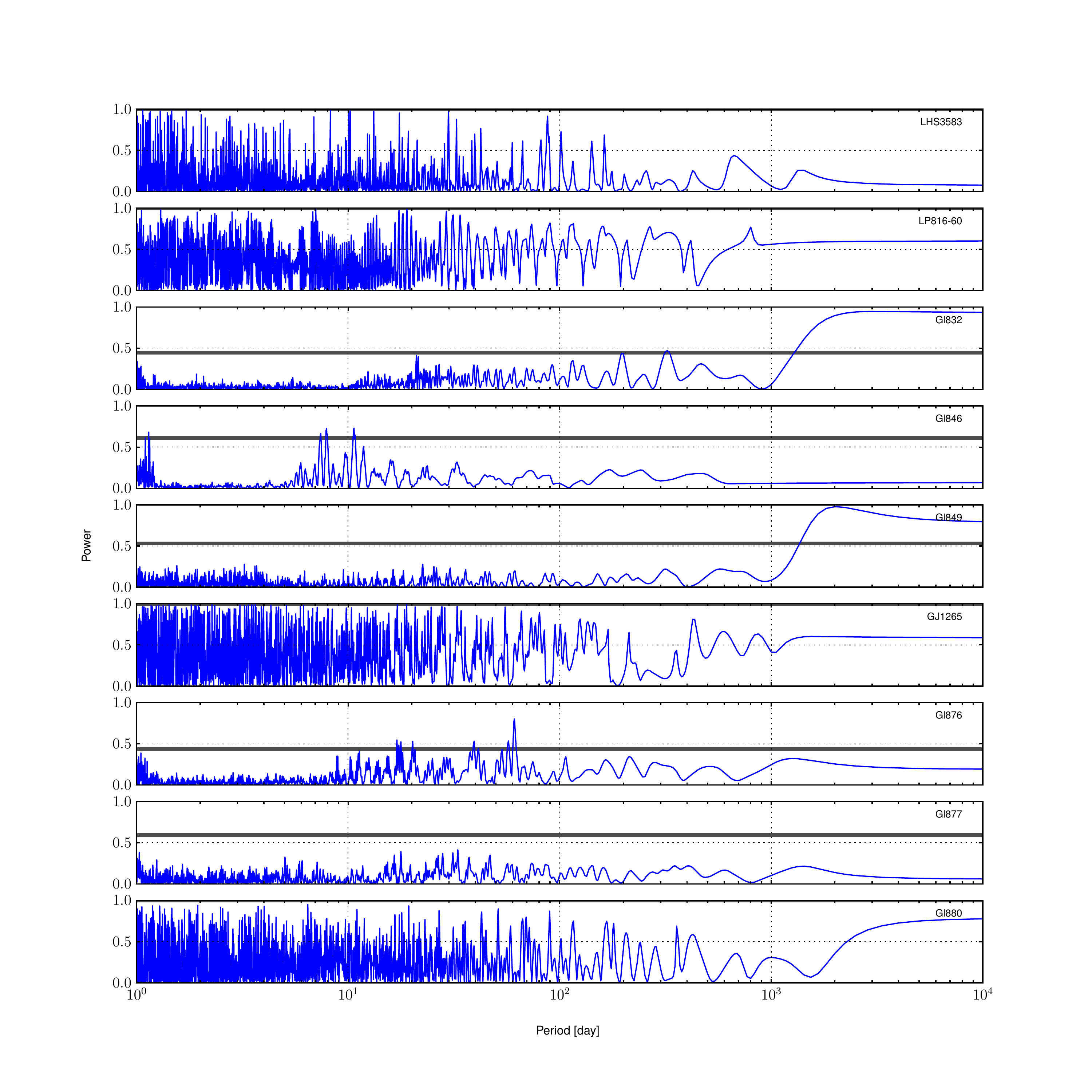}
\caption{Periodograms for RV time series (continued).}
\end{figure*}
\addtocounter{figure}{-1}
\begin{figure*}[t]
\hspace{-1.5cm}
\includegraphics[width=1.2\linewidth]{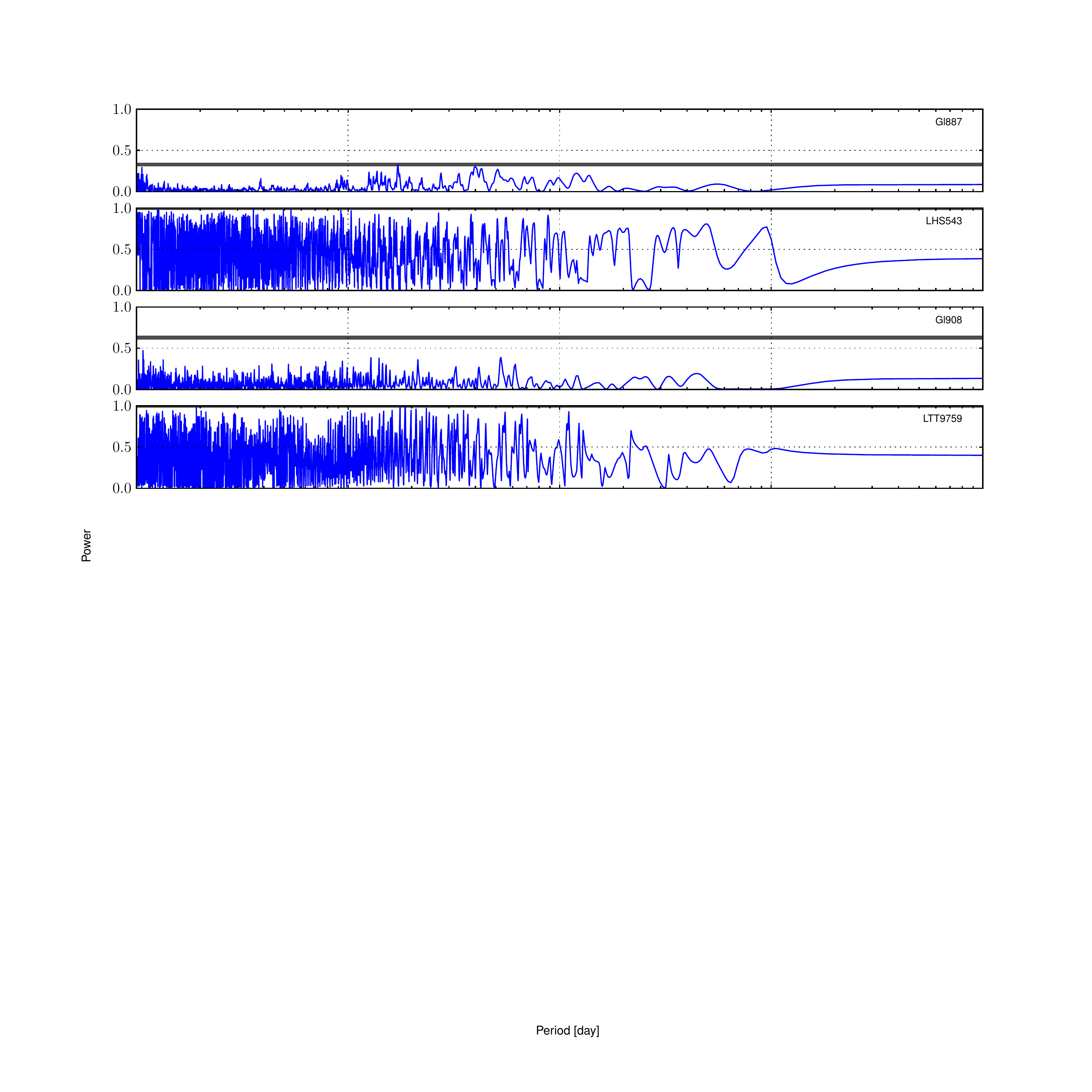}
\caption{Periodograms for RV time series (continued).}
\end{figure*}

\longtab{5}{
\begin{longtable}{lccccc}
\caption{\label{tab:periodos} Test for periodicity}\\
\hline\hline
Name    & Period     & K        & $\chi^2_{sin}$      & FAP   & FAP$_2$   \\
        & [day]      & [m/s]    &                      &       &       \\ \hline \endfirsthead
\caption{continued.}\\
\hline\hline
Name    & Period     & K        & $\chi^2_{sin}$      & FAP   & FAP$_2$   \\
        & [day]      & [m/s]    &                      &       &       \\ \hline \endhead \hline \endfoot
Gl1     &71.9   &1.50   &330    &$0.340$        &$0.236$\\
Gl12    &2.25   &4.69   & 0.0   &$0.936$        &$0.991$\\
LHS1134 &3.77   &17.43  & 0.1   &$0.935$        &$0.320$\\
Gl54.1  &6.22   &4.73   & 6.8   &$0.731$        &$0.634$\\
Gl87    &16.2   &1.53   &11.4   &$0.339$        &$0.331$\\
Gl105B  &4.04   &5.20   &14.7   &$0.713$        &$0.674$\\
CD-44-836A      &5.78   &4.43   & 0.2   &$0.804$        &$0.879$\\
LHS1481 &2.05   &6.74   & 0.1   &$0.535$        &$0.447$\\
\end{longtable}}

\begin{figure*}[t]
\hspace{-1.5cm}
\includegraphics[width=1.2\linewidth]{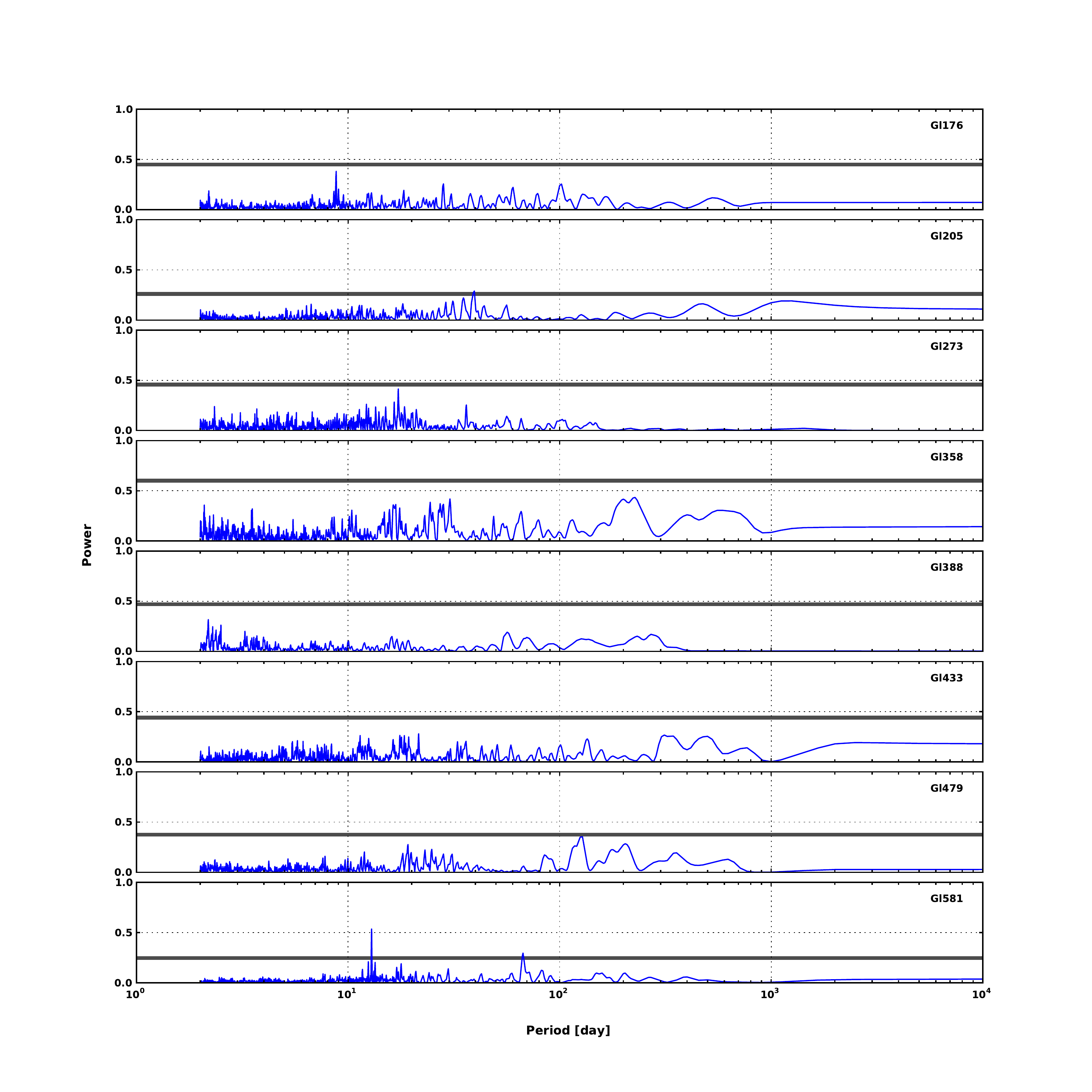}
\caption{\label{fig:periodos2} Periodograms for RV time series with 1st Keplerian signal removed.}
\end{figure*}
\addtocounter{figure}{-1}
\begin{figure*}[t]
\hspace{-1.5cm}
\includegraphics[width=1.2\linewidth]{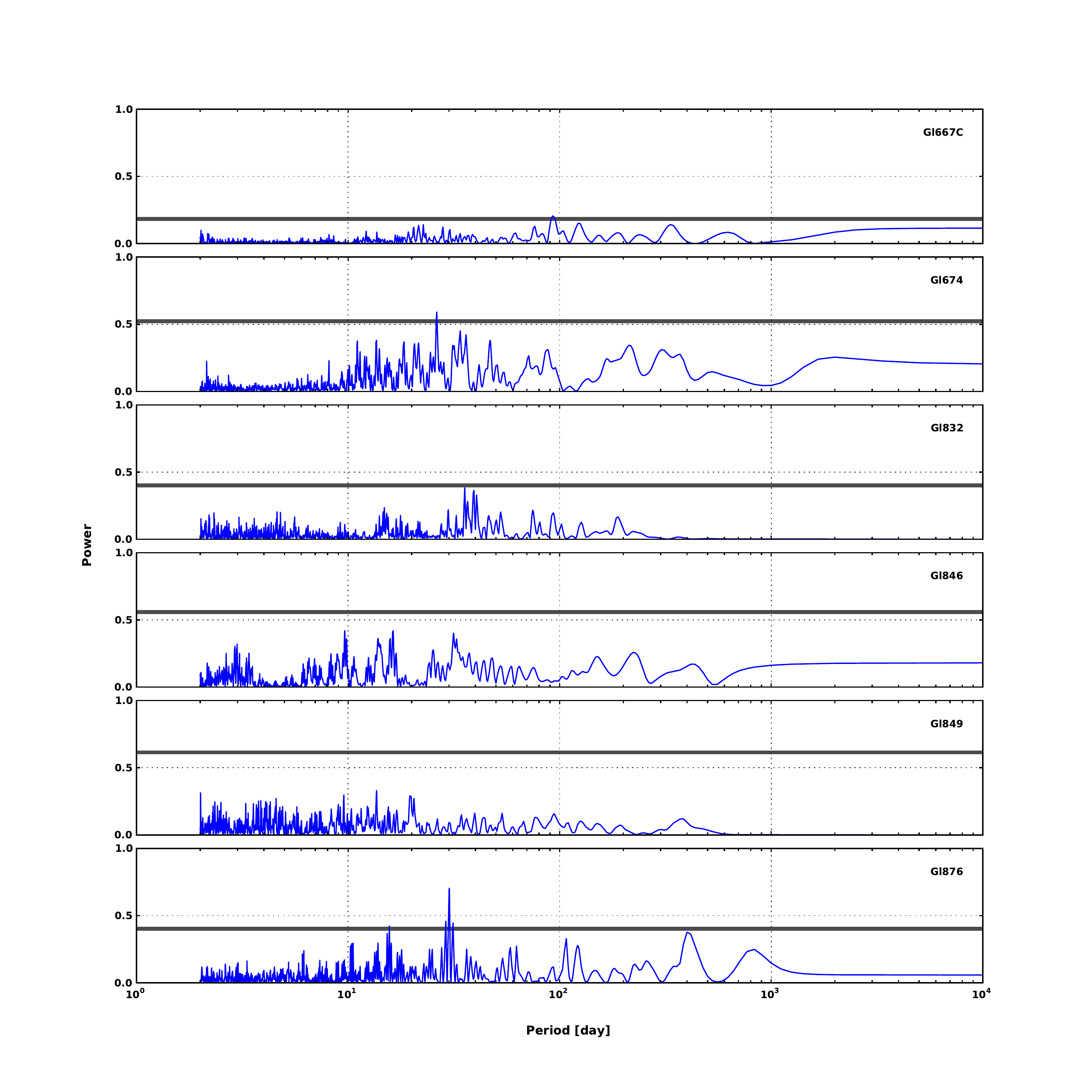}
\caption{Periodograms for RV time series with 1st Keplerian signal removed (continued).}
\end{figure*}

\longtab{6}{\begin{longtable}{lccccc}
\caption{\label{tab:per2} Test for periodicity after subtraction of the best keplerian fit}\\
\hline\hline
Name    & Period     & K        & $\chi^2_{sin}$      & FAP   \\
        & [day]      & [m/s]    &                      &       \\ \hline \endfirsthead
\caption{continued.}\\
\hline\hline
Name    & Period     & K        & $\chi^2_{sin}$      & FAP   \\
        & [day]      & [m/s]    &                      &       \\ \hline \endhead \hline \endfoot
Gl176   &40.1   &4.19   &537    &{\boldmath$<~10^{-3}$}\\
Gl205   &39.3   &1.91   &1524   &{\boldmath$<~10^{-3}$}\\
Gl273   &17.3   &1.58   &199    &$0.038$\\
Gl358   &2.08   &4.21   &158    &$0.278$\\
Gl433   &5.76   &1.61   &256    &$0.415$\\
Gl479   &11.3   &2.07   &341    &$0.040$\\
Gl581   &12.9   &3.64   &946    &{\boldmath$<~10^{-3}$}\\
Gl667C  &28.1   &1.72   &1096   &$0.012$\\
Gl674   &26.2   &3.69   &540    &{\boldmath$<~10^{-3}$}\\
Gl832   &35.7   &1.68   &324    &$0.016$\\
Gl846   &30.8   &4.73   &142    &$0.014$\\
Gl849   &19.6   &1.61   &132    &$0.999$\\
Gl876   &30.1   &47.52  &$2.6~10^{+4}$  &{\boldmath$<~10^{-3}$}\\
\end{longtable}}

\longtabL{7}{

\begin{landscape}
\begin{longtable}{llccr@{$\pm$}lr@{$\pm$}lr@{$\pm$}lr@{$\pm$}lr@{$\pm$}l}
\caption{\label{tab:paramKep} Keplerian solutions with various models}\\
\hline\hline 
Name	& Model	&$\gamma$ & Pl.	& \multicolumn{2}{c}{$P$}&\multicolumn{2}{c}{$K$}&\multicolumn{2}{c}{$T_0$}	& \multicolumn{2}{c}{$e$}&	\multicolumn{2}{c}{$\omega$}  \\
                   &                & [km/s]         &            &\multicolumn{2}{c}{[day]}&\multicolumn{2}{c}{[m/s]}&\multicolumn{2}{c}{BJD$-2,400,000.0$ [day]}&     &         &         &                \\ \hline \endfirsthead
\caption{continued.}\\ \hline\hline
Name	& Model	& $\gamma$ & Pl.	& \multicolumn{2}{c}{$P$}&\multicolumn{2}{c}{$K$}&\multicolumn{2}{c}{$T_0$}	& \multicolumn{2}{c}{$e$}&	\multicolumn{2}{c}{$\omega$}  \\
                   &                & [km/s]         &            &\multicolumn{2}{c}{[day]}&\multicolumn{2}{c}{[m/s]}&\multicolumn{2}{c}{BJD$-2,400,000.0$ [day]}&     &         &         &                \\ \hline \endhead \hline \endfoot
Gl1   & 1 pl.         & 25.7004 $\pm$ 0.0007 &  b &72.2558 & 0.3694 & 1.95 & 0.46 & 54233.8776 & 4.1981 & 0.43 & 0.18 & -66 & 29 \\
~	& 2 pl.	&  25.7004$\pm$0.0010	& b 	&  2.6469	& 0.0002	&  1.9	& 0.6	&  54208.9992	& 0.0624	&  0.7	& 0.1	&  2.7	& 15.3\\ 
~	& ~	         &  ~	                                     & c 	&  72.1756	& 0.3091	&  2.1	& 0.5	&  54236.2643	& 2.9804	&  0.5	& 0.1	&  -43.3	& 20.7\\ 
~	& 3 pl.	&  25.7009$\pm$0.8959	& b 	&  2.9756	& 0.0001	&  4.5	& 110.8	&  54206.7547	& 0.1894	&  0.9	& 1.3	&  -34.8	& 94.0\\ 
~	& ~	         &  ~	                                     & c 	&  71.6748	& 0.0281	&  35.6	& 613.0	&  54244.9922	& 0.8154	&  1.0	& 0.4	&  7.3	& 63.1\\ 
~	& ~	         & ~	                                     & d 	&  216.3889	& 1.3191	&  2.8	& 0.5	&  54368.6900	& 7.3816	&  0.5	& 0.1	&  -26.1	& 13.4\\ 
~	& 4 pl.	& 25.6995$\pm$0.8533	& b 	&  2.8253	& 0.0005	&  3.7	& 4.1	&  54207.4751	& 0.1024	&  0.9	& 0.1	&  -95.1	& 78.6\\ 
~	& ~	         & ~					& c 	&  60.1075	& 0.3006	&  15.9	& 500.1	&  54260.6315	& 7.9037	&  1.0	& 1.5	&  -153.1	& 65.9\\
\end{longtable}
\end{landscape}

}

\longtab{8}{

\begin{longtable}{llccccc}
\caption{\label{tab:fapKepd0} Model comparison based on false-alarm probabilities (FAP). 
Tested models are composed of either {\it a constant}, {\it 1}, {\it 2}, {\it 3} or {\it 4} planets. 
Their FAP are in boldface when the more complex model is found to be a statistically significant improvement over the simpler model.}\\
\hline\hline
Name    & Model  &       $\sqrt{\chi_{red}^2}$ &       Cste      &       1 planet      &       2 planets      & 3 planets \\\hline  \endfirsthead
\caption{continued.}\\ \hline\hline
Name    & Model   &       $\sqrt{\chi_{red}^2}$ &       Cste      &       1 planet      &       2 planets      & 3 planets \\\hline \endhead \hline \endfoot
Gl1     & 1 planet & 8.3        & 67.2  & ~     & ~     & ~ \\
~       & 2 planets & 4.1       & ~     & 94.6  & ~     & ~ \\
~       & 3 planets & 2.7       & 61.2  & 9.9   & ~     & ~ \\
Gl54.1  & 1 planet & 0.3        & 26.1  & ~     & ~     & ~ \\
Gl87    & 1 planet & 0.9        & 65.7  & ~     & ~     & ~ \\
Gl105B  & 1 planet & 0.7        & 35.4  & ~     & ~     & ~ \\
Gl176   & 1 planet & 25.9       & {\boldmath $0.1$}     & ~     & ~     & ~ \\
~       & 2 planets & 9.2       & ~     & {\boldmath $0.2$}     & ~     & ~ \\
\end{longtable}

}

\longtab{9}{

\begin{longtable}{llccccc}
\caption{\label{tab:fapKepd1} Model comparison based on false-alarm probabilities (FAP). 
Tested models are composed of either {\it a constant}, {\it 1}, {\it 2}, {\it 3} or {\it 4} planets. 
Their FAP are in boldface when the more complex model is found to be a statistically significant improvement over the simpler model.}\\
\hline\hline
Name    & Model  &       $\sqrt{\chi_{red}^2}$ &       Drift      &       1 pl.$+$dr.      &       2 pl.$+$dr.      & 3 pl.$+$dr. \\\hline  \endfirsthead
\caption{continued.}\\ \hline\hline
Name    & Model   &       $\sqrt{\chi_{red}^2}$ &       Drift      &       1 pl.$+$dr.      &       2 pl.$+$dr.      & 3 pl.$+$dr. \\\hline \endhead \hline \endfoot
Gl1     & 1 pl.$+$dr. & 8.1     & 75.3  & ~     & ~     & ~ \\
~       & 2 pl.$+$dr. & 4.3     & ~     & 92.3  & ~     & ~ \\
~       & 3 pl.$+$dr. & 2.0     & 37.3  & 13.3  & ~     & ~ \\
Gl54.1  & 1 pl.$+$dr. & 0.3     & 65.9  & ~     & ~     & ~ \\
Gl87    & 1 pl.$+$dr. & 0.9     & 71.1  & ~     & ~     & ~ \\
Gl105B  & 1 pl.$+$dr. & 0.7     & 57.0  & ~     & ~     & ~ \\
Gl176   & 1 pl.$+$dr. & 26.6    & {\boldmath $0.4$}     & ~     & ~     & ~ \\
~       & 2 pl.$+$dr. & 9.4     & ~     & {\boldmath $0.7$}     & ~     & ~ \\
\end{longtable}

}

\begin{figure}
\includegraphics[width=.9\linewidth]{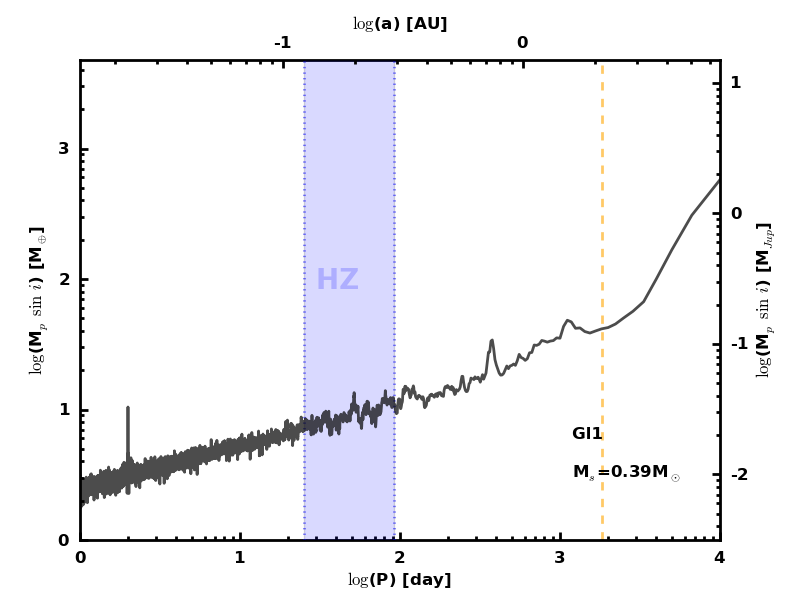}
\includegraphics[width=.9\linewidth]{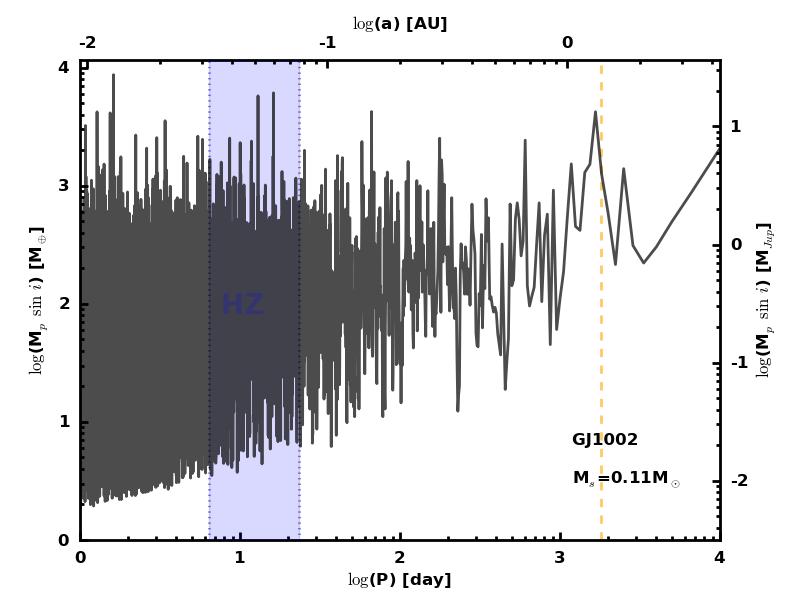}
\includegraphics[width=.9\linewidth]{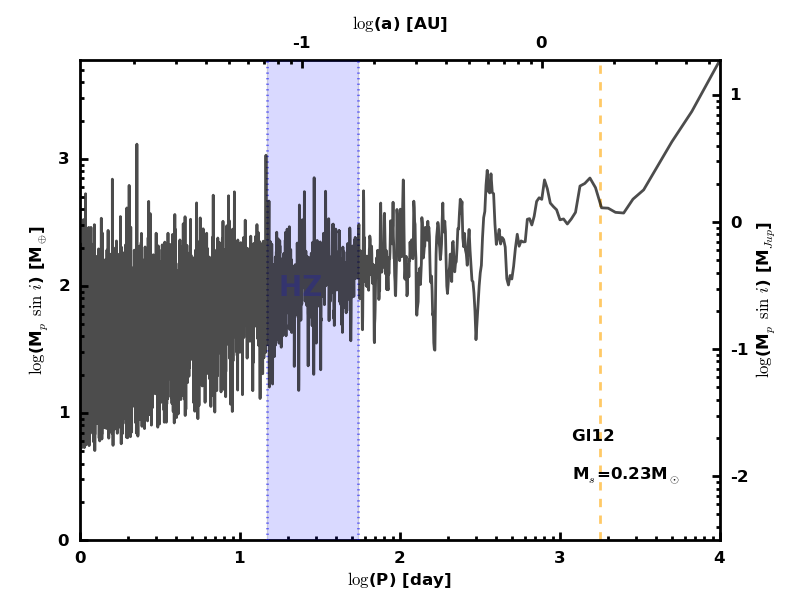}
\includegraphics[width=.9\linewidth]{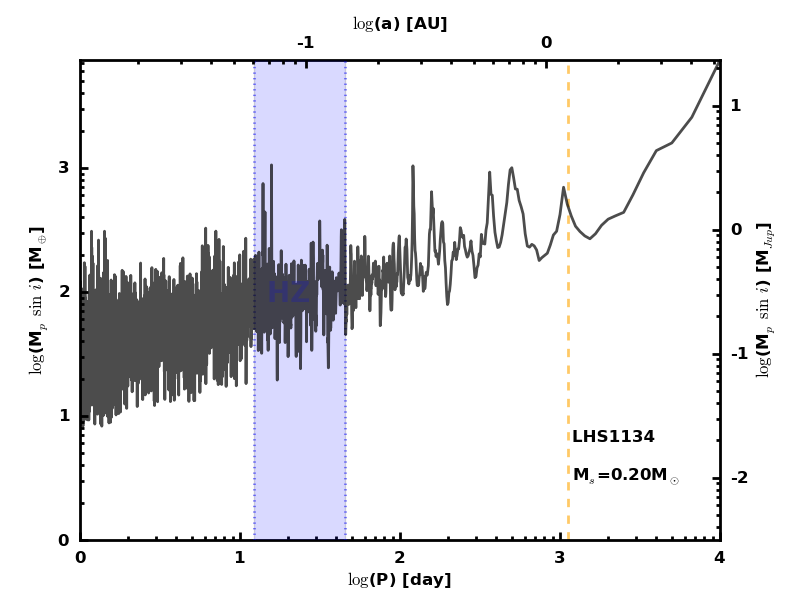}
\end{figure}\begin{figure}
\includegraphics[width=.9\linewidth]{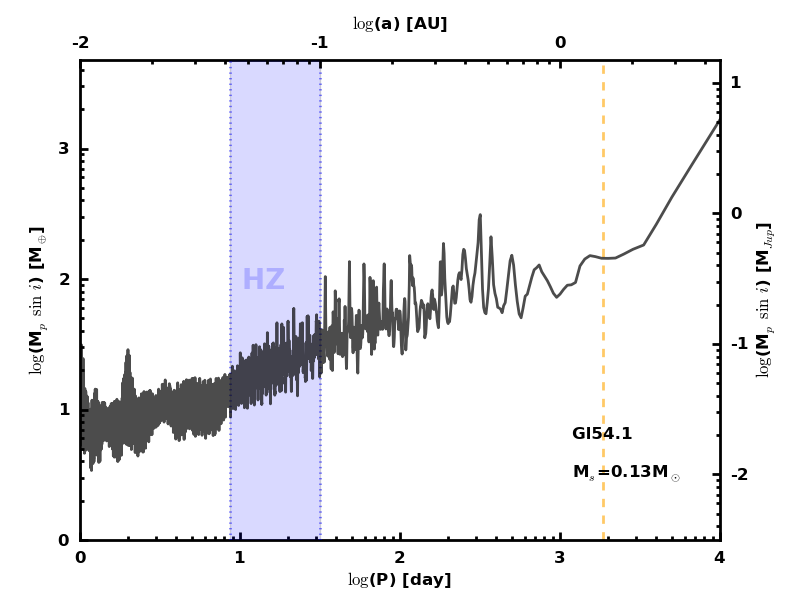}
\includegraphics[width=.9\linewidth]{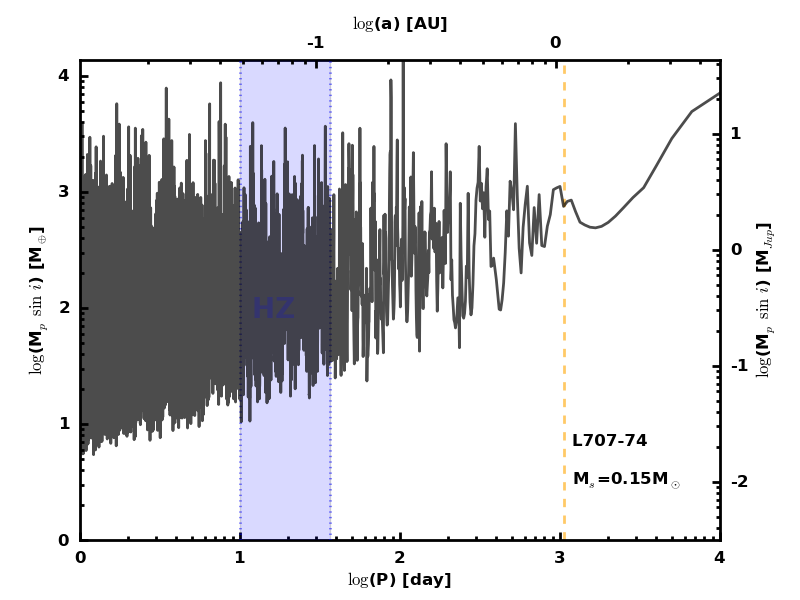}
\includegraphics[width=.9\linewidth]{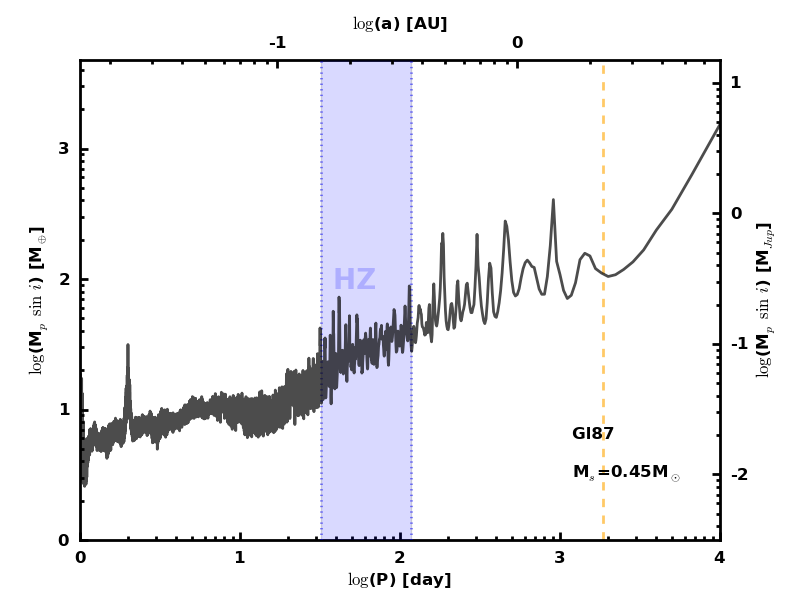}
\includegraphics[width=.9\linewidth]{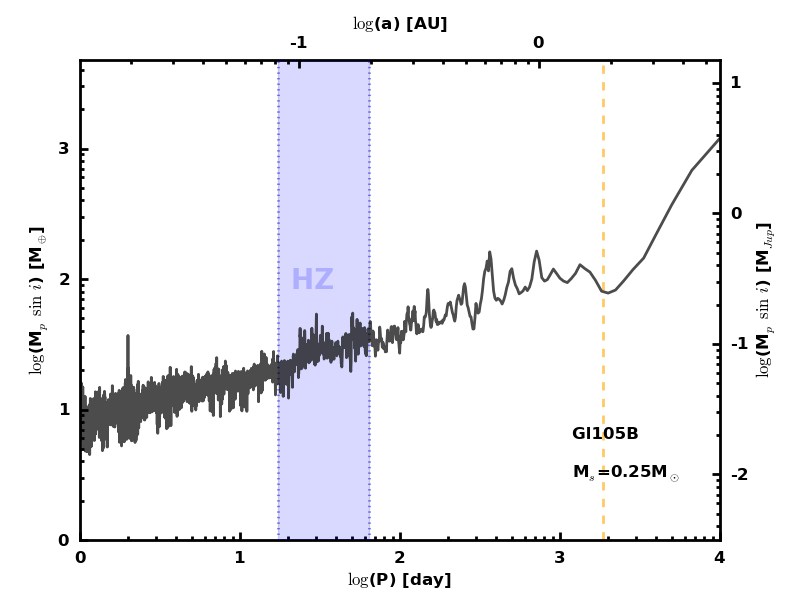}
\end{figure}\begin{figure}
\includegraphics[width=.9\linewidth]{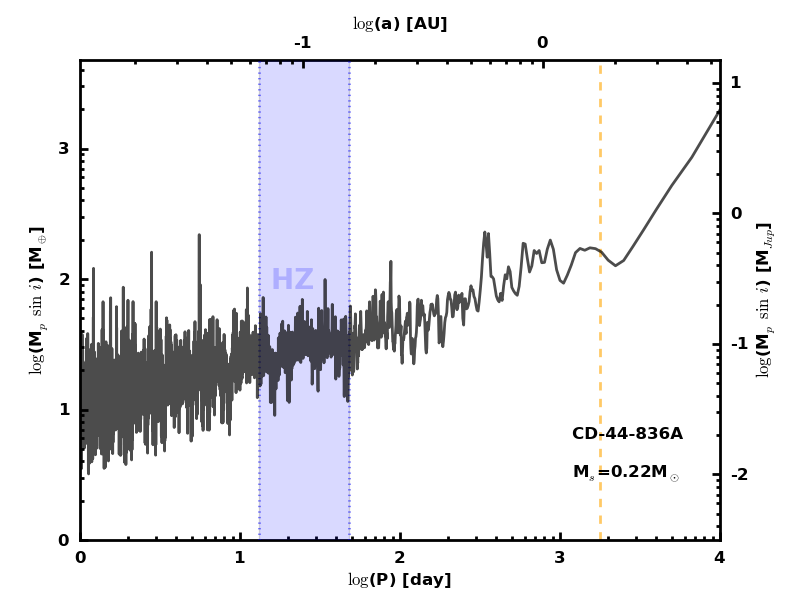}
\includegraphics[width=.9\linewidth]{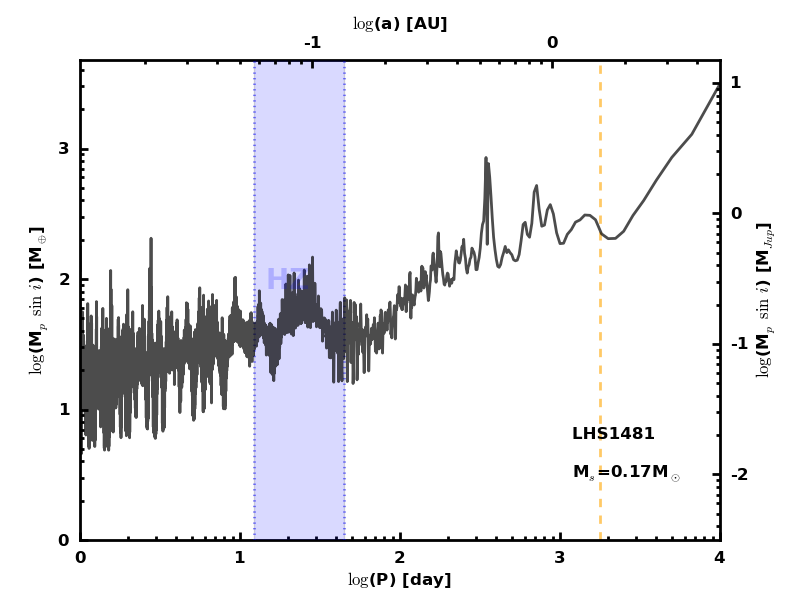}
\includegraphics[width=.9\linewidth]{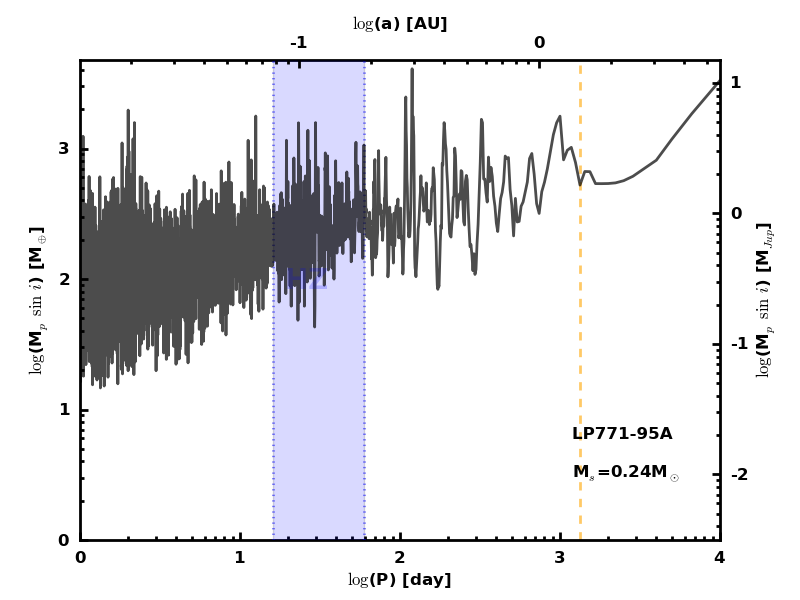}
\includegraphics[width=.9\linewidth]{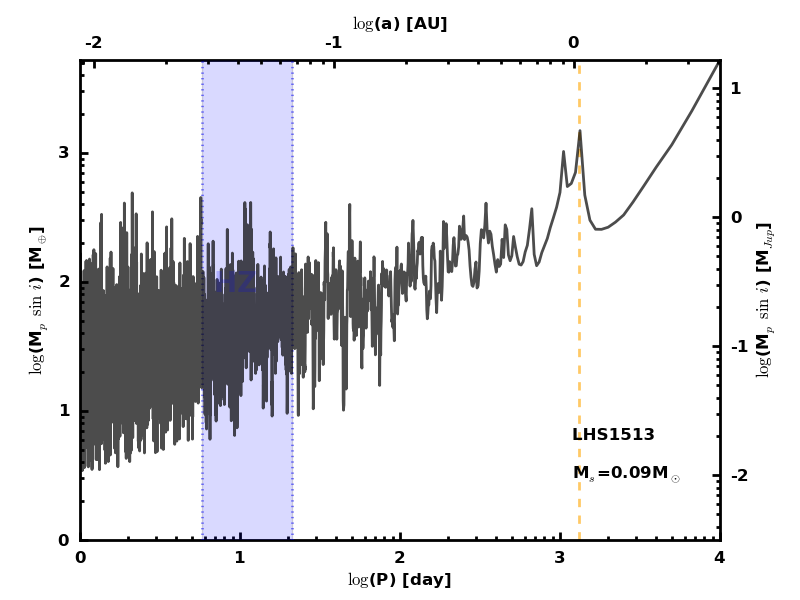}
\end{figure}\begin{figure}
\includegraphics[width=.9\linewidth]{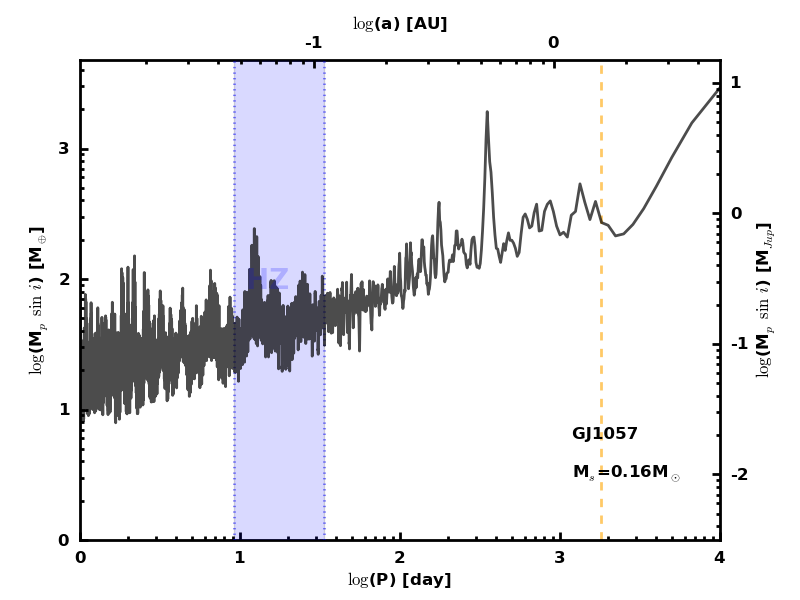}
\includegraphics[width=.9\linewidth]{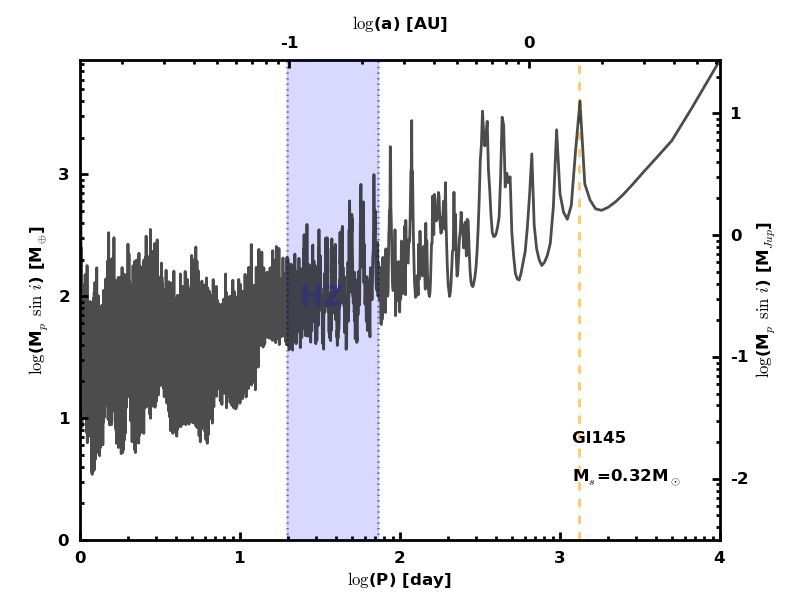}
\includegraphics[width=.9\linewidth]{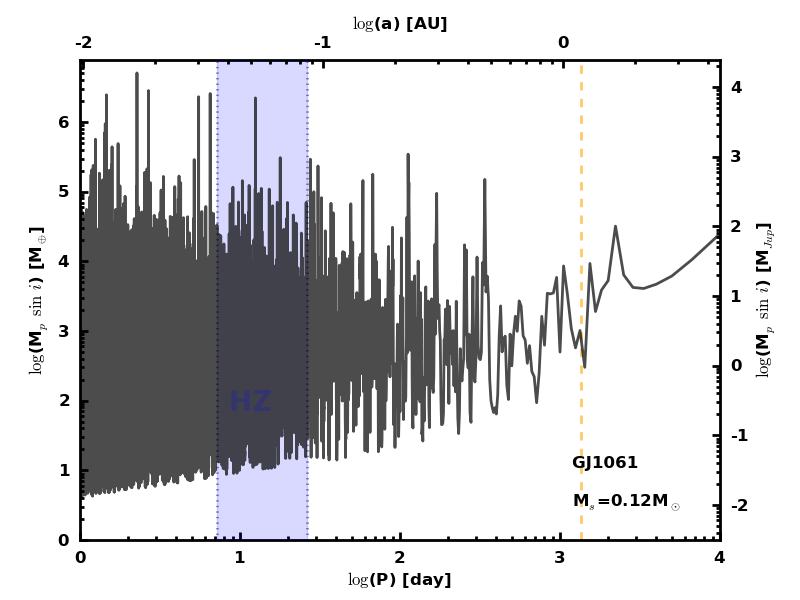}
\includegraphics[width=.9\linewidth]{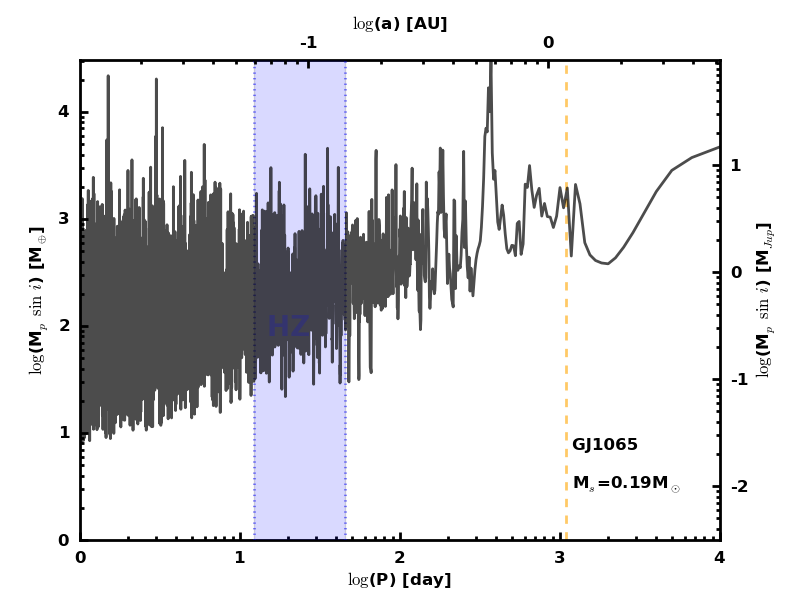}
\end{figure}\clearpage\begin{figure}
\includegraphics[width=.9\linewidth]{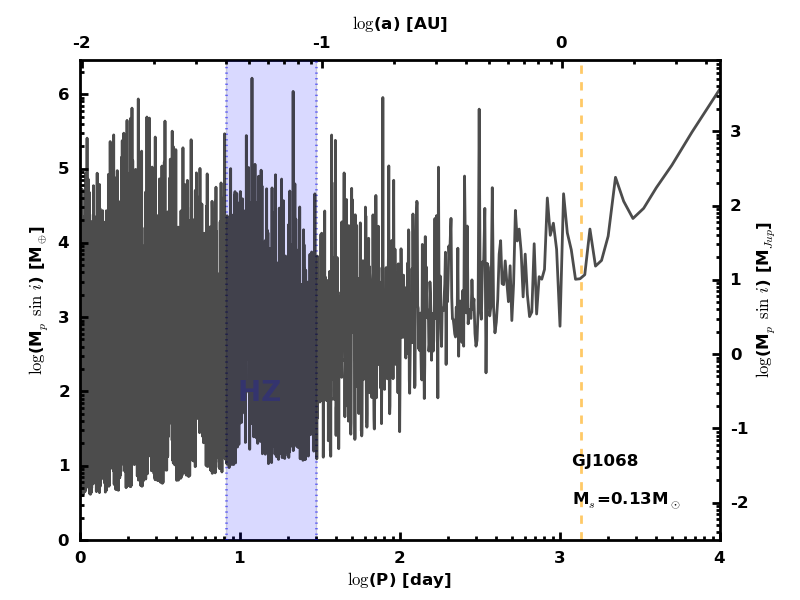}
\includegraphics[width=.9\linewidth]{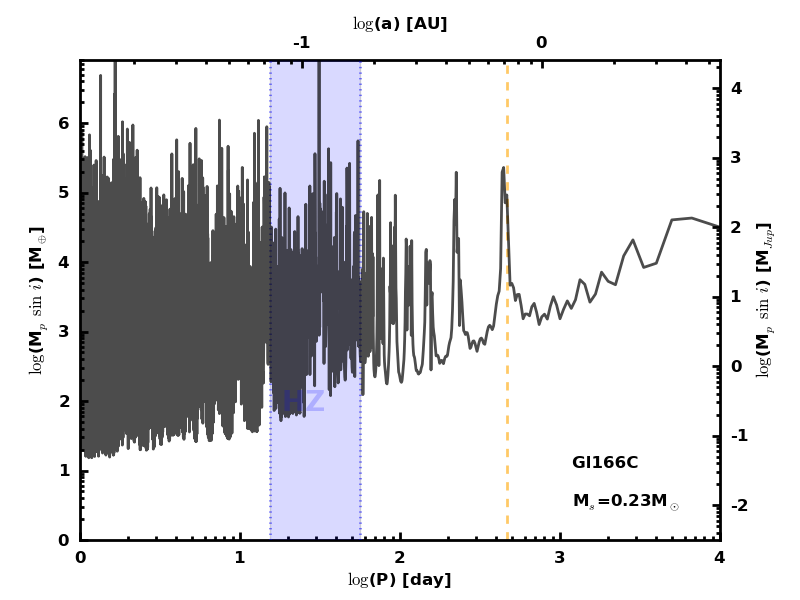}
\includegraphics[width=.9\linewidth]{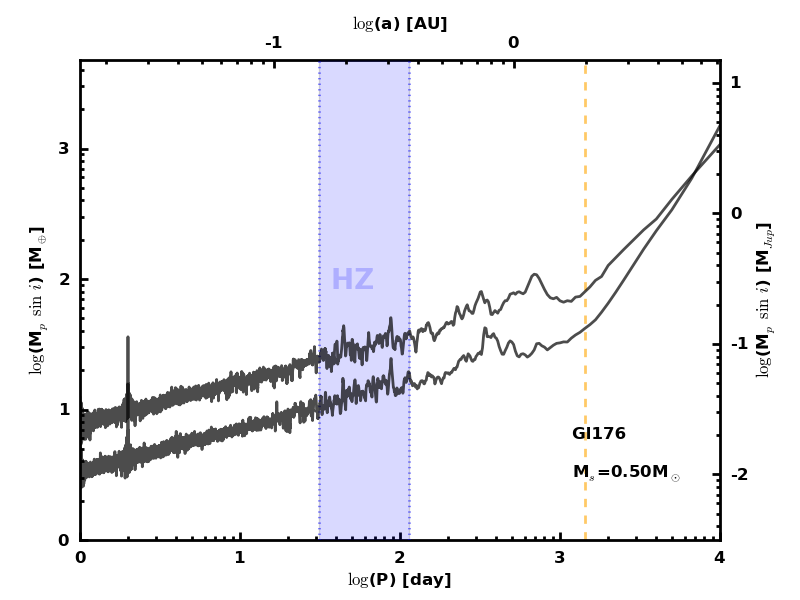}
\includegraphics[width=.9\linewidth]{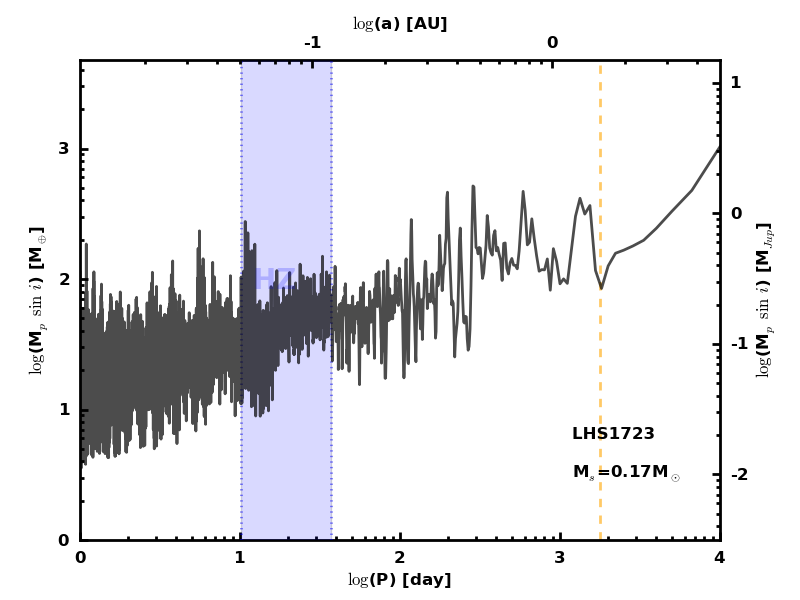}
\end{figure}\begin{figure}
\includegraphics[width=.9\linewidth]{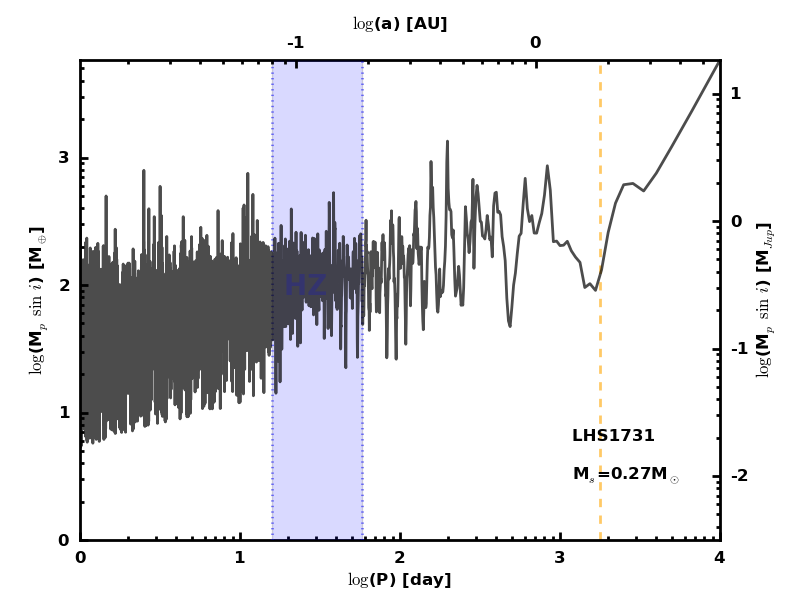}
\includegraphics[width=.9\linewidth]{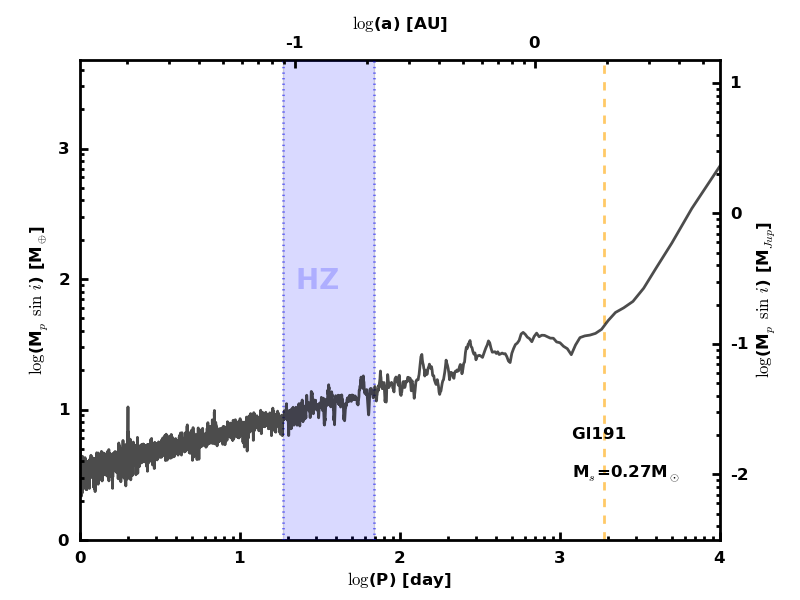}
\includegraphics[width=.9\linewidth]{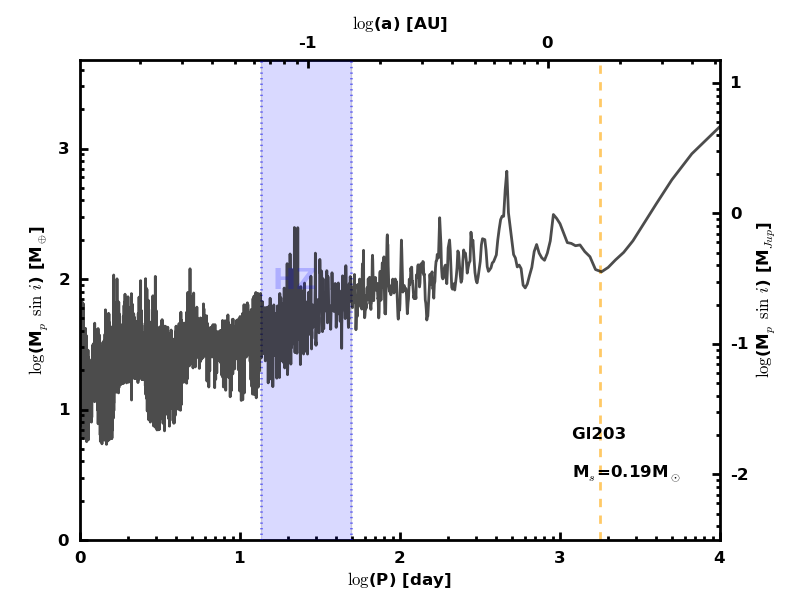}
\includegraphics[width=.9\linewidth]{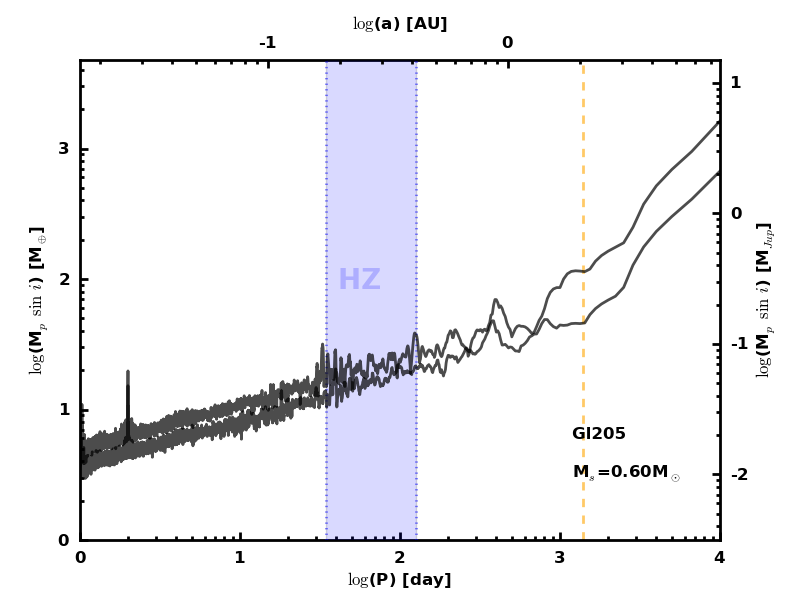}
\end{figure}\begin{figure}
\includegraphics[width=.9\linewidth]{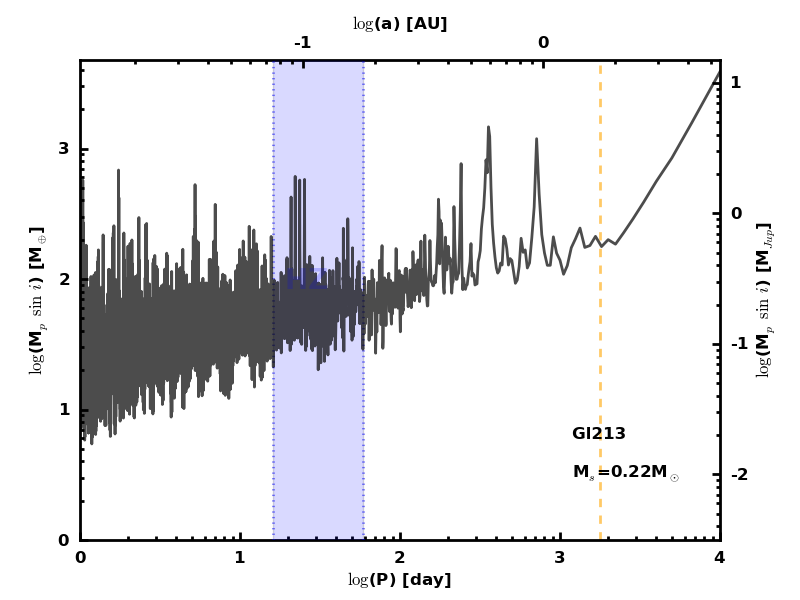}
\includegraphics[width=.9\linewidth]{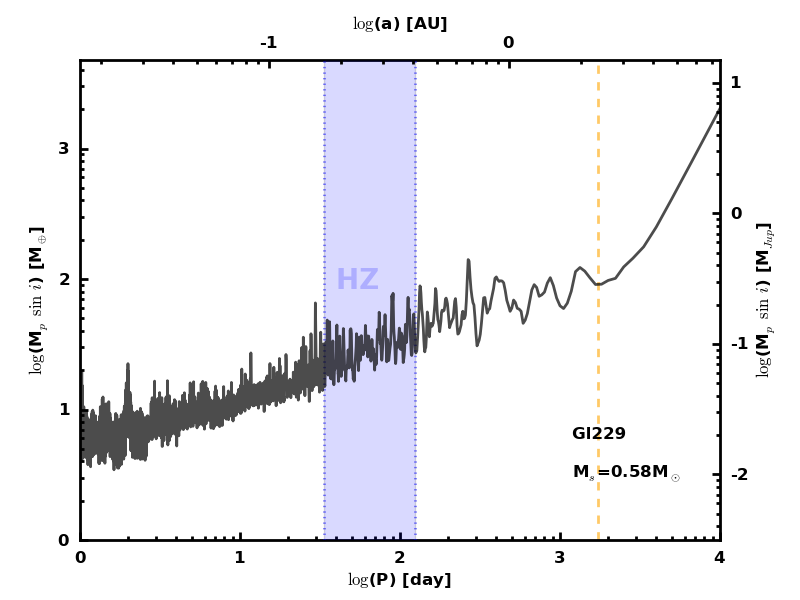}
\includegraphics[width=.9\linewidth]{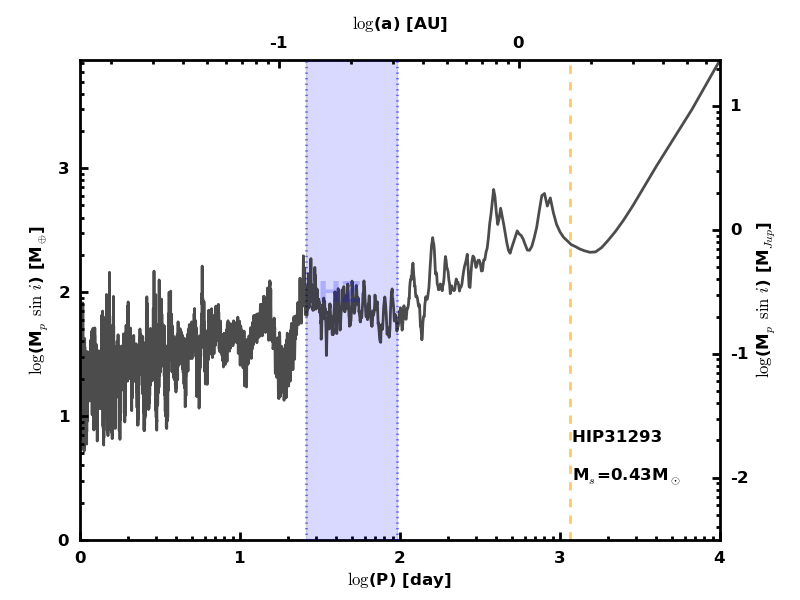}
\includegraphics[width=.9\linewidth]{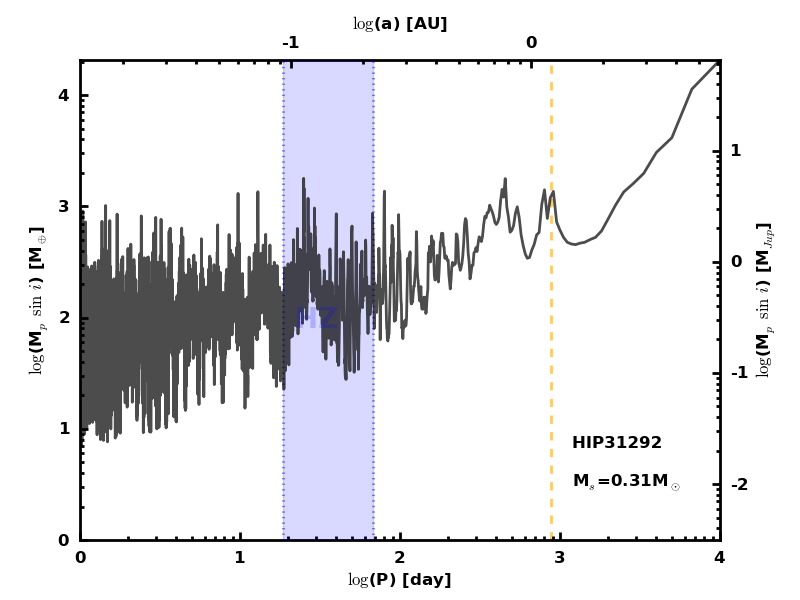}
\end{figure}\begin{figure}
\includegraphics[width=.9\linewidth]{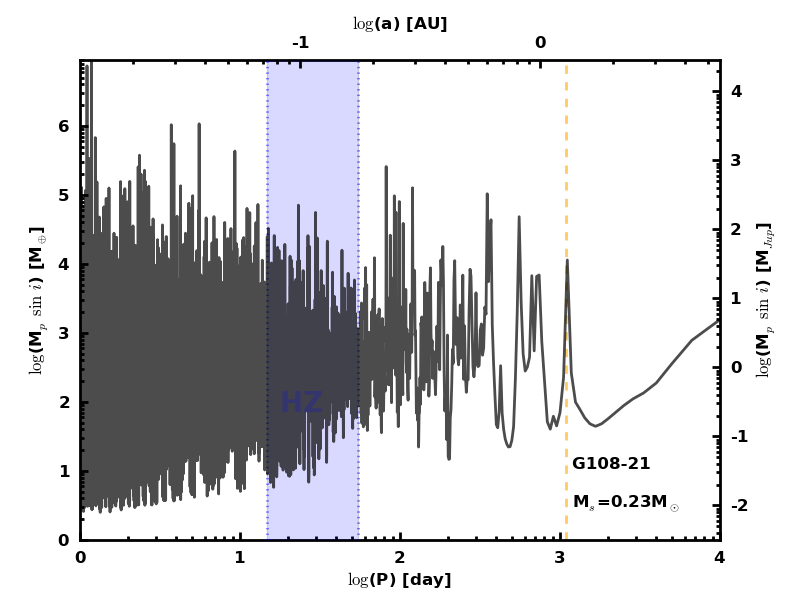}
\includegraphics[width=.9\linewidth]{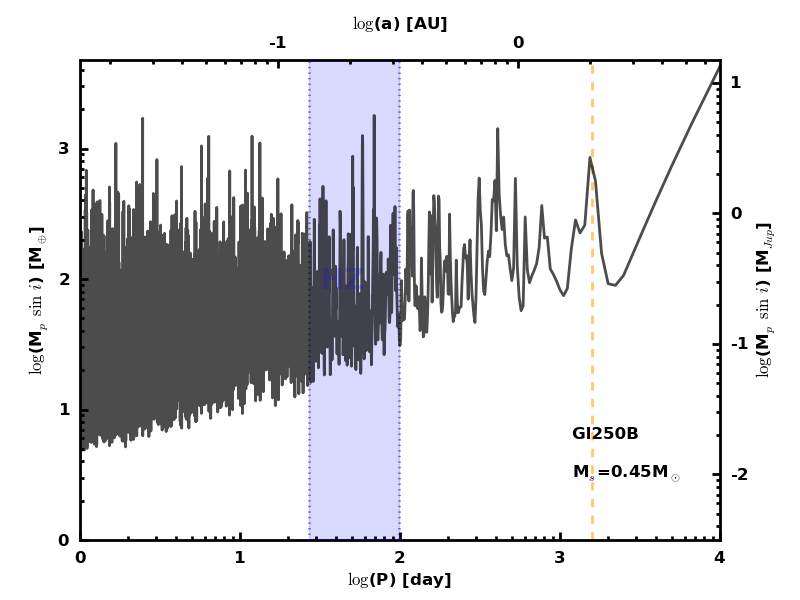}
\includegraphics[width=.9\linewidth]{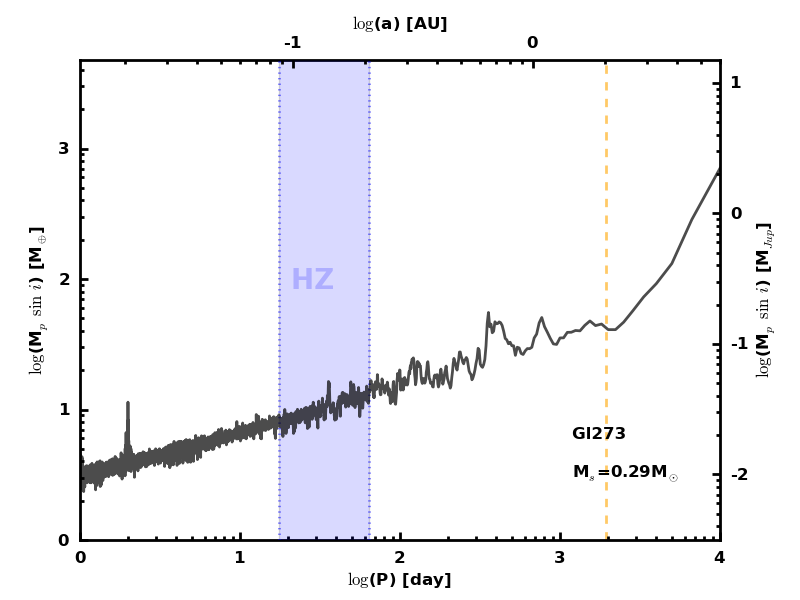}
\includegraphics[width=.9\linewidth]{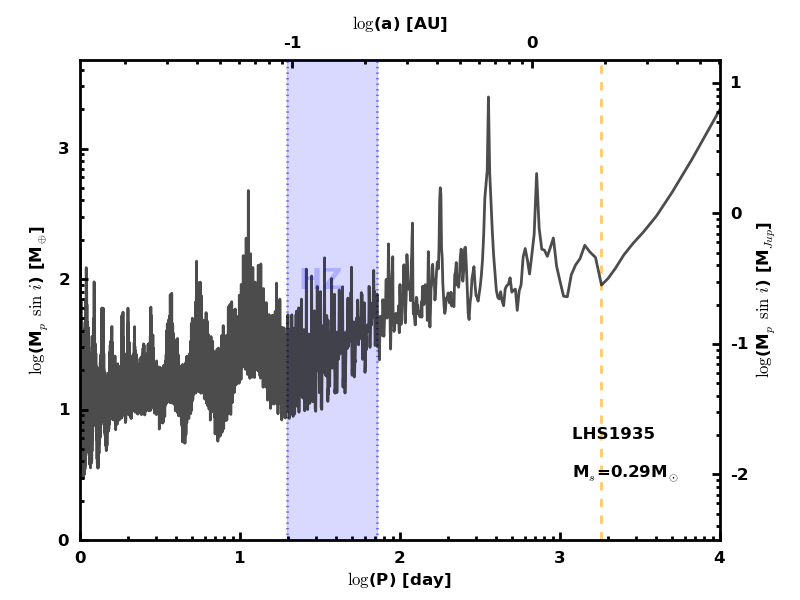}
\end{figure}\clearpage\begin{figure}
\includegraphics[width=.9\linewidth]{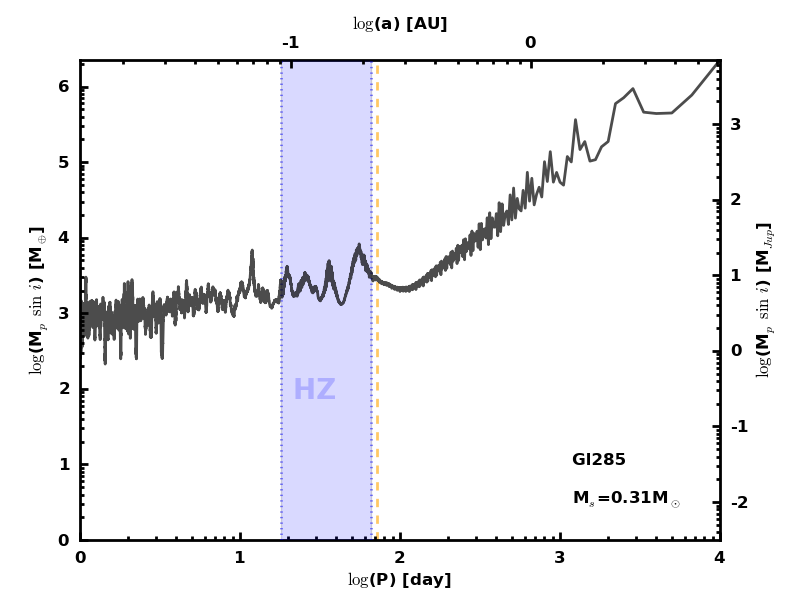}
\includegraphics[width=.9\linewidth]{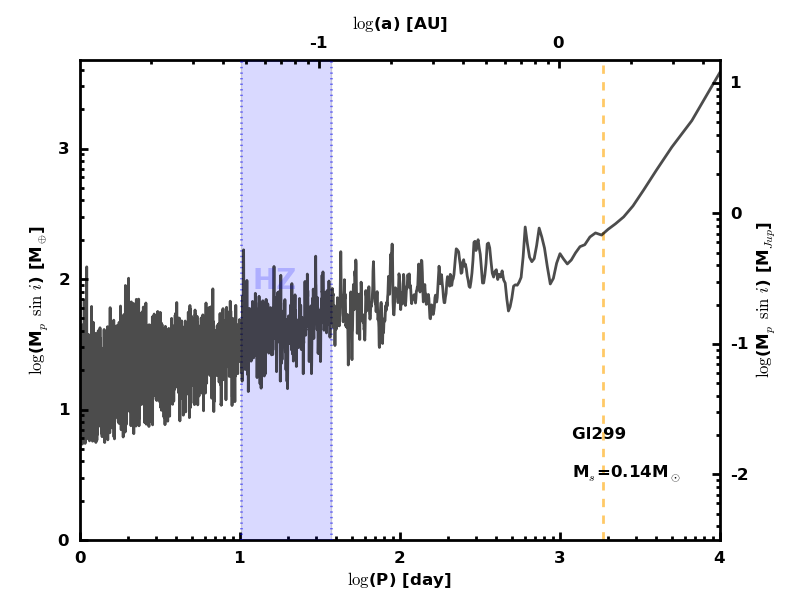}
\includegraphics[width=.9\linewidth]{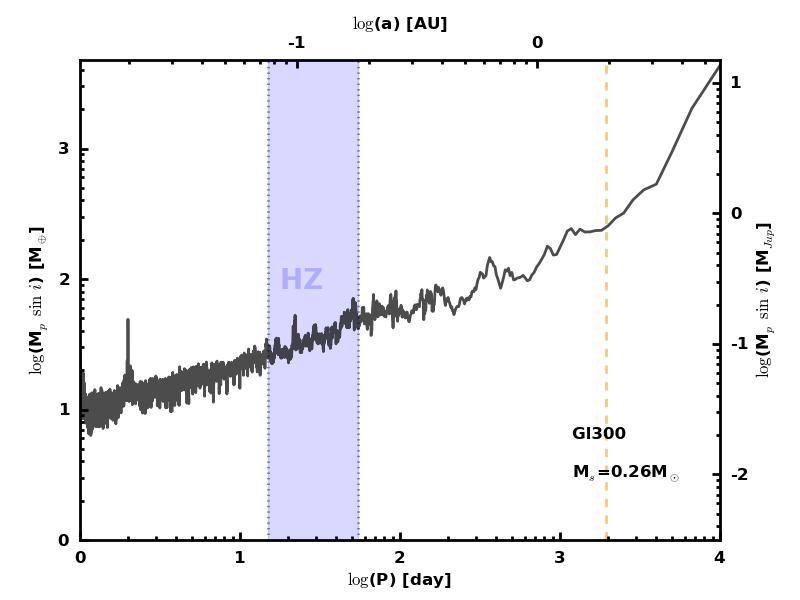}
\includegraphics[width=.9\linewidth]{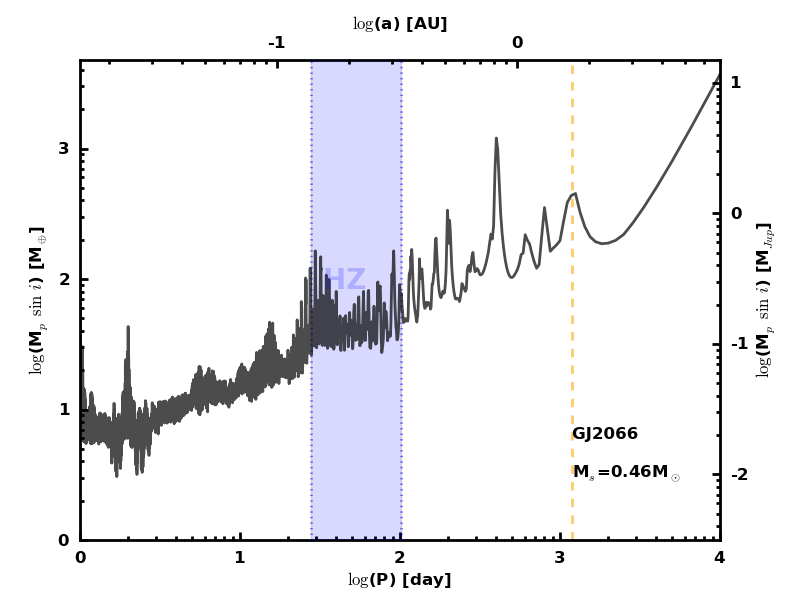}
\end{figure}\begin{figure}
\includegraphics[width=.9\linewidth]{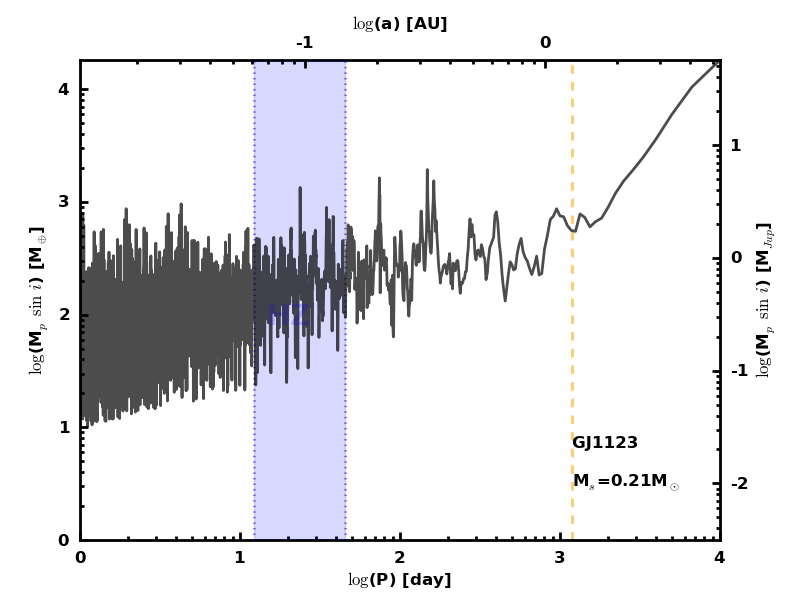}
\includegraphics[width=.9\linewidth]{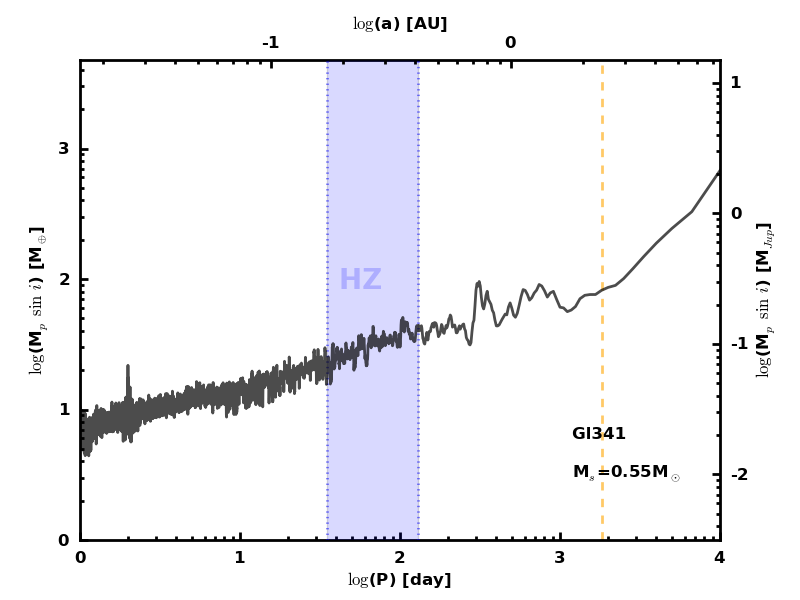}
\includegraphics[width=.9\linewidth]{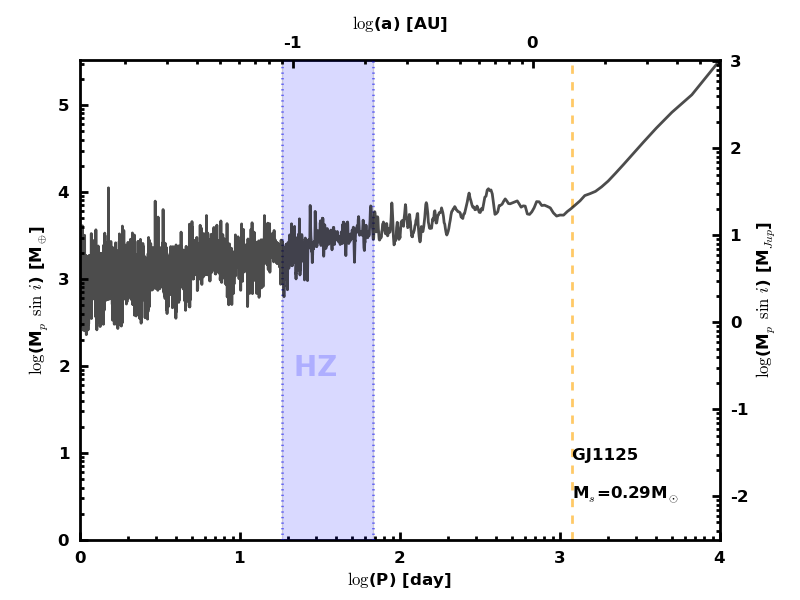}
\includegraphics[width=.9\linewidth]{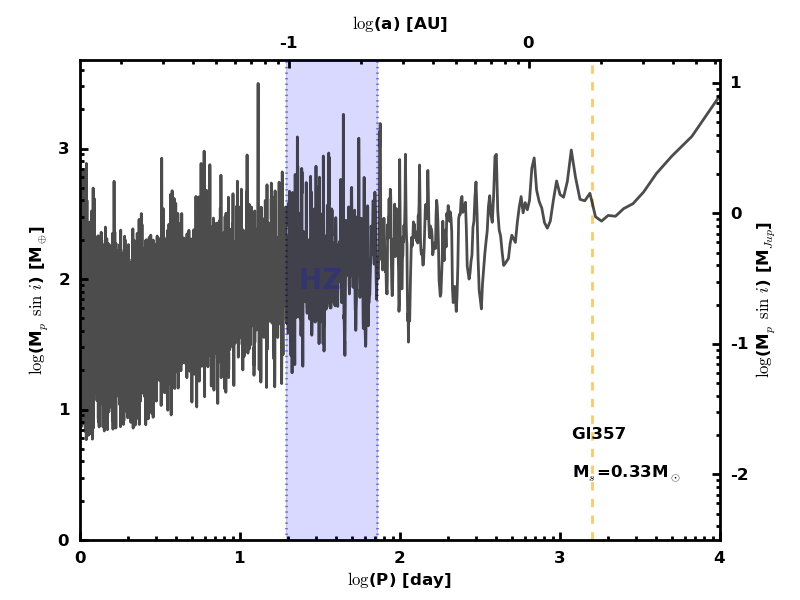}
\end{figure}\begin{figure}
\includegraphics[width=.9\linewidth]{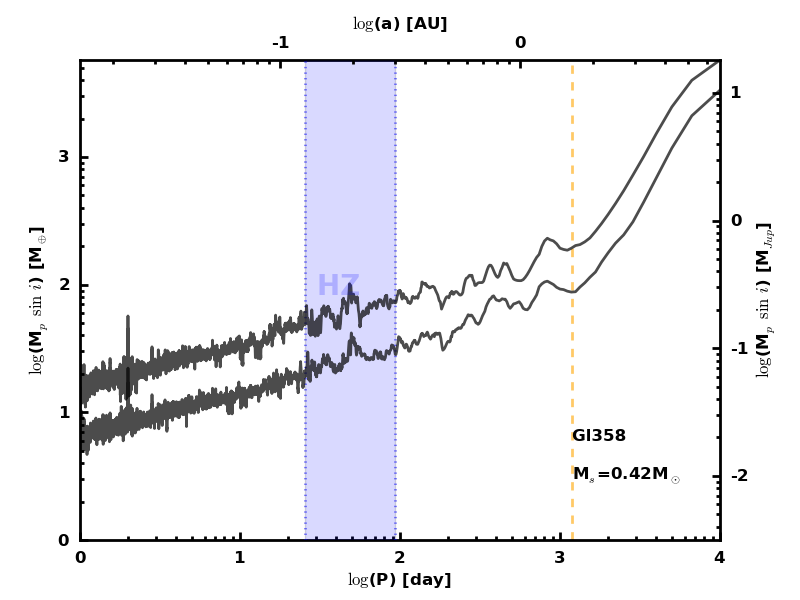}
\includegraphics[width=.9\linewidth]{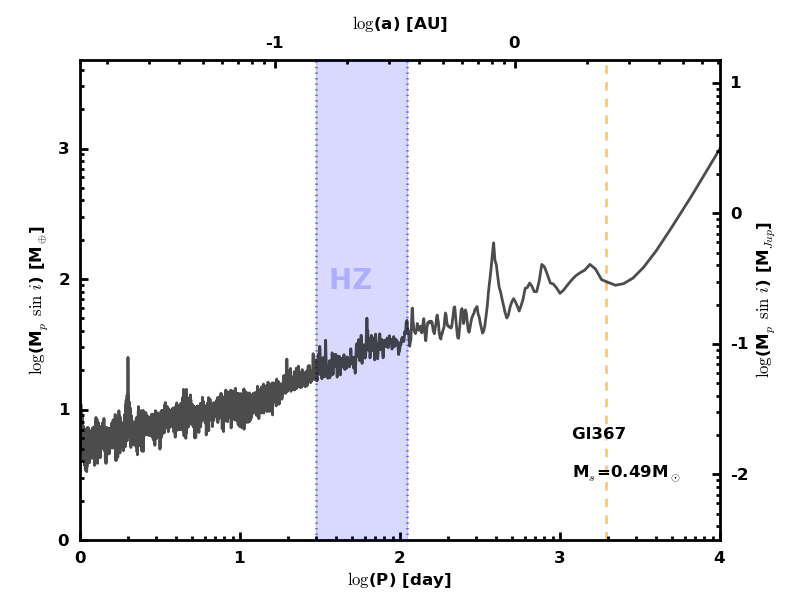}
\includegraphics[width=.9\linewidth]{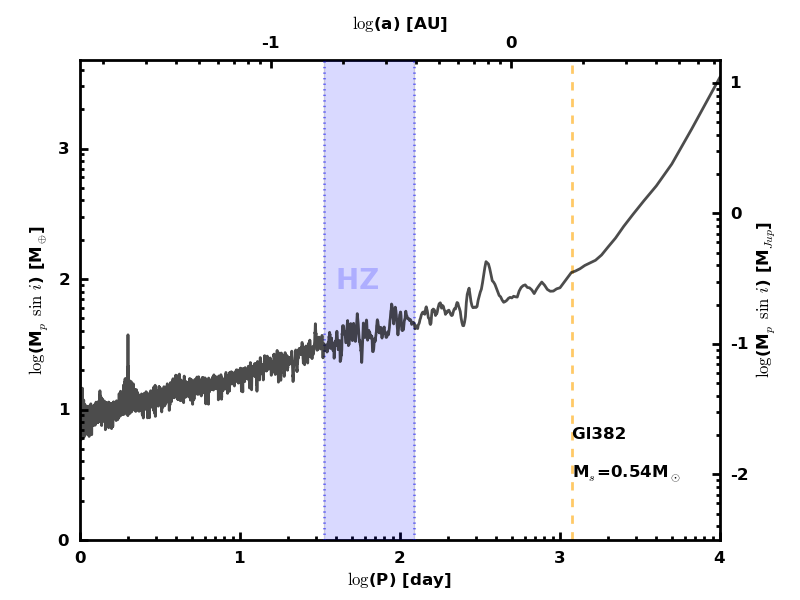}
\includegraphics[width=.9\linewidth]{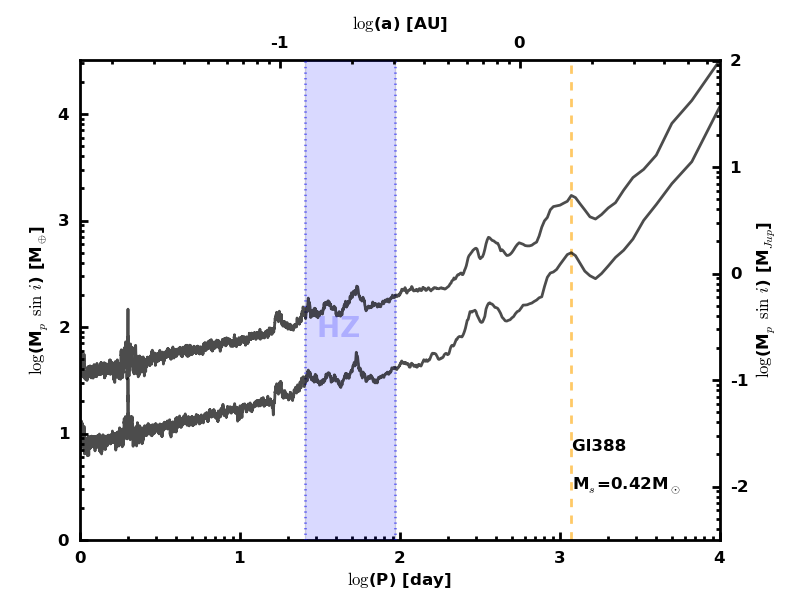}
\end{figure}\begin{figure}
\includegraphics[width=.9\linewidth]{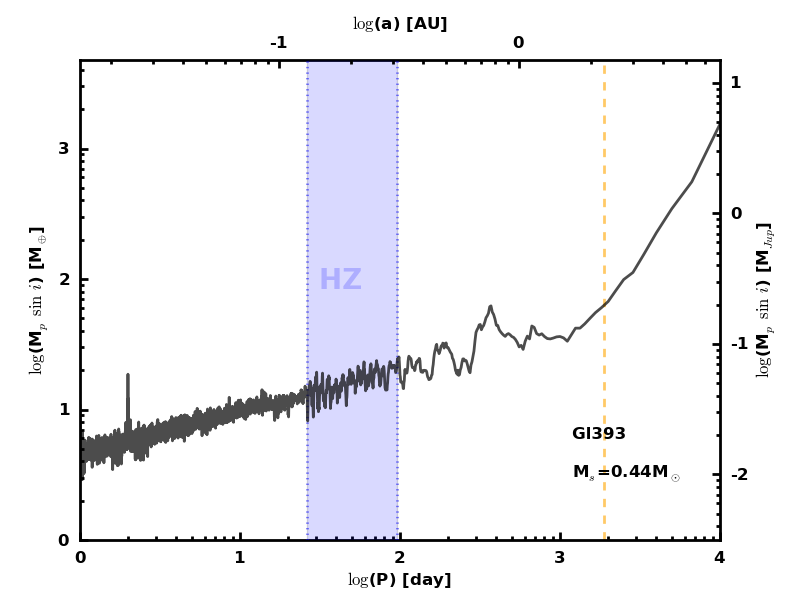}
\includegraphics[width=.9\linewidth]{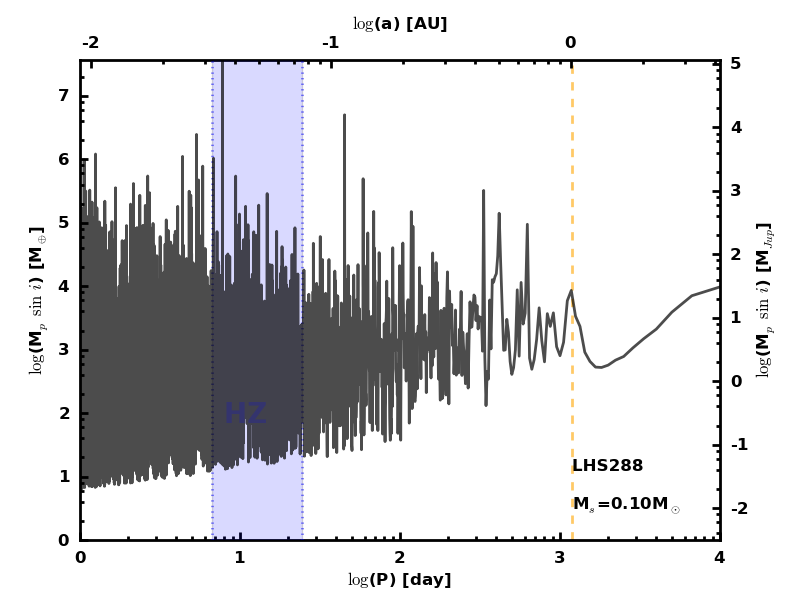}
\includegraphics[width=.9\linewidth]{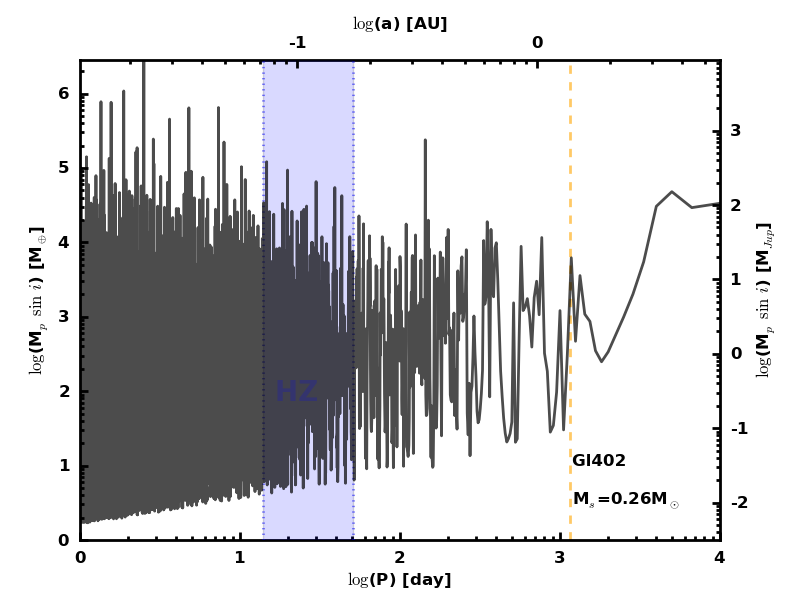}
\includegraphics[width=.9\linewidth]{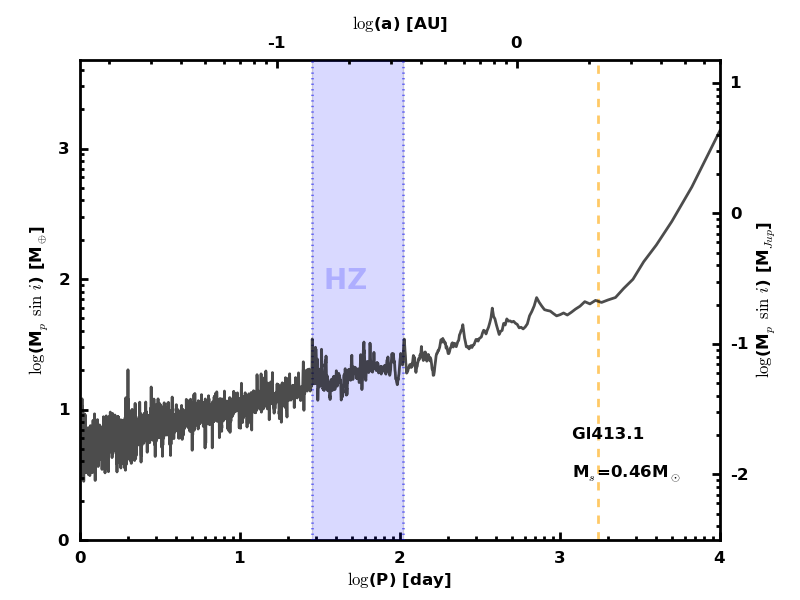}
\end{figure}\clearpage\begin{figure}
\includegraphics[width=.9\linewidth]{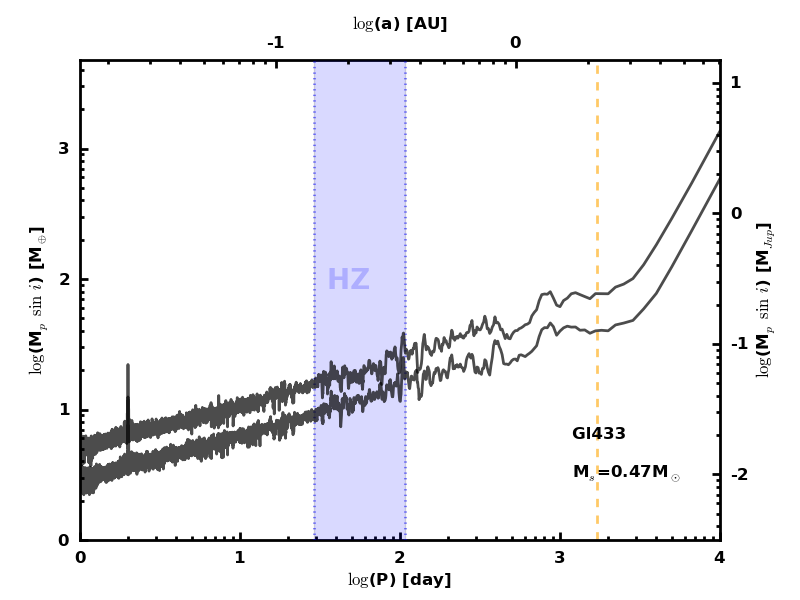}
\includegraphics[width=.9\linewidth]{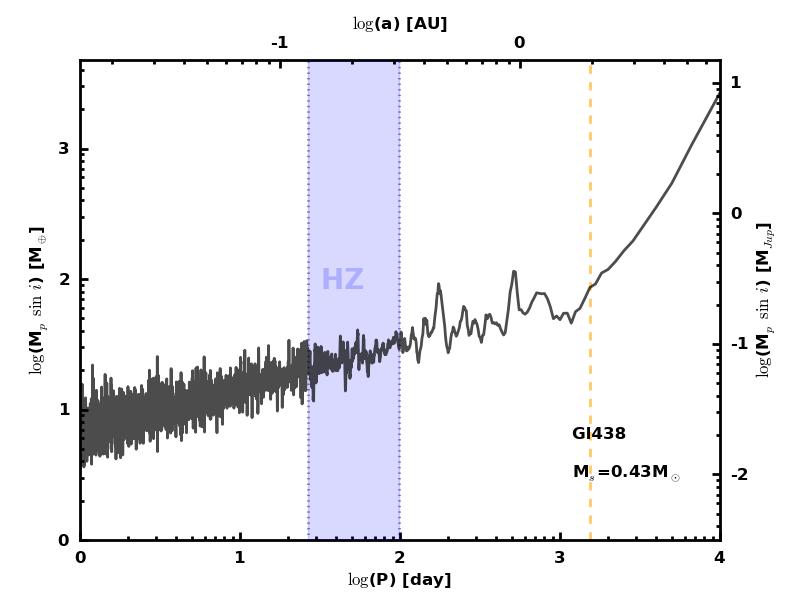}
\includegraphics[width=.9\linewidth]{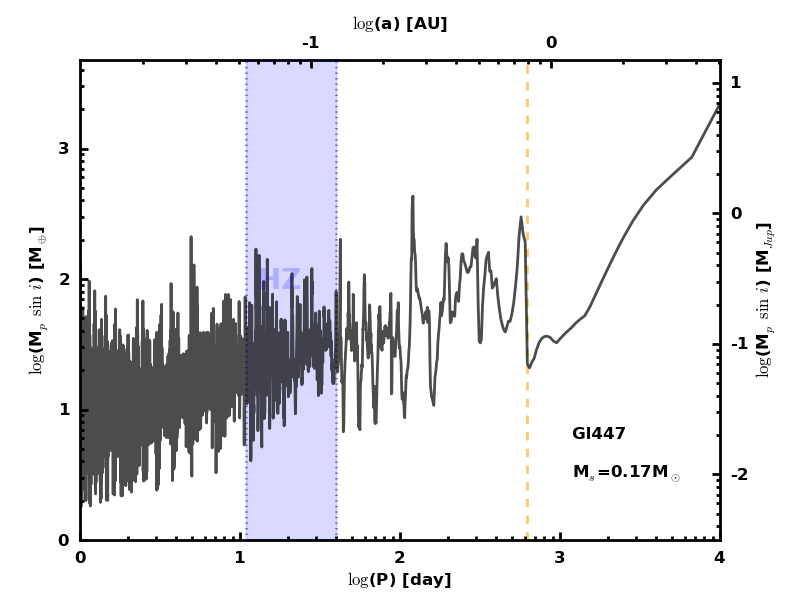}
\includegraphics[width=.9\linewidth]{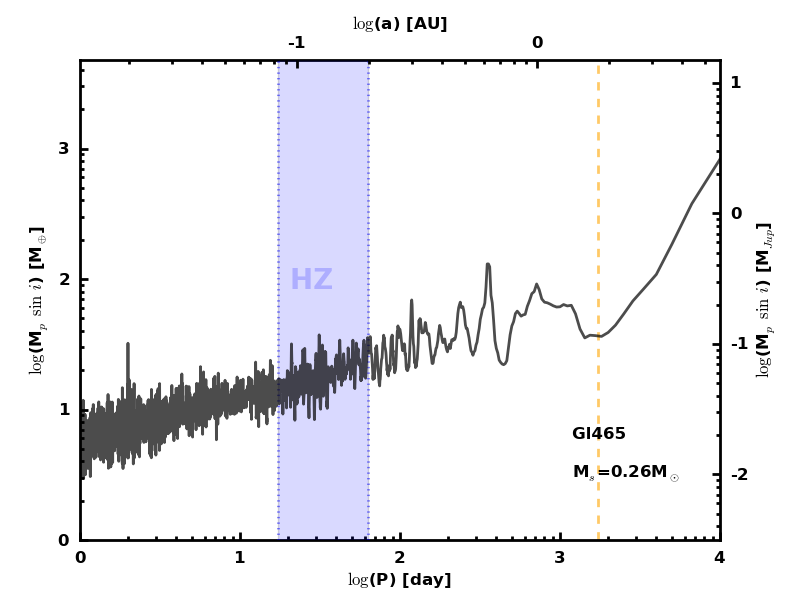}
\end{figure}\begin{figure}
\includegraphics[width=.9\linewidth]{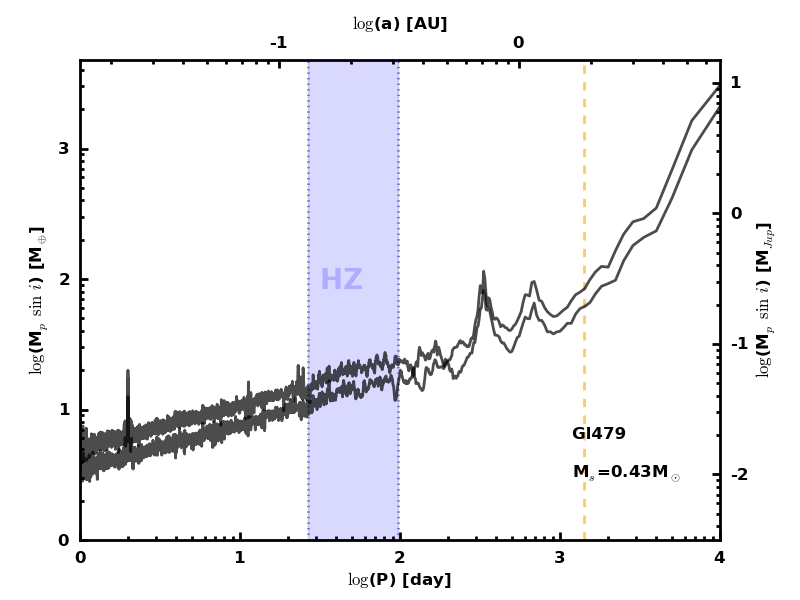}
\includegraphics[width=.9\linewidth]{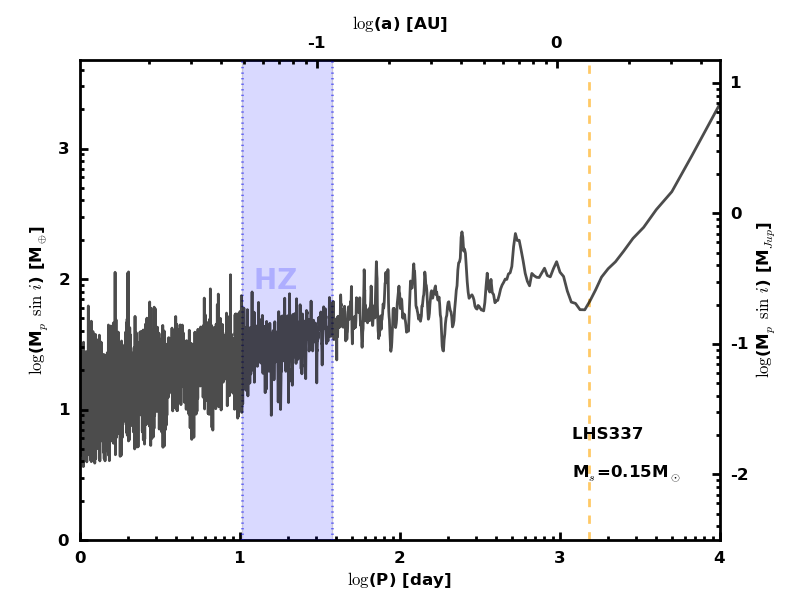}
\includegraphics[width=.9\linewidth]{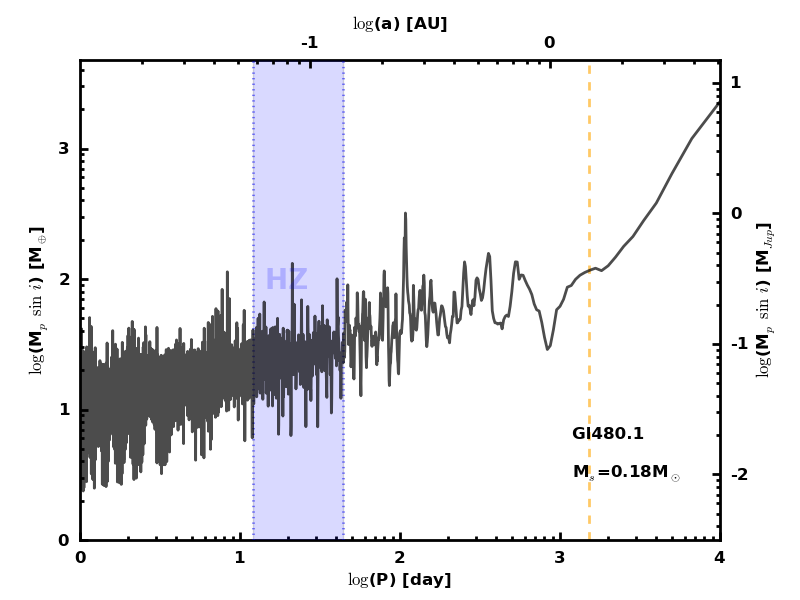}
\includegraphics[width=.9\linewidth]{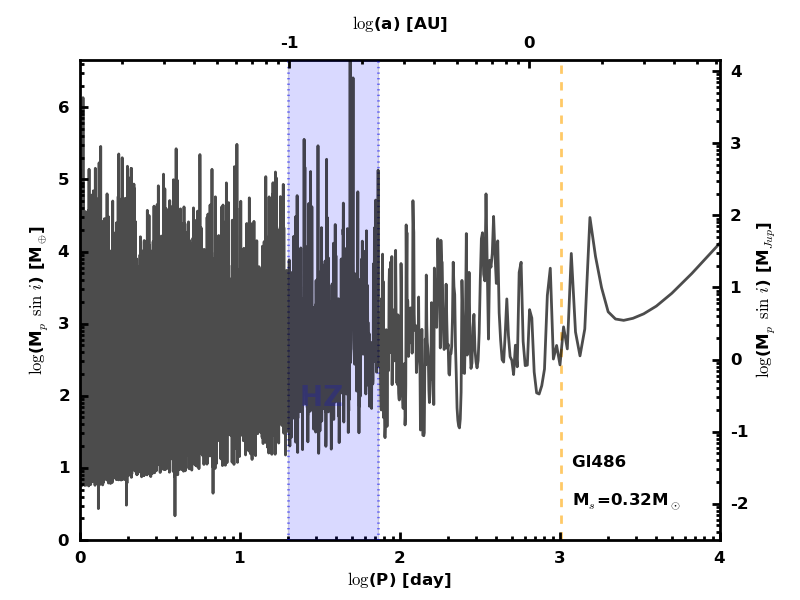}
\end{figure}\begin{figure}
\includegraphics[width=.9\linewidth]{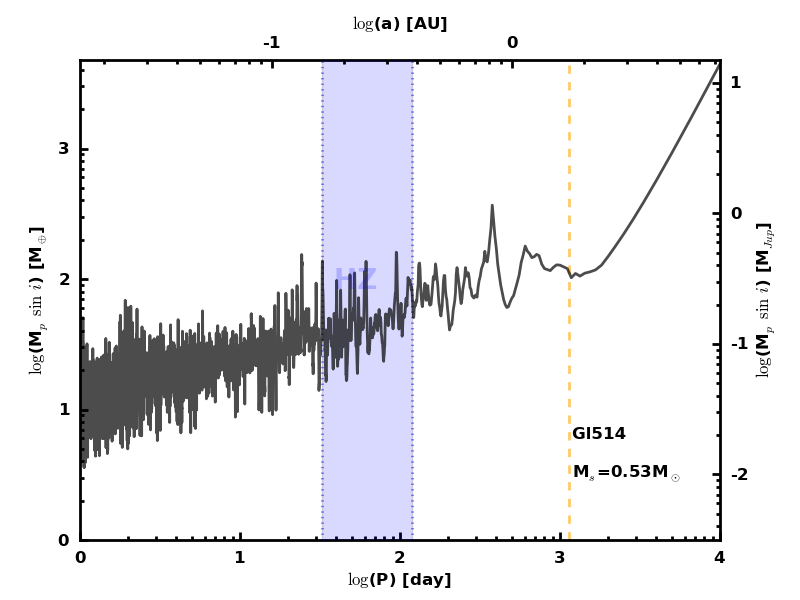}
\includegraphics[width=.9\linewidth]{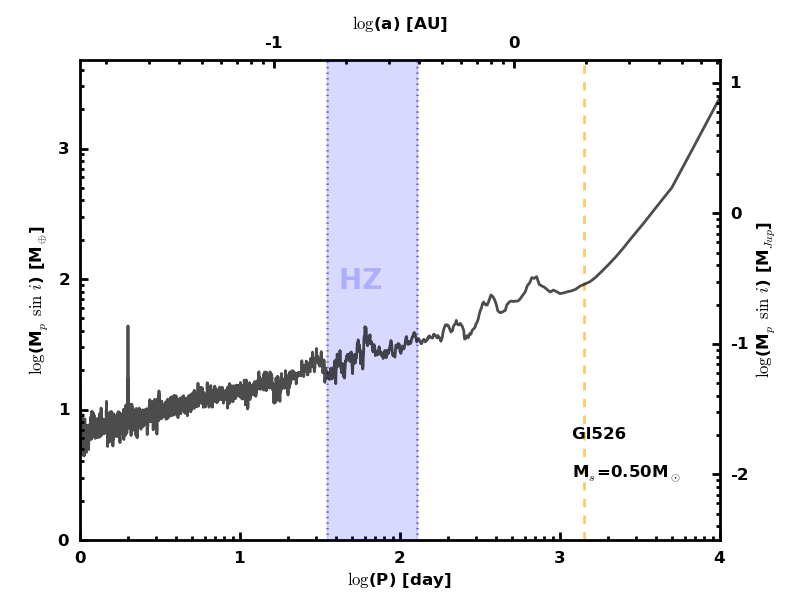}
\includegraphics[width=.9\linewidth]{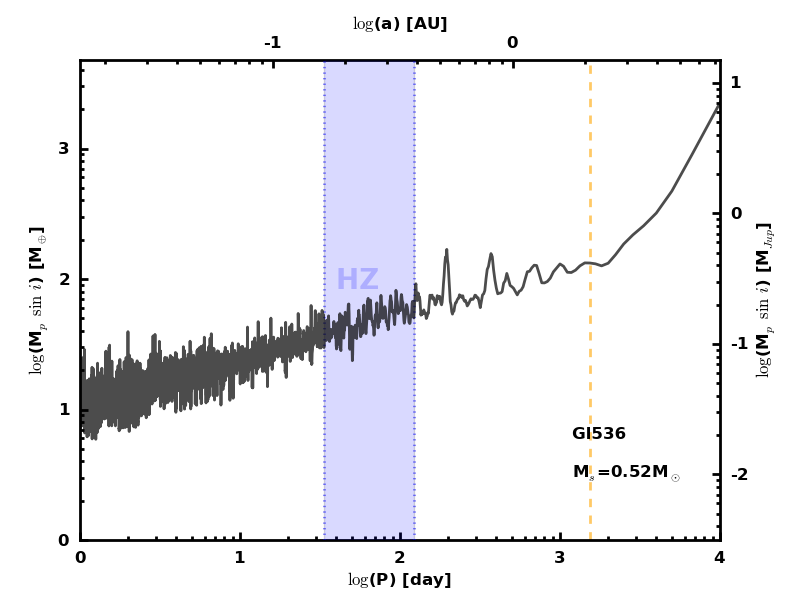}
\includegraphics[width=.9\linewidth]{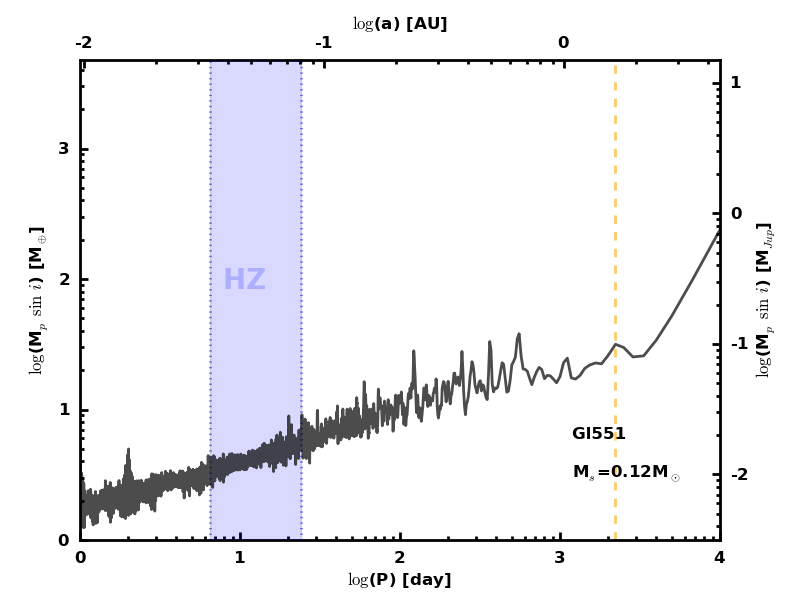}
\end{figure}\begin{figure}
\includegraphics[width=.9\linewidth]{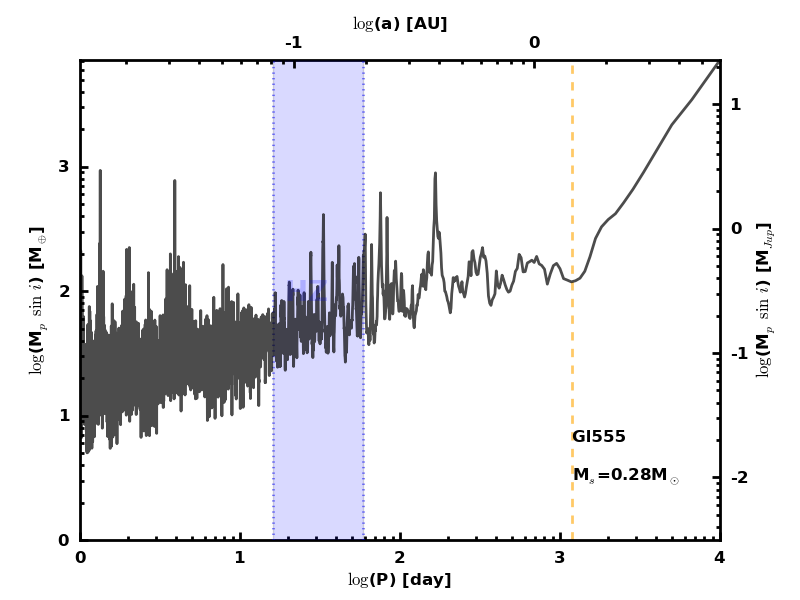}
\includegraphics[width=.9\linewidth]{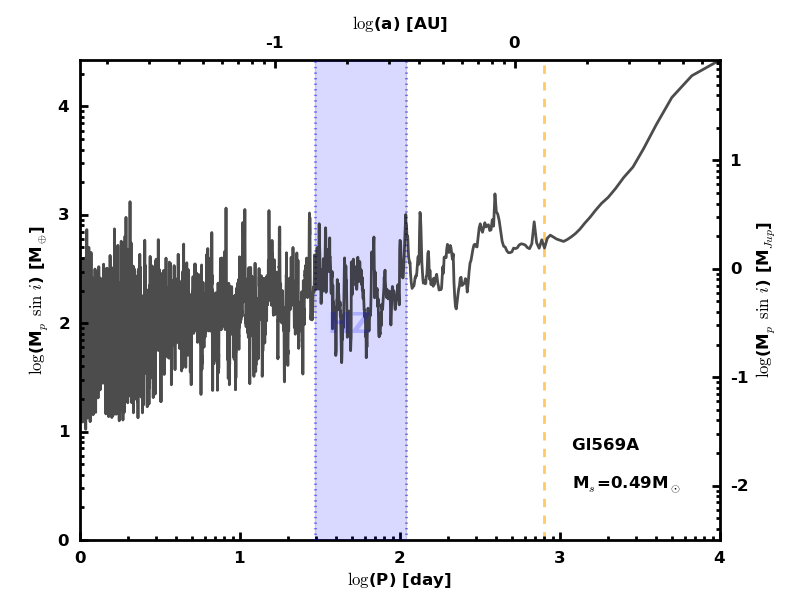}
\includegraphics[width=.9\linewidth]{Gl581_dl.png}
\includegraphics[width=.9\linewidth]{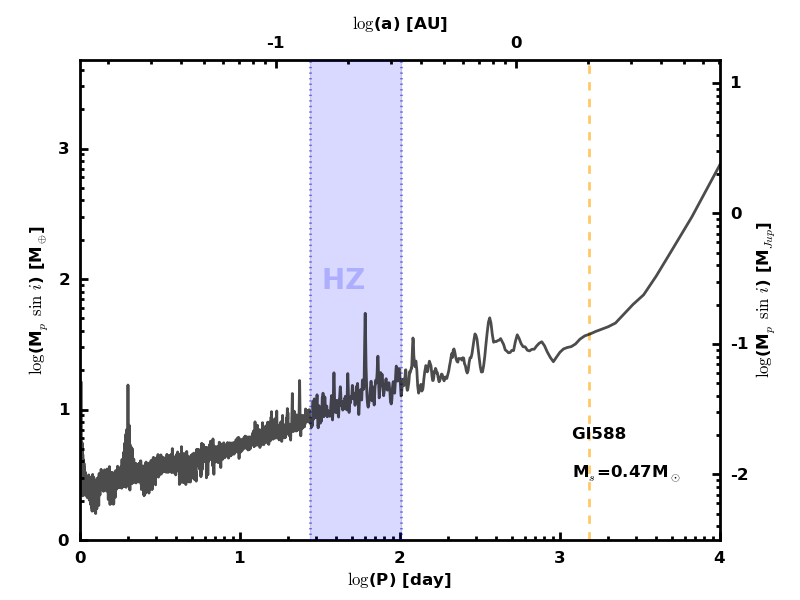}
\end{figure}\clearpage\begin{figure}
\includegraphics[width=.9\linewidth]{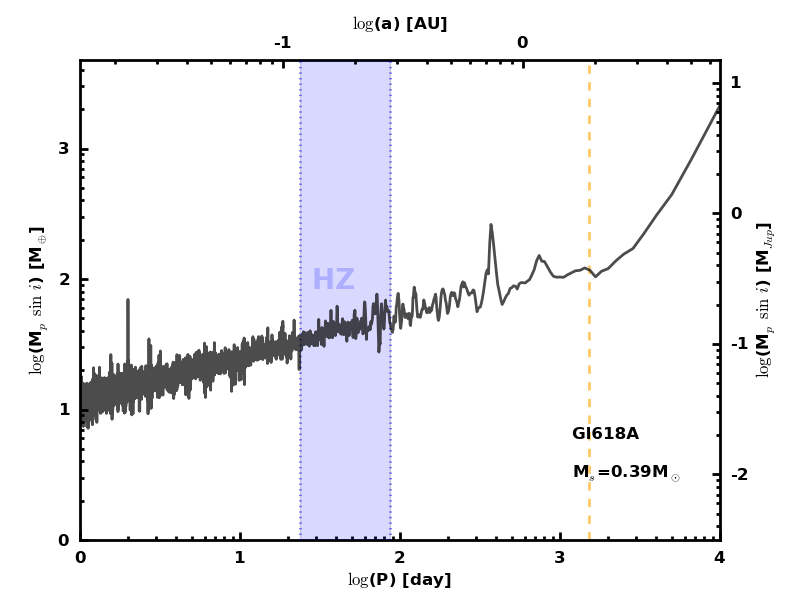}
\includegraphics[width=.9\linewidth]{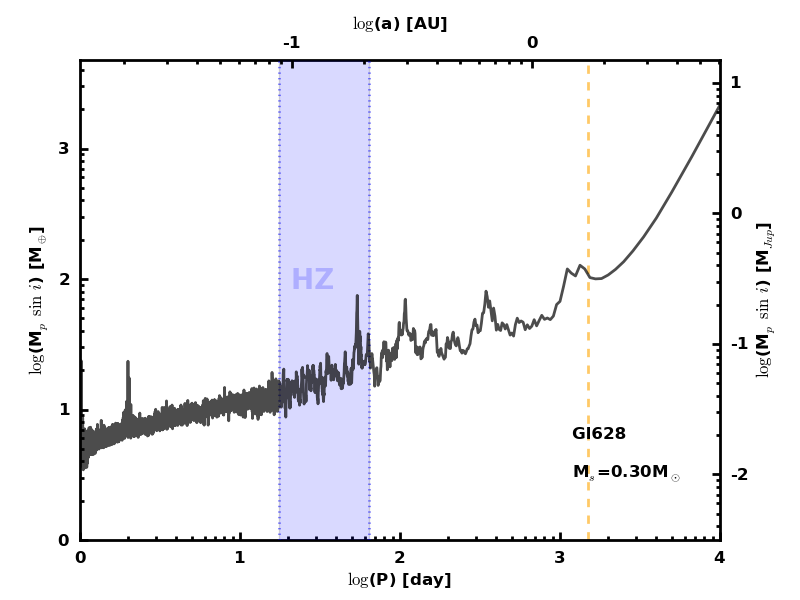}
\includegraphics[width=.9\linewidth]{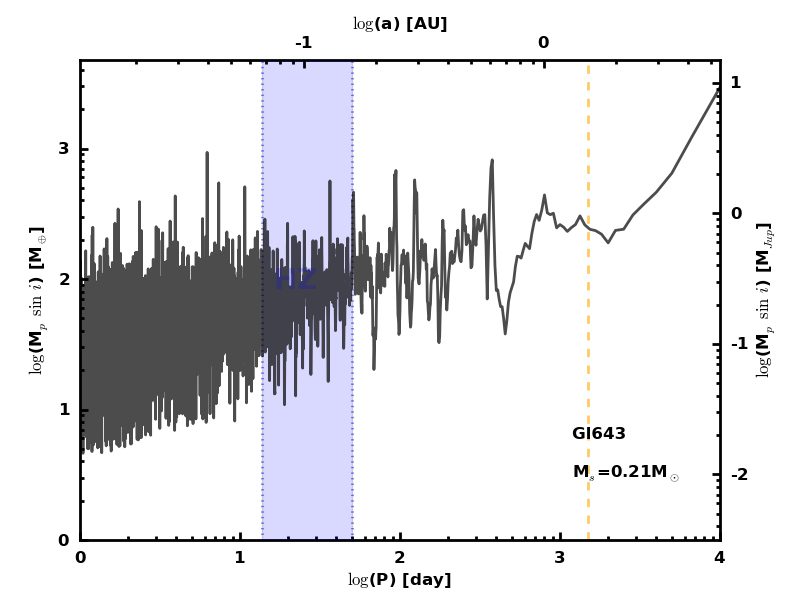}
\includegraphics[width=.9\linewidth]{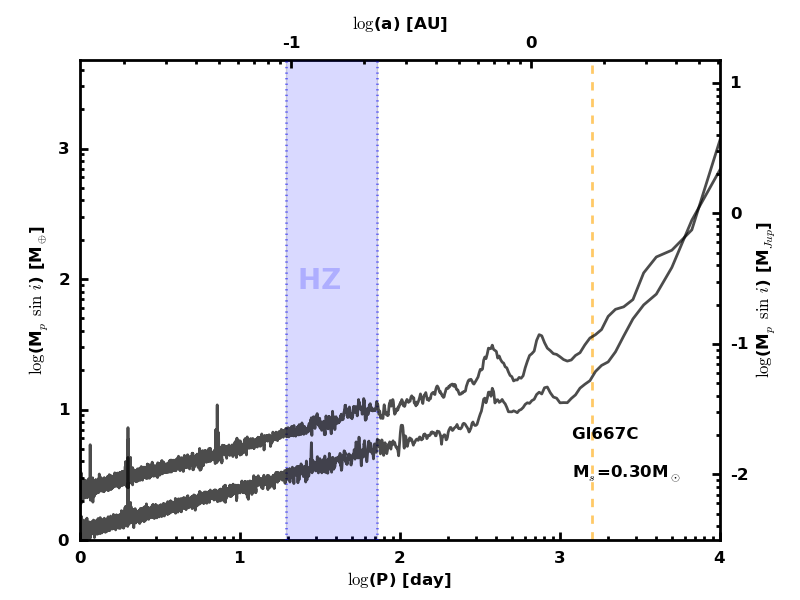}
\end{figure}\begin{figure}
\includegraphics[width=.9\linewidth]{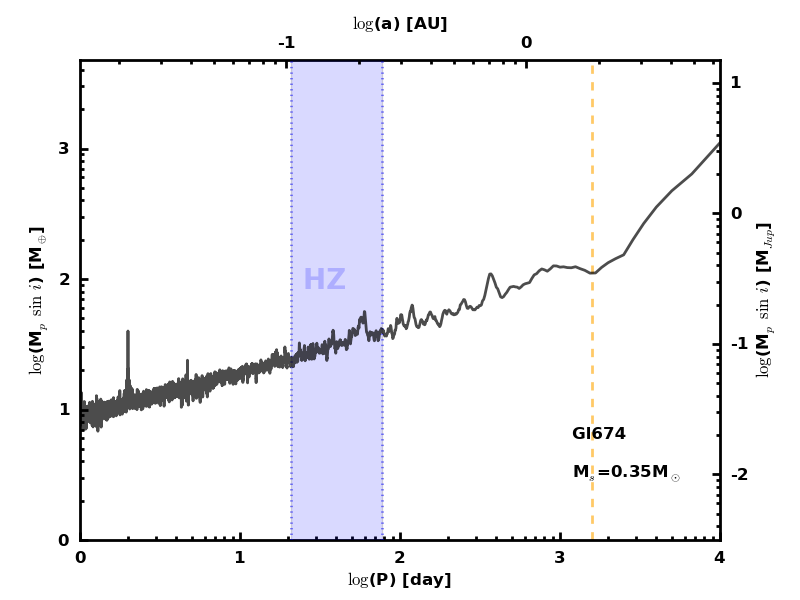}
\includegraphics[width=.9\linewidth]{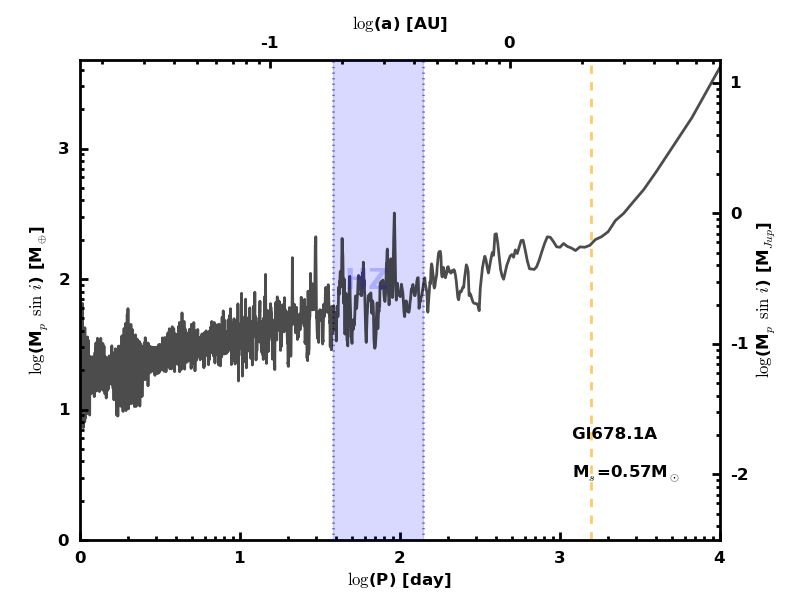}
\includegraphics[width=.9\linewidth]{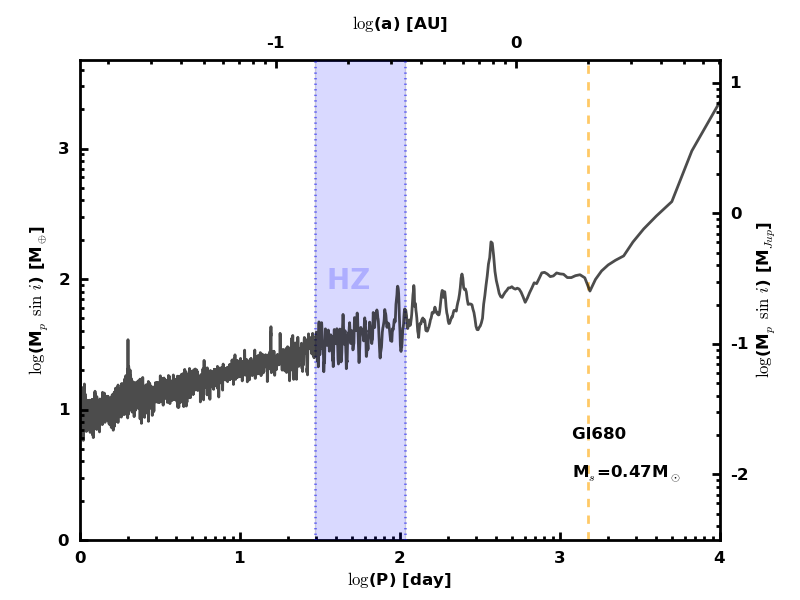}
\includegraphics[width=.9\linewidth]{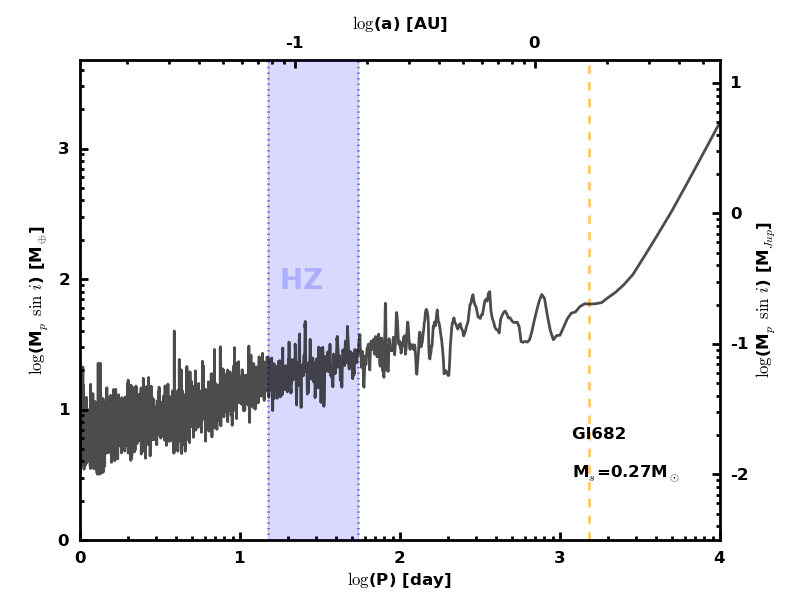}
\end{figure}\begin{figure}
\includegraphics[width=.9\linewidth]{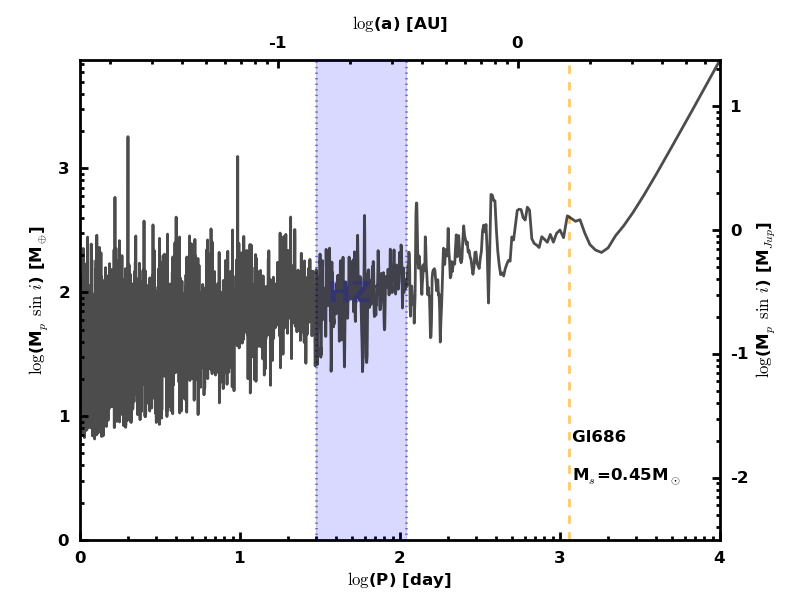}
\includegraphics[width=.9\linewidth]{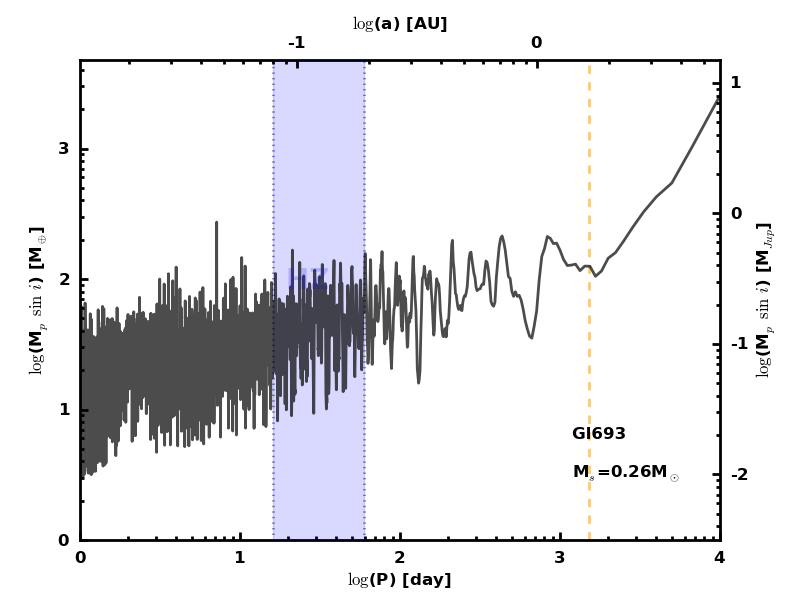}
\includegraphics[width=.9\linewidth]{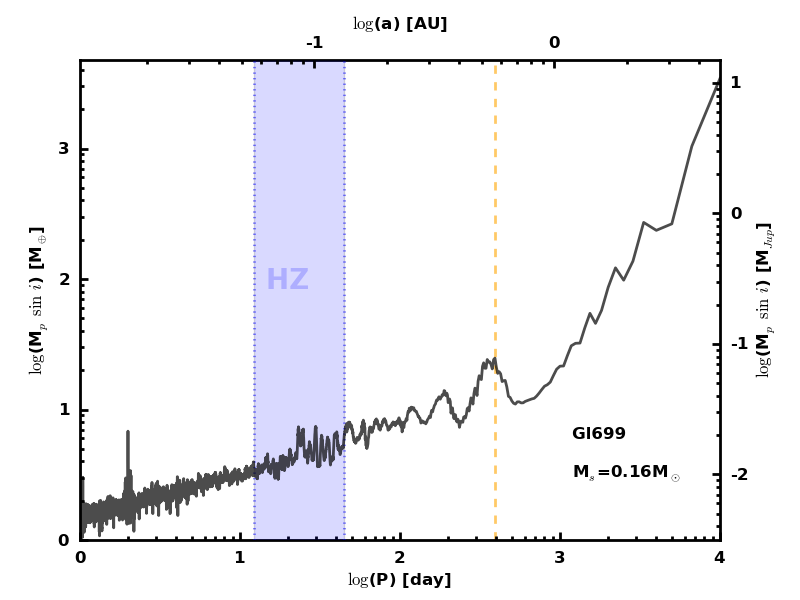}
\includegraphics[width=.9\linewidth]{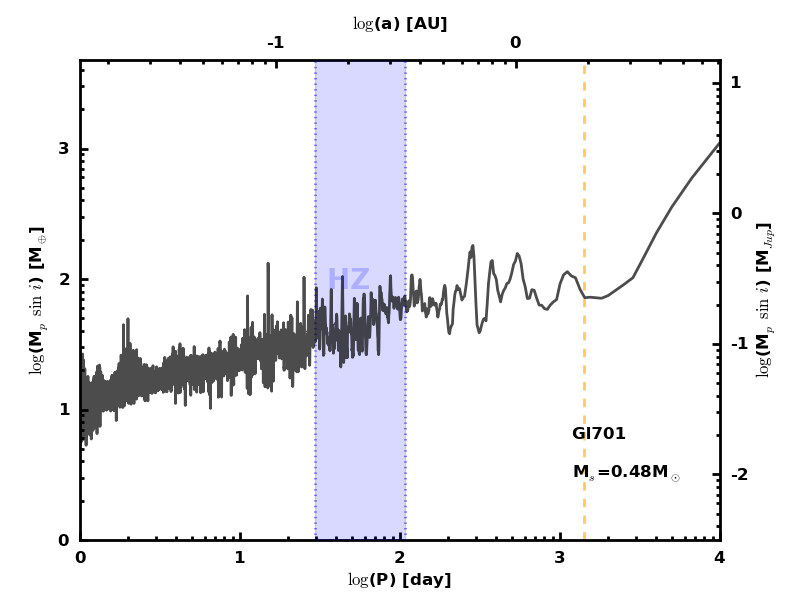}
\end{figure}\begin{figure}
\includegraphics[width=.9\linewidth]{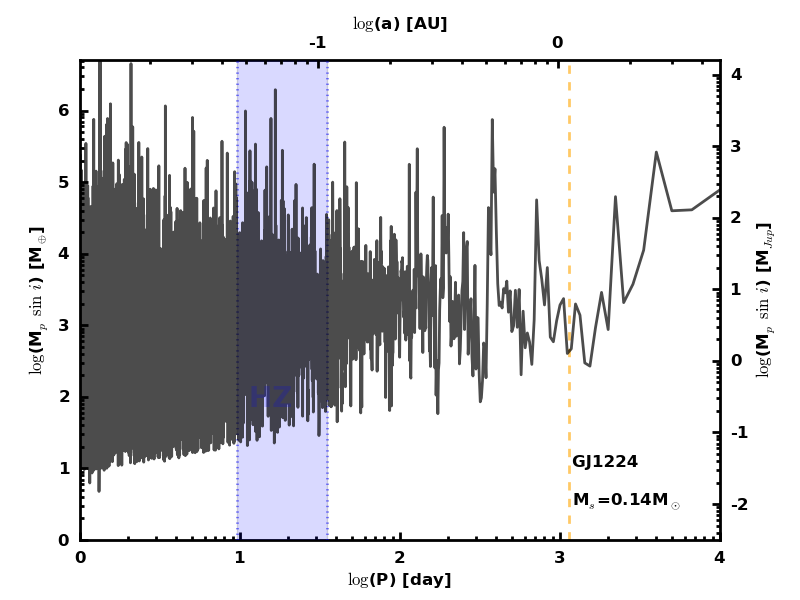}
\includegraphics[width=.9\linewidth]{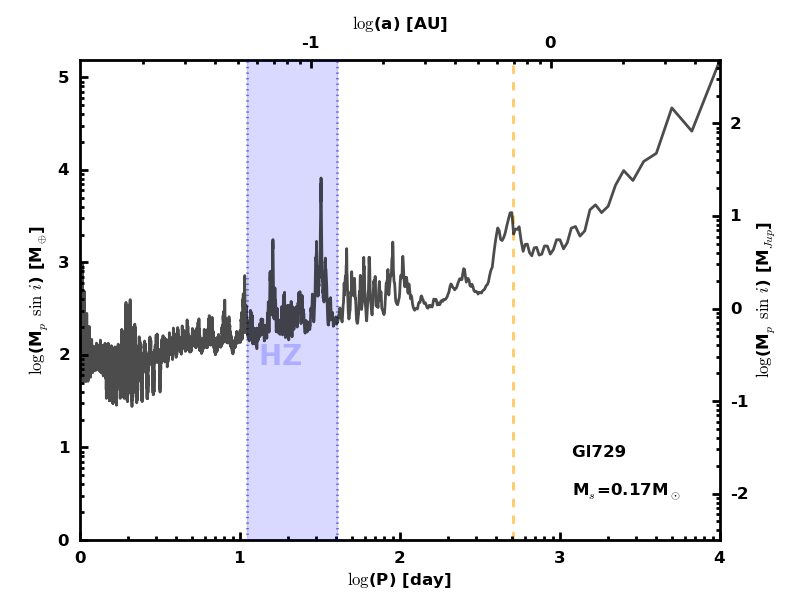}
\includegraphics[width=.9\linewidth]{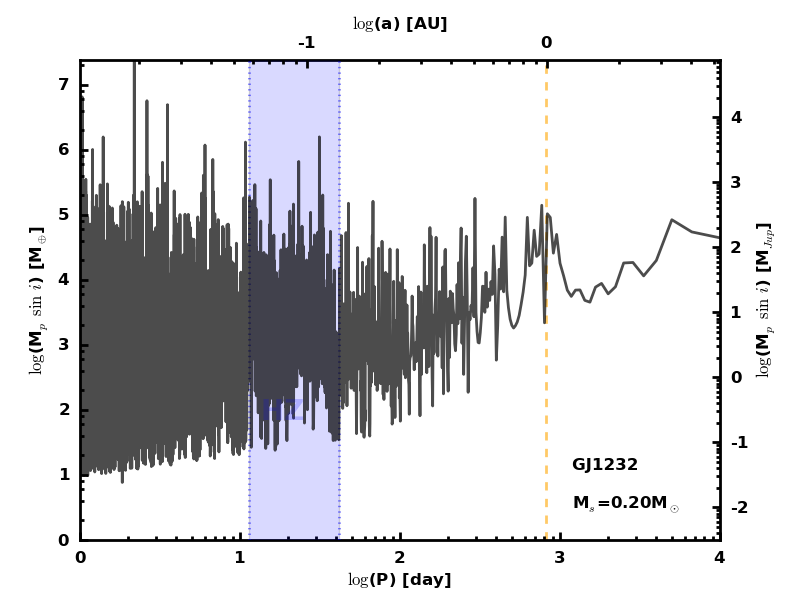}
\includegraphics[width=.9\linewidth]{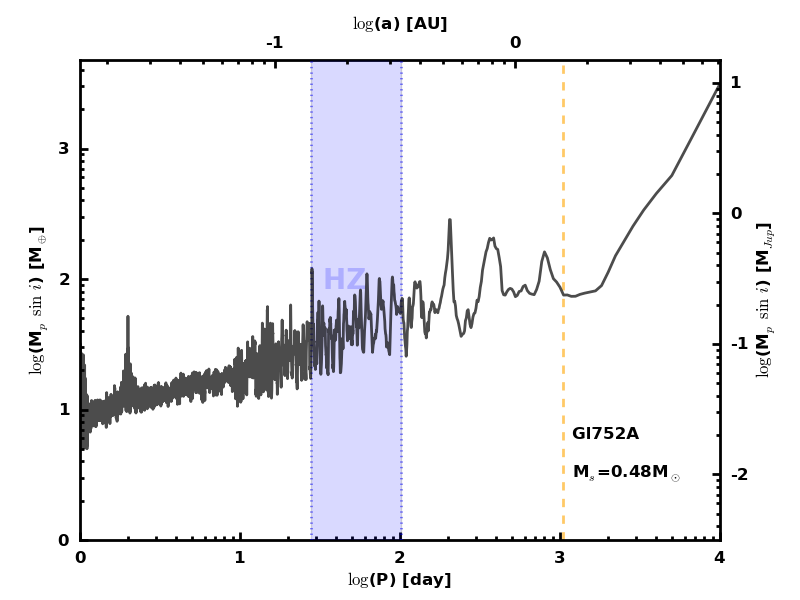}
\end{figure}\clearpage\begin{figure}
\includegraphics[width=.9\linewidth]{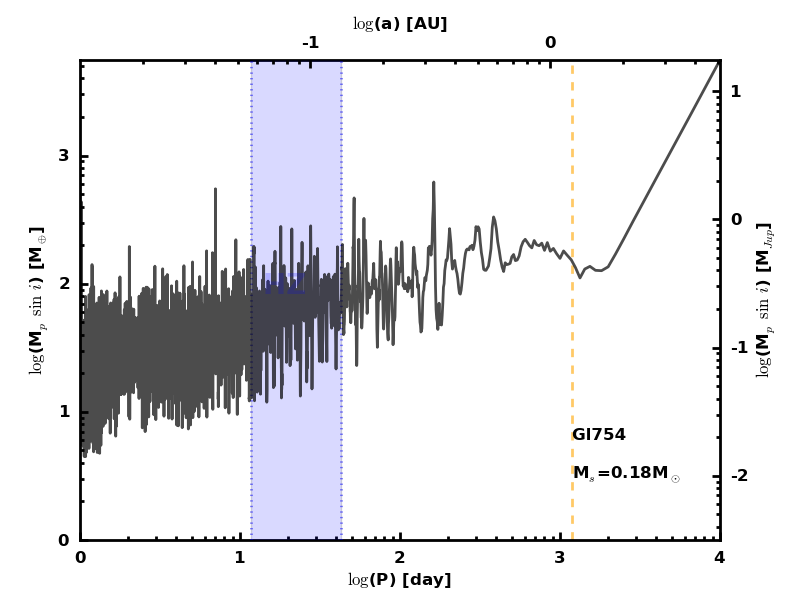}
\includegraphics[width=.9\linewidth]{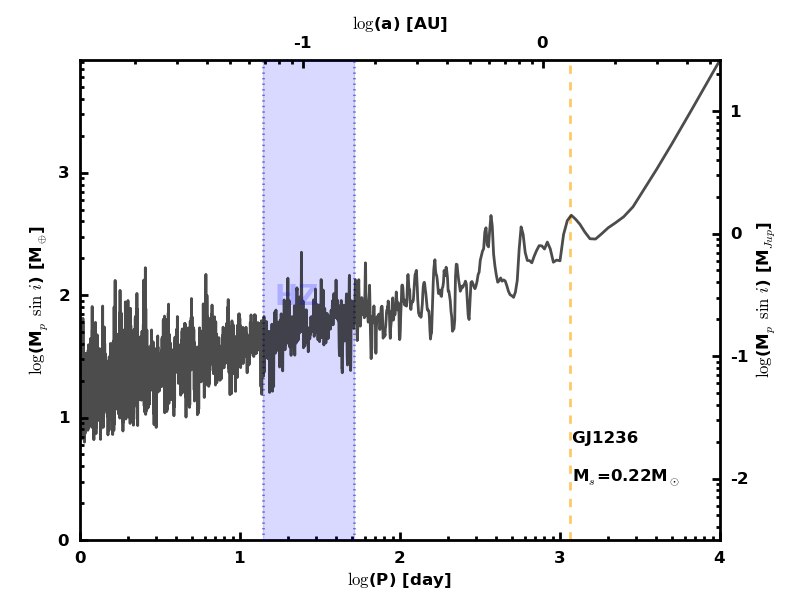}
\includegraphics[width=.9\linewidth]{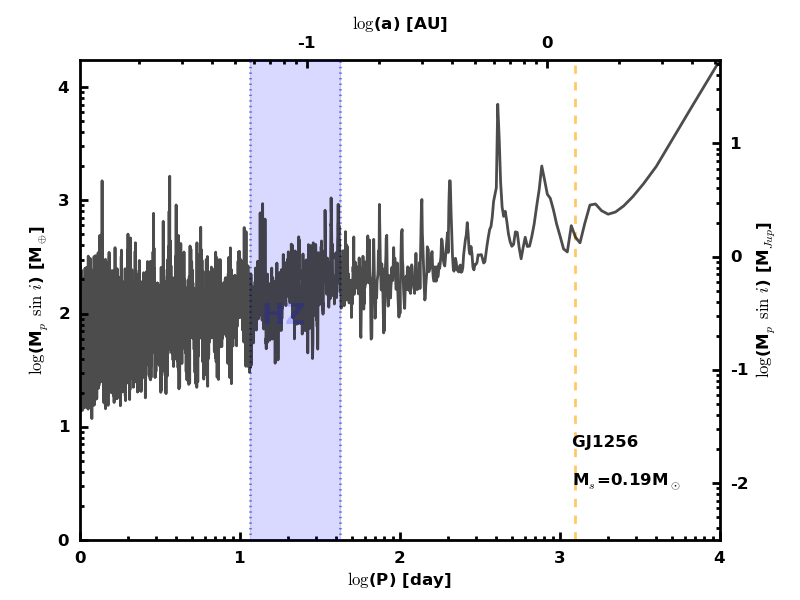}
\includegraphics[width=.9\linewidth]{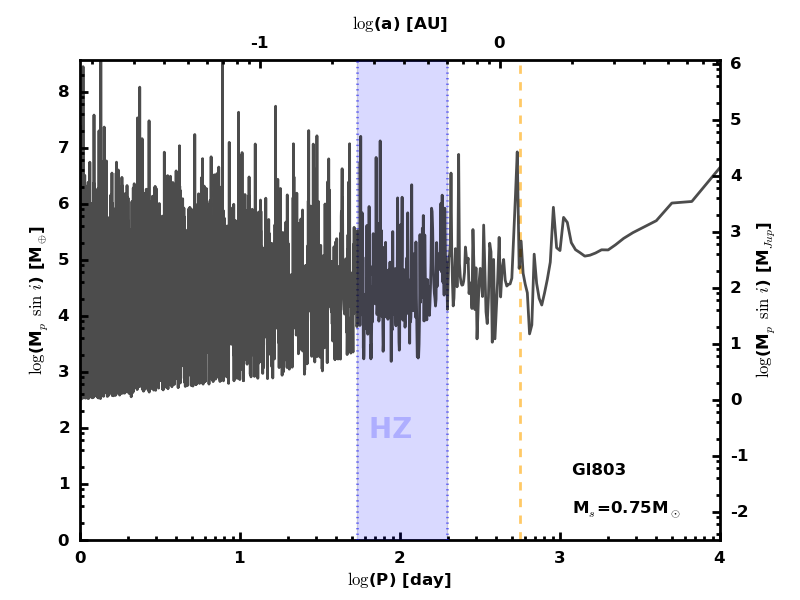}
\end{figure}\begin{figure}
\includegraphics[width=.9\linewidth]{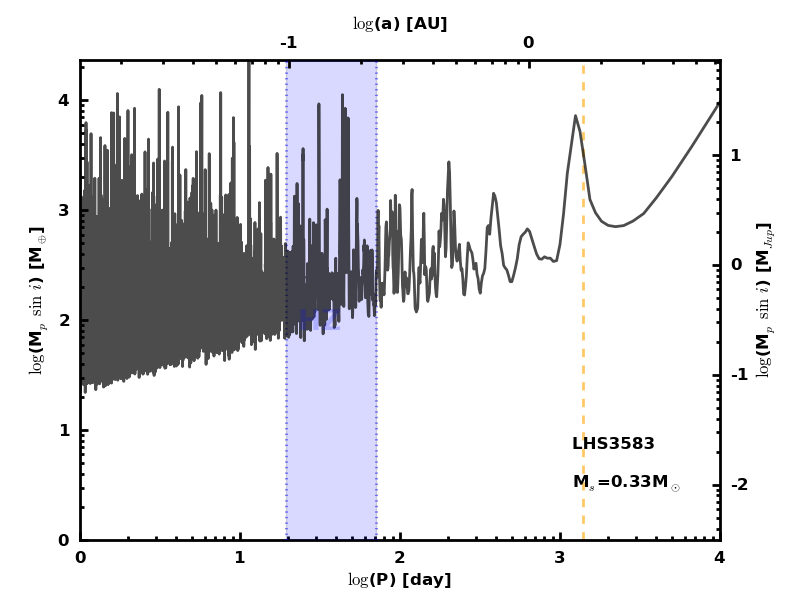}
\includegraphics[width=.9\linewidth]{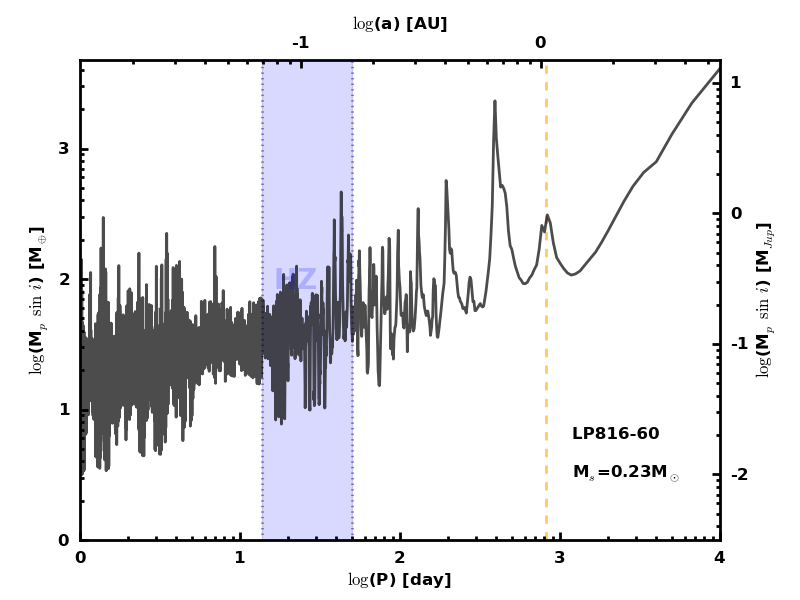}
\includegraphics[width=.9\linewidth]{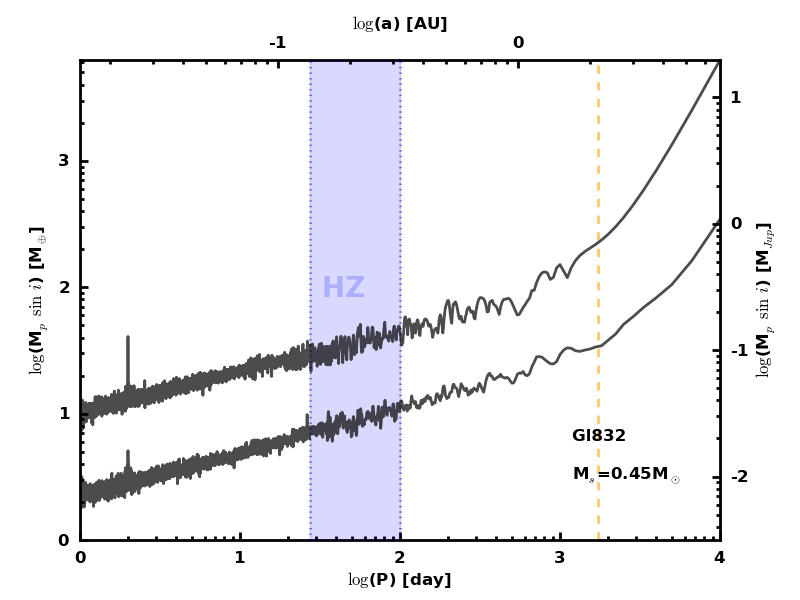}
\includegraphics[width=.9\linewidth]{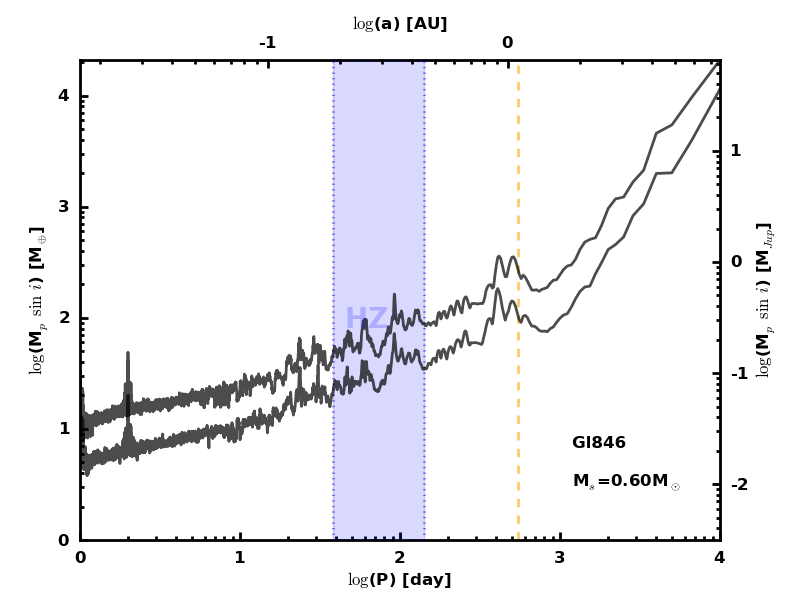}
\end{figure}\begin{figure}
\includegraphics[width=.9\linewidth]{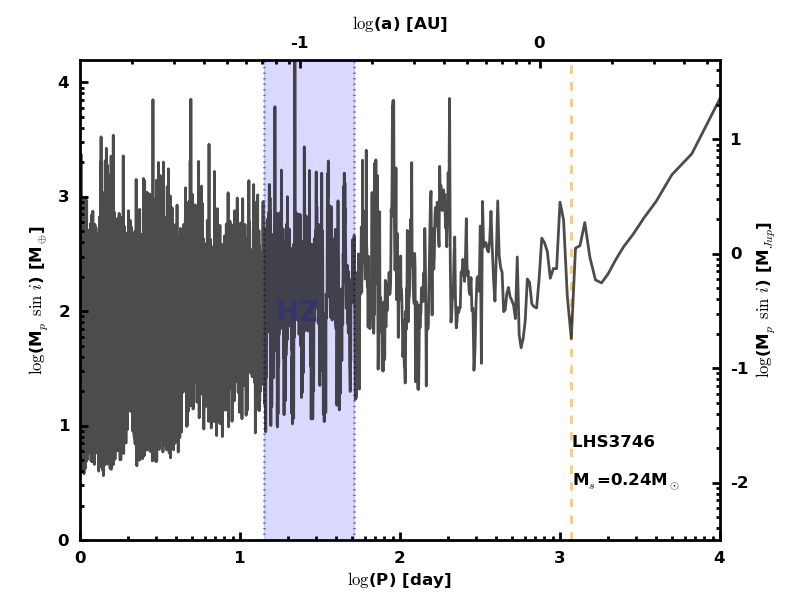}
\includegraphics[width=.9\linewidth]{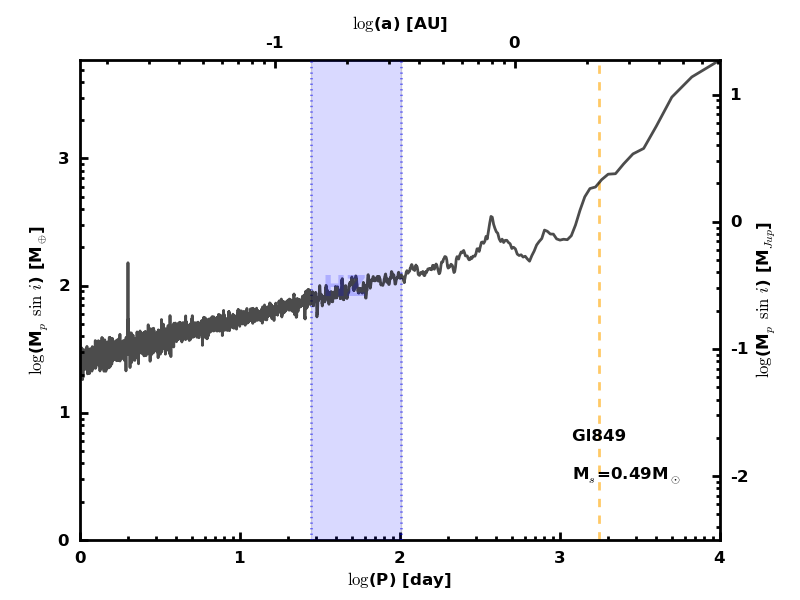}
\includegraphics[width=.9\linewidth]{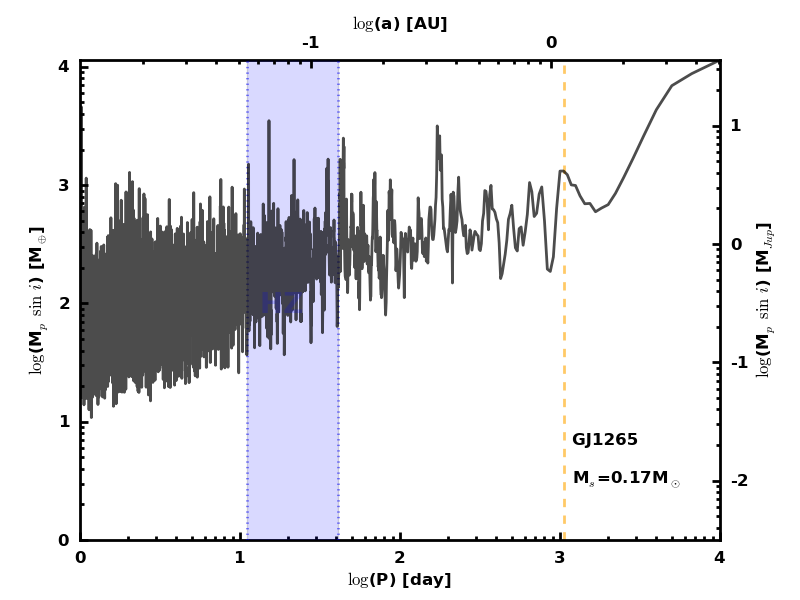}
\includegraphics[width=.9\linewidth]{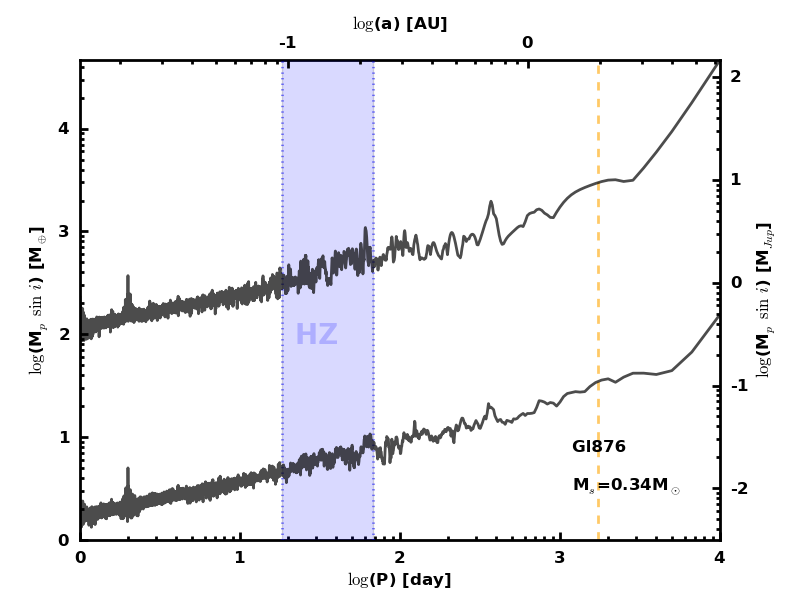}
\end{figure}\begin{figure}
\includegraphics[width=.9\linewidth]{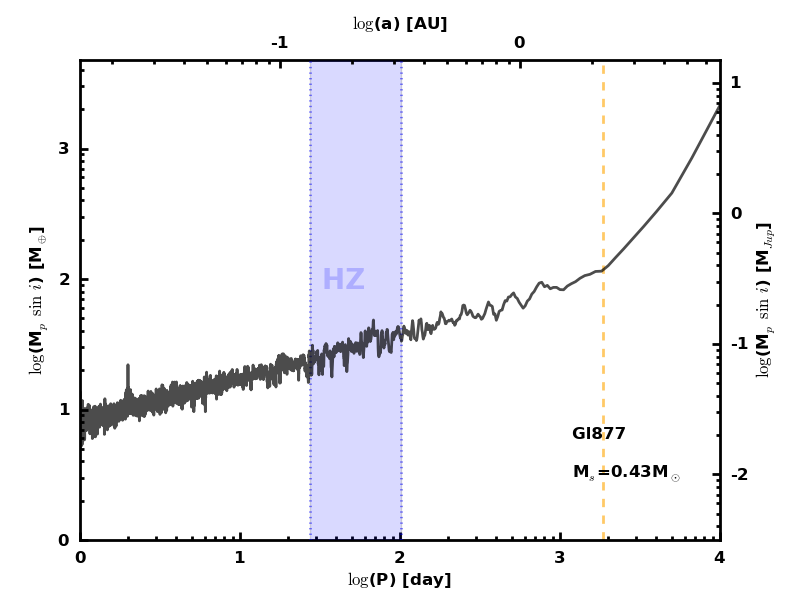}
\includegraphics[width=.9\linewidth]{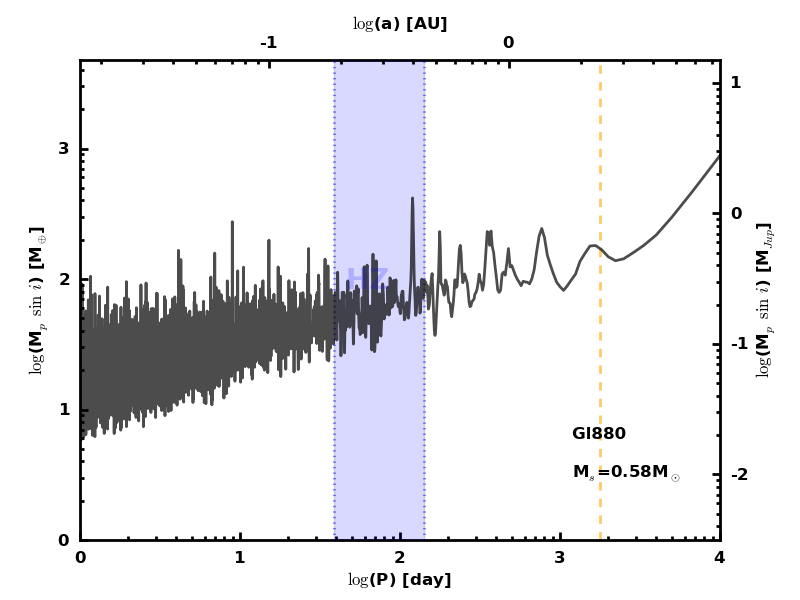}
\includegraphics[width=.9\linewidth]{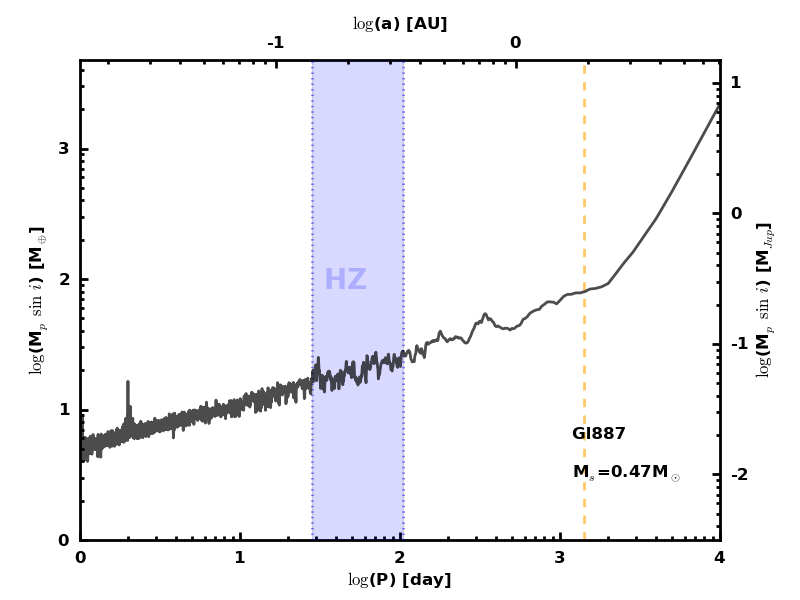}
\includegraphics[width=.9\linewidth]{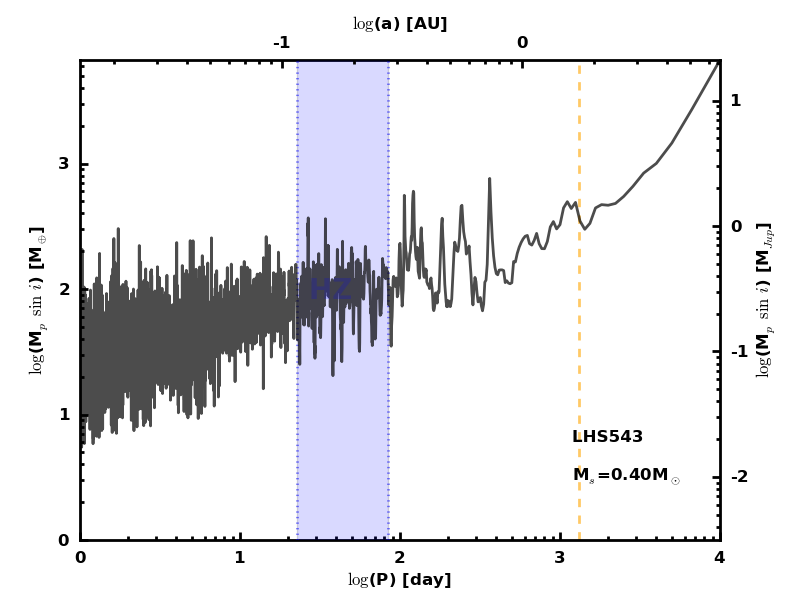}
\end{figure}\clearpage\begin{figure}
\includegraphics[width=.9\linewidth]{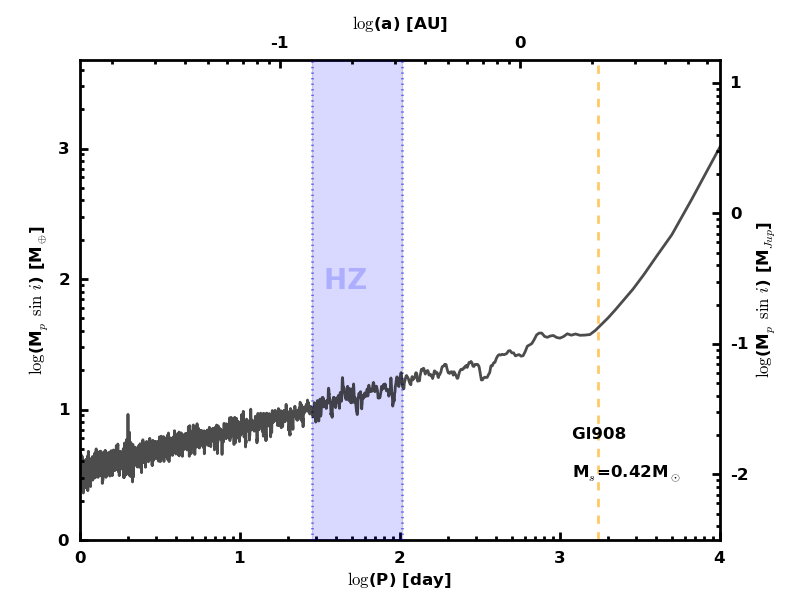}
\includegraphics[width=.9\linewidth]{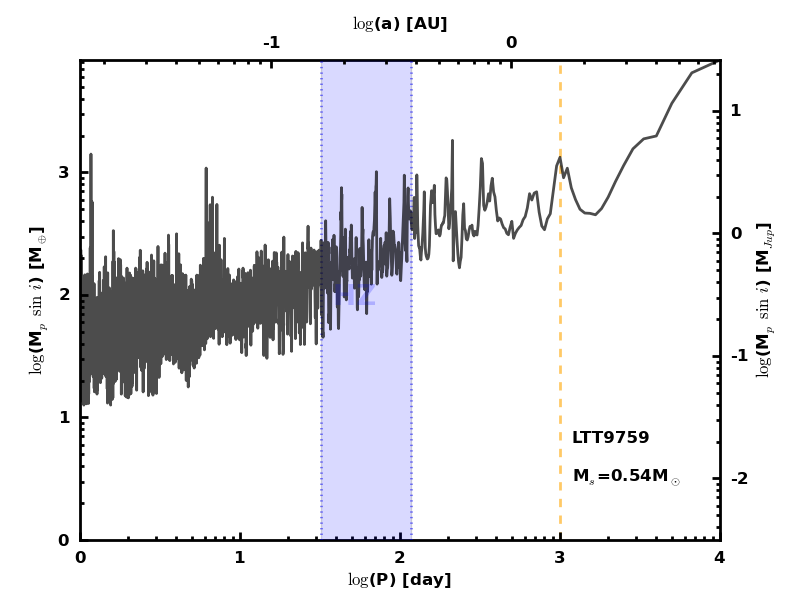}
\caption{\label{fig:limites} Conservative detection limits on $m \ sin i$ for time-series with more than 4 measurements. Planets above the limit are excluded, with a 99\% confidence level, for {\it all} 12 trial phases. Some panels appear with 2 curves : the upper one is the detection limits before any model is subtracted and the bottom one is for the residuals around a chosen model (composed of planets, linear drifts and/or simple sine function). See Sect.~\ref{sect:limits} for details.}
\end{figure}

\newpage
\begin{figure}
\includegraphics[width=.9\linewidth]{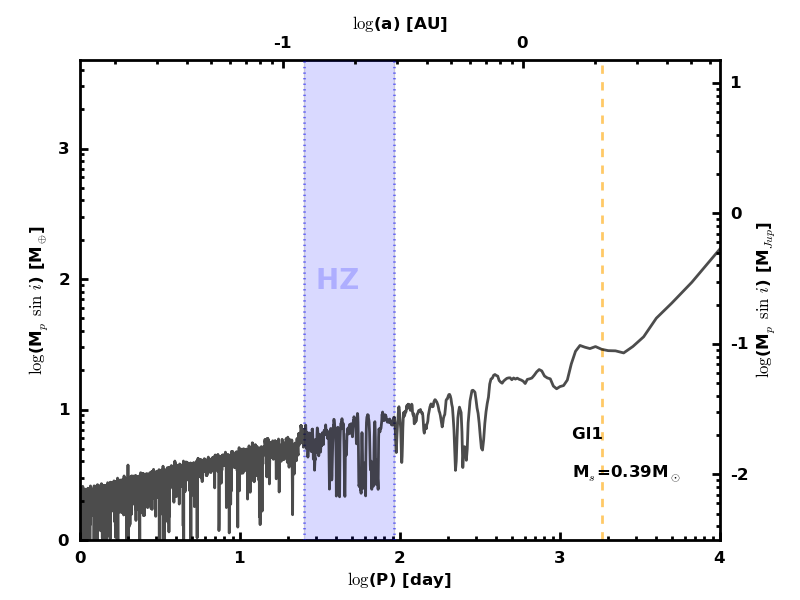}
\includegraphics[width=.9\linewidth]{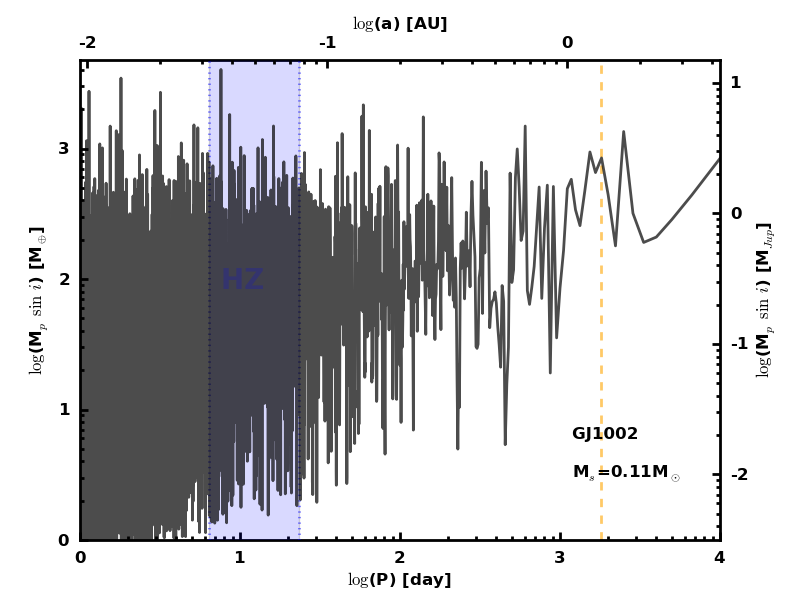}
\includegraphics[width=.9\linewidth]{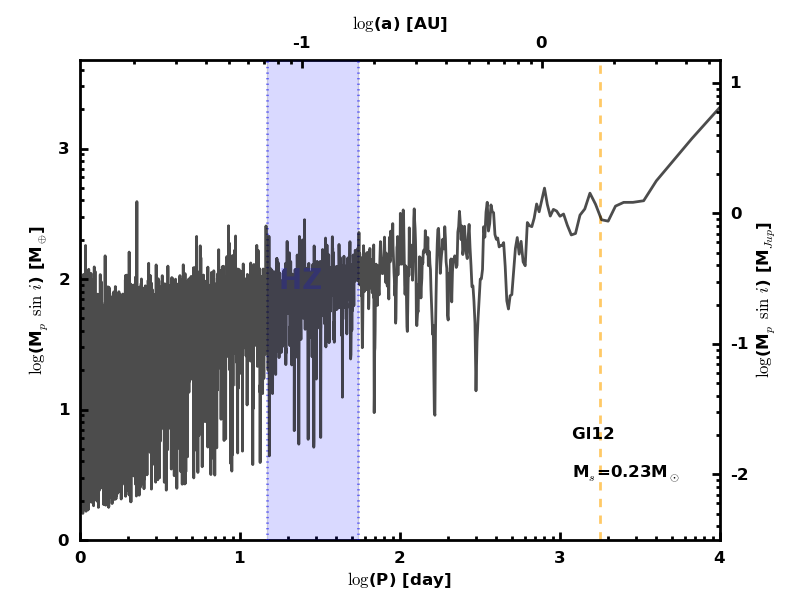}
\includegraphics[width=.9\linewidth]{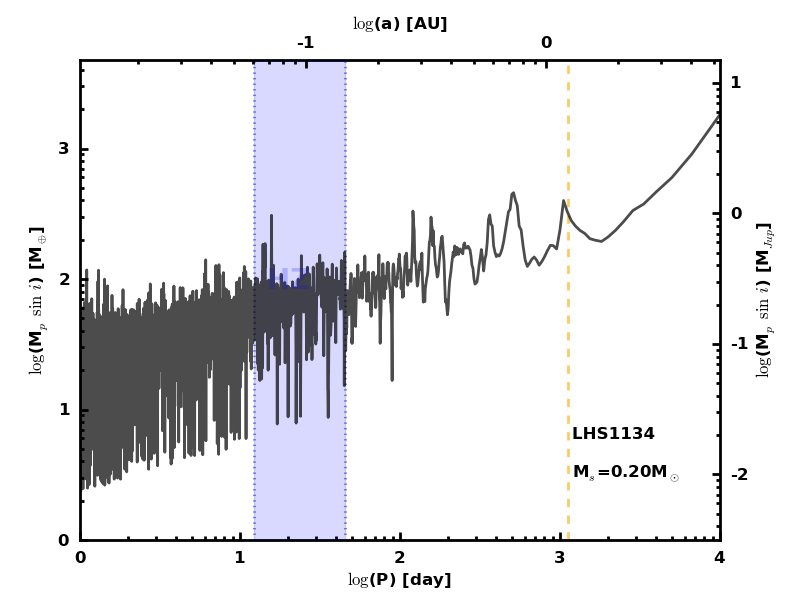}
\end{figure}\begin{figure}
\includegraphics[width=.9\linewidth]{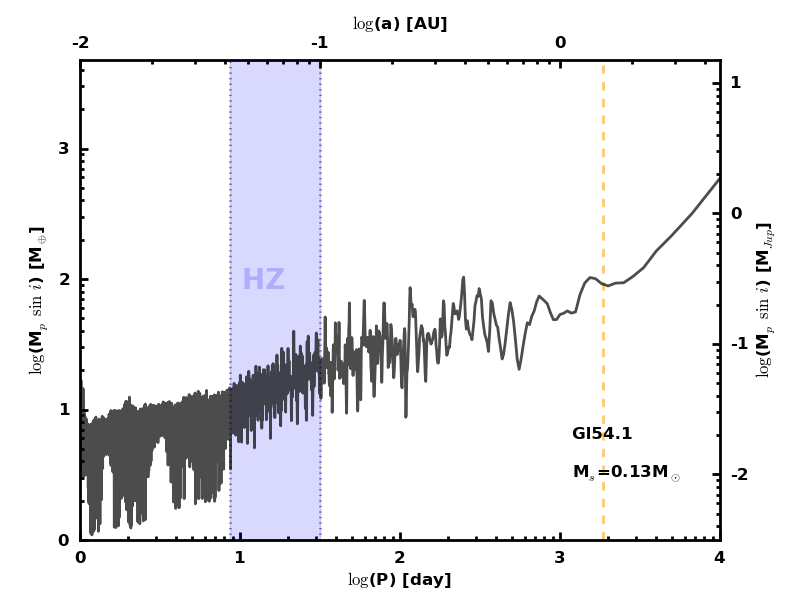}
\includegraphics[width=.9\linewidth]{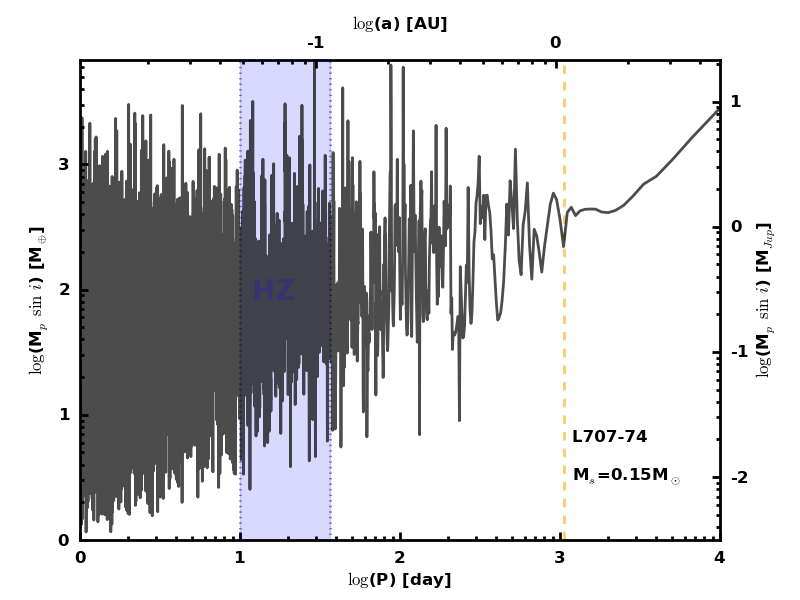}
\includegraphics[width=.9\linewidth]{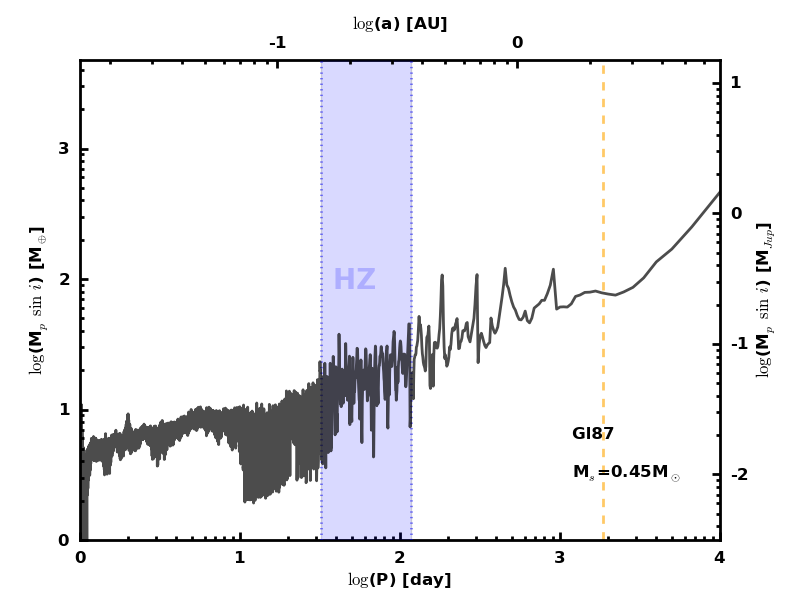}
\includegraphics[width=.9\linewidth]{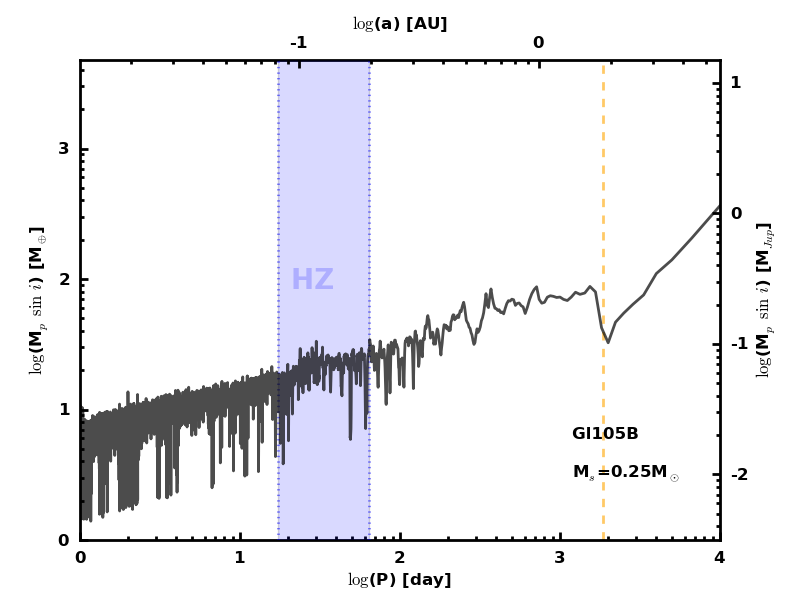}
\end{figure}\begin{figure}
\includegraphics[width=.9\linewidth]{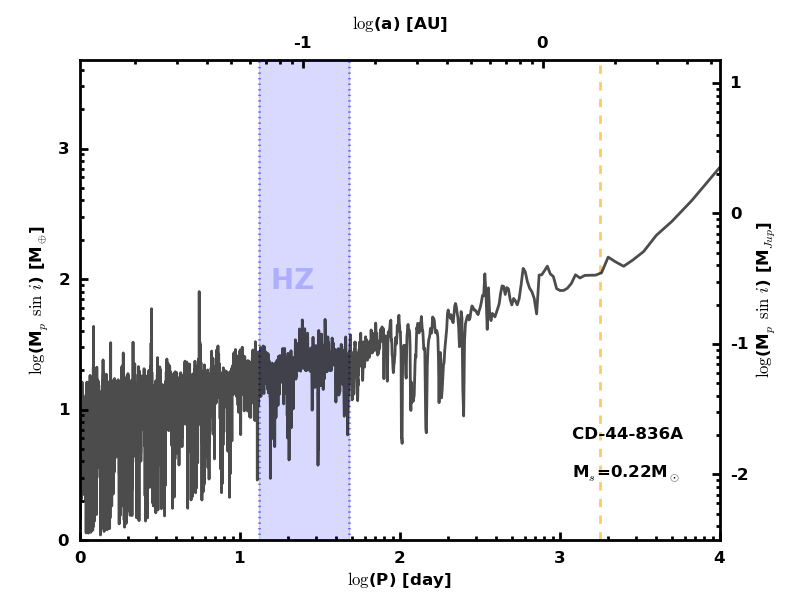}
\includegraphics[width=.9\linewidth]{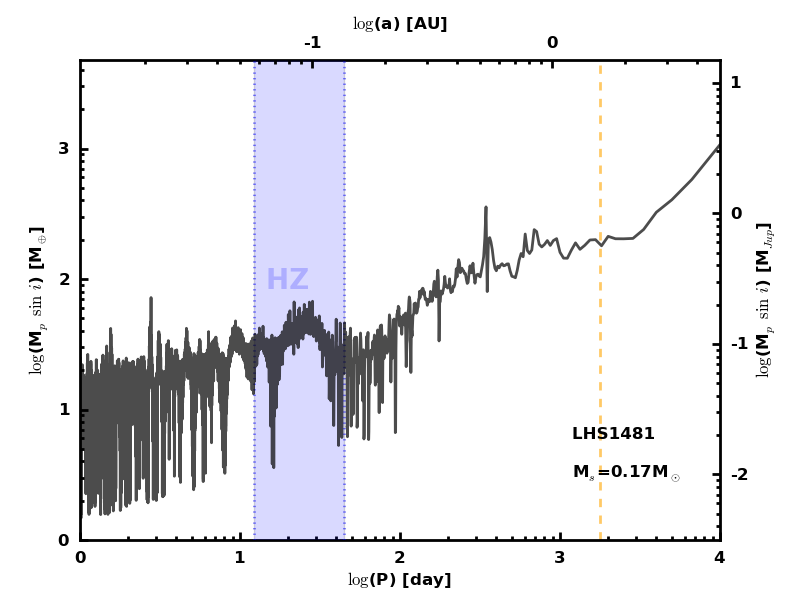}
\includegraphics[width=.9\linewidth]{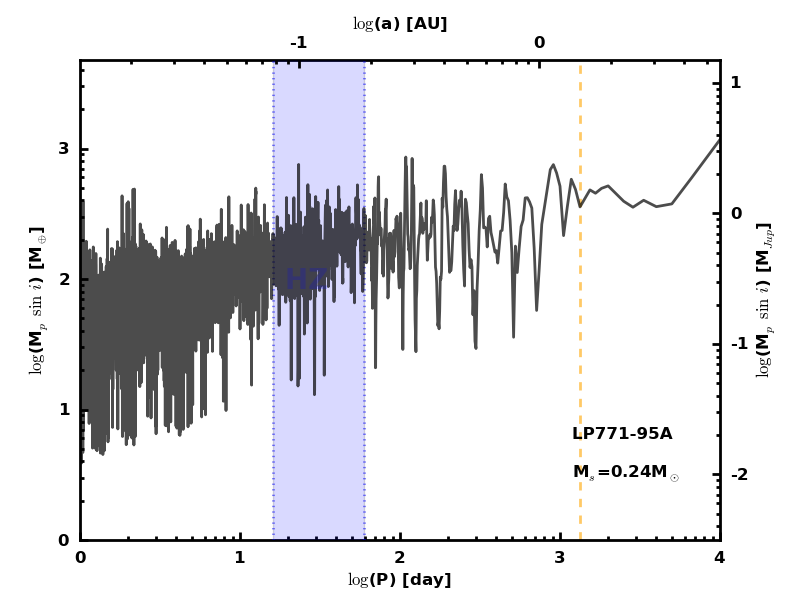}
\includegraphics[width=.9\linewidth]{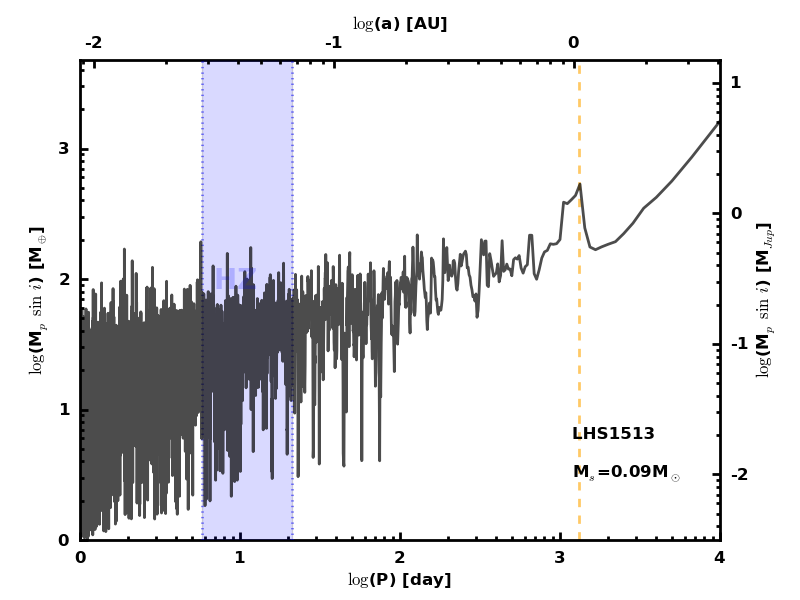}
\end{figure}\begin{figure}
\includegraphics[width=.9\linewidth]{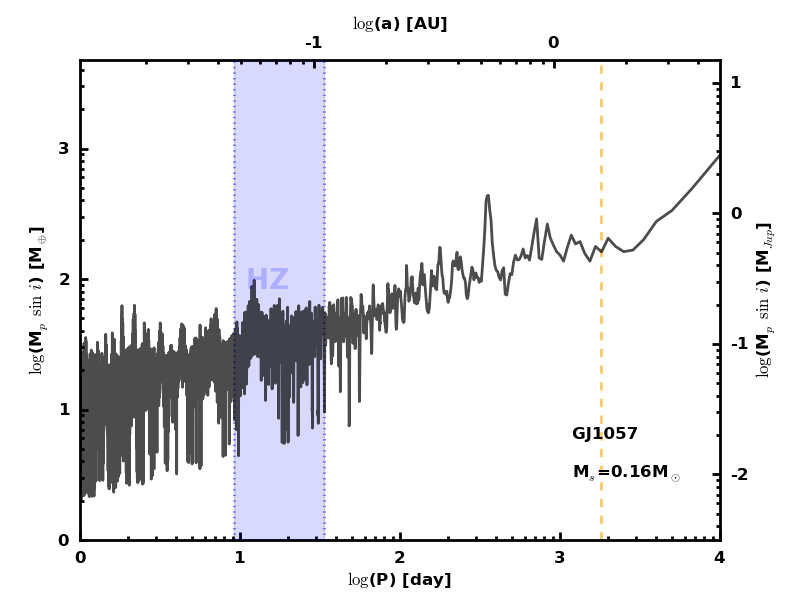}
\includegraphics[width=.9\linewidth]{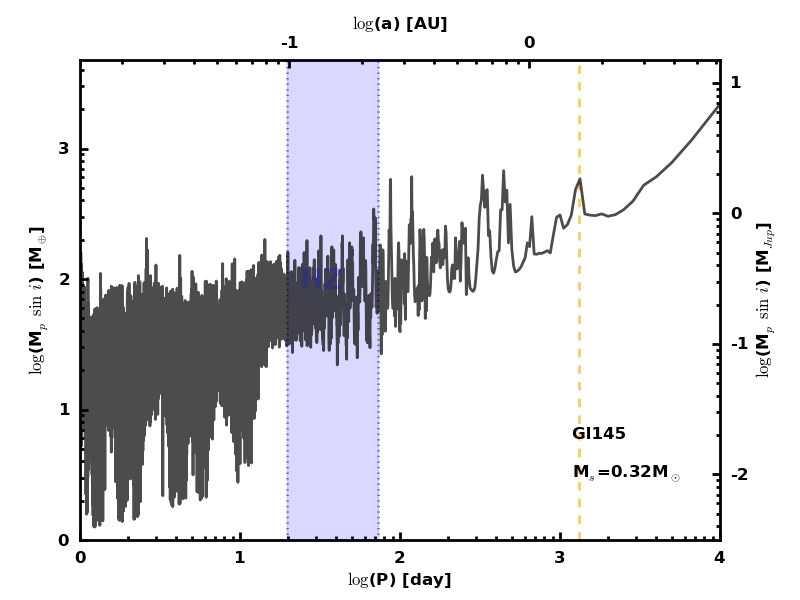}
\includegraphics[width=.9\linewidth]{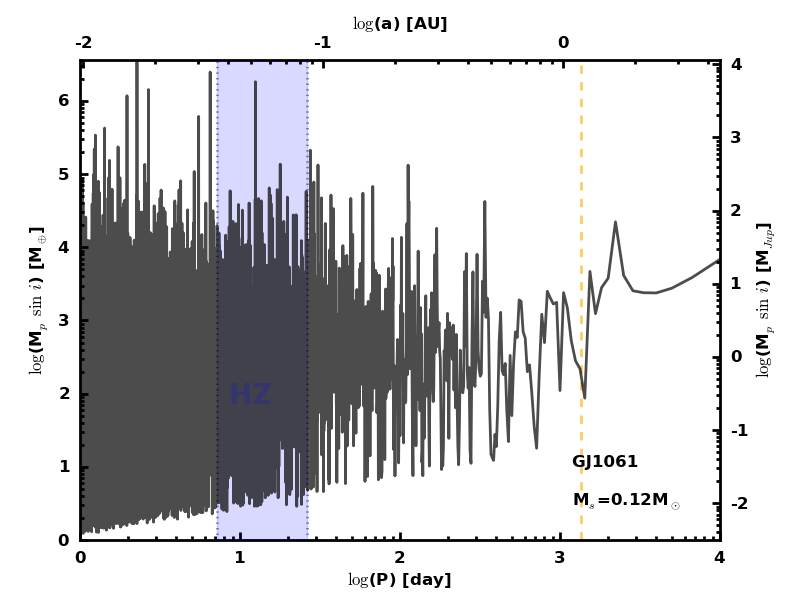}
\includegraphics[width=.9\linewidth]{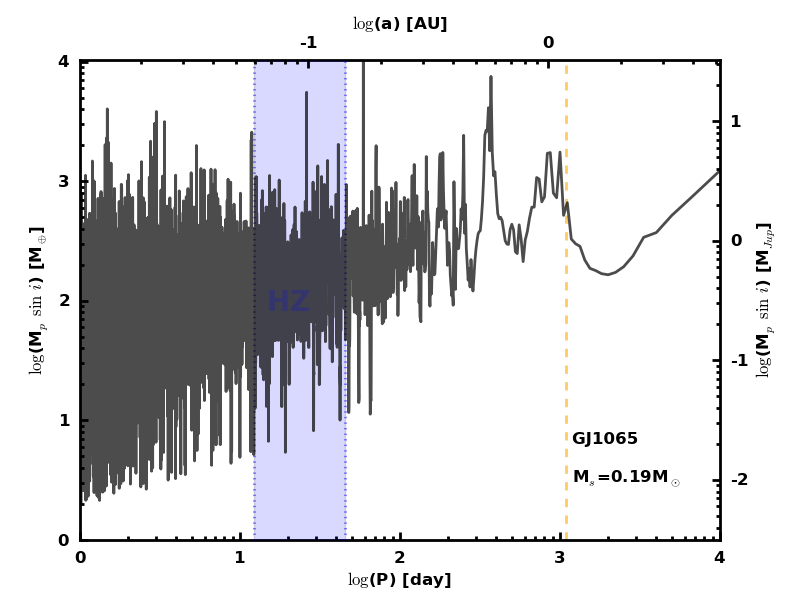}
\end{figure}\clearpage\begin{figure}
\includegraphics[width=.9\linewidth]{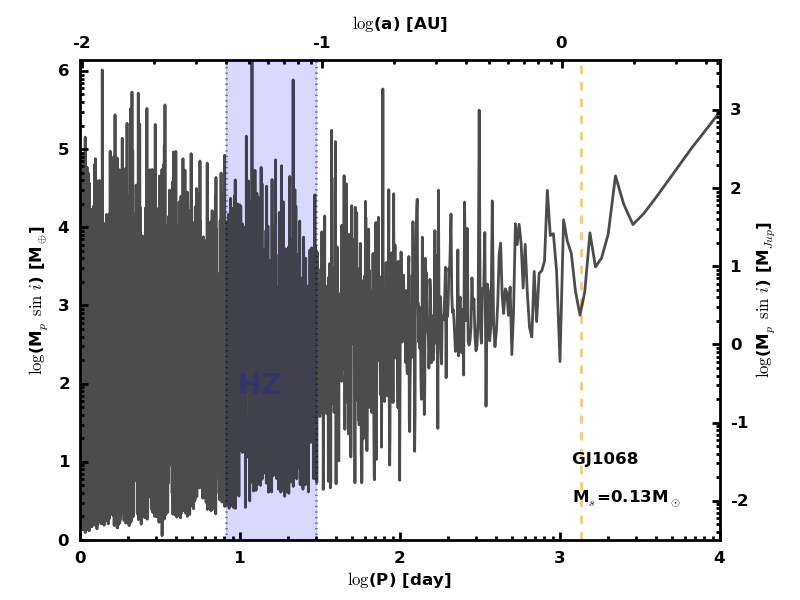}
\includegraphics[width=.9\linewidth]{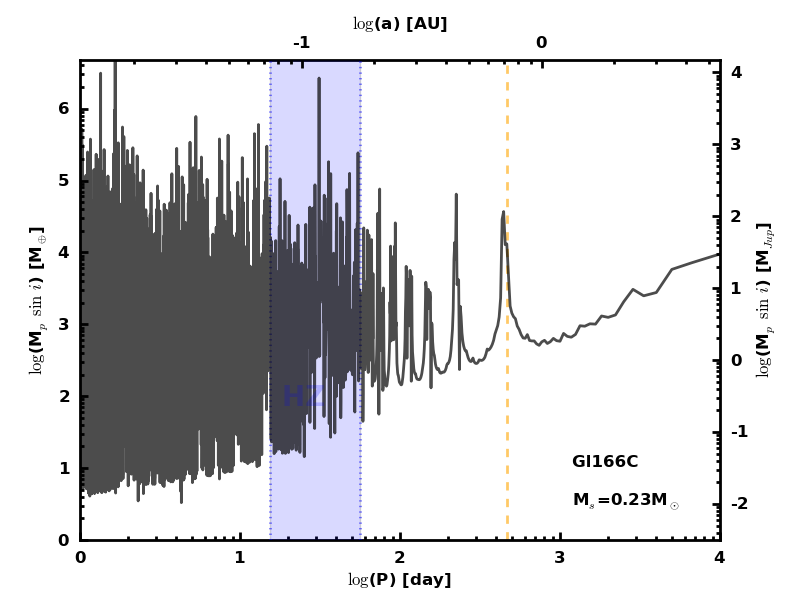}
\includegraphics[width=.9\linewidth]{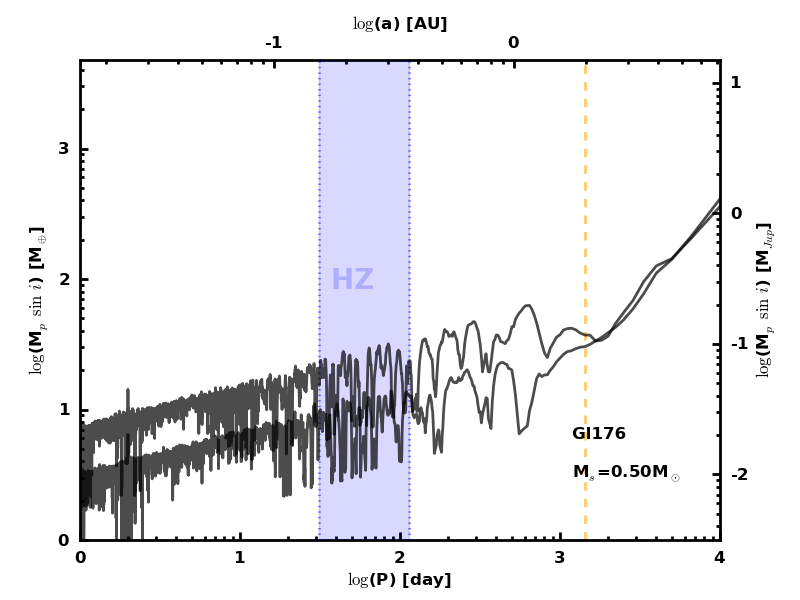}
\includegraphics[width=.9\linewidth]{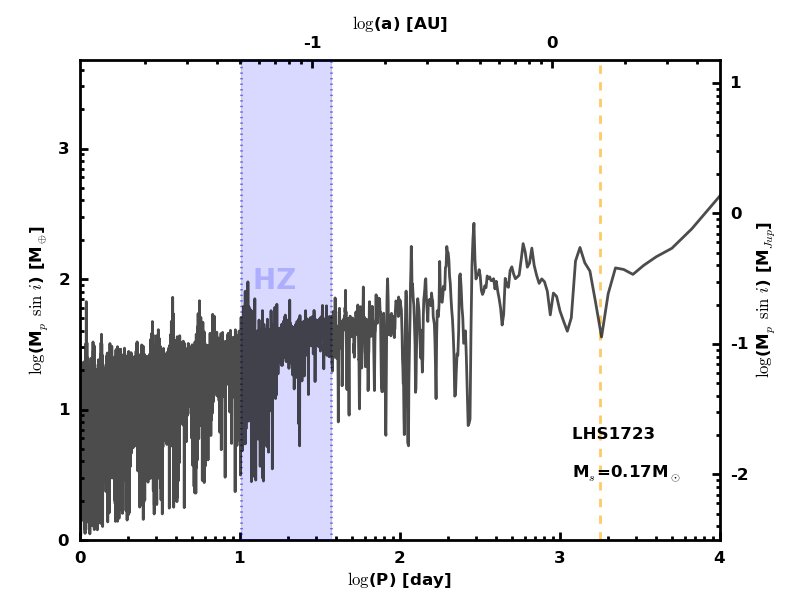}
\end{figure}\begin{figure}
\includegraphics[width=.9\linewidth]{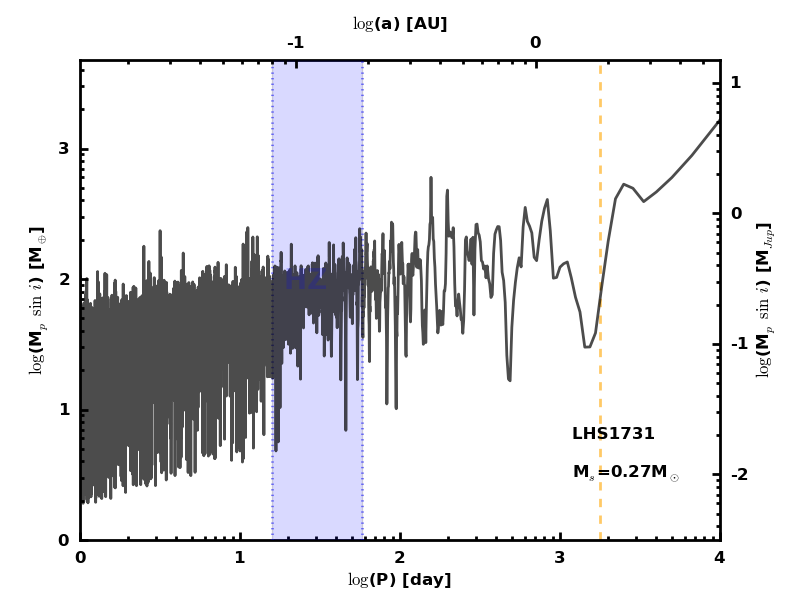}
\includegraphics[width=.9\linewidth]{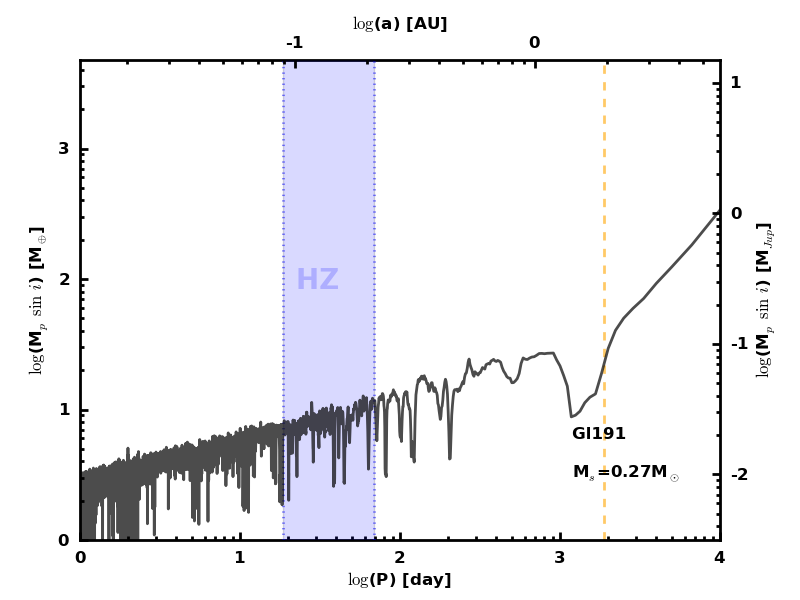}
\includegraphics[width=.9\linewidth]{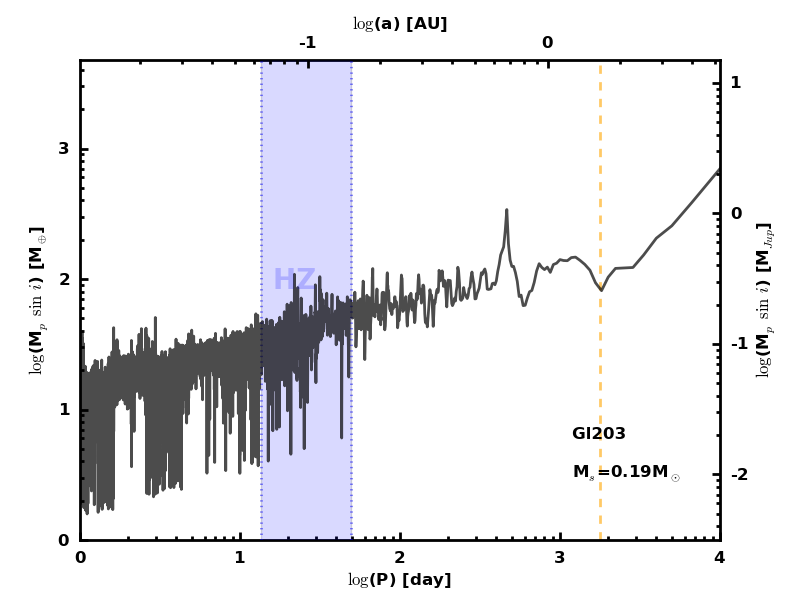}
\includegraphics[width=.9\linewidth]{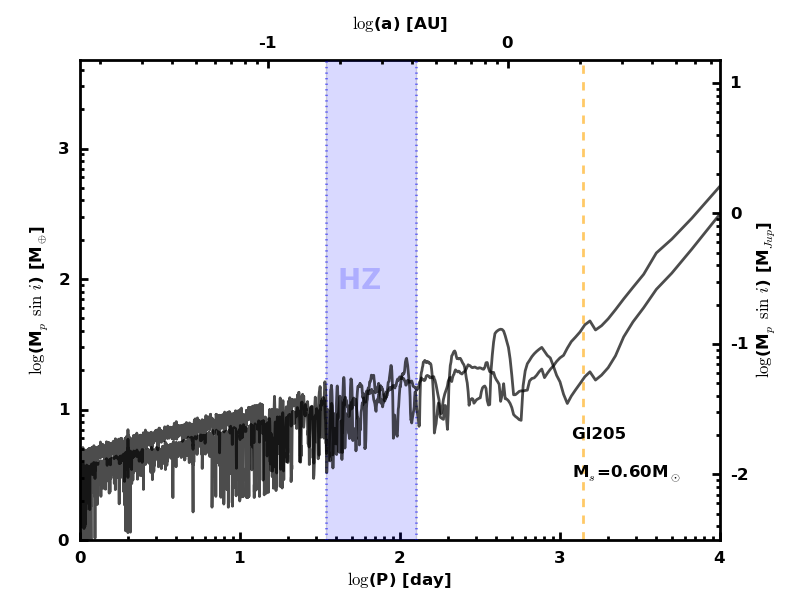}
\end{figure}\begin{figure}
\includegraphics[width=.9\linewidth]{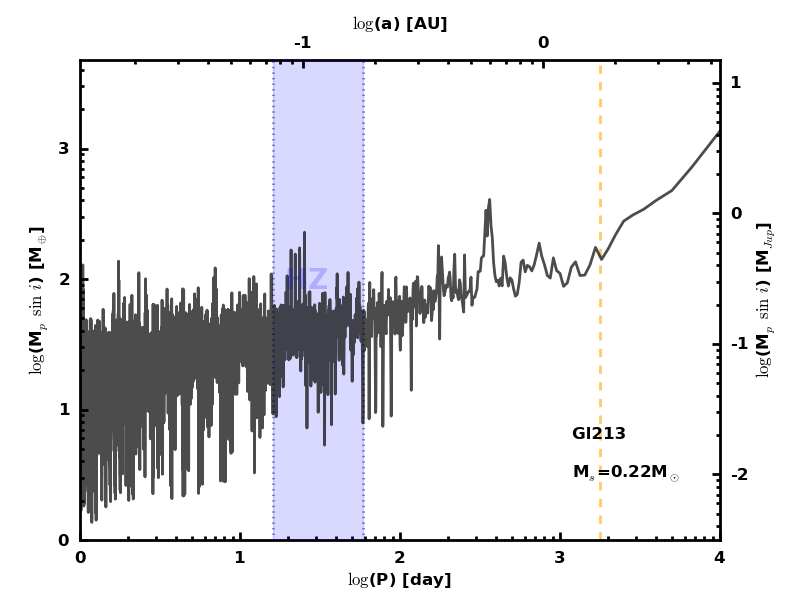}
\includegraphics[width=.9\linewidth]{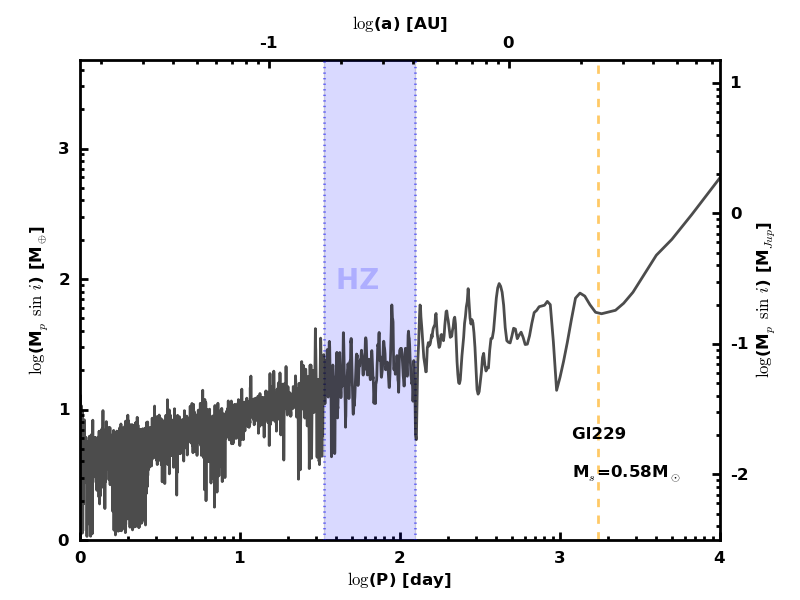}
\includegraphics[width=.9\linewidth]{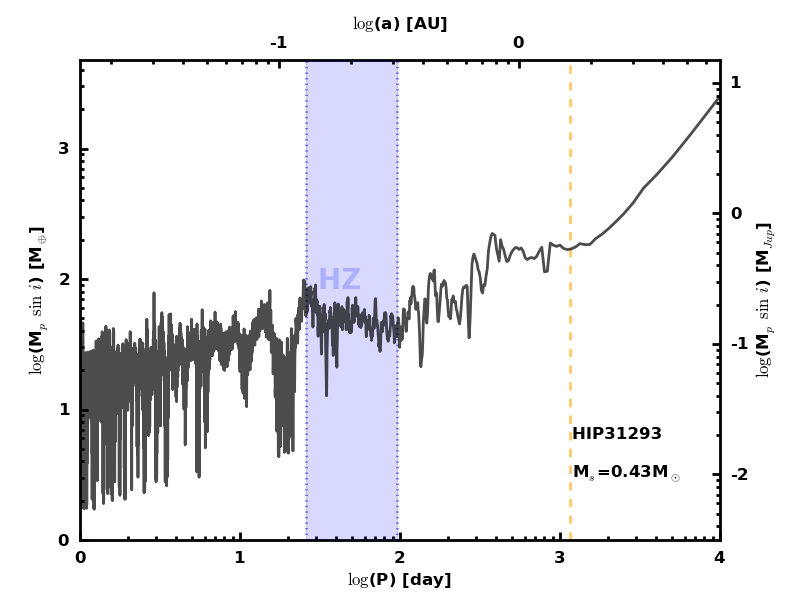}
\includegraphics[width=.9\linewidth]{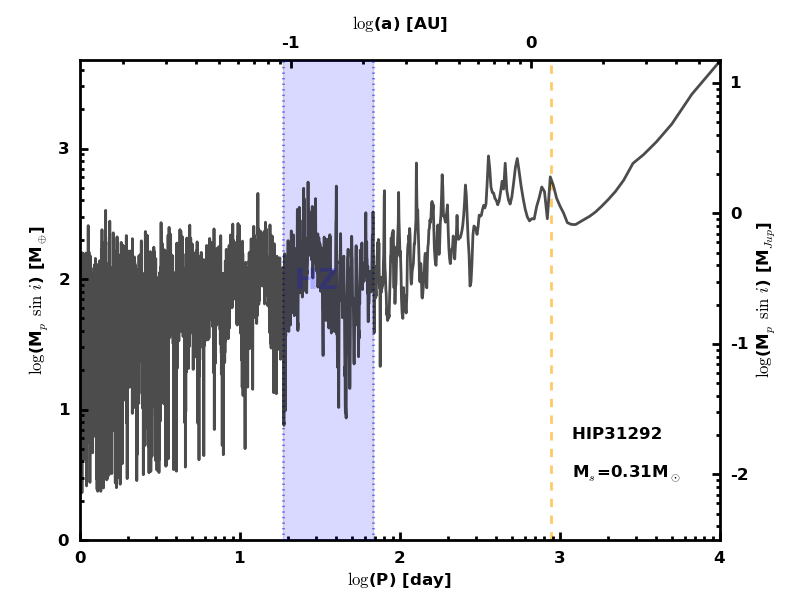}
\end{figure}\begin{figure}
\includegraphics[width=.9\linewidth]{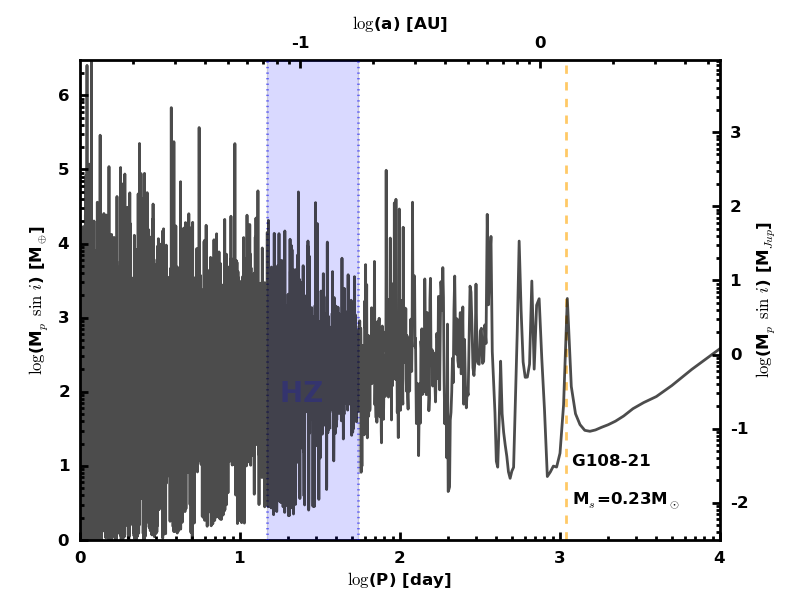}
\includegraphics[width=.9\linewidth]{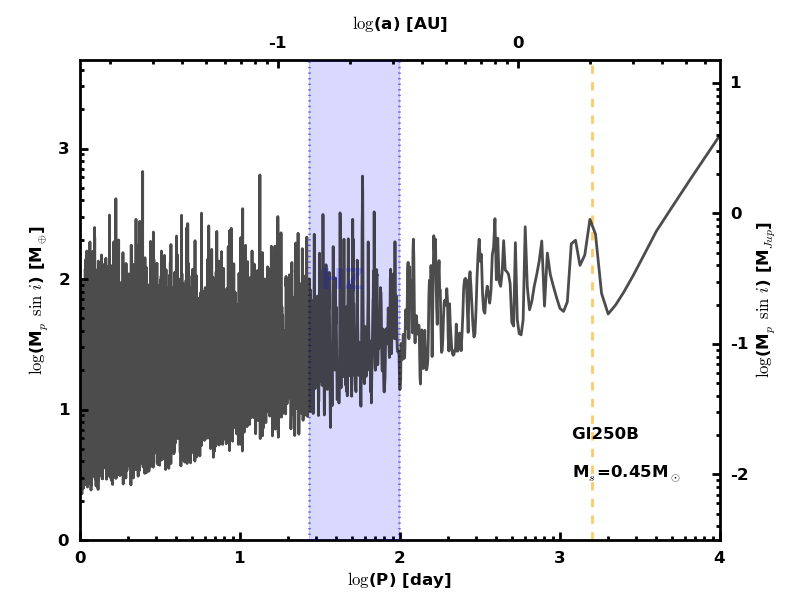}
\includegraphics[width=.9\linewidth]{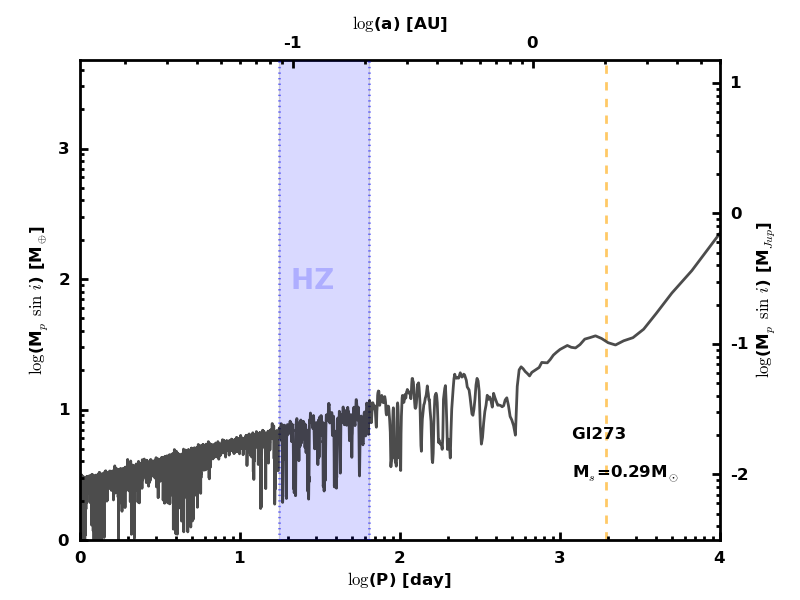}
\includegraphics[width=.9\linewidth]{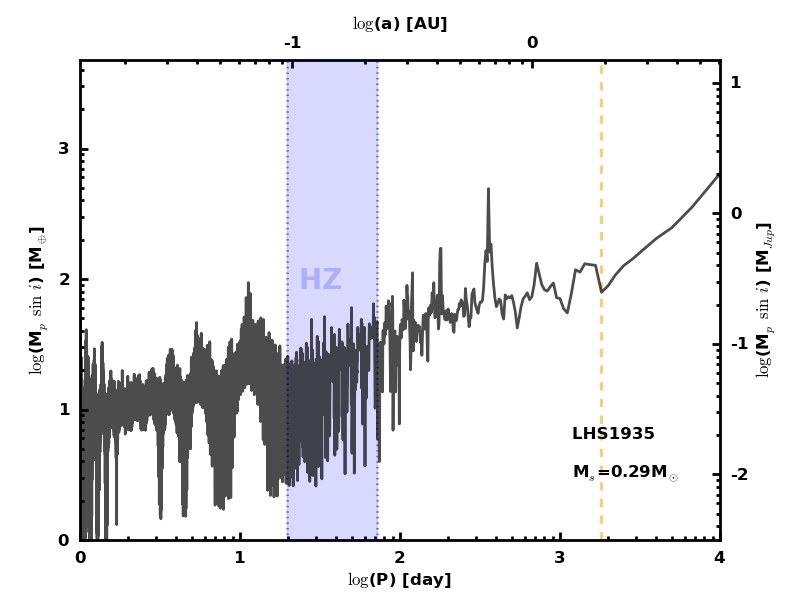}
\end{figure}\clearpage\begin{figure}
\includegraphics[width=.9\linewidth]{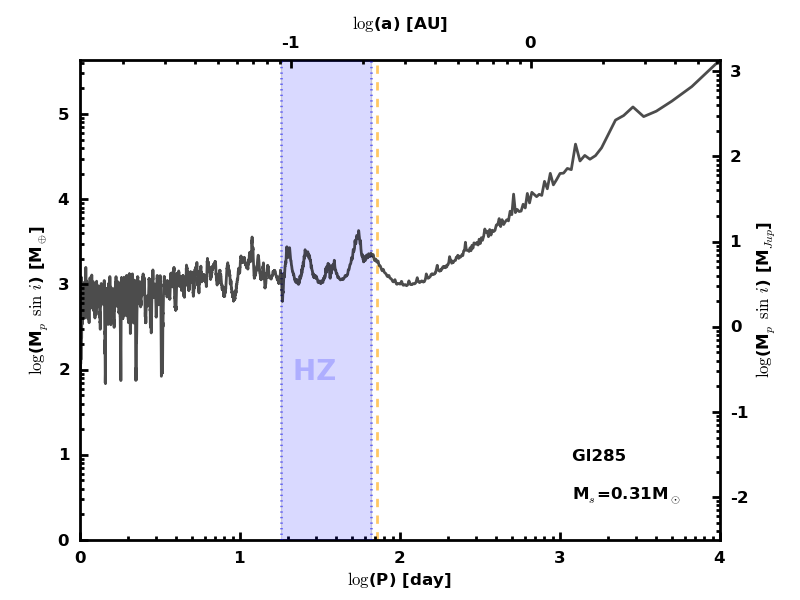}
\includegraphics[width=.9\linewidth]{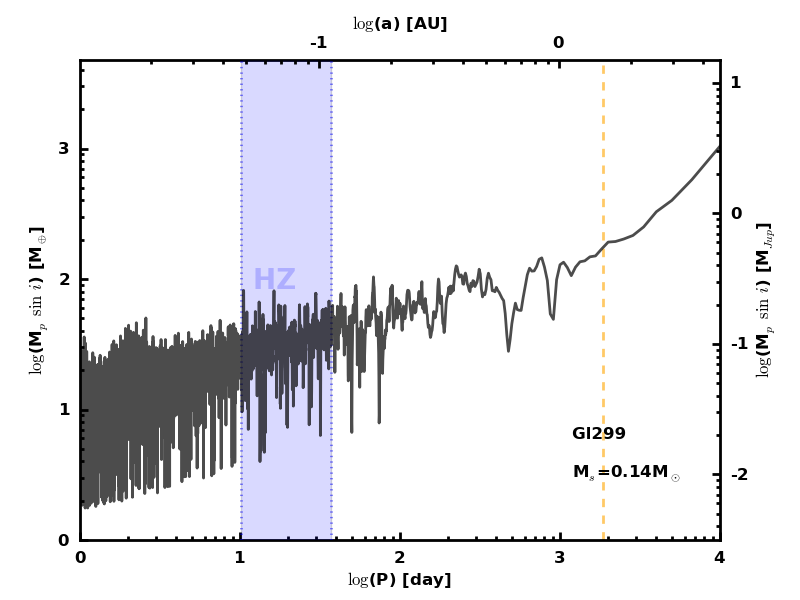}
\includegraphics[width=.9\linewidth]{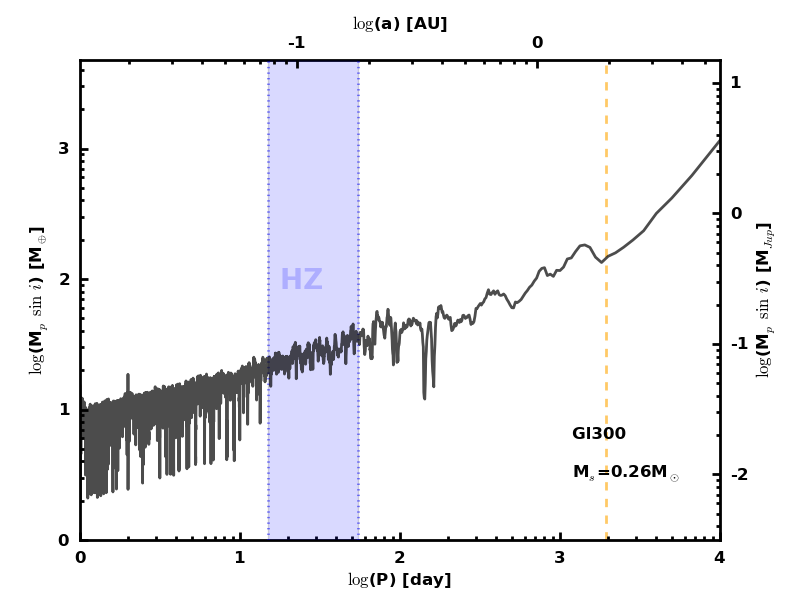}
\includegraphics[width=.9\linewidth]{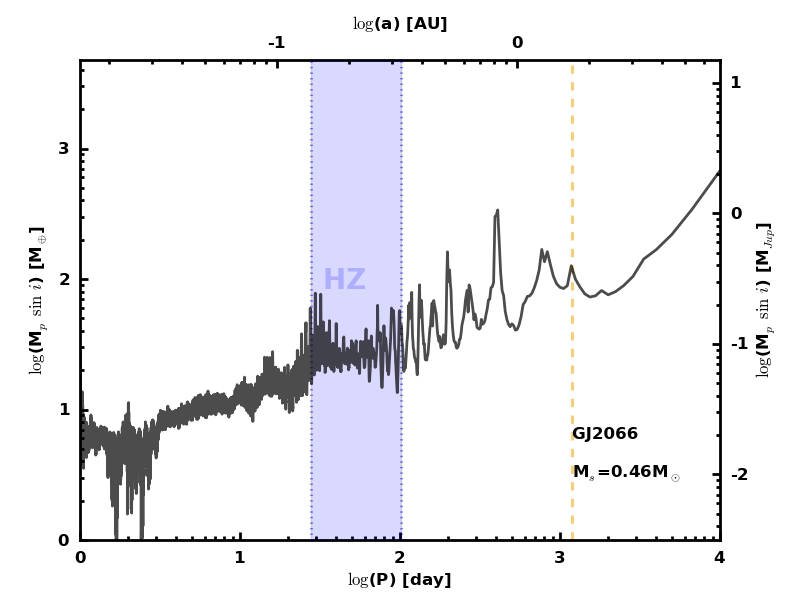}
\end{figure}\begin{figure}
\includegraphics[width=.9\linewidth]{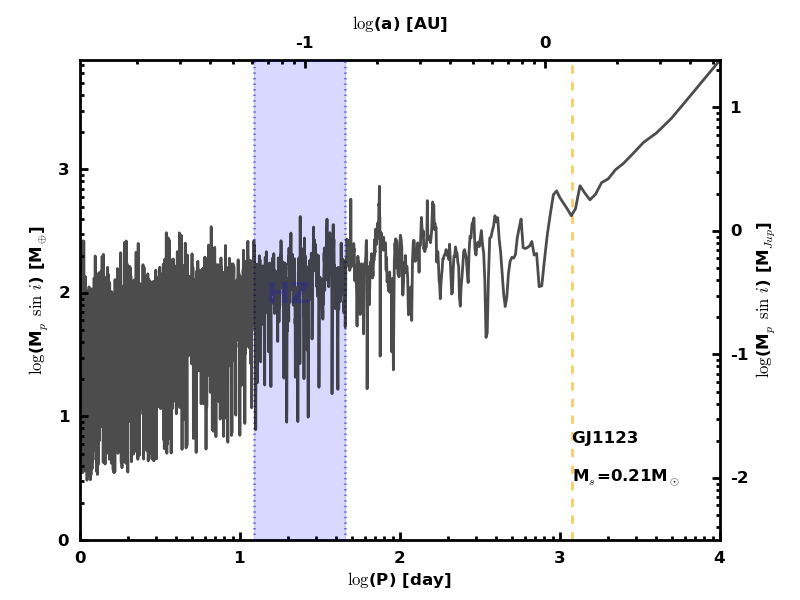}
\includegraphics[width=.9\linewidth]{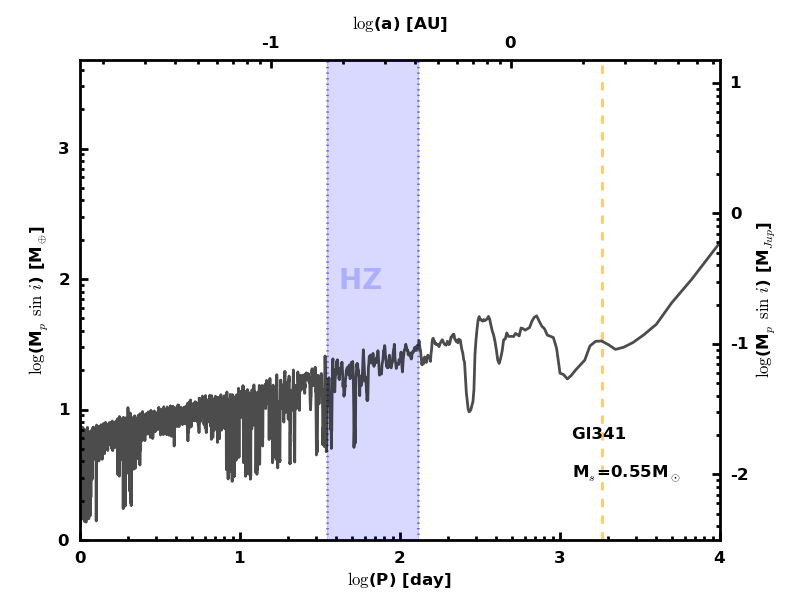}
\includegraphics[width=.9\linewidth]{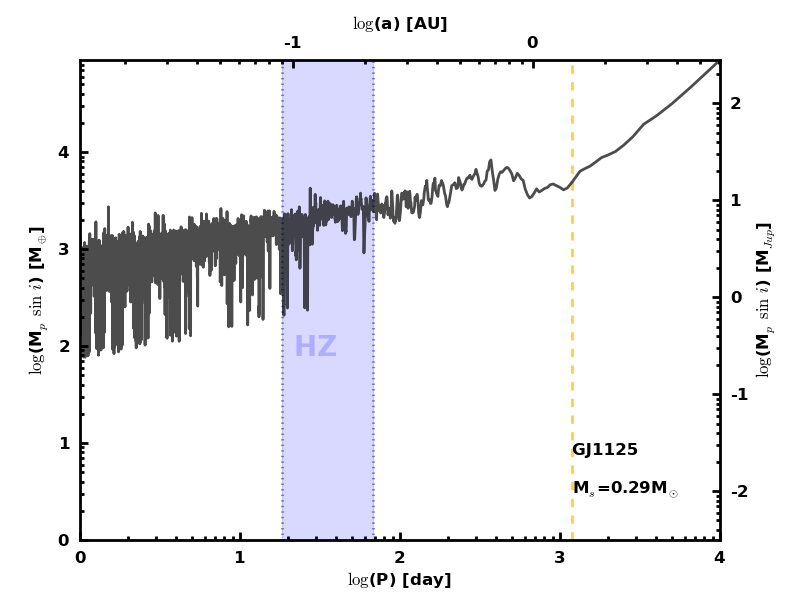}
\includegraphics[width=.9\linewidth]{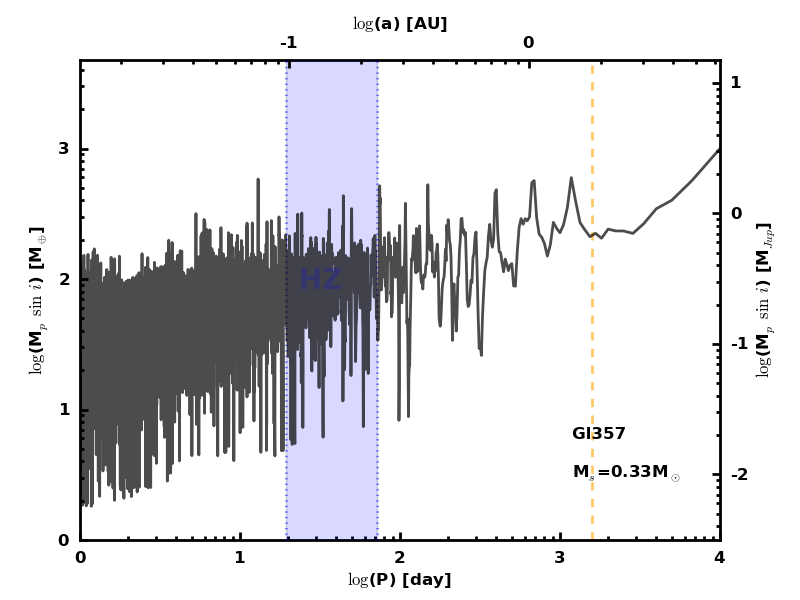}
\end{figure}\begin{figure}
\includegraphics[width=.9\linewidth]{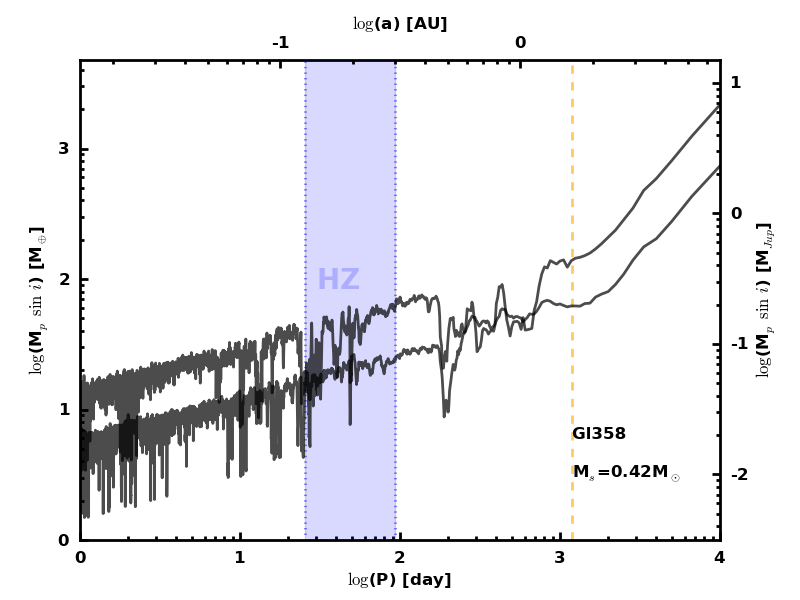}
\includegraphics[width=.9\linewidth]{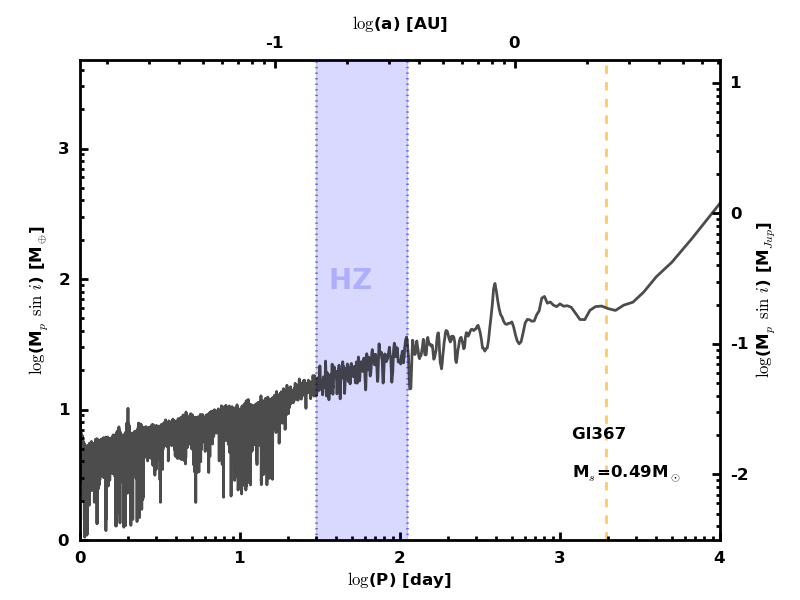}
\includegraphics[width=.9\linewidth]{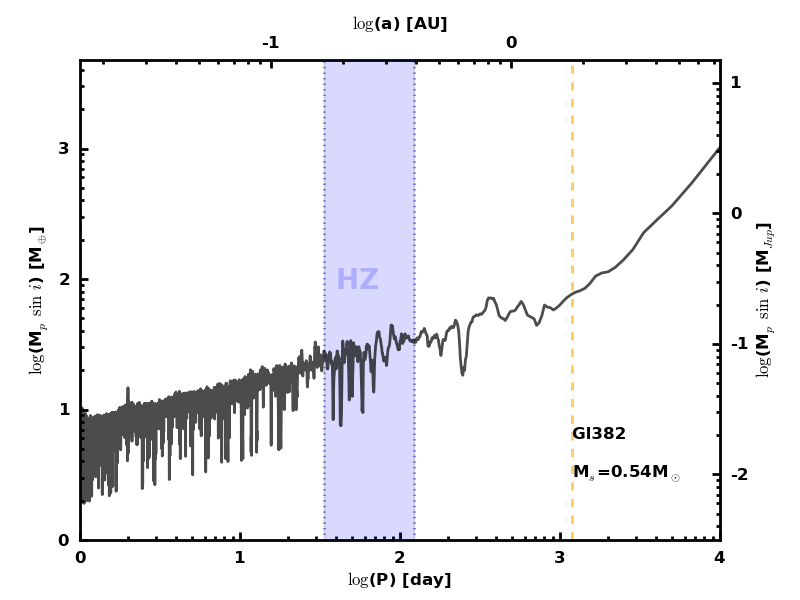}
\includegraphics[width=.9\linewidth]{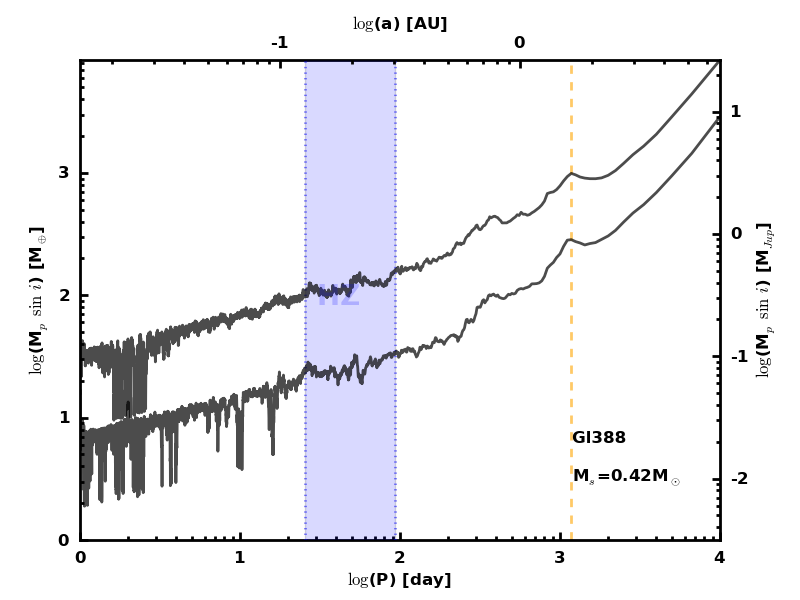}
\end{figure}\begin{figure}
\includegraphics[width=.9\linewidth]{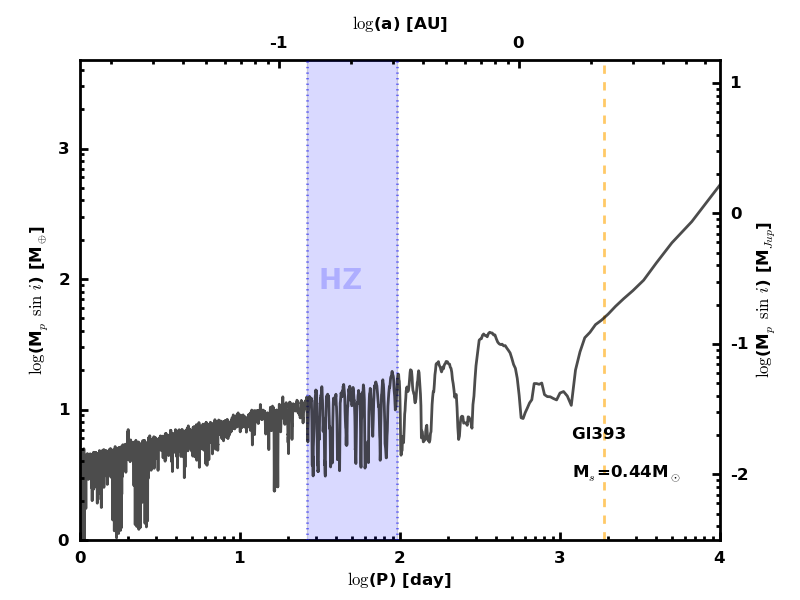}
\includegraphics[width=.9\linewidth]{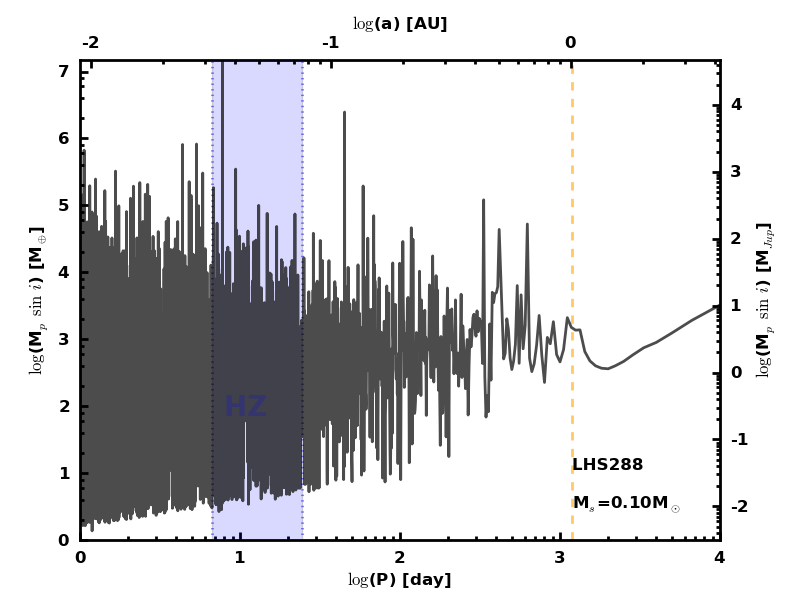}
\includegraphics[width=.9\linewidth]{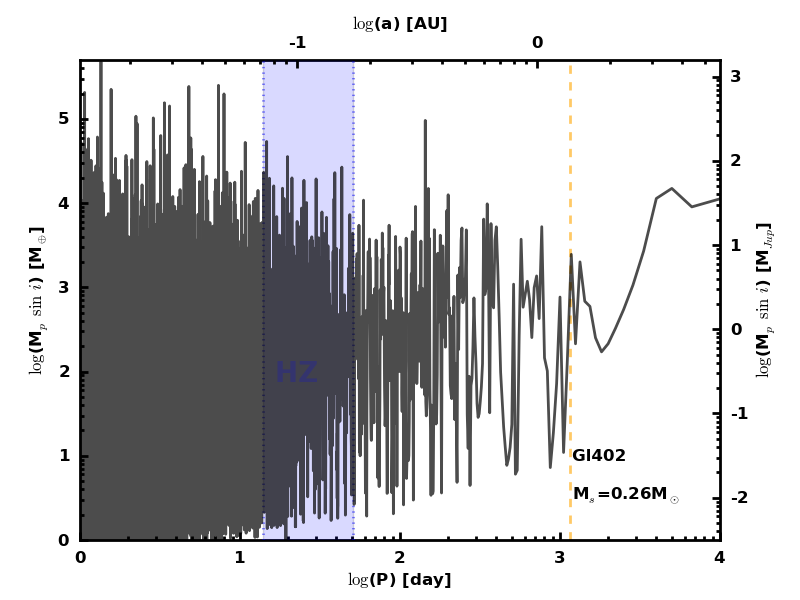}
\includegraphics[width=.9\linewidth]{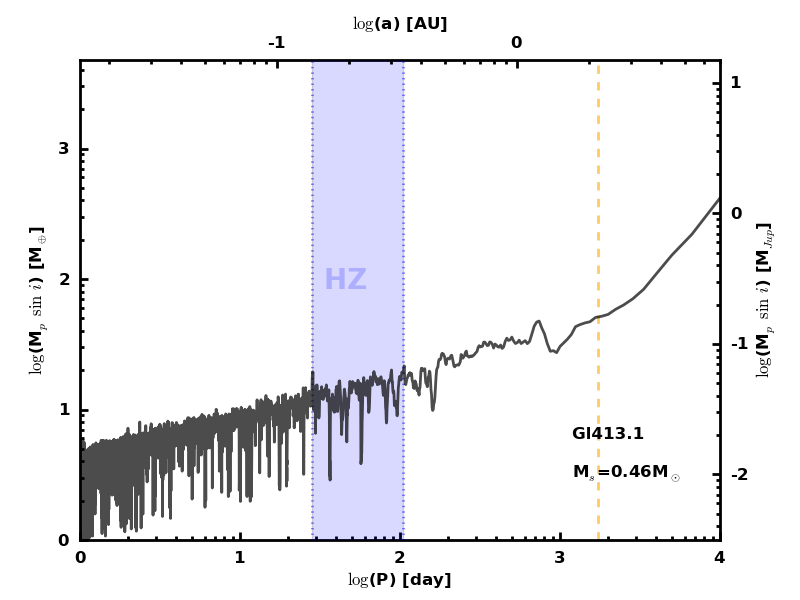}
\end{figure}\clearpage\begin{figure}
\includegraphics[width=.9\linewidth]{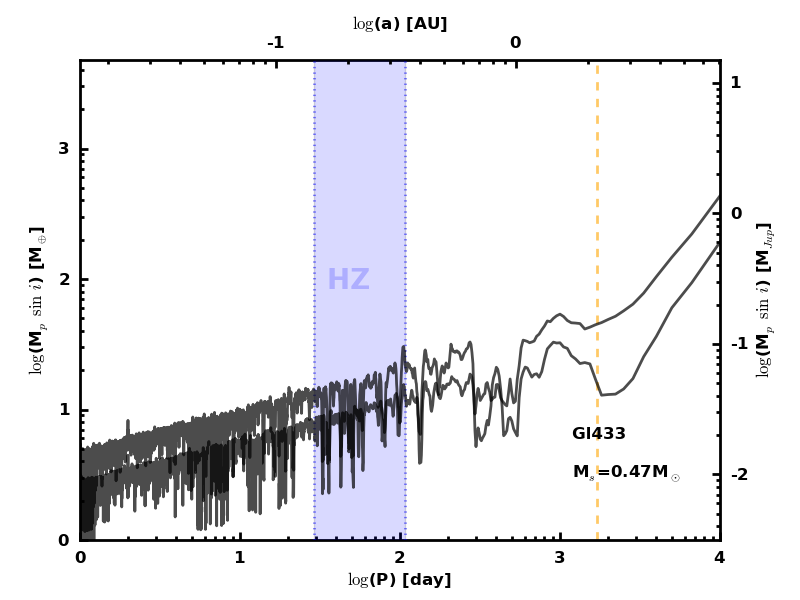}
\includegraphics[width=.9\linewidth]{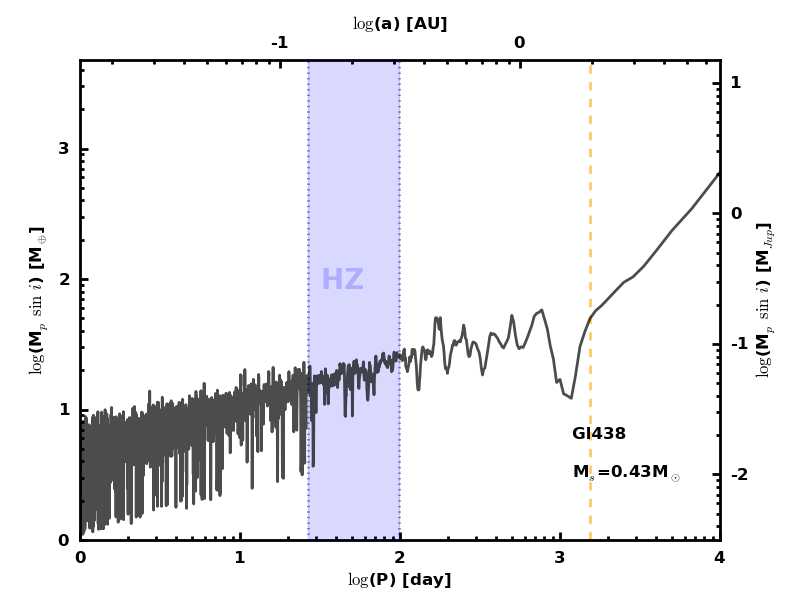}
\includegraphics[width=.9\linewidth]{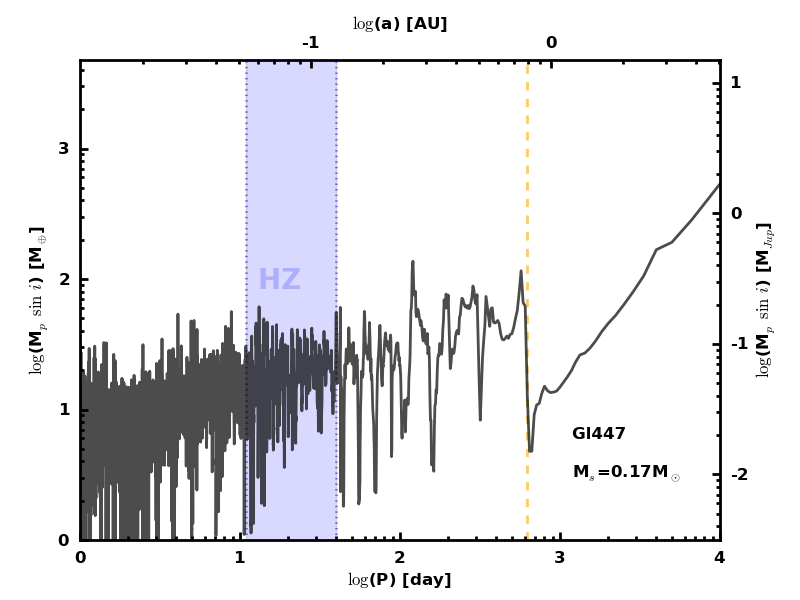}
\includegraphics[width=.9\linewidth]{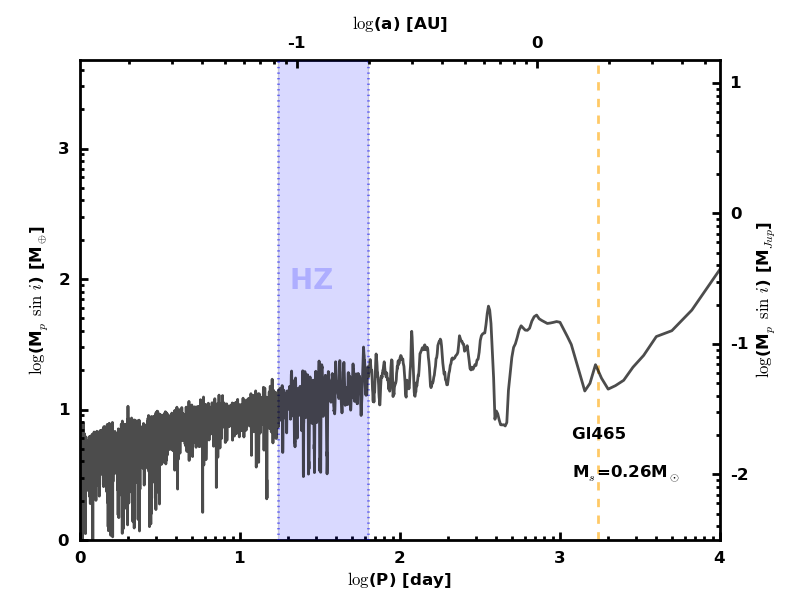}
\end{figure}\begin{figure}
\includegraphics[width=.9\linewidth]{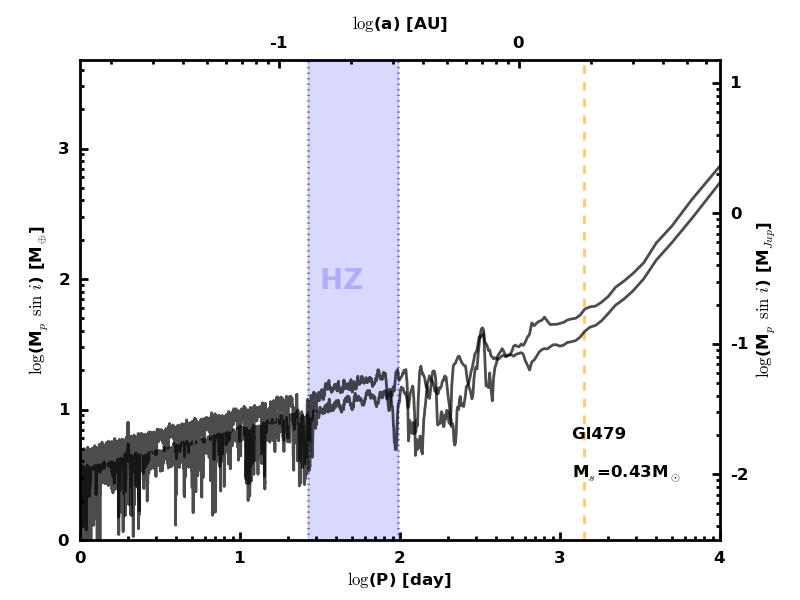}
\includegraphics[width=.9\linewidth]{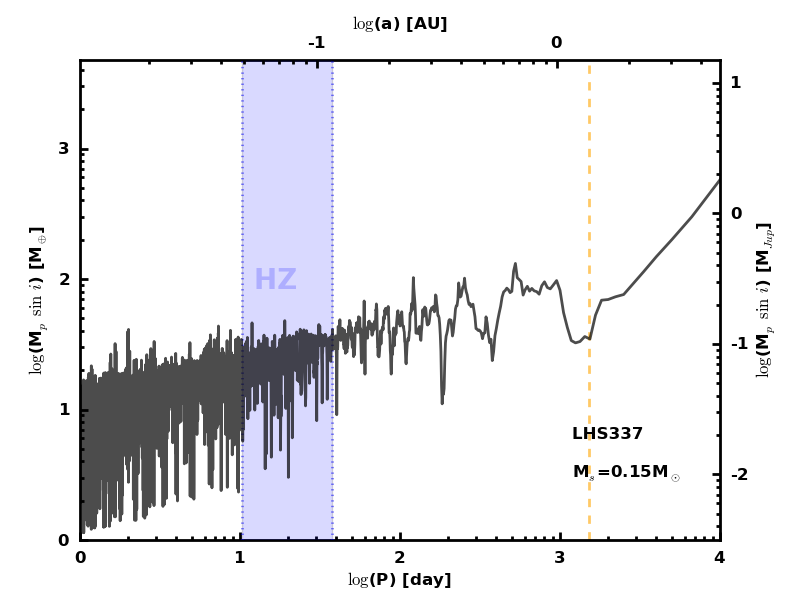}
\includegraphics[width=.9\linewidth]{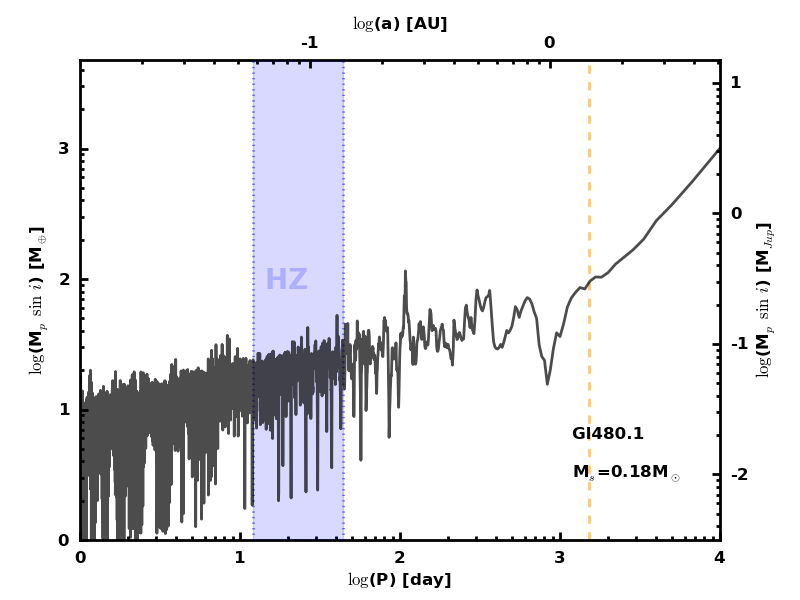}
\includegraphics[width=.9\linewidth]{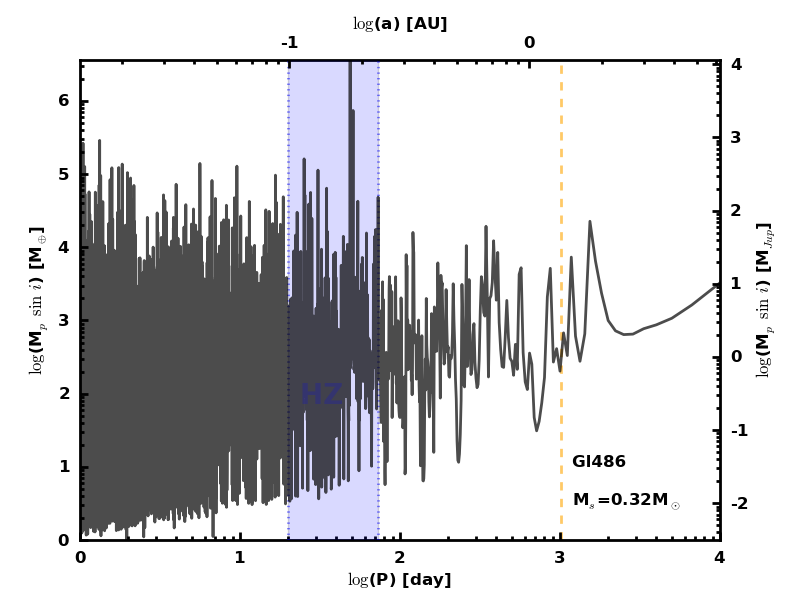}
\end{figure}\begin{figure}
\includegraphics[width=.9\linewidth]{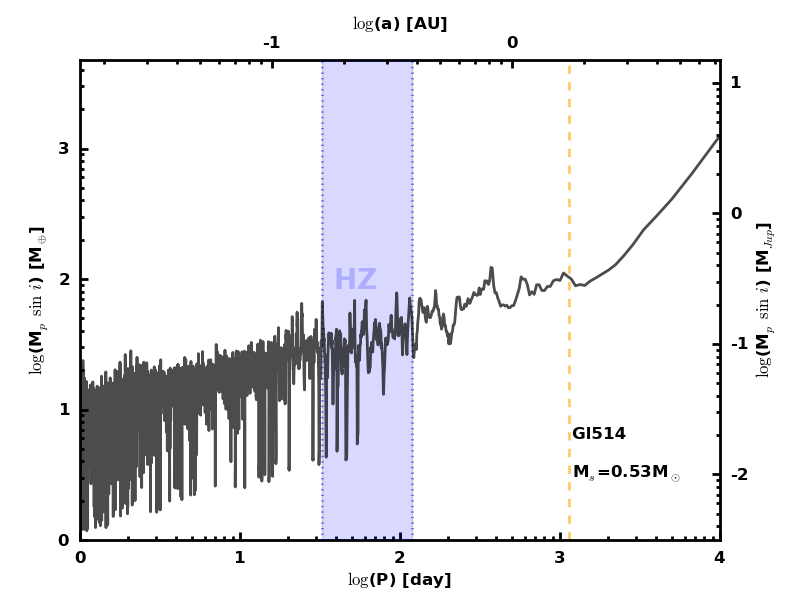}
\includegraphics[width=.9\linewidth]{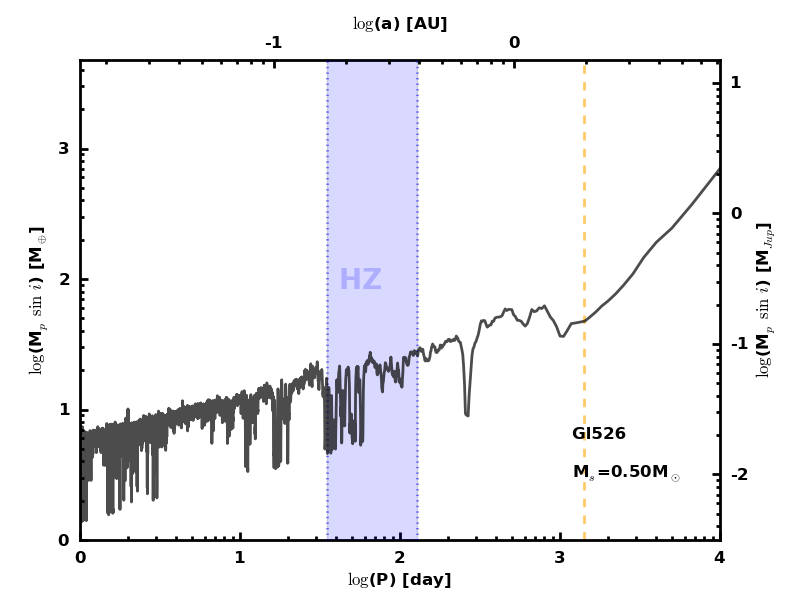}
\includegraphics[width=.9\linewidth]{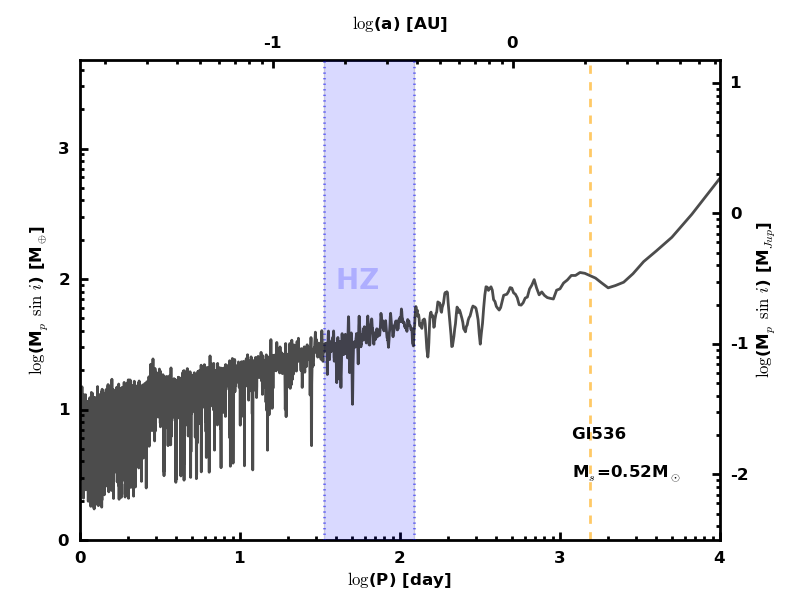}
\includegraphics[width=.9\linewidth]{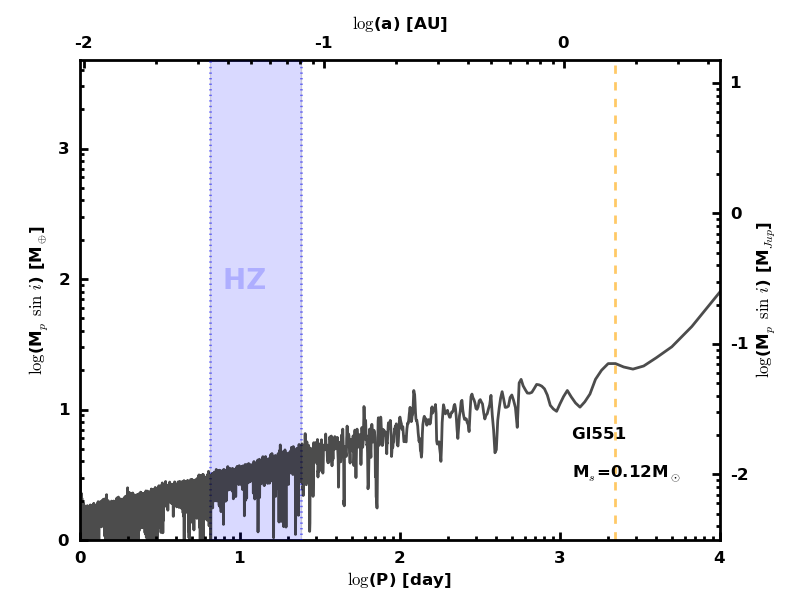}
\end{figure}\begin{figure}
\includegraphics[width=.9\linewidth]{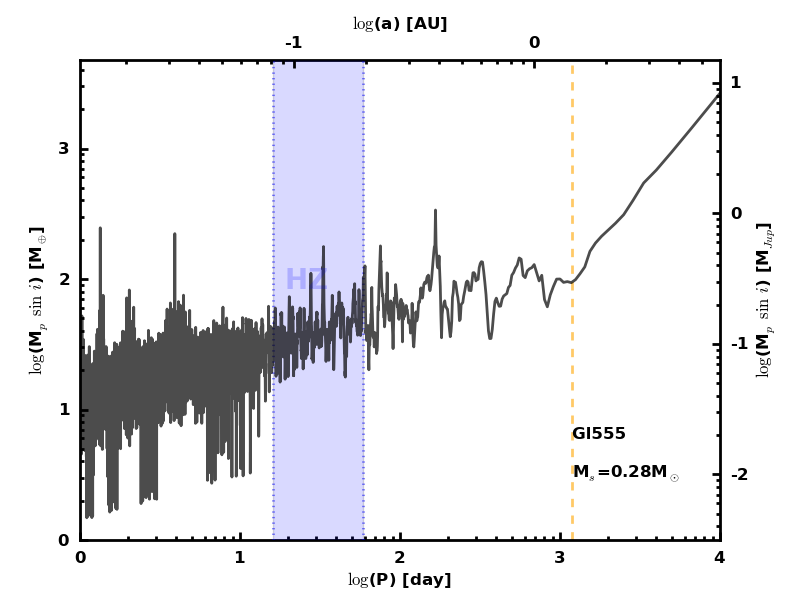}
\includegraphics[width=.9\linewidth]{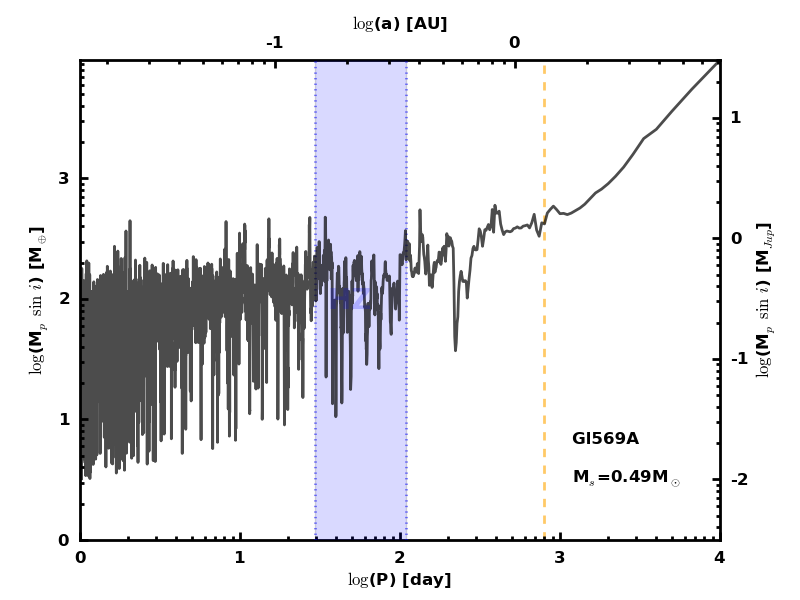}
\includegraphics[width=.9\linewidth]{Gl581_dlm2.png}
\includegraphics[width=.9\linewidth]{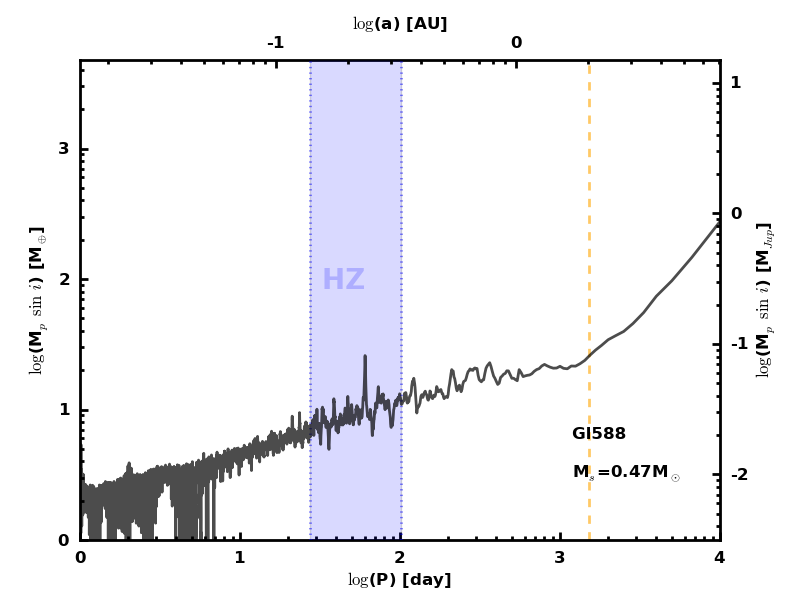}
\end{figure}\clearpage\begin{figure}
\includegraphics[width=.9\linewidth]{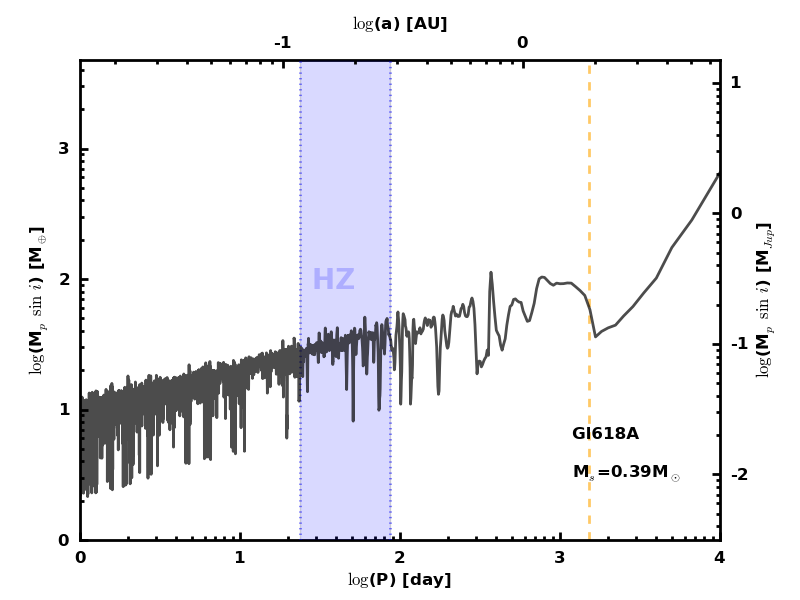}
\includegraphics[width=.9\linewidth]{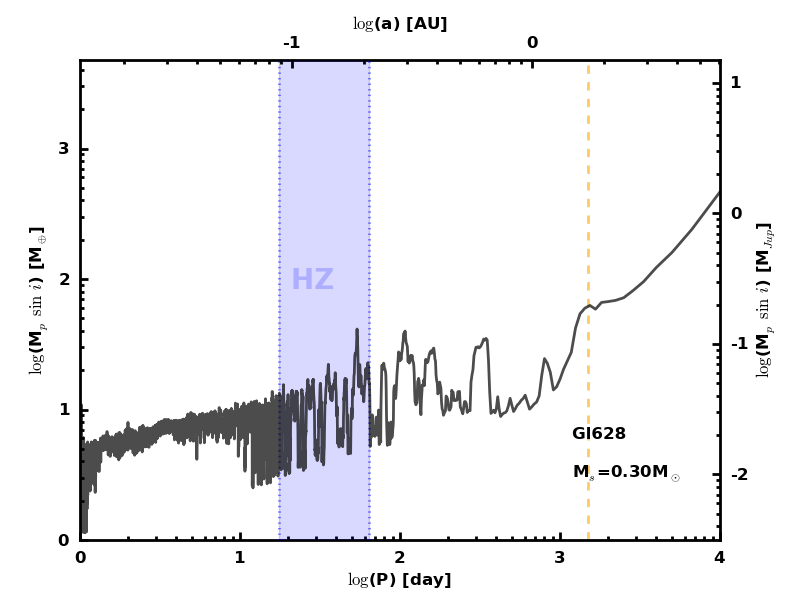}
\includegraphics[width=.9\linewidth]{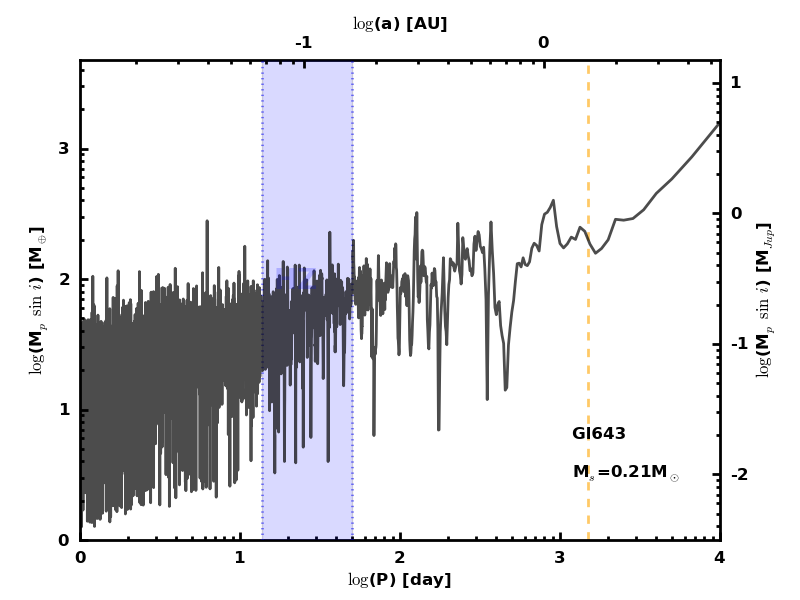}
\includegraphics[width=.9\linewidth]{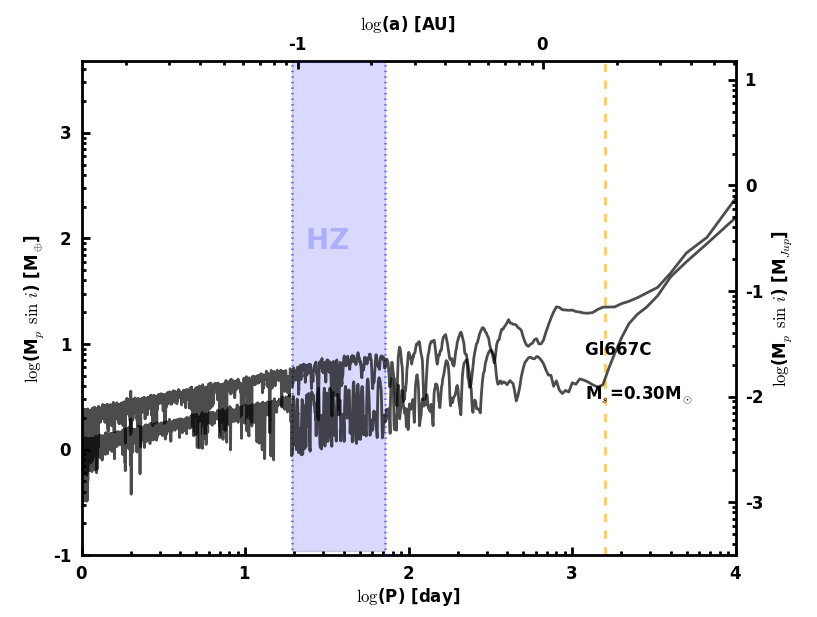}
\end{figure}\begin{figure}
\includegraphics[width=.9\linewidth]{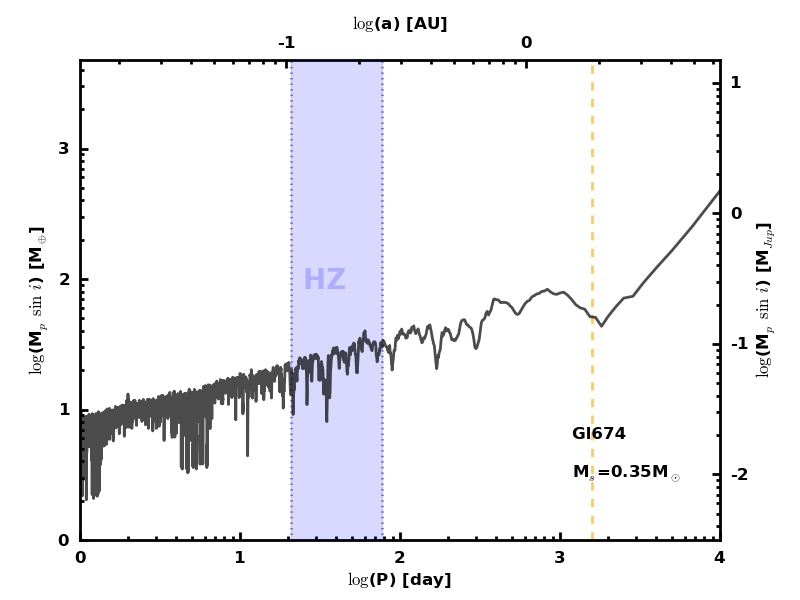}
\includegraphics[width=.9\linewidth]{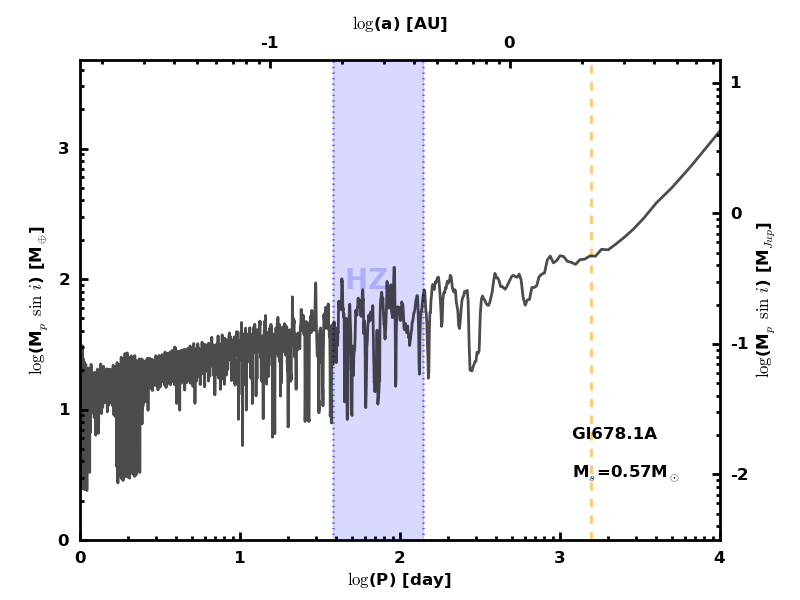}
\includegraphics[width=.9\linewidth]{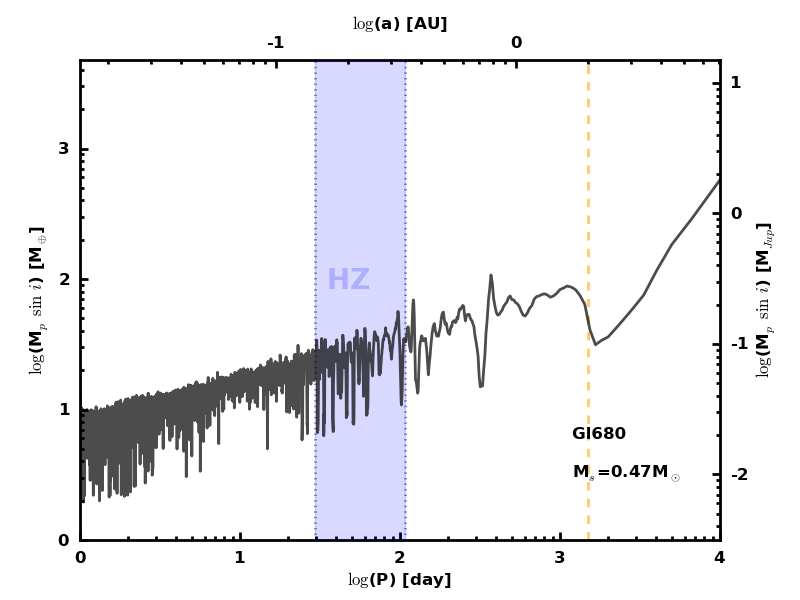}
\includegraphics[width=.9\linewidth]{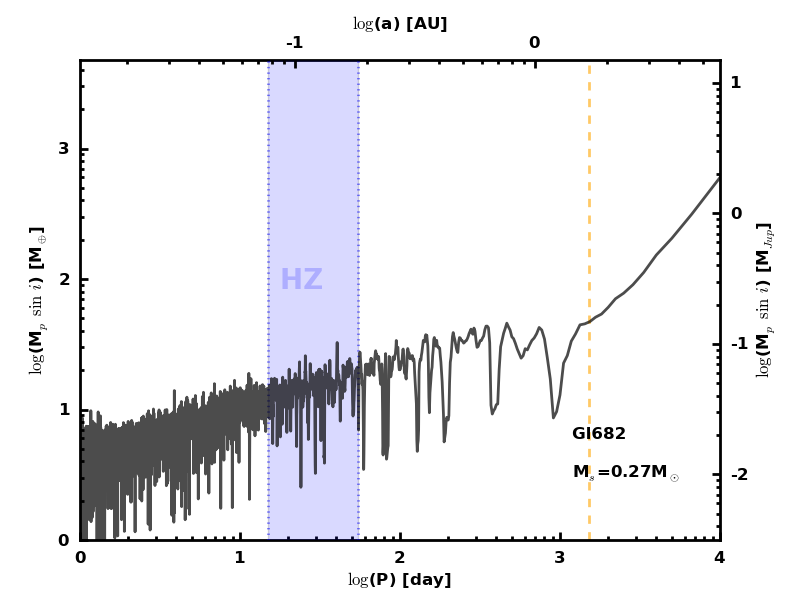}
\end{figure}\begin{figure}
\includegraphics[width=.9\linewidth]{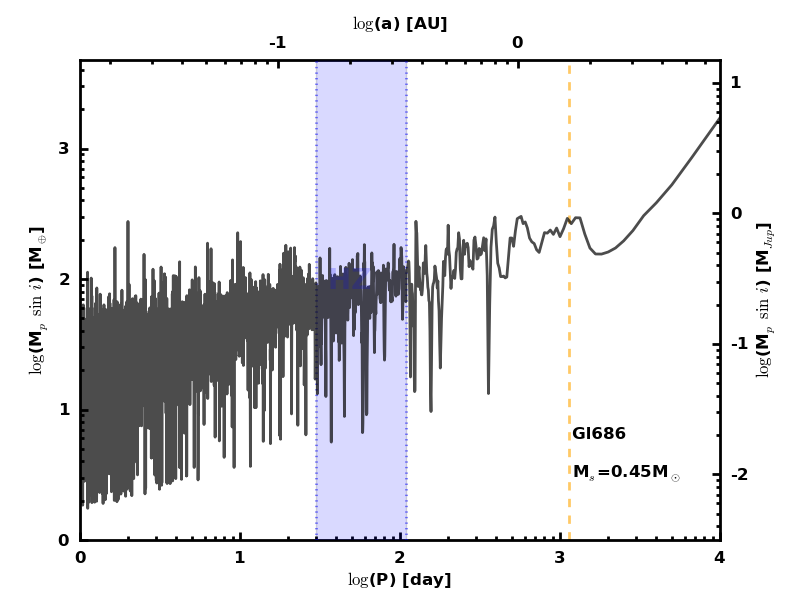}
\includegraphics[width=.9\linewidth]{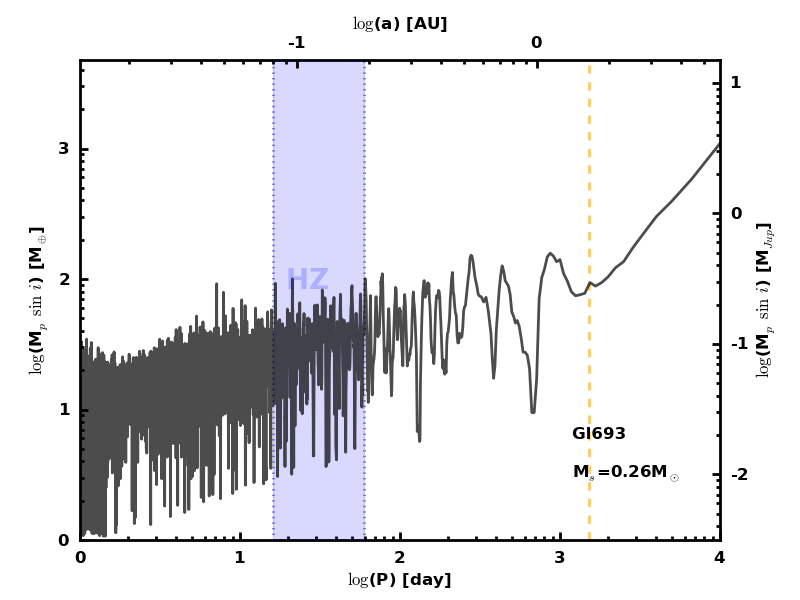}
\includegraphics[width=.9\linewidth]{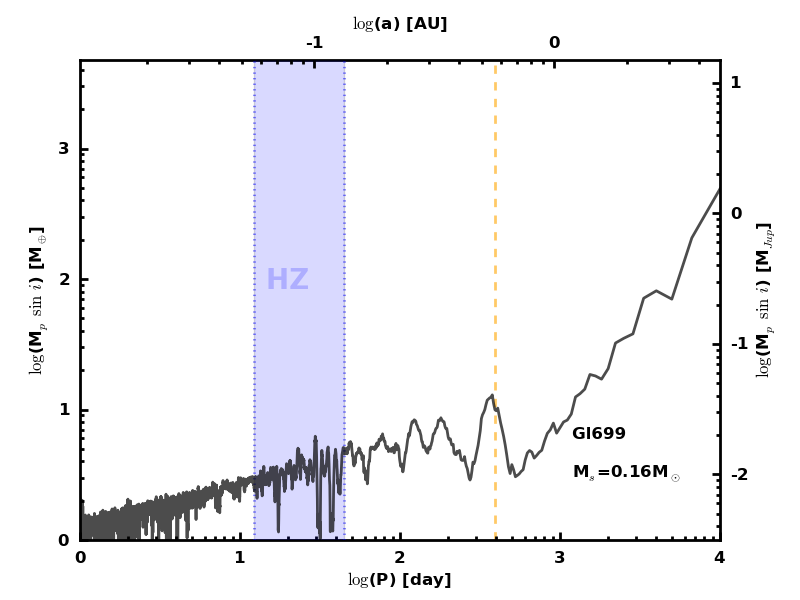}
\includegraphics[width=.9\linewidth]{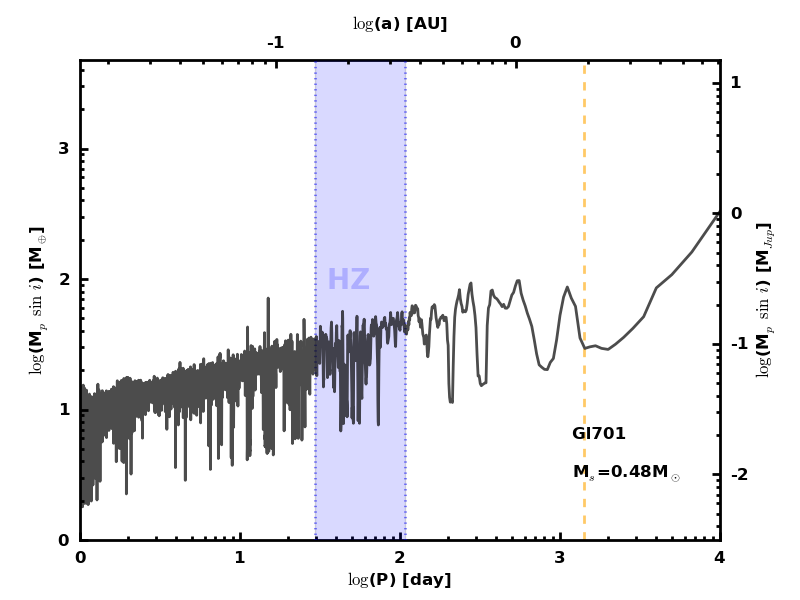}
\end{figure}\begin{figure}
\includegraphics[width=.9\linewidth]{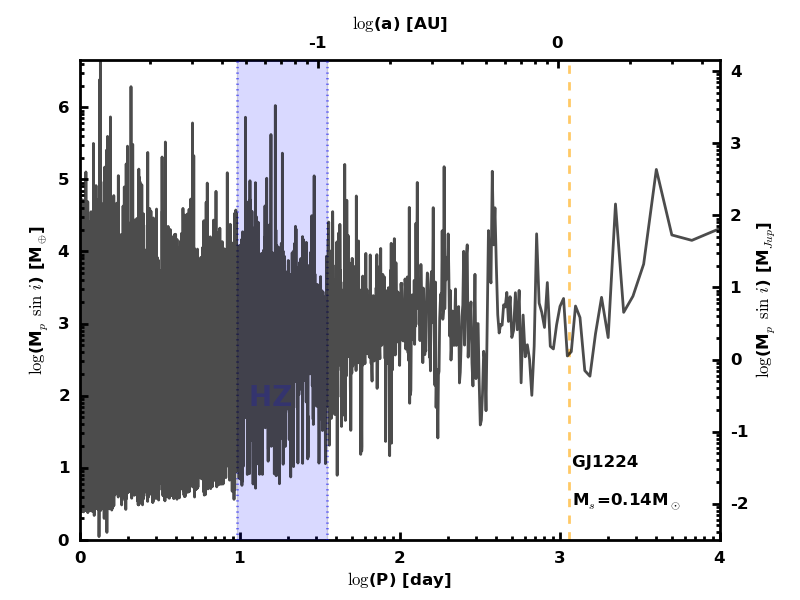}
\includegraphics[width=.9\linewidth]{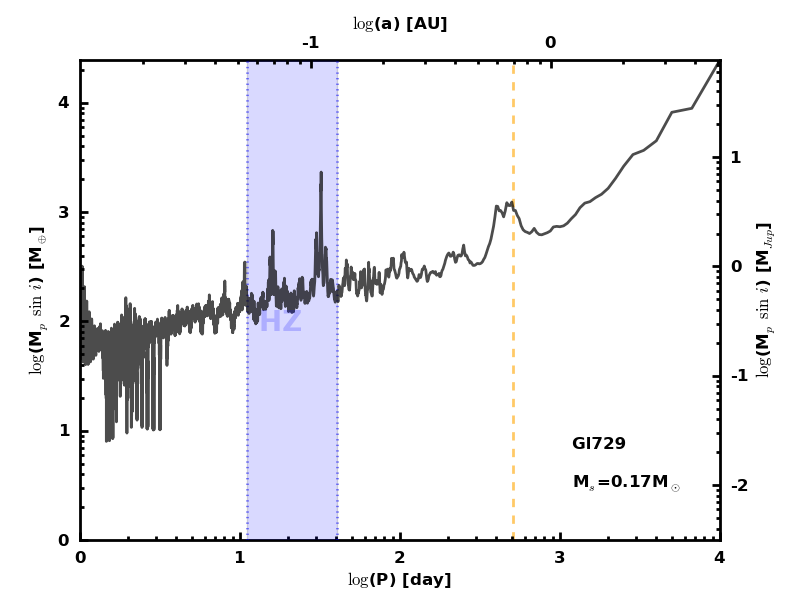}
\includegraphics[width=.9\linewidth]{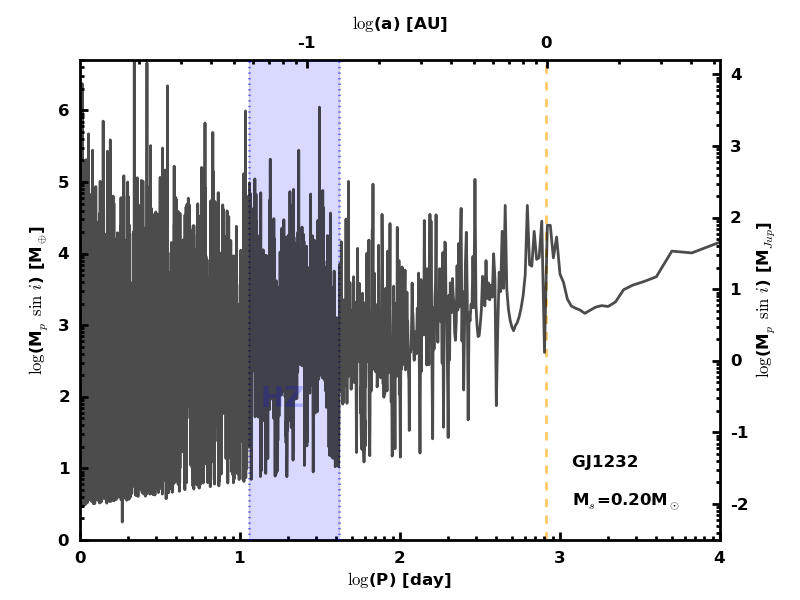}
\includegraphics[width=.9\linewidth]{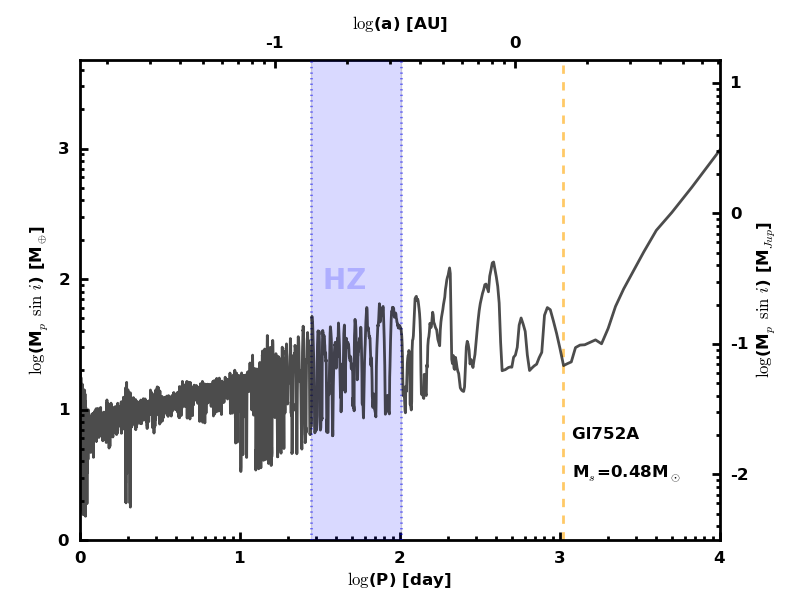}
\end{figure}\clearpage\begin{figure}
\includegraphics[width=.9\linewidth]{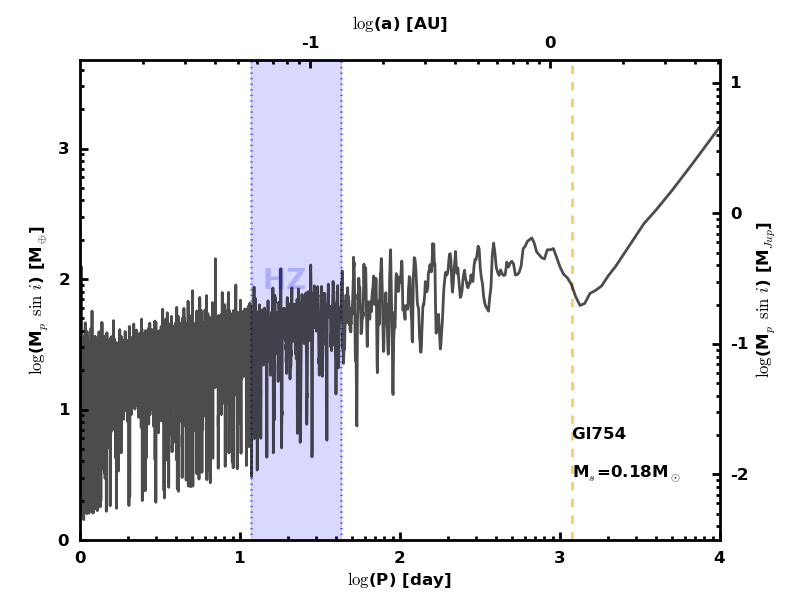}
\includegraphics[width=.9\linewidth]{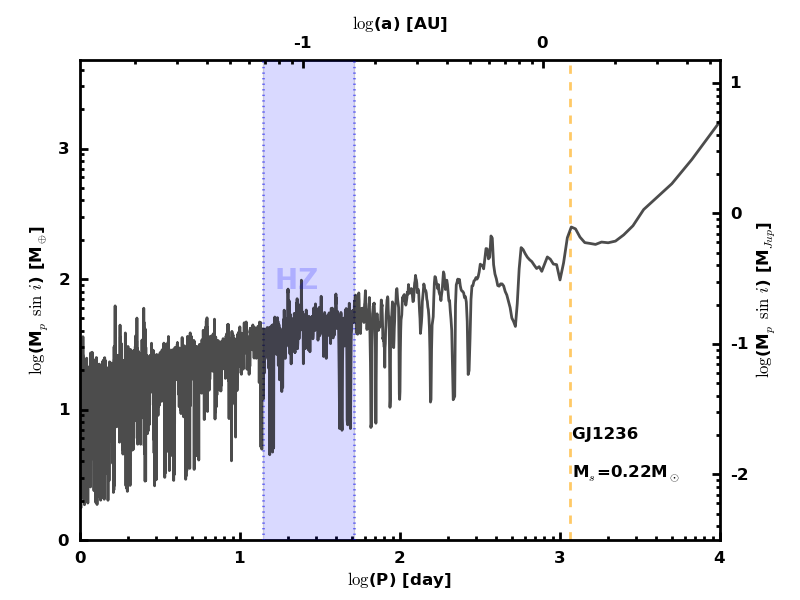}
\includegraphics[width=.9\linewidth]{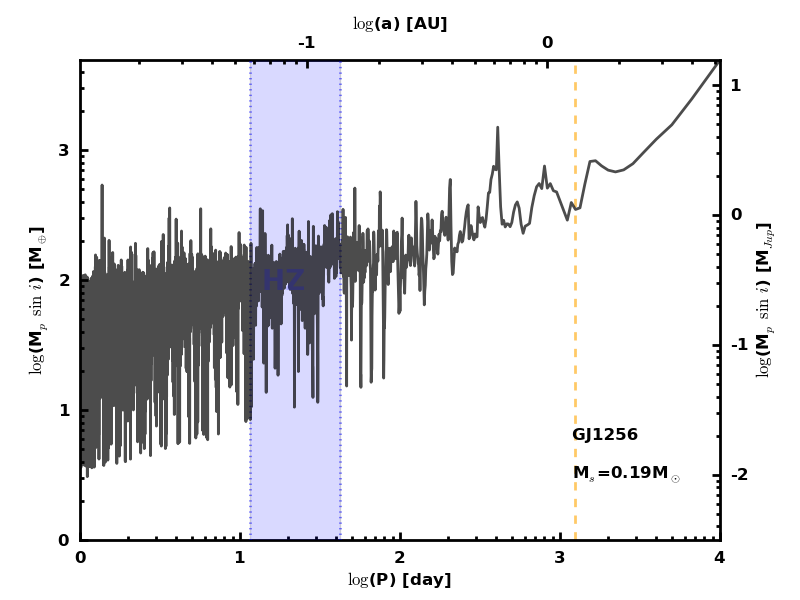}
\includegraphics[width=.9\linewidth]{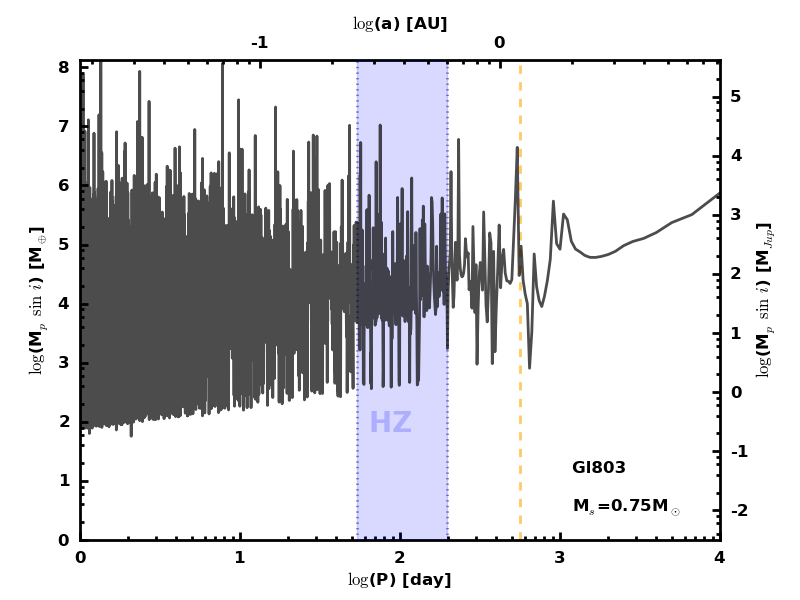}
\end{figure}\begin{figure}
\includegraphics[width=.9\linewidth]{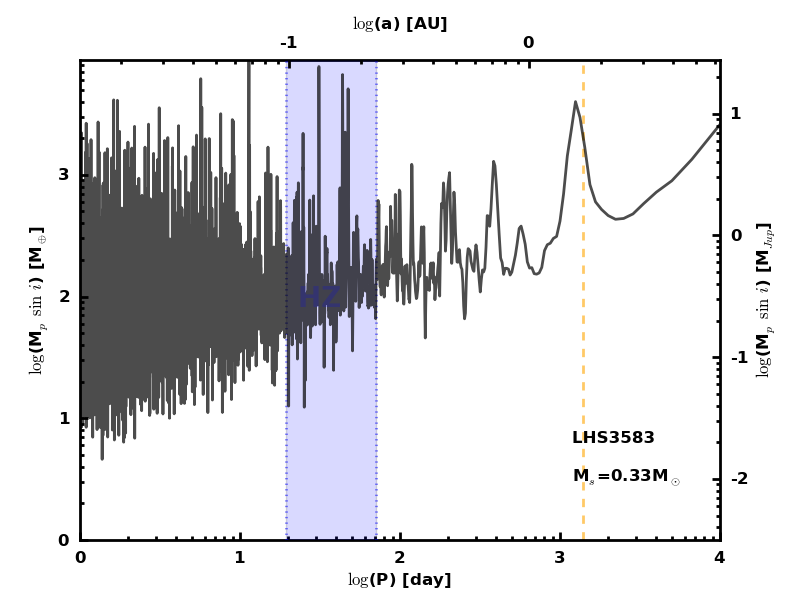}
\includegraphics[width=.9\linewidth]{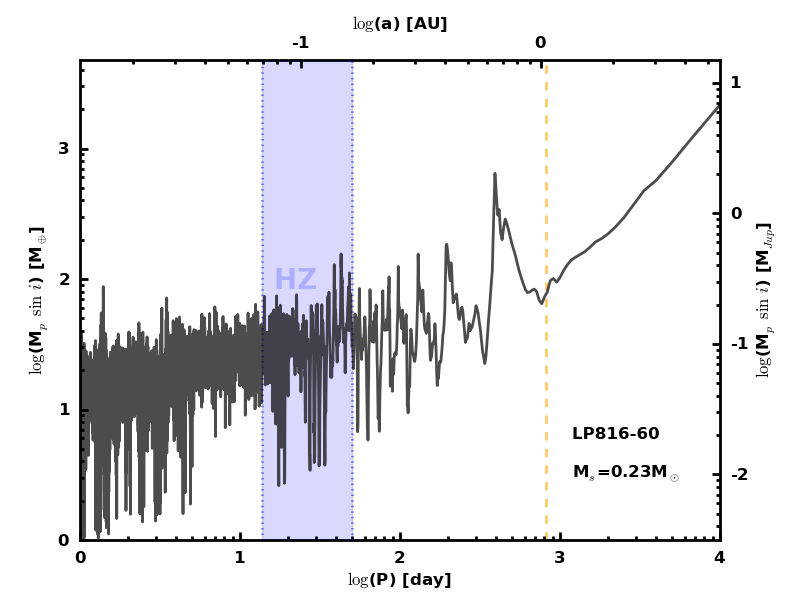}
\includegraphics[width=.9\linewidth]{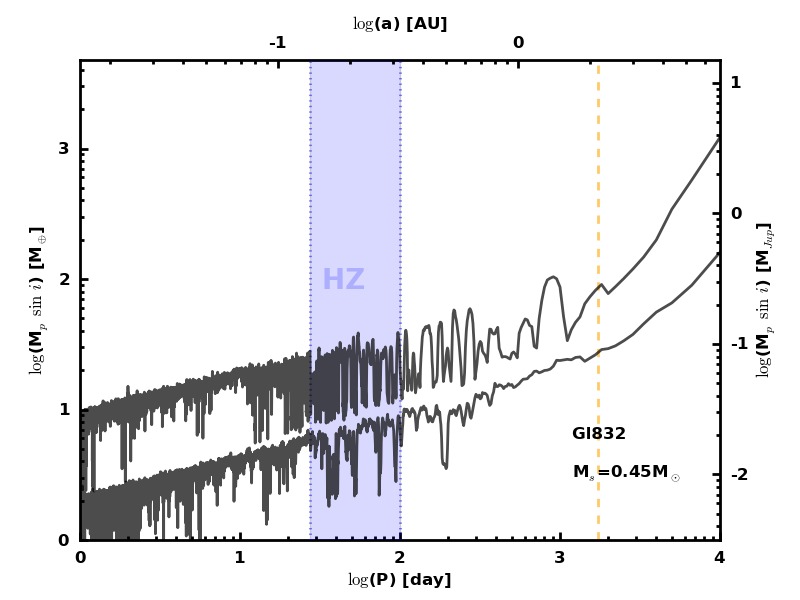}
\includegraphics[width=.9\linewidth]{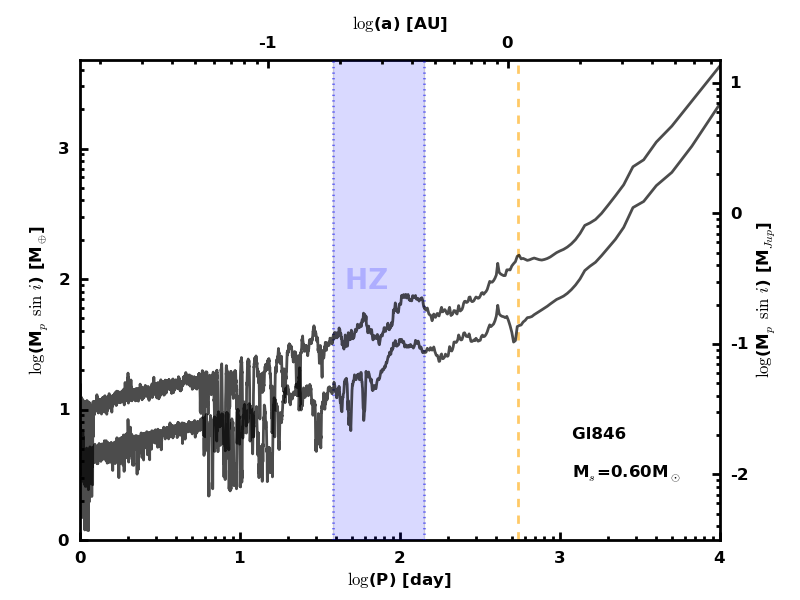}
\end{figure}\begin{figure}
\includegraphics[width=.9\linewidth]{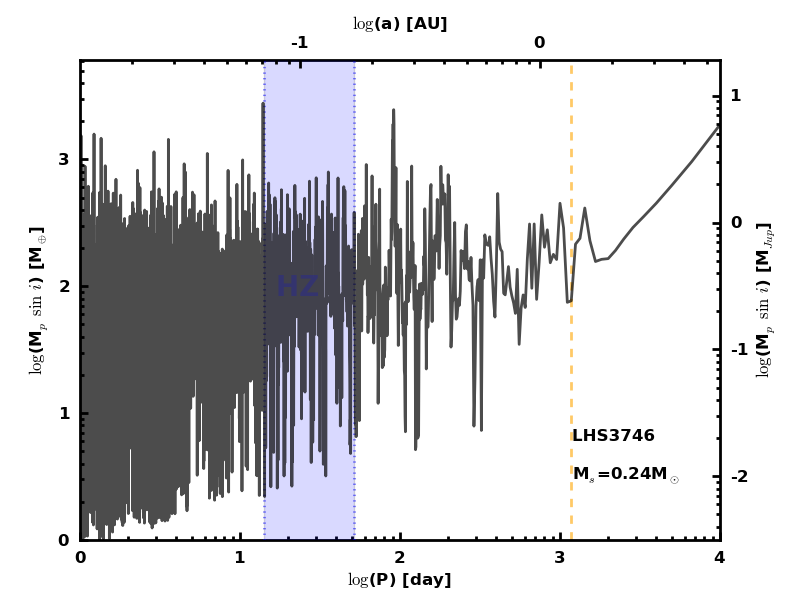}
\includegraphics[width=.9\linewidth]{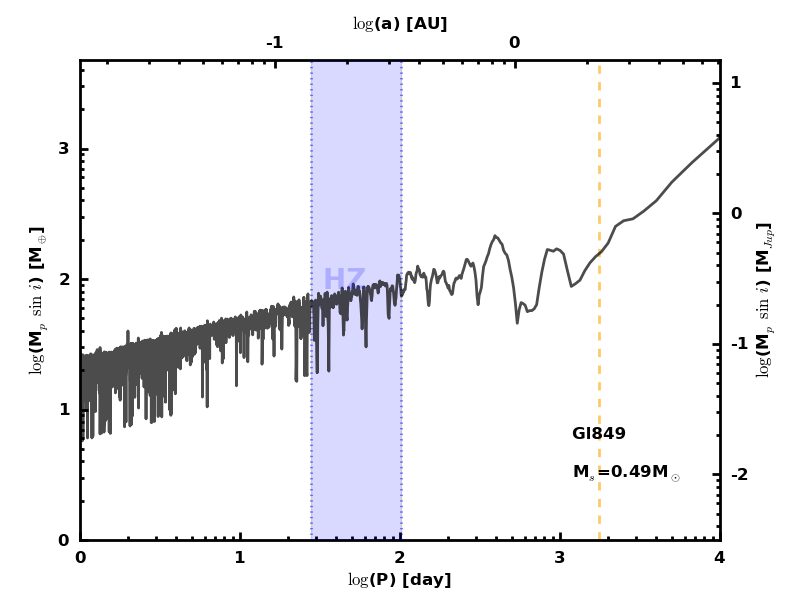}
\includegraphics[width=.9\linewidth]{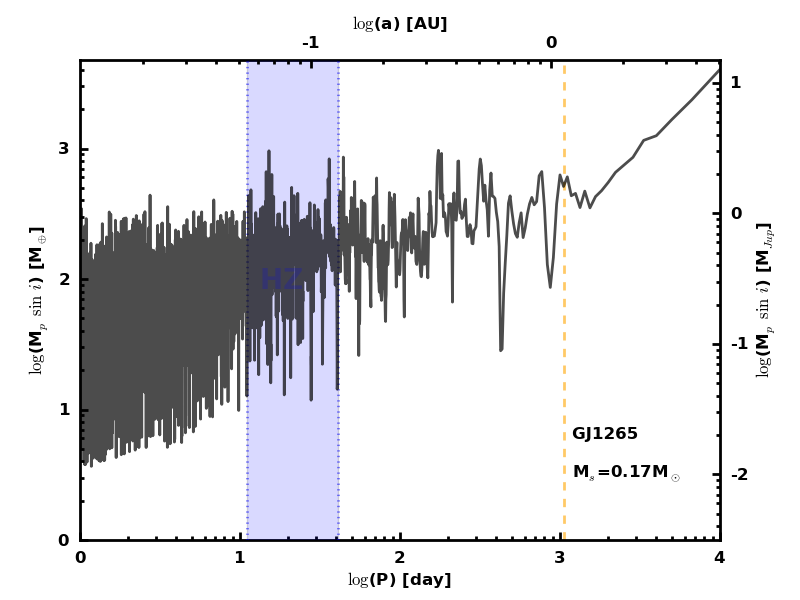}
\includegraphics[width=.9\linewidth]{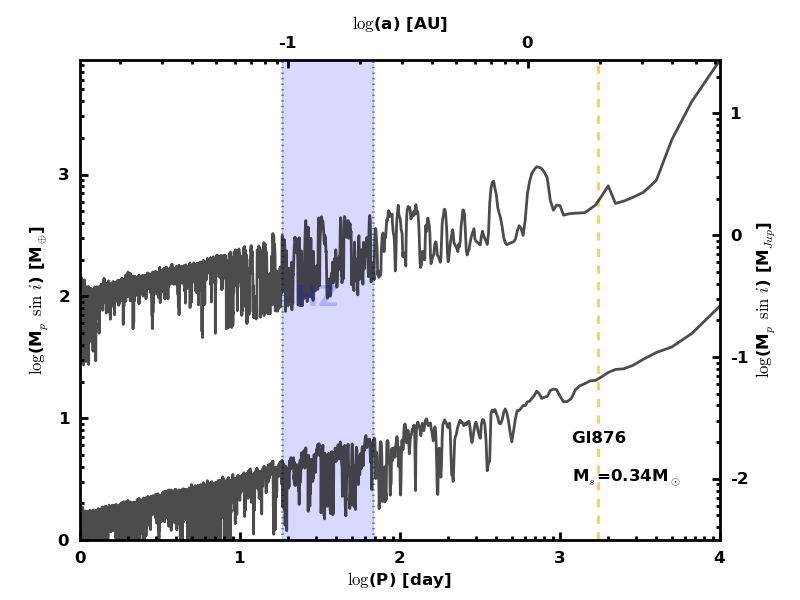}
\end{figure}\begin{figure}
\includegraphics[width=.9\linewidth]{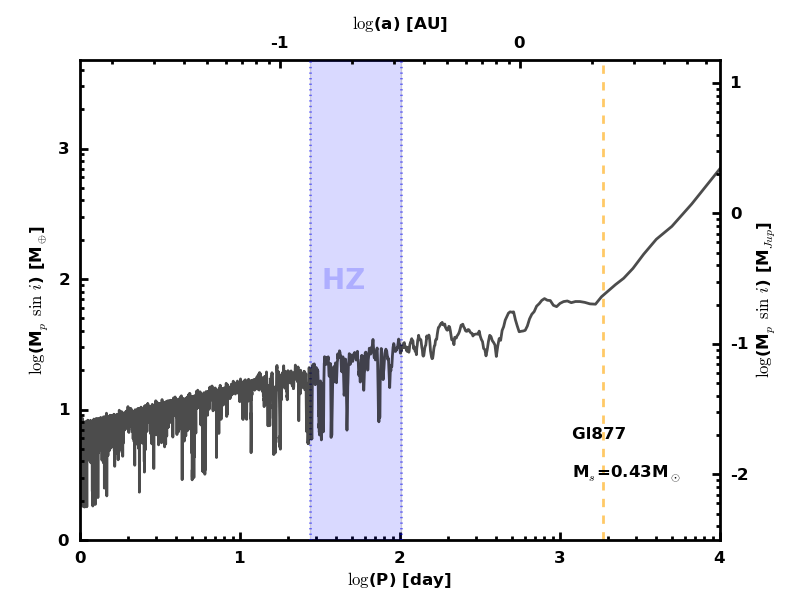}
\includegraphics[width=.9\linewidth]{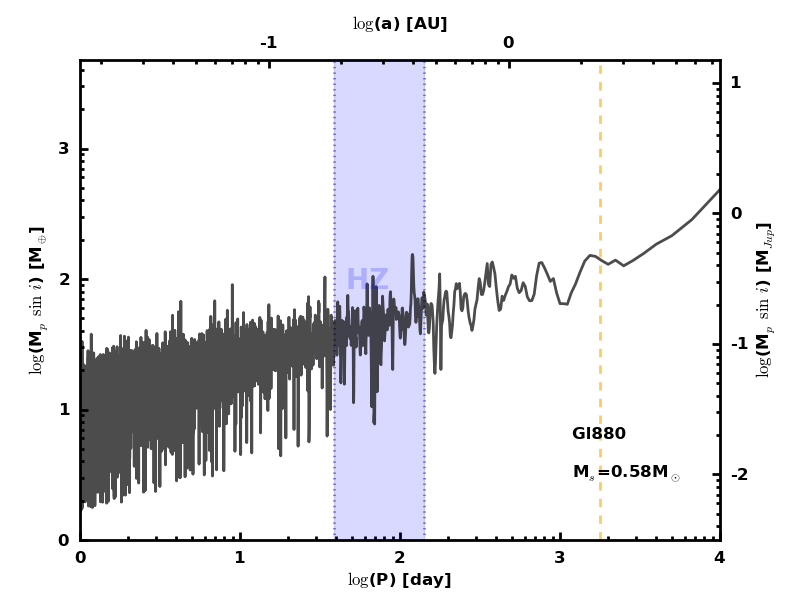}
\includegraphics[width=.9\linewidth]{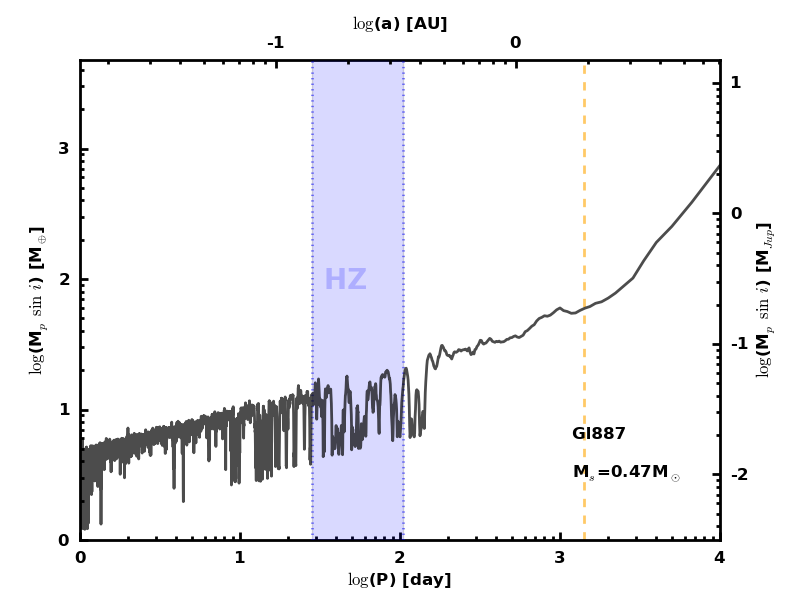}
\includegraphics[width=.9\linewidth]{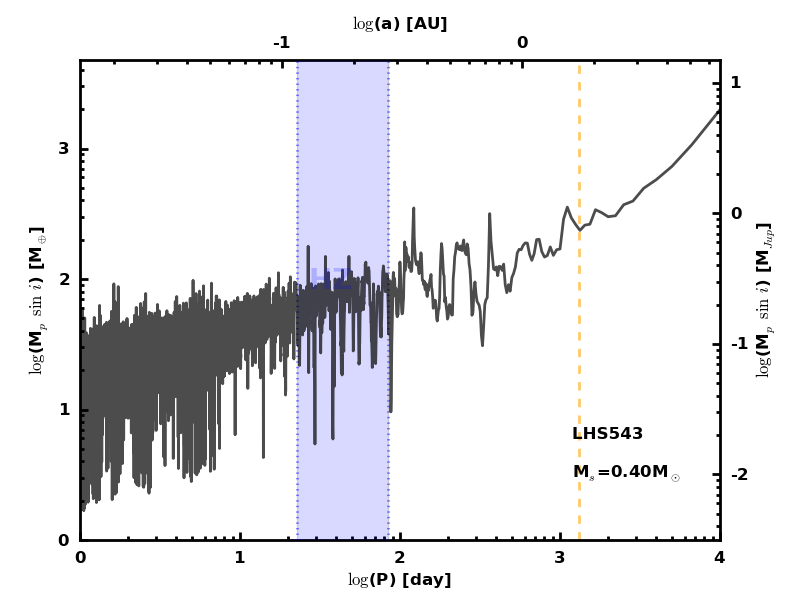}
\end{figure}\clearpage\begin{figure}
\includegraphics[width=.9\linewidth]{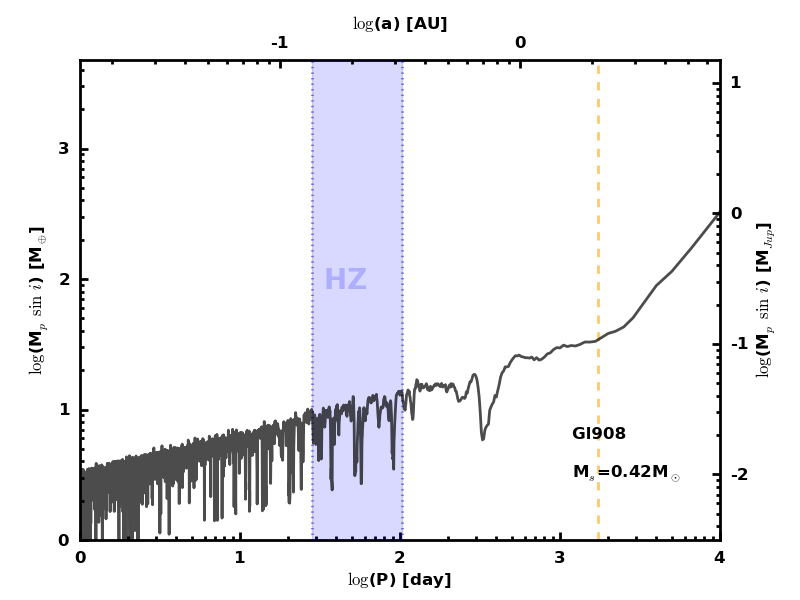}
\includegraphics[width=.9\linewidth]{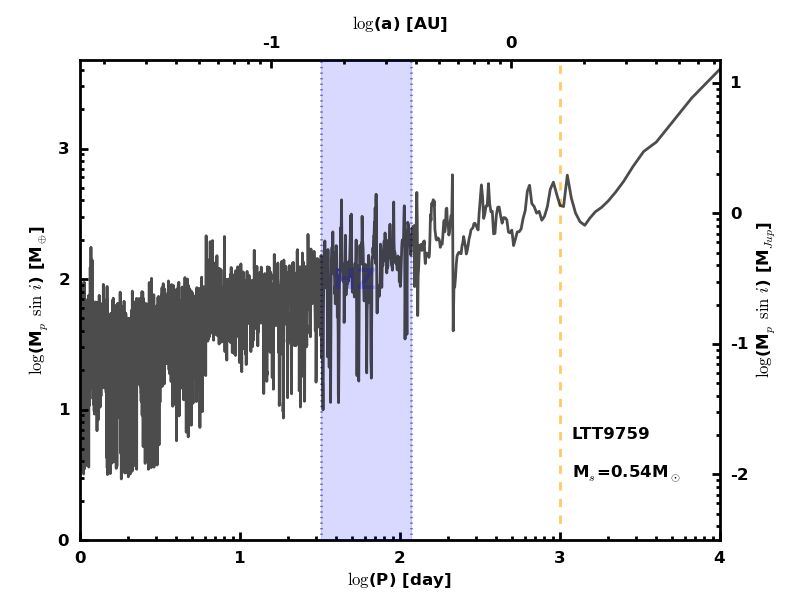}
\caption{\label{fig:limitesb} Phase-averaged detection limits on $m \ sin i$ for time-series with more than 4 measurements. Planets above the limit are statistically excluded, with a 99\% confidence level, for {\it half} the 12 trial phases. Some panels appear with 2 curves : the upper one is the detection limits before any model is subtracted and the bottom one is for the residuals around a chosen model (composed of planets, linear drifts and/or simple sine function). See Sect.~\ref{sect:limits} for details.}
\end{figure}

\appendix
\section{\label{app:comparison}Comparison with published time series}
\subsection{Variability}

Compared to other published time-series, we measured lower dispersions for all M dwarfs but Gl\,846 and known planet-host stars. Gl\,1 is not found variable in E06 and Z09 but their dispersion is limited by a higher photon noise ($\sim 2.6$ m/s, against $\sigma_e=1.9$ m/s in our case). We report variability for Gl\,229 but at a level of $<$1.9 m/s while the variability reported in E06 and Z09 implies a jitter of $3.9-4.7$ m/s. The slightly lower dispersion we observe for Gl\,357 ($\sigma_e=3.2$ m/s) against 3.7 and 5.3 m/s for E06 and Z09, respectively, might not be significant given our low number of observations (6). For Gl\,551, we measured a dispersion only slightly lower (3.3 against 3.6 m/s). We observe significantly lower variability for Gl\,682 (1.8 against 3.6 m/s) and Gl\,699 (1.7 against 3.4 and 3.3 m/s), and a higher dispersion for Gl\,846 (5.6 against 3.0 m/s). Although different time-spans, epochs of observations and activity levels at those epochs could explain different dispersions for individual stars (as it is certainly the case for Gl\,846 -- see Sect.~\ref{subsec:act}), the fact we measure a lower dispersion for most comparison stars most likely reflects the better performances of the {\textsc Harps} spectrograph.

\subsection{Trends}
Like this paper, Z09 reports non-significant slopes for Gl\,357 and Gl\,682 and significant slopes for Gl\,1, Gl\,551 and Gl\,699 (although in our case Gl699 is attributed a significant trend by the F-test only). Nonetheless, the slopes reported for Gl\,551 and Gl\,699 seem different and they moreover found a significant trend for Gl\,229 whereas we do not. Time series have also been published for Gl\,832 and Gl\,849 as they were singled out from their sample to report an orbiting planet \citep{Bailey:2009, Butler:2006}. For both stars, the planet reflex motion clearly dominates the radial velocity signal so we discard them from a quantitative comparison. In Table~\ref{tab:trendZ09}, we compare the slopes of linear fits to the time series in Z09 and to those of this paper. We note that most often the significant differences reflect a signal more complex than a simple linear drift.

\begin{table}
\begin{tabular}{lccc}
\hline
\hline
slope	     &	This paper 		&	Z09	& $\sigma$-difference\\
                        & [m/s/yr]      & [m/s/yr]                     &  \\
\hline
Gl1	&	$+0.332\pm0.212$	&	$-0.204\pm0.305$	&	0.09 \\
Gl229	&	$-0.257\pm0.296$	&	$+1.410\pm0.269$	& 4.30	\\
Gl357	&	$-1.682\pm0.710$	&	$+0.273\pm0.305$	& 2.77	\\
Gl551	&	$-0.234\pm0.162$	&	$+0.715\pm0.135$	& 4.44	\\
Gl682	&	$+0.685\pm0.490$	&	$+2.395\pm0.562$	& 2.54	\\
Gl699	&	$-3.043\pm0.646$	&	$-0.697\pm0.133$	& 3.73	\\
\hline
\hline
\end{tabular}
\caption{\label{tab:trendZ09}Linear trends for the time series of stars common to \citet[][Z09]{Zechmeister:2009b} and this paper. The fourth column reports the significance of the difference, expressed in $\sigma$ (and corresponding to the overlap of 2 gaussian distributions, evaluated with Monte Carlo trials).}
\end{table}

\subsection{Periodicity}
Among stars with identified periodicity in RV data, Gl\,832, Gl\,849 and Gl\,876 have time series published to report on detected planets. The periodicities we have found for those three stars are similar to their planets' orbital periods. Only Gl\,876d is not detected with our automated procedure because one has to do a full N-body integration to subtract properly the signal induced by planets 'b' and 'c'. Besides known planet hosts, Z09 also report on absence of periodicities for Gl\,229, Gl\,357, Gl\,433 and Gl\,682, and significant periodicities for Gl\,551 and Gl\,699. Our results and Z09 are therefore in contradiction for three stars : Gl\,433, Gl\,551 and Gl\,699. We noted in \S~\ref{subsec:plsys} that, for Gl\,433, the RVs reported by Z09 and in this paper are not incompatible provided that the merged data set is fitted by a model composed of 1 planet plus a quadratic drift. In the case of Gl\,551, the $\sim$1-year periodicity Z09 and \citet{Endl:2008b} have attributed the signal to an alias of a low frequency signal with the typical 1-year sampling. After \citet{Endl:2008b}, the low frequency signal is believed to be caused by a cluster of points that are both blue-shifted and with a higher H$\alpha$ index compared to other points in the time series. This putative activity signal might not be seen in our time series because it counts only 24 measurements, against 229 in Z09. Finally, the periodicity found for Gl\,699 is also attributed to activity, with a clear counterpart in H$\alpha$ filling factor. Again, if that activity signal is not seen in our time series, it is likely because it counts only 22 measurements, against 226 for Z09.

\end{document}